%% file: tetrahedronCondMat.tex
\documentclass[adp,fleqn]{w-art-modifiedbyrs}
\usepackage{w-thm}
\usepackage[]{graphicx}
\usepackage[]{epsfig}
\usepackage{amsmath}
\input{./Parts/mycommands}
\begin{document}
\input{./Parts/titlepage}
\input{./Parts/section1}
\input{./Parts/section2}
\input{./Parts/section3a}

\input{./Parts/section3b}
\input{./Parts/section4a}

\input{./Parts/section4b}
\input{./Parts/section5}

\input{./Parts/acknowledgement}
\input{./Parts/literature}
\newpage
\appendix
\input{./Parts/appendix1a}

\input{./Parts/appendix1b}
\input{./Parts/appendix1c}
\input{./Parts/appendix2a}

\input{./Parts/appendix2b}

\input{./Parts/appendix2c}
\newpage
\input{./Parts/figure01.tex}
\input{./Parts/figure02.tex}
\input{./Parts/figure03.tex} 
\input{./Parts/figure04_01.tex}
\input{./Parts/figure04_23.tex}
\input{./Parts/figure05.tex}
\input{./Parts/figure06.tex}
\input{./Parts/figure07.tex}
\input{./Parts/figure0809.tex}
\input{./Parts/figure10.tex}

\input{./Parts/figure11.tex}
\input{./Parts/figure11a.tex}
\input{./Parts/figure12.tex}
\input{./Parts/figure13.tex}
\input{./Parts/figure14.tex}
\input{./Parts/figure15.tex}
\input{./Parts/figure16.tex}
\newpage\end{document}

%% file: Parts/mycommands.tex
\newcommand{\beq}{\begin{eqnarray} }
\newcommand{\eeq}{\end{eqnarray}}

\newcommand{\punkt}{\, .}
\newcommand{\komma}{\, ,}
\newcommand{\summe}[1]{\sum\limits_{#1}}
\newcommand{\sumijs}{ \sum\limits_{i\neq j\sigma}}
\newcommand{\cplus}[1]{{\bf c}^+_{#1}}
\newcommand{\cminus}[1]{{\bf c}_{#1}}
\newcommand{\nsigma}[1]{{\bf n}_{#1 \sigma}}
\newcommand{\noperator}[1]{{\bf n}_{#1 }}

\newcommand{\hamilton}{{\bf H}}

\newcommand{\opS}{\mbox{\boldmath $ S $}}

\newcommand{\ket}[1]{| #1 \rangle}

\newcommand{\Aindex}[1]{A_{#1}}

\newcommand{\Cindex}[2]{C_{#1-#2}}

\newcommand{\Thetaindex}[1]{\Theta_{#1}}


%% file: Parts/titlepage.tex
\renewcommand{\copyrightyear}{2006}
\pagespan{3}{}
\Reviseddate{}
\Dateposted{}
\keywords{Hubbard model, analytical solution, nearest-neighbour interaction,cluster}
\subjclass[pacs]{85.80.+n, 73.22.-f, 71.27.+a}



\title[Extended Hubbard model on small clusters]{Analytical solution of a Hubbard model extended by 
nearest-neighbour Coulomb and exchange interaction on a triangle and tetrahedron }


\author[Schumann]{Rolf Schumann \inst{1,}%
  \footnote{Corresponding author\quad E-mail:~\textsf{schumann@theory.phy.tu-dresden.de}, 
            Phone: +49\,351\,463\,33644, 
            Fax: +49\,351\,463\,37079}}
\address[\inst{1}]{Institute for Theoretical Physics, TU Dresden, Dresden D-01062, Germany}
\begin{abstract}
The Hubbard model extended by either
nearest-neighbour Coulomb correlation and/or nearest neighbour
Heisenberg exchange is solved analytically for a triangle and tetrahedron. 
All eigenvalues and eigenvectors are given as functions of the model parameters 
in a closed form.
The groundstate crossings and degeneracies are discussed both for the canonical and grand-canonical energy levels.
The grand canonical potential $\Omega (\mu,T,h)$ and the electron occupation $N(\mu,T,h)$ 
of the related cluster gases were calculated for arbitrary values 
(attractive and repulsive) of the three interaction constants.
In the pure Hubbard model we found various steps in $N(\mu,T=0,h)$ higher than one.
It is shown that the various degeneracies of the grand-canonical energy levels are partially lifted by an antiferromagnetic exchange interaction, whereas a moderate ferromagnetic exchange
modifies only slightly the results of the pure Hubbard model. A repulsive
nn Coulomb correlation lifts these degeneracies completely.
The relation of the cluster gas results to extended systems is discussed.

\end{abstract}
\maketitle                   






%% file: Parts/section1.tex
\section{Introduction}
Besides the fact that the analytical solution of non-trivial models of strong correlation
is a task of high pedagogical value, the renewed interest in the analytical solution stems mainly
from cluster methods, which were successfully developed within the context of strong electron correlation during the last decade \cite{Senechal00,Maier05,Wang05} and applied to problems of high-$T_c$ superconductivity and from modelling of electron transport through quantum dots, see
e.g. \cite{Kikoin06} and the references therein.
Both topics have in common that a detailed knowledge of the cluster physics is a key ingredient.
The by far most employed model of strong correlated electrons is the Hubbard model.
Despite its simplicity, the only analytical solutions available up to now are the groundstate 
for n-site rings by help of the Bethe ansatz or transfer matrix method \cite{Lieb68,Korepin00} 
and the complete solution on small clusters, i.e. the triangle, the tetrahedron \cite{Falicov84} and the square \cite{Schumann02}. For models containing five and more sites only approximate or numerical solutions exist. 
A further topic of increasing interest is the interplay between frustration
and strong electron correlation - a topic which acquired a lot of interest due to the discovery of a superconducting phase in the hydrate of ${\rm Na_xCoO_2}$. For a theoretical survey
to that topic I refer to \cite{Merino06} and the literature therein. 
Furthermore, the latter paper demonstrates explicitly the use one can make from the exact knowledge of small clusters in understanding the results of more elaborate many body methods.
Although I have up to now not dealt personally with attractive Hubbard models this point
seems to be also of some interest \cite{Perez96,Nakano06}. Thus researchers in this field may probably also benefit
from our work, since the results are valid for arbitrary values of the model parameters.
The present paper is very related to our former analytical solution 
of the Hubbard model on
a square and the related cluster gas \cite{Schumann02}, thus we will extensively refer to it.
One of the main points made in \cite{Schumann02} was the discovery of the degeneration of cluster states with different particle numbers and spin, which has surely a great influence of the coupling of spin- and 
charge degrees of freedom, and indeed, the degeneracy of the groundstates for
the 4-site cluster occupied either with two or with four electrons, is responsible for
the charge separation and inhomogeneities in the pseudogap phase of the cuprates, whereas
their bosonic character accounts for the superconducting phase \cite{Kocharian05,Kocharian06}. Also the d-wave symmetry of
the gap seems to be a direct consequence of the symmetry of the 4-site cluster groundstates \cite{Tsai06}.
Usually the addition of further terms to the pure Hubbard Hamiltonian destroys this degeneracy and makes the system unsolvable due to a reduced number of global symmetries, with the consequence that we have a reduced number of analytical solutions and exact numerical results 
\cite{Zhang89}.
In the present paper we study the influence
of  additional nearest-neighbour (nn) Coulomb interaction, exchange term in the case of both the triangle and the tetrahedron or next-nearest neighbour (nnn) hopping 
in the case of the square respectively, which are added to the standard Hubbard model. The latter is used in the notation of \cite{Schumann02}. 
A common feature of the considered models is frustration, which is present due to geometry
in the triangle and tetrahedron, even without the additional terms, and in the
square due to the nnn-hopping, what becomes evident since
for $t'=t$ the model is equivalent to the pure Hubbard model on the tetrahedron. 
For the triangular and tetrahedral geometry
the spatial and spin symmetries are enough to get a complete solution.
For the square geometry any of the additional terms breaks the global pseudospin symmetry, 
what prevents closed form solutions for
all eigenvalues and eigenstates.
Nevertheless, there exist non-trivial parameter sets, where analytical solutions exist at least for the canonical partition sums. In the following we abstain to discuss the square geometry
since the most important solvable case $t=t'$ is contained in our results for the tetrahedron.
In the next section we shall present the extensions of the models and its closed form
solutions, which we shifted to the appendices due to its unusual large page volume.
In the third section the low lying levels and its parameter dependencies are discussed, where
emphasis is given to their dependence on the additional interaction (hopping) parameters.
A lot of degeneracies and ground state level crossings are presented, which may give
rise to a likely rich physics in the ``cluster gas'' or extended systems built from these
clusters as it was observed in the relation of the 4-site Hubbard model on a square to the square lattice.
In a concluding section I will discuss the results.
At least I want to beg for the readers pardon for the unusual volume of this article, which is 
due to the closed form expressions of the eigenstates of the models. The reason
is that in response to Ref. \cite{Schumann02} a lively interest arose in the closed form 
of the eigenstates, which I have published on my website only, whereas the thermodynamical results of the model contained in the Ref. \cite{Schumann02} obviously did not find as much interest. Unfortunately I can not
guarantee for the future existence of this website, therefore I decided
to include the eigenstates as appendix to provide a hard copy to the interested
reader this time.
Whereas preliminary results on thermodynamics and magnetics were published
in Ref. \cite{Schumann06},
more detailed data on these topics will be published in a forthcoming paper.

%% file: Parts/section2.tex
\section{The Hamiltonian}
In the following we consider the model
\beq
\hamilton&=&\hamilton_H+\hamilton_C+\hamilton_J
\eeq
with $\hamilton_H$ being the pure Hubbard model.
\beq
\!\!\!\!
\hamilton_H &=&  t \sumijs \cplus{i \sigma}\cminus{j \sigma} + 
\summe{i \sigma} \left(
\frac{U}{2}\nsigma{i}\noperator{i-\sigma}-(\mu+\sigma h) \nsigma{i} \right ) \komma
\label{hamilton}
\eeq
$\hamilton_C$  is the nn-Coulomb repulsion 
\beq
\hamilton_C=W\,\sum_{\langle i \neq j \rangle} \noperator{i} \noperator{j} \punkt
\eeq
Furthermore a nn-Heisenberg exchange term $\hamilton_J$ 
\beq
\hamilton_J=J\,\sum_{\langle i \neq j \rangle} \opS_i \opS_j
\eeq
was added.
Here $\cplus{i \sigma}$ and $\cminus{i \sigma}$ are the creation and 
destruction operators of electrons at site $i$ with spin $\sigma$,
$\noperator{i \sigma}=\cplus{i \sigma}\cminus{i \sigma}$ and
$\noperator{i}=\sum_{i\sigma} \noperator{i \sigma}$. $\opS_i$ indicates
the local spin operator at site $i$.
For a detailed physical reasoning of the additional terms in the Hamiltonian see 
e.g. \cite{FuldeBuch}. Our Hamiltonian is an extension of the Hamiltonian used
e.g. in \cite{Davoudi06} (Although up to now an erata was not published, we think that the $\sigma$ in the hopping term of eq. 1 of that paper is simply a misprint), 
since we treat the nearest-neighbour 
correlation $W$ and the nearest-neighbour exchange constant $J$ as independent parameters.
This is done mainly to cover pure Heisenberg type models by setting $W/t=0,\, J/t\ne0$.
The chemical potential $\mu$  and the magnetic 
field $h$ in $z$-direction are introduced
to take into account the effects of doping and applying external magnetic fields.
Please note, that the signs in front of the hopping and exchange term are positive,
thus one has to be carefully while comparing with papers where other conventions
are used.
Regarding the method of calculating the eigenvalues and eigenvectors we refer to \cite{Falicov84, Schumann02}.
The analytical solution, i.e. the closed form expressions for the eigenvalues
and eigenvectors in dependence on the parameters $U$, $J$, and $W$ are shifted 
to the appendices, where they are numbered. These numbers are occasionally used for 
referencing in the following sections.
Since I did not found any analytical solution of that model, it was hard to prove
the results. For $W/t=J/t=0$ we recover the results of \cite{Falicov84}, i.e. the
characteristical polynomials and characteristics of the eigenstates, which themself are not
given explicitely. For the triangle we recover, again setting $W/t=J/t=0$, the
analytical result for the groundstate energy and the wave function with 4 electrons
given in \cite{Merino06}.
Furthermore for small electron occupation or some limiting cases the results could 
be proved by a pencil calculation.
Nevertheless, the main confidence stems from our experience with the Hubbard model on
the square. A very difficult point was also the transcription of the analytical
computer results into a human readable form. Since this had to be done automatically
also, additional progamming was necessary, and finally I had to manhandle some
of the expression to format it for publication, a further chance 
to introduce mistakes.

%% file: Parts/section3a.tex
\section{The triangle and the related cluster gas}
\subsection{The spectrum of the triangle}
 In Fig. \ref{triangleCanonicalSpectra} we study the dependence of
the canonical eigenvalues on the parameters $U'$, $J'$, and $W'$ separately. The new
parameters are defined according to
\beq
U'&=&\frac{U}{4t+|U|} \label{Uprime}\\
J'&=&\frac{J}{t+|J|} \label{Jprime}\\
W'&=&\frac{W}{t+|W|} \label{Wprime} \punkt
\eeq
in order to map the parameter space from the interval $(-\infty,+\infty)$ to $(-1,1)$.
The mapping of the on-site correlation was chosen in a different way, since we are mainly interested 
in the strong interacting case, and thus, roughly speaking, we achieved that the weak interaction region is limited to the interval $(-0.5,0.5)$ and the strong correlated
case outside. To avoid infinite energies we map the energy in the same way
\beq
E'&=&\frac{E}{t+|E|}\label{Eprime}
\eeq
A multitude of level crossings in dependence of the different interaction parameters is found. Although  level crossings of higher states may be important for spectroscopy, we focus to the level crossings of the canonical groundstates, due to its relation to quantum phase transitions in extended systems. 
First of all there is only one groundstate level crossing in dependence of $U$ if $J$ and $W$
is set to zero. This occurs if the cluster is occupied with two electrons and $U$ changes sign.
In order to study the dependence of the groundstate on the other parameters we fix $U/t$ to four.
The dependence on $W$ is also very simple, since there is one levelcrossing for two electrons  and $W=U$ only. Otherwise, the dependence on the exchange parameter is very interesting. In Fig. \ref{triangleLevelCrossings} we show all the groundstate levelcrossings. The insets show the $J$ dependence of the complete canonical spectra. The levels are denoted in the legend by their quantum numbers.
if the cluster is occupied with two electrons we find an $U$ independent crossing from a 
twofold degenerate groundstate with spin one and spin projection one (no. 8, 20) to the non-degenerate eigenstate (no. 13) with spin one and spin projection zero. 
These states belong both to the one-dimensional irreducible representation $\Gamma_2$.
At the second crossing point, a crossover to two eigenstates of spin zero (no. 14, 17) occurs. 
The symmetry
changes to $\Gamma_3$ resulting again in a twofold degenerate groundstate, due to the two-dimensionality of the irred. representation. Whereas the first transition happens always
at $J_c/t=0$, the second one is dependent on the on-site correlation according to 
\beq
\frac{J_c}{t}&=&\frac{U}{2t+U} \punkt
\eeq
For three electrons there is only one ground state level crossing, changing from
fourfold degenerate level (no. 27, 30, 36, 39)  with spin one half and irred. representation $\Gamma_3$ to 
the two-fold degenerate states(no. 23, 46) with spin three half and irred. representation $\Gamma_2$. The crossing point $J_c$ is implicitely given as function on $U$ by
\beq
\!\!\!\!\!\!\!\!\!\!\!\!\!\!\!\!\!\!\!\!\!\!\!\!\
0&=&11\,J - 4\,U  
   + 4\,{\sqrt{27 \,t^2  + J^2 + 2\,J\,U + U^2}} 
  \cos \left ( \frac{1}{3}\arccos (\frac{{\left( J + U \right) }^3}
         {{\left( 27\,t^2  + J^2 + 2\,J\,U + U^2 \right) }^{\frac{3}{2}}})
\right )\punkt
\eeq
Fig. \ref{figureJcvonUN3} shows the solution for the primed values.
A crossover from a fourfold degenerated groundstate (no. 44, 45, 56, 57) 
of symmetry $\Gamma_3$ with spin one and spin projection $\pm 1$ to a singlet (no. 46) with $\Gamma_1$ and spin zero takes place for negative values
of $J$ (remember: $J<0$ favourites parallel spins in our notation) if the electron occupation is four. 
The $U$ dependence of the
critical exchange parameter is given by 
\beq
J_c &=& \frac{1}{3}(2\,t+3\,U - {\sqrt{112\,t^2 + 24\,U\,t + 9\,U^2}}) \punkt
\eeq

%% file: Parts/section3b.tex
\subsection{The cluster gas} 
In the following the most interesting aspects of an ensemble of
triangular clusters (clustergas) will be discussed.
In the left panel of Fig. \ref{triangleNvonMueprimeUprime} we show the dependence of the electron occupation
on the chemical potential for the pure model for different values of the on-site
electron correlation. It is obvious, that the 16-fold degeneracy of states of one, two,
three, and four electron states is lifted by the on-site electron correlation.
But the more interesting feature is the jump of the occupation number from 0 to 2. 
The reason is, that 
a $N=1$ state is never the (grand-canonical) ground state. This feature is persistent 
even for arbitrary large values of $U$, since the $N=2$ groundstate avoids 
correlation. This situation is analog to the
four-site model, where states with two, three, and four electrons are degenerated  for certain values of the chemical potential. It was conjectured, that this may be one of the main issues for both the superconduction and the stripe structure if the high-$T_C$-cuprates are modeled 
by a square lattice Hubbard model. Since the triangular cluster gas is related in the same sense
to the triangular lattice, it seems reasonable to conjecture that we have analogous
peculiarities in systems, which are modeled by a Hubbard model on a triangular lattice.
The recently reported superconductivity in ${\rm Na_xCoO_2}$ may be a hint
in this direction. Although we are mainly interested in the repulsive model we show also the curves for the attractive case ($U<0$) for seek of completeness.
The right panel depicts the groundstate "phase diagram", i.e. the eigenvalues of the groundstates in dependence on the chemical potential and the electron correlation.
In order to cover the whole parameter range we used again eq. (\ref{Uprime}) and introduced 
$\mu'$ according to
\beq
\mu'&=&\frac{\mu}{t+|\mu|}\label{Mueprime }  \punkt
\eeq
The meaning of the colours is explained by help of the palette shown in Fig. \ref{palette3site},
whereas the left picture shows the groundstate of the complete $\mu'$-$h'$ plane, with
\beq
h'&=&\frac{h}{t+|h|}\label{hprime }  \punkt
\eeq
and $U/t$ fixed to four. This $U$ value was chosen since it is the region, where in extendeded
systems the cross-over from weak correlation to strong correlation happens, what means
that perturbation methods can not be applied .
The influence of an ferromagnetic ($J<0$) and antiferromagnetic
exchange ($J>0$) interaction is studied in Fig. \ref{triangleU4NvonMueprimeJprime}.
This interaction is not able to lift the mentioned degeneracy of
the groundstates. And the same holds for an attractive
nearest-neighbour Coulomb interaction, whereas the more realistic case of 
repulsive interaction does, as may be seen in 
Fig. \ref{triangleU4NvonMueprimeWprime}.
Regarding the grand canonical spectrum of the triangle the most 
interesting physics results for $\mu=-1/2$. In the left part of Fig. \ref{triangleLowestGrandCanonicalLevels} 
we depicted the lowest part of the grand-canonical 
spectrum. At this point, an empty state is degenerated with the four single-electron states 
belonging to irreducible representation $\Gamma_3$ and furthermore with a two-electron triplet  of symmetry $\Gamma_2$, if the nn-interactions are neglected. A small external magnetic field lifts this degeneracy partially.
If the exchange interaction is taken into account this degeneracy is also lifted partially
as may be seen from the right panel of Fig. \ref{triangleLowestGrandCanonicalLevels}, 
where we depicted the grand-canonical level system for both a ferromagnetic
and an antiferromagnetic exchange integral.

%% file: Parts/section4a.tex
\section{Results for the Tetrahedron and the related cluster gas}
\subsection{The spectrum of the tetrahedron}
In Fig. \ref{primedewsTetraU} we depict the dependence of
the canonical eigenvalues on the parameter $U'$.
The degeneracy can be seen from the legends, since
the irreducible representations
$\Gamma_1$, $\Gamma_2$ are one-dimensional, $\Gamma_3$, is two dimensional, and $\Gamma_4$, $\Gamma_5$ are three-dimensional. Furthermore these degeneracies have to be multiplied by
the normal spin degeneration.
For the pure Hubbard model there is a groundstate crossing at $U=0$ for an electron 
occupation of 2, 3 and 4 but more interesting is its absence for $N=5$ where the groundstate
$\Gamma_4$, $S=1/2$ is sixfold degenerated and for $N=6$ where we find a nondegenerated $\Gamma_1, \,S=0$ groundstate for the whole parameter range.
The groundstate in dependence on U and J is plotted for the cases, where level crossings 
in the ground state appear. Note that we used again primed parameter values to cover the whole plane. The meaning of the colours is explained in the palette shown at the down-right corner.
The dependence on the exchange-parameter $J'$ is shown in Fig. \ref{primedewsTetraJ}.
Due to the exchange interaction the degeneration is lifted and the degeneration 
factor reduces to two, besides the $\Gamma$-multiplicity.
Whereas the groundstate crossing for the $N=2$ system was rather expected, since it is coupled to the sign change of $J$, we find a more complex behaviour for three and four electrons on the cluster. As may be seen from Fig. \ref{lowEwsU4N3J} we find a change from a sixfold degenerated groundstate with $\Gamma_5$, $S=1/2$ to a twofold degenerated groundstate with $\Gamma_2$, $S=1/2$ at a positive value of $J$ for reasonabel $U$ values.
\beq
J_c &=&
\frac{10\,t  + 3\,U - {\sqrt{100\,t^2 + 48\,t\,U + 9\,U^2}}}{3}
\eeq
what amounts for $U=4t$ to $J_c/t=(22 - 2\,{\sqrt{109}})/3=0.373129$.
Another crossover happens when $J$ changes the sign, since for negative values
a $S=3/2$ state with $S_z=3/2$ is groundstate. Since in the nature 
the high degeneration of the $\Gamma_5$ groundstate will be lifted by a Jahn-Teller
distortion, one has to expect the system to undergo a structural transition at
positive values of J followed by a magnetic one when J changes sign.
A similar situation happens, if the cluster is occupied with four electrons, see
Fig. \ref{lowEwsU4N4J}. Now, we have transitions for negative values of $J$ only 
the first one, which is structural and magnetic, happens for $J_{c,1}(U=4t)=-0.19106\,t$
and the second one, which is purely magnetic, for $J_{c,2}(U=4t)=-0.575885\,t$.
The dependence of $J_c$ from the
Hubbard parameter $U$ one gets from equating the eigenvalues No. 115 with No. 102 and No. 102 with No. 94
respectively (for the eigenvalues see appendix \ref{appendix2a}). These equations
contain the parameter $J$ in an essentially non-algebraic way and have to be solved
numerically. The degeneracy of the groundstate steps from four to six in the first
crossover and remains six in the second one.
For 5 electrons on the cluster we find only one crossing point $J_{c}(U=4t)=-3.02201\,t$
from equating the eigenvalues No. 164 with No 174.
If the cluster is occupied with six electrons a crossover from a non-degenerated non-magnetic
state ($\Gamma_1$, $S=0$) to a sixfold degenerated groundstate happens at
\beq
J_c &=& \frac{4\,t + 3\,U - {\sqrt{208\,t^2 + 48\,t\,U + 9\,U^2}}}{3} \komma
\eeq
what yields $-2.44127\,t$ if $U=4t$ is used.\\
In Fig. \ref{primedewsTetraW} we plotted the canonical levels in dependence on the nearest neighbour Coulomb interaction $W$ scaled to the primed value $W'$ for $J=0$.
Again, we have chosen $U=4t$. Here we find one levelcrossing for $W_c=U/2$ in each of the cases of 2, 3, and 4 electrons on the cluster and none for the other occupation numbers.
At the first glance this seems to be not very interesting, but if one looks closer, one recognises a transition from a singlet state with $\Gamma_1$, $S=0$ to an eleven-fold (!)
degenerated groundstate (nine with $\Gamma_5$, $S=1$ and two with $\Gamma_3$, $S=0$)
for $N=2$ and $W<U/2$. For $N=4$ the transition is to a doublet with $\Gamma_3$ and $S=0$.
For three electrons we have for $W>U/2$ a $\Gamma_1$ $S=1/2$ doublet as groundstate and below the transition a $\Gamma_3$ spin quartet.

%% file: Parts/section4b.tex
\subsection{The tetrahedral cluster gas} 
The most interesting part of this paper is surely the ensemble of
tetrahedral clusters (clustergas), which are coupled to a common electron reservoir.
In Fig \ref{NvonMueU} we study the electron occupation  in dependence of the
the chemical potential for different values of the Hubbard correlation energy.
In the trivial case of $U=0$ we have a two step function only, due to the huge degeneration
of the grand-canonical groundstate. Of course this degeneration is partially lifted
if the on-site electron correlation is switched on. Nevertheless there remains a huge 
degeneration of grand-canonical groundstates with 0, 1, 2, and 3 electrons at $\mu=-t$.
Indeed the empty state is degenerated with the six $N=1$ $\Gamma_{4}$ states, 
further with the two $N=2$ $S=0$ $\Gamma_{3}$ and nine  $N=2$ $S=1$ $\Gamma_{5}$  states and 
finally with the four $N=3$ $S=3/2$ $\Gamma_{2}$ states. Of course this behaviour is due
to the fact, that for such low numbers of electrons uncorrelated many-particle states
exist without correlation. 
This situation is drawn in detail in Fig. \ref{lowlevelsN0N3}.
For other values of the correlation energy $U$
this degeneration happens at
\beq
\!\!\!\!\!\!\!\!\!\!\!\!\!\!\!\!
\mu&=&-t + \frac{3\,U}{4} - \frac{{\sqrt{64\,t^2 + 8\,t\,U + U^2}}}{4} + 
      \frac{{\sqrt{16\,t^2 + U^2}}}{\sqrt{3}}\,
\cos \left (\frac{1}{3}\arccos (\frac{12\,{\sqrt{3}}\,t^2\,U}
              {{\left( 16\,t^2 + U^2 \right) }^{\frac{3}{2}}})\right )
\eeq
The situation is completely different for $\mu/t=2.4056$,
where we find a degeneration of the two $N=4$ $S=0$ $\Gamma_{3}$ states with the
single $N=6$ $S=0$ $\Gamma_{1}$ state, as shown in Fig. \ref{lowlevelsN4N6}.
A $N=5$ state is never a grand-canonical groundstate. The situation resembles closely
the Hubbard square, where we also found a step from
$N=2$ to $N=4$ and from $N=4$ to $N=6$ \cite{Schumann02}. 
In Fig. \ref{gcLevelsU4} we depicted the
mentioned grand-canonical groundstate energies over the chemical potential 
in very detail.
Now, taking an exchange interaction into account the situation is
changed according to Fig. \ref{NvonMueJU4}. Whereas a ferromagnetic 
exchange does not change the qualitative picture, an antiferromagnetic does.
For that case the system avoids an odd number of electrons to stay with total spin zero.
For an extended system this means that for an odd mean electron occupation only
clusters with even electron occupation will exist in the groundstate.
Of course $|J/t|=1$ is usually a to large value. We have chosen it to make the plots
clearer. Nevertheless the picture is not qualitatively changed if smaller values are considered as one can see from the right panel in Fig. \ref{NvonMueJU4}.
As one can see from Fig. \ref{NvonMueWU4} a repulsive nearest-neighbour interaction 
lifts the degeneracy of levels with different electron occupation completely. Otherwise,
a large attractive interaction would result in a one-step function.

%% file: Parts/section5.tex
\section{Discussion}
Since my original interest for these clusters stems from extended systems 
I will focus the discussion to the consequences of the cluster behaviour
for crystals. Although the eigensystem and the shown figures cover the whole parameter space
from $-\infty$ to $+\infty$ for all model parameters I will not discuss the 
cases with attractive Coulomb interaction, since I believe they are of minor interest with
respect to real electron systems only.
Nevertheless papers dealing with attractive Hubbard models are published from time to
time, thus the presented results may be interesting to that community also.\\
The two (or three if the square with diagonal hopping $t'=t$ is taken as a separate
model) clusters are considered usually as "frustrated" systems. The reason is mainly that for large on-site coulomb interaction the Hubbard model may be mapped onto an anti-ferromagnetic 
Heisenberg model with the consequence that the localised magnetic moments can not be arranged
in such a configuration that all localised spins are antiparallel.
The reader may wonder that the conception of frustration was never touched in the paper.
That is due to the fact, that thinking in localised spins means thinking in one-particle
states, which do not play any role here, since the basis states, e.g. $\ket{2,u,0,d}$, are
many particle states from the beginning. Of course, if for instance the electron or spin density
as explicite functions of the position are wanted, one has to specify the action of the creation operators, what is done usually by means of determinants or one-particle-orbitals in clusters or Wannier functions in crystals.
Nevertheless, in my opinion the consequences of frustration are reflected in some sense by the some times large degeneracy of the ground states and the degeneration of ground states with
different occupation numbers. The ``frustration'' increases the groundstate energy for some
occupation numbers in such a manner that adjacent ``non-frustrated'' ground states have lower
grand canonical energies, i.e. exhibit higher (or equal) weights in the grand-canonical partition sum.
This feature, first observed in the Hubbard model on the square, is much more pronounced
in the Hubbard model on the triangle and tetrahedron. 
In a periodical arrangement of such clusters with a weak inter-cluster hopping, these
degeneration will result in patterns built up from clusters with different electron
occupation, alike the stripe structures observed in the copper-oxide planes.
But in the case of the triangular or tetrahedral array these patterns will be much more richer, since we have for the latter e.g. steps from zero to three (for the pure Hubbard model at $T=0$)
in the underdoped region or from four to six in the overdoped region.
Indeed, an material which can be described by an array of tetrahedra with dominating
on-site correlation will show very intersting features if it is doped to five electrons
per tetrahedron. If there is a small intercluster hopping, the three groundstates (see Fig. \ref{lowlevelsN4N6}) are broadened to small bands with the bandwidth of the order of
the (small!) intercluster hopping. Since the $N=4$ and $N=6$ bands will be magnetically inactive
for instance a small magnetic field will split the $N=5$ groundstate shifting the lower one 
below the $N=4$ or $N=6$ groundstates.
An equal effect could be reached by a Jahn-Teller scenario, but since a distortion
of the tetrahedral symmetry will also split the two fold degeneration of the $N=4$ groundstate,
whereas the non-degenerated ground state with six electrons will remain in position,
it is not clear, which state will be the groundstate, but this could be decided easily by
an experiment determining the ground-state spin density, or by including
the elastic energy into the Hamiltonian and treating it perturbatively.
Before one deals with such subtle features, we had to answer the question, to wether
degree a nearest-neighbour exchange or coulomb correlation changes the spectra of the
pure Hubbard model. A repulsive nearest neighbour coulomb interaction destroys
the degeneration of groundstates with different electron occupation. But if it is smaller
or in the order of the width of the bands formed from the cluster states, if an array
is considered, the subtle interplay of groundstates with different occupation numbers 
survives probably. But very interesting is the fact that an exchange interaction will
not necessarily lift these degeneracies. For example a ferromagnetic exchange does not
change the step curve of the pure Hubbard model qualitatively, whereas a relative
big antiferromagnetic exchange results in a step function avoiding odd number
occupation.
Furthermore, a detailed view of the spectra in dependence of the parameters 
offers several ground
state level crossings in dependence on the parameters $U$, $J$ and $W$. 
For extended systems these level crossings will result in  ``quantum phase transitions''.
If the chemical potential is far from the ``degeneration points'' a weak-coupled cluster
array will contain clusters with a well fixed integer number of electrons.
The ``quantum phase diagrams'' of such a system will be very alike to the related
``groundstate diagram'' of the cluster, calculated in the same way as it was done for
Figs. \ref{NvonMueU}, \ref{NvonMueJU4} and \ref{NvonMueWU4}.\\

To relate the results of the present paper to arrays of triangular or tetragonal clusters
it may be helpful to recall what we learned from the analytical 
solution of the square-cluster gas applied to the Hubbard model on a square lattice.
since for that case numerical cluster methods provided additional insight 
\cite{Senechal00,Maier05,Tremblay05,Maier06}.
Due to the inter-cluster hopping the three degenerated groundstates with N=2,3,4 gain a
bandwidth and will be probably shifted slightly in position, nevertheless 
an overlap in some region remains. 
If the chemical potential is within this region the consequence is that we
have degenerated non-magnetic and magnetic band states stemming from the N=2,4 and the N=3 
cluster states. Since I see no reason, why the effective mass of these states should be
the same, it will be not astonishing that the two charged (and bosonic!) states
(remember: for a dotation of N=3 to the
cuprates neutrality is restored by the ions not considered in the square Hubbard model) and the uncharged magnetic (and fermionic!) state will propagate
with different velocity (of course this is the spinon-holon picture extended to
the cluster states) resulting in the so-called spin-charge separation \cite{Senechal00}.
This feature is to expect within triangular- or tetragonal cluster arrays also,
since the main reason, i.e the degeneration of spinless charged and uncharged magnetic grand canonical groundstates, exists as well.

Finally I want to speculate about the relation of my results to
(high-$T_c$-) superconductivity in arrays of triangular or tetragonal clusters. 
It was shown \cite{Kocharian05,Kocharian06,Tsai06}, that the clustergas already 
accounts for a lot of the features of the phase diagram of the high-$T_c$ cuprates. 
Both the position of the superconducting domes (for electron and hole dotation) for low
temperature and the stripe phase at higher temperatures are roughly spoken in the 
vicinity of the degeneration 
points of the related cluster gas, i.e. for 3 or 5 electrons per 4-site cluster.
The slight asymmetry in the phase diagram for hole and electron doping within the
cuprates hints to a broken electron-hole symmetry, a feature which can not be
modeled by help of the pure Hubbard model. 
Otherwise the models considered here lack this symmetry at all. 
The most interesting physics for electron dotation will 
be e.g. in an array of tetrahedrons if five electrons per cluster are present,
whereas for hole dotation it will be a much broader region between one and three electrons 
per cluster.\\
Regarding superconductivity I believe that a good approximation for the superconducting groundstate is 
a covering of the two dimensional plane by a mixture of 
degenerated cluster states.
For instance for an array of squares coupled to an electron bath a mean occupation of five electrons per cluster would result in a ground state containing $x$ \% $N=4$, $y$ \% $N=6$, and $1-x-y$ \% N=5 cluster states. Since these three cluster ground states are degenerated, interchanging of an $N=4$ cluster and an adjacent $N=5$ or $N=6$ cluster will need no additional energy. 
Of course this 
interchange is forbidden if the clusters are isolated.
But introducing a small hopping, as it is done in the
checkerboard model \cite{Tsai06} will then, besides selecting one state out of the vast number of degenerated coverings, allow for that process, thereby propagating one 
electron (or two if an N=4/N=6 interchange is also allowed) from one cluster to another.
In the square lattice such an interchange is only possible in (1,0) and (0,1) direction, but forbidden over the diagonal (as may be seen easily for the checkerboard covering). 
If one adopts this simple picture the d symmetry of the charge propagation turns out to be a simple geometrical effect. 
Also the fact that the results calculated 
within the numerical cluster methods by help of a 4-site cluster do not show any
qualitative new features if the cluster size is increased hints to the
central role of the four-site square cluster \cite{Senechal00,Maier05}.
If an array of triangles or tetrahedrons is considered the same simple mechanism 
will take place if the system is doped to a chemical potential in the vicinity of the 
relevant degeneration points, i.e. where $N(\mu)$ steps at least by two, 
of course different geometrical constraints will be effective.
Apart from the introductorily mentioned but not discussed topics, 
where the presented cluster solution may be useful, there is, due the fast development of nanotechnology, also some chance
that two-dimensional arrays of triangular or tetragonal clusters with various geometries will be available and utilised for
experiments in the near future.

%% file: Parts/acknowledgement.tex
\begin{acknowledgement}
I am grateful to K. Kikoin and D. Efremov
for discussion on Kondo tunneling through 
small clusters.
\end{acknowledgement}

%% file: Parts/appendix1a.tex
\section{Eigensystem of the triangle}
\label{appendix1}
In the following tables the first column gives the number of the state for referencing
in the text. The second column labels the eigenstate by the eigenvalues of the U-independent
symmetry operators, i.e. the electron occupation number  $N_e$, the spin-projection 
$\opS_z$ in z-direction $m_s$, the eigenvalues $s(s+1)$ of $\opS^2$ and the spatial
symmetry is indicated by $\Gamma_{i,j}$, where the first index labels the irreducible
representation of the tetrahedral group and the second numbers the partner.
The notation is based on Ref. \cite{CornwellBook}.
The third column gives the enery eigenvalues in abbreviated form. The abbreviations
are listed subsequently to the tables. In the last column a numerical value is given for example.
For comparison we have chosen the same parameters as in Ref. \cite{Schumann02}
corresponding to a pure Hubbard-model with $U=5t$, $W=J=0t$. If the grand-canonical 
energy levels in an applied magnetic field are needed one has to substract $\mu \, N_e+h\,m_s$.
\subsection{\bf The eigenvalues of the triangle}
\parindent0cm
\begin{tabular}[t]{|r|l|c|c|}
\multicolumn{4}{c}{\large \bf \boldmath Eigenkets and eigenvalues for ${\rm  N_e}$=0 and   ${\rm m_s}$= $0$. } \\ \hline
\parbox[c]{1cm}{No}  & \parbox[c]{2.5cm}{\begin{center} Eigenstate \end{center}}  & \parbox[c]{7.5cm}{ \begin{center}   Energy \end{center}}  & \parbox[c]{2cm}{ \begin{center} Example \end{center}}  \\ \hline 
\hline 
1 & $\ket{0,0,0,\Gamma_1}$  & $0$   & 0. \\ 
\hline
\end{tabular} \\[2ex]
\begin{tabular}[t]{|r|l|c|c|}
\multicolumn{4}{c}{\large \bf \boldmath Eigenkets and eigenvalues for ${\rm  N_e}$=1 and   ${\rm m_s}$= $- 1 \over 2 $. } \\ \hline
\parbox[c]{1cm}{No}  & \parbox[c]{2.5cm}{\begin{center} Eigenstate \end{center}}  & \parbox[c]{7.5cm}{ \begin{center}   Energy \end{center}}  & \parbox[c]{2cm}{ \begin{center} Example \end{center}}  \\ \hline 
\hline 
2 & $\ket{1,- {1 \over 2} , {3 \over 4} ,\Gamma_1}$  & $2\,t$   & 2. \\ 
3 & $\ket{1,- {1 \over 2} , {3 \over 4} ,\Gamma_{3,1}}$  & $-t$   & -1. \\ 
4 & $\ket{1,- {1 \over 2} , {3 \over 4} ,\Gamma_{3,2}}$  & $-t$   & -1. \\ 
\hline
\end{tabular} \\[2ex]
\begin{tabular}[t]{|r|l|c|c|}
\multicolumn{4}{c}{\large \bf \boldmath Eigenkets and eigenvalues for ${\rm  N_e}$=1 and   ${\rm m_s}$= $ {1 \over 2} $. } \\ \hline
\parbox[c]{1cm}{No}  & \parbox[c]{2.5cm}{\begin{center} Eigenstate \end{center}}  & \parbox[c]{7.5cm}{ \begin{center}   Energy \end{center}}  & \parbox[c]{2cm}{ \begin{center} Example \end{center}}  \\ \hline 
\hline 
5 & $\ket{1, {1 \over 2} , {3 \over 4} ,\Gamma_1}$  & $2\,t$   & 2. \\ 
6 & $\ket{1, {1 \over 2} , {3 \over 4} ,\Gamma_{3,1}}$  & $-t$   & -1. \\ 
7 & $\ket{1, {1 \over 2} , {3 \over 4} ,\Gamma_{3,2}}$  & $-t$   & -1. \\ 
\hline
\end{tabular} \\[2ex]
\begin{tabular}[t]{|r|l|c|c|}
\multicolumn{4}{c}{\large \bf \boldmath Eigenkets and eigenvalues for ${\rm  N_e}$=2 and   ${\rm m_s}$= $-1$. } \\ \hline
\parbox[c]{1cm}{No}  & \parbox[c]{2.5cm}{\begin{center} Eigenstate \end{center}}  & \parbox[c]{7.5cm}{ \begin{center}   Energy \end{center}}  & \parbox[c]{2cm}{ \begin{center} Example \end{center}}  \\ \hline 
\hline 
8 & $\ket{2,-1,2,\Gamma_2}$  & $\frac{J}{2} - 2\,t + W$   & -2. \\ 
9 & $\ket{2,-1,2,\Gamma_{3,1}}$  & $\frac{J}{2} + t + W$   & 1. \\ 
10 & $\ket{2,-1,2,\Gamma_{3,2}}$  & $\frac{J}{2} + t + W$   & 1. \\ 
\hline
\end{tabular} \\[2ex]
\begin{tabular}[t]{|r|l|c|c|}
\multicolumn{4}{c}{\large \bf \boldmath Eigenkets and eigenvalues for ${\rm  N_e}$=2 and   ${\rm m_s}$= $0$. } \\ \hline
\parbox[c]{1cm}{No}  & \parbox[c]{2.5cm}{\begin{center} Eigenstate \end{center}}  & \parbox[c]{7.5cm}{ \begin{center}   Energy \end{center}}  & \parbox[c]{2cm}{ \begin{center} Example \end{center}}  \\ \hline 
\hline 
11 & $\ket{2,0,0,\Gamma_1}$  & $\frac{-{\sqrt{\Aindex{5}}} - J + 2\,t + U + W}{2}$   & 0. \\ 
12 & $\ket{2,0,0,\Gamma_1}$  & $\frac{{\sqrt{\Aindex{5}}} - J + 2\,t + U + W}{2}$   & 6. \\ 
13 & $\ket{2,0,2,\Gamma_2}$  & $-2\,t + W$   & -2. \\ 
14 & $\ket{2,0,0,\Gamma_{3,1}}$  & $\frac{-{\sqrt{\Aindex{1}}} - J - t + U + W}{2}$   & -1.37228 \\ 
15 & $\ket{2,0,0,\Gamma_{3,1}}$  & $\frac{{\sqrt{\Aindex{1}}} - J - t + U + W}{2}$   & 4.37228 \\ 
16 & $\ket{2,0,2,\Gamma_{3,1}}$  & $t + W$   & 1. \\ 
17 & $\ket{2,0,0,\Gamma_{3,2}}$  & $\frac{-{\sqrt{\Aindex{1}}} - J - t + U + W}{2}$   & -1.37228 \\ 
18 & $\ket{2,0,0,\Gamma_{3,2}}$  & $\frac{{\sqrt{\Aindex{1}}} - J - t + U + W}{2}$   & 4.37228 \\ 
19 & $\ket{2,0,2,\Gamma_{3,2}}$  & $t + W$   & 1. \\ 
\hline
\end{tabular} \\[2ex]
\begin{tabular}[t]{|r|l|c|c|}
\multicolumn{4}{c}{\large \bf \boldmath Eigenkets and eigenvalues for ${\rm  N_e}$=2 and   ${\rm m_s}$= $1$. } \\ \hline
\parbox[c]{1cm}{No}  & \parbox[c]{2.5cm}{\begin{center} Eigenstate \end{center}}  & \parbox[c]{7.5cm}{ \begin{center}   Energy \end{center}}  & \parbox[c]{2cm}{ \begin{center} Example \end{center}}  \\ \hline 
\hline 
20 & $\ket{2,1,2,\Gamma_2}$  & $\frac{J}{2} - 2\,t + W$   & -2. \\ 
21 & $\ket{2,1,2,\Gamma_{3,1}}$  & $\frac{J}{2} + t + W$   & 1. \\ 
22 & $\ket{2,1,2,\Gamma_{3,2}}$  & $\frac{J}{2} + t + W$   & 1. \\ 
\hline
\end{tabular} \\[2ex]
\begin{tabular}[t]{|r|l|c|c|}
\multicolumn{4}{c}{\large \bf \boldmath Eigenkets and eigenvalues for ${\rm  N_e}$=3 and   ${\rm m_s}$= $- {3 \over 2} $. } \\ \hline
\parbox[c]{1cm}{No}  & \parbox[c]{2.5cm}{\begin{center} Eigenstate \end{center}}  & \parbox[c]{7.5cm}{ \begin{center}   Energy \end{center}}  & \parbox[c]{2cm}{ \begin{center} Example \end{center}}  \\ \hline 
\hline 
23 & $\ket{3,- {3 \over 2} , {15 \over 4} ,\Gamma_2}$  & $\frac{3\,\left( J + 2\,W \right) }{2}$   & 0. \\ 
\hline
\end{tabular} \\[2ex]
\begin{tabular}[t]{|r|l|c|c|}
\multicolumn{4}{c}{\large \bf \boldmath Eigenkets and eigenvalues for ${\rm  N_e}$=3 and   ${\rm m_s}$= $- {1 \over 2} $. } \\ \hline
\parbox[c]{1cm}{No}  & \parbox[c]{2.5cm}{\begin{center} Eigenstate \end{center}}  & \parbox[c]{7.5cm}{ \begin{center}   Energy \end{center}}  & \parbox[c]{2cm}{ \begin{center} Example \end{center}}  \\ \hline 
\hline 
24 & $\ket{3,- {1 \over 2} , {3 \over 4} ,\Gamma_1}$  & $U + 2\,W$   & 4. \\ 
25 & $\ket{3,- {1 \over 2} , {3 \over 4} ,\Gamma_2}$  & $U + 2\,W$   & 4. \\ 
26 & $\ket{3,- {1 \over 2} , {15 \over 4} ,\Gamma_2}$  & $\frac{J + 6\,W}{2}$   & 0. \\ 
27 & $\ket{3,- {1 \over 2} , {3 \over 4} ,\Gamma_{3,1}}$  & $\frac{-J + 2\,U + 7\,W - 2\,{\sqrt{\Aindex{2}}}\,\cos (\Thetaindex{1})}{3}$   & -1.27492 \\ 
28 & $\ket{3,- {1 \over 2} , {3 \over 4} ,\Gamma_{3,1}}$  & $\frac{\Aindex{8}}{3}$   & 6.27492 \\ 
29 & $\ket{3,- {1 \over 2} , {3 \over 4} ,\Gamma_{3,1}}$  & $\frac{\Aindex{7}}{3}$   & 3. \\ 
30 & $\ket{3,- {1 \over 2} , {3 \over 4} ,\Gamma_{3,2}}$  & $\frac{-J + 2\,U + 7\,W - 2\,{\sqrt{\Aindex{2}}}\,\cos (\Thetaindex{1})}{3}$   & -1.27492 \\ 
31 & $\ket{3,- {1 \over 2} , {3 \over 4} ,\Gamma_{3,2}}$  & $\frac{\Aindex{8}}{3}$   & 6.27492 \\ 
32 & $\ket{3,- {1 \over 2} , {3 \over 4} ,\Gamma_{3,2}}$  & $\frac{\Aindex{7}}{3}$   & 3. \\ 
\hline
\end{tabular} \\[2ex]
\begin{tabular}[t]{|r|l|c|c|}
\multicolumn{4}{c}{\large \bf \boldmath Eigenkets and eigenvalues for ${\rm  N_e}$=3 and   ${\rm m_s}$= $ {1 \over 2} $. } \\ \hline
\parbox[c]{1cm}{No}  & \parbox[c]{2.5cm}{\begin{center} Eigenstate \end{center}}  & \parbox[c]{7.5cm}{ \begin{center}   Energy \end{center}}  & \parbox[c]{2cm}{ \begin{center} Example \end{center}}  \\ \hline 
\hline 
33 & $\ket{3, {1 \over 2} , {3 \over 4} ,\Gamma_1}$  & $U + 2\,W$   & 4. \\ 
34 & $\ket{3, {1 \over 2} , {3 \over 4} ,\Gamma_2}$  & $U + 2\,W$   & 4. \\ 
35 & $\ket{3, {1 \over 2} , {15 \over 4} ,\Gamma_2}$  & $\frac{J + 6\,W}{2}$   & 0. \\ 
36 & $\ket{3, {1 \over 2} , {3 \over 4} ,\Gamma_{3,1}}$  & $\frac{-J + 2\,U + 7\,W - 2\,{\sqrt{\Aindex{2}}}\,\cos (\Thetaindex{1})}{3}$   & -1.27492 \\ 
37 & $\ket{3, {1 \over 2} , {3 \over 4} ,\Gamma_{3,1}}$  & $\frac{\Aindex{8}}{3}$   & 6.27492 \\ 
38 & $\ket{3, {1 \over 2} , {3 \over 4} ,\Gamma_{3,1}}$  & $\frac{\Aindex{7}}{3}$   & 3. \\ 
39 & $\ket{3, {1 \over 2} , {3 \over 4} ,\Gamma_{3,2}}$  & $\frac{-J + 2\,U + 7\,W - 2\,{\sqrt{\Aindex{2}}}\,\cos (\Thetaindex{1})}{3}$   & -1.27492 \\ 
40 & $\ket{3, {1 \over 2} , {3 \over 4} ,\Gamma_{3,2}}$  & $\frac{\Aindex{8}}{3}$   & 6.27492 \\ 
41 & $\ket{3, {1 \over 2} , {3 \over 4} ,\Gamma_{3,2}}$  & $\frac{\Aindex{7}}{3}$   & 3. \\ 
\hline
\end{tabular} \\[2ex]
\begin{tabular}[t]{|r|l|c|c|}
\multicolumn{4}{c}{\large \bf \boldmath Eigenkets and eigenvalues for ${\rm  N_e}$=3 and   ${\rm m_s}$= $ {3 \over 2} $. } \\ \hline
\parbox[c]{1cm}{No}  & \parbox[c]{2.5cm}{\begin{center} Eigenstate \end{center}}  & \parbox[c]{7.5cm}{ \begin{center}   Energy \end{center}}  & \parbox[c]{2cm}{ \begin{center} Example \end{center}}  \\ \hline 
\hline 
42 & $\ket{3, {3 \over 2} , {15 \over 4} ,\Gamma_2}$  & $\frac{3\,\left( J + 2\,W \right) }{2}$   & 0. \\ 
\hline
\end{tabular} \\[2ex]
\begin{tabular}[t]{|r|l|c|c|}
\multicolumn{4}{c}{\large \bf \boldmath Eigenkets and eigenvalues for ${\rm  N_e}$=4 and   ${\rm m_s}$= $-1$. } \\ \hline
\parbox[c]{1cm}{No}  & \parbox[c]{2.5cm}{\begin{center} Eigenstate \end{center}}  & \parbox[c]{7.5cm}{ \begin{center}   Energy \end{center}}  & \parbox[c]{2cm}{ \begin{center} Example \end{center}}  \\ \hline 
\hline 
43 & $\ket{4,-1,2,\Gamma_2}$  & $\frac{J}{2} + 2\,t + U + 5\,W$   & 6. \\ 
44 & $\ket{4,-1,2,\Gamma_{3,1}}$  & $\frac{J}{2} - t + U + 5\,W$   & 3. \\ 
45 & $\ket{4,-1,2,\Gamma_{3,2}}$  & $\frac{J}{2} - t + U + 5\,W$   & 3. \\ 
\hline
\end{tabular} \\[2ex]
\begin{tabular}[t]{|r|l|c|c|}
\multicolumn{4}{c}{\large \bf \boldmath Eigenkets and eigenvalues for ${\rm  N_e}$=4 and   ${\rm m_s}$= $0$. } \\ \hline
\parbox[c]{1cm}{No}  & \parbox[c]{2.5cm}{\begin{center} Eigenstate \end{center}}  & \parbox[c]{7.5cm}{ \begin{center}   Energy \end{center}}  & \parbox[c]{2cm}{ \begin{center} Example \end{center}}  \\ \hline 
\hline 
46 & $\ket{4,0,0,\Gamma_1}$  & $\frac{-{\sqrt{\Aindex{3}}} - J - 2\,t + 3\,U + 9\,W}{2}$   & 0.876894 \\ 
47 & $\ket{4,0,0,\Gamma_1}$  & $\frac{{\sqrt{\Aindex{3}}} - J - 2\,t + 3\,U + 9\,W}{2}$   & 9.12311 \\ 
48 & $\ket{4,0,2,\Gamma_2}$  & $2\,t + U + 5\,W$   & 6. \\ 
49 & $\ket{4,0,0,\Gamma_{3,1}}$  & $\frac{-{\sqrt{\Aindex{4}}} - J + t + 3\,U + 9\,W}{2}$   & 4.43845 \\ 
50 & $\ket{4,0,0,\Gamma_{3,1}}$  & $\frac{{\sqrt{\Aindex{4}}} - J + t + 3\,U + 9\,W}{2}$   & 8.56155 \\ 
51 & $\ket{4,0,2,\Gamma_{3,1}}$  & $-t + U + 5\,W$   & 3. \\ 
52 & $\ket{4,0,0,\Gamma_{3,2}}$  & $\frac{-{\sqrt{\Aindex{4}}} - J + t + 3\,U + 9\,W}{2}$   & 4.43845 \\ 
53 & $\ket{4,0,0,\Gamma_{3,2}}$  & $\frac{{\sqrt{\Aindex{4}}} - J + t + 3\,U + 9\,W}{2}$   & 8.56155 \\ 
54 & $\ket{4,0,2,\Gamma_{3,2}}$  & $-t + U + 5\,W$   & 3. \\ 
\hline
\end{tabular} \\[2ex]
\begin{tabular}[t]{|r|l|c|c|}
\multicolumn{4}{c}{\large \bf \boldmath Eigenkets and eigenvalues for ${\rm  N_e}$=4 and   ${\rm m_s}$= $1$. } \\ \hline
\parbox[c]{1cm}{No}  & \parbox[c]{2.5cm}{\begin{center} Eigenstate \end{center}}  & \parbox[c]{7.5cm}{ \begin{center}   Energy \end{center}}  & \parbox[c]{2cm}{ \begin{center} Example \end{center}}  \\ \hline 
\hline 
55 & $\ket{4,1,2,\Gamma_2}$  & $\frac{J}{2} + 2\,t + U + 5\,W$   & 6. \\ 
56 & $\ket{4,1,2,\Gamma_{3,1}}$  & $\frac{J}{2} - t + U + 5\,W$   & 3. \\ 
57 & $\ket{4,1,2,\Gamma_{3,2}}$  & $\frac{J}{2} - t + U + 5\,W$   & 3. \\ 
\hline
\end{tabular} \\[2ex]
\begin{tabular}[t]{|r|l|c|c|}
\multicolumn{4}{c}{\large \bf \boldmath Eigenkets and eigenvalues for ${\rm  N_e}$=5 and   ${\rm m_s}$= $- {1 \over 2} $. } \\ \hline
\parbox[c]{1cm}{No}  & \parbox[c]{2.5cm}{\begin{center} Eigenstate \end{center}}  & \parbox[c]{7.5cm}{ \begin{center}   Energy \end{center}}  & \parbox[c]{2cm}{ \begin{center} Example \end{center}}  \\ \hline 
\hline 
58 & $\ket{5,- {1 \over 2} , {3 \over 4} ,\Gamma_1}$  & $-2\,t + 2\,U + 8\,W$   & 6. \\ 
59 & $\ket{5,- {1 \over 2} , {3 \over 4} ,\Gamma_{3,1}}$  & $t + 2\,U + 8\,W$   & 9. \\ 
60 & $\ket{5,- {1 \over 2} , {3 \over 4} ,\Gamma_{3,2}}$  & $t + 2\,U + 8\,W$   & 9. \\ 
\hline
\end{tabular} \\[2ex]
\begin{tabular}[t]{|r|l|c|c|}
\multicolumn{4}{c}{\large \bf \boldmath Eigenkets and eigenvalues for ${\rm  N_e}$=5 and   ${\rm m_s}$= $ {1 \over 2} $. } \\ \hline
\parbox[c]{1cm}{No}  & \parbox[c]{2.5cm}{\begin{center} Eigenstate \end{center}}  & \parbox[c]{7.5cm}{ \begin{center}   Energy \end{center}}  & \parbox[c]{2cm}{ \begin{center} Example \end{center}}  \\ \hline 
\hline 
61 & $\ket{5, {1 \over 2} , {3 \over 4} ,\Gamma_1}$  & $-2\,t + 2\,U + 8\,W$   & 6. \\ 
62 & $\ket{5, {1 \over 2} , {3 \over 4} ,\Gamma_{3,1}}$  & $t + 2\,U + 8\,W$   & 9. \\ 
63 & $\ket{5, {1 \over 2} , {3 \over 4} ,\Gamma_{3,2}}$  & $t + 2\,U + 8\,W$   & 9. \\ 
\hline
\end{tabular} \\[2ex]
\begin{tabular}[t]{|r|l|c|c|}
\multicolumn{4}{c}{\large \bf \boldmath Eigenkets and eigenvalues for ${\rm  N_e}$=6 and   ${\rm m_s}$= $0$. } \\ \hline
\parbox[c]{1cm}{No}  & \parbox[c]{2.5cm}{\begin{center} Eigenstate \end{center}}  & \parbox[c]{7.5cm}{ \begin{center}   Energy \end{center}}  & \parbox[c]{2cm}{ \begin{center} Example \end{center}}  \\ \hline 
\hline 
64 & $\ket{6,0,0,\Gamma_1}$  & $3\,\left( U + 4\,W \right) $   & 12. \\ 
\hline
\end{tabular} \\

%% file: Parts/appendix1b.tex
\parindent0cm
\subsection{\bf List of abbreviations }
\parindent0cm
\subsection*{ List of abbreviations }
\beq 
\Aindex{1} &=& 4\,\left( 2\,t^2 + t\,U + U\,\left( J - W \right)  \right)  + {\left( J + t - U - W \right) }^2
 \nonumber \\
\Aindex{2} &=& J^2 + 27\,t^2 + 2\,J\,\left( U - W \right)  + {\left( U - W \right) }^2
 \nonumber \\
\Aindex{3} &=& J^2 + 36\,t^2 + 4\,t\,\left( U - W \right)  + {\left( U - W \right) }^2 + 2\,J\,\left( 2\,t + U - W \right) 
 \nonumber \\
\Aindex{4} &=& J^2 + 9\,t^2 - 2\,t\,\left( U - W \right)  + {\left( U - W \right) }^2 - 2\,J\,\left( t - U + W \right) 
 \nonumber \\
\Aindex{5} &=& 4\,\left( 8\,t^2 - 2\,t\,U + U\,\left( J - W \right)  \right)  + {\left( -J + 2\,t + U + W \right) }^2
 \nonumber \\
\Aindex{6} &=& J - 2\,U - 7\,W + 2\,{\sqrt{\Aindex{2}}}\,\cos (\Thetaindex{1})
 \nonumber \\
\Aindex{7} &=& -J + 2\,U + 7\,W + {\sqrt{\Aindex{2}}}\,\left( \cos (\Thetaindex{1}) - {\sqrt{3}}\,\sin (\Thetaindex{1}) \right) 
 \nonumber \\
\Aindex{8} &=& -J + 2\,U + 7\,W + {\sqrt{\Aindex{2}}}\,\left( \cos (\Thetaindex{1}) + {\sqrt{3}}\,\sin (\Thetaindex{1}) \right) 
 \nonumber \\
\eeq 
\beq
\Thetaindex{1} &=& \frac{\arccos (\frac{{\left( J + U - W \right) }^3}{{\Aindex{2}}^{\frac{3}{2}}})}{3}
\nonumber \eeq 

%% file: Parts/appendix1c.tex
\subsection{\bf The eigenvectors}
In the following the unnormalised eigenvectors are given in abbreviated form, whereby
the indices give the number of the state in accordance with the numbering of
the energy levels given above.
For getting the normalised eigenvectors the coefficients have to be divided by the
normalisation constant $N_i$, which is given for the vectors  containing more than one coefficient.
\mathindent0cm
{\subsection*{\boldmath Unnormalized eigenvectors for ${\rm  N_e}=0$ and   ${\rm m_s}$= $0$.}
\beq
\ket{\Psi}_{1}& = &\ket{0,0,0,\Gamma_1} \nonumber \\ 
&=& 1
 \left ( \ket{000}\right) \nonumber 
\eeq
{\subsection*{\boldmath Unnormalized eigenvectors for ${\rm  N_e}=1$ and   ${\rm m_s}$= $- {1 \over 2} $.}
\beq
\ket{\Psi}_{2}& = &\ket{1,- {1 \over 2} , {3 \over 4} ,\Gamma_1} \nonumber \\ 
&=& \frac{1}{{\sqrt{3}}}
 \left ( \ket{00d} + \ket{0d0} + \ket{d00}\right) \nonumber 
\eeq
\beq
\ket{\Psi}_{3}& = &\ket{1,- {1 \over 2} , {3 \over 4} ,\Gamma_{3,1}} \nonumber \\ 
&=& \frac{1}{{\sqrt{2}}}
 \left ( \ket{00d} - \ket{0d0}\right) \nonumber 
\eeq
\beq
\ket{\Psi}_{4}& = &\ket{1,- {1 \over 2} , {3 \over 4} ,\Gamma_{3,2}} \nonumber \\ 
& = &
\Cindex{4}{1} \left ( 
\ket{00d} + \ket{0d0}\right) 
 \nonumber \\
& + &
\Cindex{4}{2} \left ( 
\ket{d00}\right) 
\nonumber \eeq
\beq
C_{4-1} &=&
-\left( \frac{1}{{\sqrt{6}}} \right)  \nonumber \\
C_{4-2} &=&
{\sqrt{\frac{2}{3}}} \nonumber \\
 N_{4} &=& {\sqrt{2\,{\Cindex{4}{1}}^2 + {\Cindex{4}{2}}^2}} \nonumber \eeq 
{\subsection*{\boldmath Unnormalized eigenvectors for ${\rm  N_e}=1$ and   ${\rm m_s}$= $ {1 \over 2} $.}
\beq
\ket{\Psi}_{5}& = &\ket{1, {1 \over 2} , {3 \over 4} ,\Gamma_1} \nonumber \\ 
&=& \frac{1}{{\sqrt{3}}}
 \left ( \ket{00u} + \ket{0u0} + \ket{u00}\right) \nonumber 
\eeq
\beq
\ket{\Psi}_{6}& = &\ket{1, {1 \over 2} , {3 \over 4} ,\Gamma_{3,1}} \nonumber \\ 
&=& \frac{1}{{\sqrt{2}}}
 \left ( \ket{00u} - \ket{0u0}\right) \nonumber 
\eeq
\beq
\ket{\Psi}_{7}& = &\ket{1, {1 \over 2} , {3 \over 4} ,\Gamma_{3,2}} \nonumber \\ 
& = &
\Cindex{7}{1} \left ( 
\ket{00u} + \ket{0u0}\right) 
 \nonumber \\
& + &
\Cindex{7}{2} \left ( 
\ket{u00}\right) 
\nonumber \eeq
\beq
C_{7-1} &=&
-\left( \frac{1}{{\sqrt{6}}} \right)  \nonumber \\
C_{7-2} &=&
{\sqrt{\frac{2}{3}}} \nonumber \\
 N_{7} &=& {\sqrt{2\,{\Cindex{7}{1}}^2 + {\Cindex{7}{2}}^2}} \nonumber \eeq 
{\subsection*{\boldmath Unnormalized eigenvectors for ${\rm  N_e}=2$ and   ${\rm m_s}$= $-1$.}
\beq
\ket{\Psi}_{8}& = &\ket{2,-1,2,\Gamma_2} \nonumber \\ 
&=& \frac{1}{{\sqrt{3}}}
 \left ( \ket{0dd} - \ket{d0d} + \ket{dd0}\right) \nonumber 
\eeq
\beq
\ket{\Psi}_{9}& = &\ket{2,-1,2,\Gamma_{3,1}} \nonumber \\ 
& = &
\Cindex{9}{1} \left ( 
\ket{0dd}\right) 
 \nonumber \\
& + &
\Cindex{9}{2} \left ( 
\ket{d0d} - \ket{dd0}\right) 
\nonumber \eeq
\beq
C_{9-1} &=&
-{\sqrt{\frac{2}{3}}} \nonumber \\
C_{9-2} &=&
-\left( \frac{1}{{\sqrt{6}}} \right)  \nonumber \\
 N_{9} &=& {\sqrt{{\Cindex{9}{1}}^2 + 2\,{\Cindex{9}{2}}^2}} \nonumber \eeq 
\beq
\ket{\Psi}_{10}& = &\ket{2,-1,2,\Gamma_{3,2}} \nonumber \\ 
&=& \frac{1}{{\sqrt{2}}}
 \left ( \ket{d0d} + \ket{dd0}\right) \nonumber 
\eeq
{\subsection*{\boldmath Unnormalized eigenvectors for ${\rm  N_e}=2$ and   ${\rm m_s}$= $0$.}
\beq
\ket{\Psi}_{11}& = &\ket{2,0,0,\Gamma_1} \nonumber \\ 
& = &
\Cindex{11}{1} \left ( 
\ket{002} + \ket{020} + \ket{200}\right) 
 \nonumber \\
& + &
\Cindex{11}{2} \left ( 
\ket{0du} - \ket{0ud} + \ket{d0u} + \ket{du0} - \ket{u0d} - \ket{ud0}\right) 
\nonumber \eeq
\beq
C_{11-1} &=&
2\,{\sqrt{\frac{2}{3}}}\,t \nonumber \\
C_{11-2} &=&
\frac{1}{2\,{\sqrt{6}}}
\, \left ( {\sqrt{\Aindex{5}}} + J - 2\,t + U - W  \right ) \nonumber \\
 N_{11} &=& {\sqrt{3\,{\Cindex{11}{1}}^2 + 6\,{\Cindex{11}{2}}^2}} \nonumber \eeq 
\beq
\ket{\Psi}_{12}& = &\ket{2,0,0,\Gamma_1} \nonumber \\ 
& = &
\Cindex{12}{1} \left ( 
\ket{002} + \ket{020} + \ket{200}\right) 
 \nonumber \\
& + &
\Cindex{12}{2} \left ( 
\ket{0du} - \ket{0ud} + \ket{d0u} + \ket{du0} - \ket{u0d} - \ket{ud0}\right) 
\nonumber \eeq
\beq
C_{12-1} &=&
2\,{\sqrt{\frac{2}{3}}}\,t \nonumber \\
C_{12-2} &=&
\frac{-1}{2\,{\sqrt{6}}}
\, \left ( {\sqrt{\Aindex{5}}} - J + 2\,t - U + W  \right ) \nonumber \\
 N_{12} &=& {\sqrt{3\,{\Cindex{12}{1}}^2 + 6\,{\Cindex{12}{2}}^2}} \nonumber \eeq 
\beq
\ket{\Psi}_{13}& = &\ket{2,0,2,\Gamma_2} \nonumber \\ 
&=& \frac{1}{{\sqrt{6}}}
 \left ( \ket{0du} + \ket{0ud} - \ket{d0u} + \ket{du0} - \ket{u0d} + \ket{ud0}\right) \nonumber 
\eeq
\beq
\ket{\Psi}_{14}& = &\ket{2,0,0,\Gamma_{3,1}} \nonumber \\ 
& = &
\Cindex{14}{1} \left ( 
\ket{002} - \ket{020}\right) 
 \nonumber \\
& + &
\Cindex{14}{2} \left ( 
\ket{d0u} - \ket{du0} - \ket{u0d} + \ket{ud0}\right) 
\nonumber \eeq
\beq
C_{14-1} &=&
t \nonumber \\
C_{14-2} &=&
\frac{1}{4}
\, \left ( {\sqrt{\Aindex{1}}} + J + t + U - W  \right ) \nonumber \\
 N_{14} &=& {\sqrt{2\,{\Cindex{14}{1}}^2 + 4\,{\Cindex{14}{2}}^2}} \nonumber \eeq 
\beq
\ket{\Psi}_{15}& = &\ket{2,0,0,\Gamma_{3,1}} \nonumber \\ 
& = &
\Cindex{15}{1} \left ( 
\ket{002} - \ket{020}\right) 
 \nonumber \\
& + &
\Cindex{15}{2} \left ( 
\ket{d0u} - \ket{du0} - \ket{u0d} + \ket{ud0}\right) 
\nonumber \eeq
\beq
C_{15-1} &=&
t \nonumber \\
C_{15-2} &=&
\frac{1}{4}
\, \left ( -{\sqrt{\Aindex{1}}} + J + t + U - W  \right ) \nonumber \\
 N_{15} &=& {\sqrt{2\,{\Cindex{15}{1}}^2 + 4\,{\Cindex{15}{2}}^2}} \nonumber \eeq 
\beq
\ket{\Psi}_{16}& = &\ket{2,0,2,\Gamma_{3,1}} \nonumber \\ 
& = &
\Cindex{16}{1} \left ( 
\ket{0du} + \ket{0ud}\right) 
 \nonumber \\
& + &
\Cindex{16}{2} \left ( 
\ket{d0u} - \ket{du0} + \ket{u0d} - \ket{ud0}\right) 
\nonumber \eeq
\beq
C_{16-1} &=&
-\left( \frac{1}{{\sqrt{3}}} \right)  \nonumber \\
C_{16-2} &=&
\frac{-1}{2\,{\sqrt{3}}} \nonumber \\
 N_{16} &=& {\sqrt{2\,{\Cindex{16}{1}}^2 + 4\,{\Cindex{16}{2}}^2}} \nonumber \eeq 
\beq
\ket{\Psi}_{17}& = &\ket{2,0,0,\Gamma_{3,2}} \nonumber \\ 
& = &
\Cindex{17}{1} \left ( 
\ket{002} + \ket{020}\right) 
 \nonumber \\
& + &
\Cindex{17}{2} \left ( 
\ket{0du} - \ket{0ud}\right) 
 \nonumber \\
& + &
\Cindex{17}{3} \left ( 
\ket{200}\right) 
 \nonumber \\
& + &
\Cindex{17}{4} \left ( 
\ket{d0u} + \ket{du0} - \ket{u0d} - \ket{ud0}\right) 
\nonumber \eeq
\beq
C_{17-1} &=&
\frac{1}{2\,{\sqrt{6}}}
\, \left ( {\sqrt{\Aindex{1}}} - J - t - U + W  \right ) \nonumber \\
C_{17-2} &=&
{\sqrt{\frac{2}{3}}}\,t \nonumber \\
C_{17-3} &=&
-\left( \frac{1}{{\sqrt{6}}} \right) 
\, \left ( {\sqrt{\Aindex{1}}} - J - t - U + W  \right ) \nonumber \\
C_{17-4} &=&
-\left( \frac{t}{{\sqrt{6}}} \right)  \nonumber \\
 N_{17} &=& {\sqrt{2\,{\Cindex{17}{1}}^2 + 2\,{\Cindex{17}{2}}^2 + {\Cindex{17}{3}}^2 + 4\,{\Cindex{17}{4}}^2}} \nonumber \eeq 
\beq
\ket{\Psi}_{18}& = &\ket{2,0,0,\Gamma_{3,2}} \nonumber \\ 
& = &
\Cindex{18}{1} \left ( 
\ket{002} + \ket{020}\right) 
 \nonumber \\
& + &
\Cindex{18}{2} \left ( 
\ket{0du} - \ket{0ud}\right) 
 \nonumber \\
& + &
\Cindex{18}{3} \left ( 
\ket{200}\right) 
 \nonumber \\
& + &
\Cindex{18}{4} \left ( 
\ket{d0u} + \ket{du0} - \ket{u0d} - \ket{ud0}\right) 
\nonumber \eeq
\beq
C_{18-1} &=&
\frac{-1}{2\,{\sqrt{6}}}
\, \left ( {\sqrt{\Aindex{1}}} + J + t + U - W  \right ) \nonumber \\
C_{18-2} &=&
{\sqrt{\frac{2}{3}}}\,t \nonumber \\
C_{18-3} &=&
\frac{1}{{\sqrt{6}}}
\, \left ( {\sqrt{\Aindex{1}}} + J + t + U - W  \right ) \nonumber \\
C_{18-4} &=&
-\left( \frac{t}{{\sqrt{6}}} \right)  \nonumber \\
 N_{18} &=& {\sqrt{2\,{\Cindex{18}{1}}^2 + 2\,{\Cindex{18}{2}}^2 + {\Cindex{18}{3}}^2 + 4\,{\Cindex{18}{4}}^2}} \nonumber \eeq 
\beq
\ket{\Psi}_{19}& = &\ket{2,0,2,\Gamma_{3,2}} \nonumber \\ 
&=& \frac{1}{2}
 \left ( \ket{d0u} + \ket{du0} + \ket{u0d} + \ket{ud0}\right) \nonumber 
\eeq
{\subsection*{\boldmath Unnormalized eigenvectors for ${\rm  N_e}=2$ and   ${\rm m_s}$= $1$.}
\beq
\ket{\Psi}_{20}& = &\ket{2,1,2,\Gamma_2} \nonumber \\ 
&=& \frac{1}{{\sqrt{3}}}
 \left ( \ket{0uu} - \ket{u0u} + \ket{uu0}\right) \nonumber 
\eeq
\beq
\ket{\Psi}_{21}& = &\ket{2,1,2,\Gamma_{3,1}} \nonumber \\ 
& = &
\Cindex{21}{1} \left ( 
\ket{0uu}\right) 
 \nonumber \\
& + &
\Cindex{21}{2} \left ( 
\ket{u0u} - \ket{uu0}\right) 
\nonumber \eeq
\beq
C_{21-1} &=&
-{\sqrt{\frac{2}{3}}} \nonumber \\
C_{21-2} &=&
-\left( \frac{1}{{\sqrt{6}}} \right)  \nonumber \\
 N_{21} &=& {\sqrt{{\Cindex{21}{1}}^2 + 2\,{\Cindex{21}{2}}^2}} \nonumber \eeq 
\beq
\ket{\Psi}_{22}& = &\ket{2,1,2,\Gamma_{3,2}} \nonumber \\ 
&=& \frac{1}{{\sqrt{2}}}
 \left ( \ket{u0u} + \ket{uu0}\right) \nonumber 
\eeq
{\subsection*{\boldmath Unnormalized eigenvectors for ${\rm  N_e}=3$ and   ${\rm m_s}$= $- {3 \over 2} $.}
\beq
\ket{\Psi}_{23}& = &\ket{3,- {3 \over 2} , {15 \over 4} ,\Gamma_2} \nonumber \\ 
&=& 1
 \left ( \ket{ddd}\right) \nonumber 
\eeq
{\subsection*{\boldmath Unnormalized eigenvectors for ${\rm  N_e}=3$ and   ${\rm m_s}$= $- {1 \over 2} $.}
\beq
\ket{\Psi}_{24}& = &\ket{3,- {1 \over 2} , {3 \over 4} ,\Gamma_1} \nonumber \\ 
&=& \frac{1}{{\sqrt{6}}}
 \left ( \ket{02d} + \ket{0d2} + \ket{20d} + \ket{2d0} + \ket{d02} + \ket{d20}\right) \nonumber 
\eeq
\beq
\ket{\Psi}_{25}& = &\ket{3,- {1 \over 2} , {3 \over 4} ,\Gamma_2} \nonumber \\ 
&=& \frac{1}{{\sqrt{6}}}
 \left ( \ket{02d} - \ket{0d2} - \ket{20d} + \ket{2d0} + \ket{d02} - \ket{d20}\right) \nonumber 
\eeq
\beq
\ket{\Psi}_{26}& = &\ket{3,- {1 \over 2} , {15 \over 4} ,\Gamma_2} \nonumber \\ 
&=& \frac{1}{{\sqrt{3}}}
 \left ( \ket{ddu} + \ket{dud} + \ket{udd}\right) \nonumber 
\eeq
\beq
\ket{\Psi}_{27}& = &\ket{3,- {1 \over 2} , {3 \over 4} ,\Gamma_{3,1}} \nonumber \\ 
& = &
\Cindex{27}{1} \left ( 
\ket{02d} - \ket{0d2}\right) 
 \nonumber \\
& + &
\Cindex{27}{2} \left ( 
\ket{20d} - \ket{2d0}\right) 
 \nonumber \\
& + &
\Cindex{27}{3} \left ( 
\ket{d02} - \ket{d20}\right) 
 \nonumber \\
& + &
\Cindex{27}{4} \left ( 
\ket{ddu} + \ket{dud}\right) 
 \nonumber \\
& + &
\Cindex{27}{5} \left ( 
\ket{udd}\right) 
\nonumber \eeq
\beq
C_{27-1} &=&
-\left( \frac{t}{{\sqrt{6}}} \right) 
\, \left ( J - 3\,t + U - W + 2\,{\sqrt{\Aindex{2}}}\,\cos (\Thetaindex{1})  \right ) \nonumber \\
C_{27-2} &=&
-\left( \frac{t}{{\sqrt{6}}} \right) 
\, \left ( J + 3\,t + U - W + 2\,{\sqrt{\Aindex{2}}}\,\cos (\Thetaindex{1})  \right ) \nonumber \\
C_{27-3} &=&
-\left( {\sqrt{6}}\,t^2 \right)  \nonumber \\
C_{27-4} &=&
\frac{-1}{9\,{\sqrt{6}}}
\,\left ( {\Aindex{6}}^2 - 27\,t^2 + 6\,J\,U - 3\,U^2 + 12\,J\,W - 30\,U\,W - 48\,W^2\right . \nonumber \\
&& \hspace{1cm} 
 + 
 \left . 12\,{\sqrt{\Aindex{2}}}\,U\,\cos (\Thetaindex{1}) + 24\,{\sqrt{\Aindex{2}}}\,W\,\cos (\Thetaindex{1})
\right  )  \nonumber \\
C_{27-5} &=&
\frac{{\sqrt{\frac{2}{3}}}}{9}
\,\left ( {\Aindex{6}}^2 - 27\,t^2 + 6\,J\,U - 3\,U^2 + 12\,J\,W - 30\,U\,W - 48\,W^2\right . \nonumber \\
&& \hspace{1cm} 
 + 
 \left . 12\,{\sqrt{\Aindex{2}}}\,U\,\cos (\Thetaindex{1}) + 24\,{\sqrt{\Aindex{2}}}\,W\,\cos (\Thetaindex{1})
\right  )  \nonumber \\
 N_{27} &=& {\sqrt{2\,{\Cindex{27}{1}}^2 + 2\,{\Cindex{27}{2}}^2 + 2\,{\Cindex{27}{3}}^2 + 2\,{\Cindex{27}{4}}^2 + {\Cindex{27}{5}}^2}} \nonumber \eeq 
\beq
\ket{\Psi}_{28}& = &\ket{3,- {1 \over 2} , {3 \over 4} ,\Gamma_{3,1}} \nonumber \\ 
& = &
\Cindex{28}{1} \left ( 
\ket{02d} - \ket{0d2}\right) 
 \nonumber \\
& + &
\Cindex{28}{2} \left ( 
\ket{20d} - \ket{2d0}\right) 
 \nonumber \\
& + &
\Cindex{28}{3} \left ( 
\ket{d02} - \ket{d20}\right) 
 \nonumber \\
& + &
\Cindex{28}{4} \left ( 
\ket{ddu} + \ket{dud}\right) 
 \nonumber \\
& + &
\Cindex{28}{5} \left ( 
\ket{udd}\right) 
\nonumber \eeq
\beq
C_{28-1} &=&
\frac{-t}{3\,{\sqrt{2}}}
\, \left ( {\sqrt{3}}\,J - 3\,{\sqrt{3}}\,t + {\sqrt{3}}\,U - {\sqrt{3}}\,W - {\sqrt{3}}\,{\sqrt{\Aindex{2}}}\,\cos (\Thetaindex{1}) - 3\,{\sqrt{\Aindex{2}}}\,\sin (\Thetaindex{1})  \right ) \nonumber \\
C_{28-2} &=&
\frac{-t}{3\,{\sqrt{2}}}
\, \left ( {\sqrt{3}}\,J + 3\,{\sqrt{3}}\,t + {\sqrt{3}}\,U - {\sqrt{3}}\,W - {\sqrt{3}}\,{\sqrt{\Aindex{2}}}\,\cos (\Thetaindex{1}) - 3\,{\sqrt{\Aindex{2}}}\,\sin (\Thetaindex{1})  \right ) \nonumber \\
C_{28-3} &=&
-\left( {\sqrt{6}}\,t^2 \right)  \nonumber \\
C_{28-4} &=&
\frac{-1}{9\,{\sqrt{6}}}
\,\left ( {\Aindex{8}}^2 - 27\,t^2 - 6\,\Aindex{8}\,U\right . \nonumber \\
&& \hspace{1cm} 
 + 
 \left . 9\,U^2 - 12\,\Aindex{8}\,W + 36\,U\,W + 36\,W^2
\right  )  \nonumber \\
C_{28-5} &=&
\frac{{\sqrt{\frac{2}{3}}}}{9}
\,\left ( {\Aindex{8}}^2 - 27\,t^2 - 6\,\Aindex{8}\,U\right . \nonumber \\
&& \hspace{1cm} 
 + 
 \left . 9\,U^2 - 12\,\Aindex{8}\,W + 36\,U\,W + 36\,W^2
\right  )  \nonumber \\
 N_{28} &=& {\sqrt{2\,{\Cindex{28}{1}}^2 + 2\,{\Cindex{28}{2}}^2 + 2\,{\Cindex{28}{3}}^2 + 2\,{\Cindex{28}{4}}^2 + {\Cindex{28}{5}}^2}} \nonumber \eeq 
\beq
\ket{\Psi}_{29}& = &\ket{3,- {1 \over 2} , {3 \over 4} ,\Gamma_{3,1}} \nonumber \\ 
& = &
\Cindex{29}{1} \left ( 
\ket{02d} - \ket{0d2}\right) 
 \nonumber \\
& + &
\Cindex{29}{2} \left ( 
\ket{20d} - \ket{2d0}\right) 
 \nonumber \\
& + &
\Cindex{29}{3} \left ( 
\ket{d02} - \ket{d20}\right) 
 \nonumber \\
& + &
\Cindex{29}{4} \left ( 
\ket{ddu} + \ket{dud}\right) 
 \nonumber \\
& + &
\Cindex{29}{5} \left ( 
\ket{udd}\right) 
\nonumber \eeq
\beq
C_{29-1} &=&
\frac{-t}{3\,{\sqrt{2}}}
\, \left ( {\sqrt{3}}\,J - 3\,{\sqrt{3}}\,t + {\sqrt{3}}\,U - {\sqrt{3}}\,W - {\sqrt{3}}\,{\sqrt{\Aindex{2}}}\,\cos (\Thetaindex{1}) + 3\,{\sqrt{\Aindex{2}}}\,\sin (\Thetaindex{1})  \right ) \nonumber \\
C_{29-2} &=&
\frac{-t}{3\,{\sqrt{2}}}
\, \left ( {\sqrt{3}}\,J + 3\,{\sqrt{3}}\,t + {\sqrt{3}}\,U - {\sqrt{3}}\,W - {\sqrt{3}}\,{\sqrt{\Aindex{2}}}\,\cos (\Thetaindex{1}) + 3\,{\sqrt{\Aindex{2}}}\,\sin (\Thetaindex{1})  \right ) \nonumber \\
C_{29-3} &=&
-\left( {\sqrt{6}}\,t^2 \right)  \nonumber \\
C_{29-4} &=&
\frac{-1}{9\,{\sqrt{6}}}
\,\left ( {\Aindex{7}}^2 - 27\,t^2 - 6\,\Aindex{7}\,U\right . \nonumber \\
&& \hspace{1cm} 
 + 
 \left . 9\,U^2 - 12\,\Aindex{7}\,W + 36\,U\,W + 36\,W^2
\right  )  \nonumber \\
C_{29-5} &=&
\frac{{\sqrt{\frac{2}{3}}}}{9}
\,\left ( {\Aindex{7}}^2 - 27\,t^2 - 6\,\Aindex{7}\,U\right . \nonumber \\
&& \hspace{1cm} 
 + 
 \left . 9\,U^2 - 12\,\Aindex{7}\,W + 36\,U\,W + 36\,W^2
\right  )  \nonumber \\
 N_{29} &=& {\sqrt{2\,{\Cindex{29}{1}}^2 + 2\,{\Cindex{29}{2}}^2 + 2\,{\Cindex{29}{3}}^2 + 2\,{\Cindex{29}{4}}^2 + {\Cindex{29}{5}}^2}} \nonumber \eeq 
\beq
\ket{\Psi}_{30}& = &\ket{3,- {1 \over 2} , {3 \over 4} ,\Gamma_{3,2}} \nonumber \\ 
& = &
\Cindex{30}{1} \left ( 
\ket{02d} + \ket{0d2}\right) 
 \nonumber \\
& + &
\Cindex{30}{2} \left ( 
\ket{20d} + \ket{2d0}\right) 
 \nonumber \\
& + &
\Cindex{30}{3} \left ( 
\ket{d02} + \ket{d20}\right) 
 \nonumber \\
& + &
\Cindex{30}{4} \left ( 
\ket{ddu} - \ket{dud}\right) 
\nonumber \eeq
\beq
C_{30-1} &=&
\frac{-t}{6}
\, \left ( -2\,J + 9\,t - 2\,U + 2\,W + 2\,{\sqrt{\Aindex{2}}}\,\cos (\Thetaindex{1})  \right ) \nonumber \\
C_{30-2} &=&
\frac{1}{18}
\,\left ( -{\Aindex{6}}^2 + 3\,J^2 - 12\,J\,t + 27\,t^2 - 12\,t\,U + 6\,U^2 - 18\,J\,W + 12\,t\,W + 24\,U\,W\right . \nonumber \\
&& \hspace{1cm} 
 + 
 \left . 51\,W^2 + 6\,{\sqrt{\Aindex{2}}}\,J\,\cos (\Thetaindex{1}) + 12\,{\sqrt{\Aindex{2}}}\,t\,\cos (\Thetaindex{1})
\right .  \nonumber \\
&& \hspace{1cm} 
 \left . -6\,{\sqrt{\Aindex{2}}}\,U\,\cos (\Thetaindex{1}) - 30\,{\sqrt{\Aindex{2}}}\,W\,\cos (\Thetaindex{1})
\right )  \nonumber \\
C_{30-3} &=&
\frac{1}{18}
\,\left ( {\Aindex{6}}^2 - 3\,J^2 + 6\,J\,t + 6\,t\,U - 6\,U^2 + 18\,J\,W - 6\,t\,W - 24\,U\,W\right . \nonumber \\
&& \hspace{1cm} 
 \left . -51\,W^2 - 6\,{\sqrt{\Aindex{2}}}\,J\,\cos (\Thetaindex{1}) - 6\,{\sqrt{\Aindex{2}}}\,t\,\cos (\Thetaindex{1})
\right .  \nonumber \\
&& \hspace{1cm} 
 + 
 \left . 6\,{\sqrt{\Aindex{2}}}\,U\,\cos (\Thetaindex{1}) + 30\,{\sqrt{\Aindex{2}}}\,W\,\cos (\Thetaindex{1})
\right )  \nonumber \\
C_{30-4} &=&
\frac{-t}{2}
\, \left ( -J + 3\,t - U + W - 2\,{\sqrt{\Aindex{2}}}\,\cos (\Thetaindex{1})  \right ) \nonumber \\
 N_{30} &=& {\sqrt{2}}\,{\sqrt{{\Cindex{30}{1}}^2 + {\Cindex{30}{2}}^2 + {\Cindex{30}{3}}^2 + {\Cindex{30}{4}}^2}} \nonumber \eeq 
\beq
\ket{\Psi}_{31}& = &\ket{3,- {1 \over 2} , {3 \over 4} ,\Gamma_{3,2}} \nonumber \\ 
& = &
\Cindex{31}{1} \left ( 
\ket{02d} + \ket{0d2}\right) 
 \nonumber \\
& + &
\Cindex{31}{2} \left ( 
\ket{20d} + \ket{2d0}\right) 
 \nonumber \\
& + &
\Cindex{31}{3} \left ( 
\ket{d02} + \ket{d20}\right) 
 \nonumber \\
& + &
\Cindex{31}{4} \left ( 
\ket{ddu} - \ket{dud}\right) 
\nonumber \eeq
\beq
C_{31-1} &=&
\frac{t}{6}
\, \left ( 2\,J - 9\,t + 2\,U - 2\,W + {\sqrt{\Aindex{2}}}\,\cos (\Thetaindex{1}) + {\sqrt{3}}\,{\sqrt{\Aindex{2}}}\,\sin (\Thetaindex{1})  \right ) \nonumber \\
C_{31-2} &=&
\frac{1}{36}
\,\left ( -2\,{\Aindex{8}}^2 + 6\,J^2 - 9\,\Aindex{8}\,t - 33\,J\,t
 + 
  54\,t^2 + 6\,\Aindex{8}\,U + 6\,J\,U - 6\,t\,U 
\right .  \nonumber \\
&& \hspace{1cm} + \left . 30\,\Aindex{8}\,W - 6\,J\,W + 87\,t\,W - 54\,U\,W
\right .  \nonumber \\
&& \hspace{1cm} 
 \left . -108\,W^2 - 6\,{\sqrt{\Aindex{2}}}\,J\,\cos (\Thetaindex{1}) - 3\,{\sqrt{\Aindex{2}}}\,t\,\cos (\Thetaindex{1})\right .\nonumber \\
&& \hspace{1cm} \left . - 6\,{\sqrt{3}}\,{\sqrt{\Aindex{2}}}\,J\,\sin (\Thetaindex{1}) - 3\,{\sqrt{3}}\,{\sqrt{\Aindex{2}}}\,t\,\sin (\Thetaindex{1})
\right )  \nonumber \\
C_{31-3} &=&
\frac{1}{36}
\,\left ( 2\,{\Aindex{8}}^2 - 6\,J^2 + 9\,\Aindex{8}\,t + 21\,J\,t - 6\,\Aindex{8}\,U - 6\,J\,U - 6\,t\,U \right . \nonumber \\
&& \hspace{1cm}  \left . - 30\,\Aindex{8}\,W + 6\,J\,W - 75\,t\,W + 54\,U\,W
+ 108\,W^2 + 6\,{\sqrt{\Aindex{2}}}\,J\,\cos (\Thetaindex{1}) 
\right .  \nonumber \\
&& \hspace{1cm} 
\left . 
- 3\,{\sqrt{\Aindex{2}}}\,t\,\cos (\Thetaindex{1})
+ 6\,{\sqrt{3}}\,{\sqrt{\Aindex{2}}}\,J\,\sin (\Thetaindex{1}) - 3\,{\sqrt{3}}\,{\sqrt{\Aindex{2}}}\,t\,\sin (\Thetaindex{1})
\right )  \nonumber \\
C_{31-4} &=&
\frac{-t}{2}
\, \left ( -J + 3\,t - U + W + {\sqrt{\Aindex{2}}}\,\cos (\Thetaindex{1}) + {\sqrt{3}}\,{\sqrt{\Aindex{2}}}\,\sin (\Thetaindex{1})  \right ) \nonumber \\
 N_{31} &=& {\sqrt{2}}\,{\sqrt{{\Cindex{31}{1}}^2 + {\Cindex{31}{2}}^2 + {\Cindex{31}{3}}^2 + {\Cindex{31}{4}}^2}} \nonumber \eeq 
\beq
\ket{\Psi}_{32}& = &\ket{3,- {1 \over 2} , {3 \over 4} ,\Gamma_{3,2}} \nonumber \\ 
& = &
\Cindex{32}{1} \left ( 
\ket{02d} + \ket{0d2}\right) 
 \nonumber \\
& + &
\Cindex{32}{2} \left ( 
\ket{20d} + \ket{2d0}\right) 
 \nonumber \\
& + &
\Cindex{32}{3} \left ( 
\ket{d02} + \ket{d20}\right) 
 \nonumber \\
& + &
\Cindex{32}{4} \left ( 
\ket{ddu} - \ket{dud}\right) 
\nonumber \eeq
\beq
C_{32-1} &=&
\frac{t}{6}
\, \left ( 2\,J - 9\,t + 2\,U - 2\,W + {\sqrt{\Aindex{2}}}\,\cos (\Thetaindex{1}) - {\sqrt{3}}\,{\sqrt{\Aindex{2}}}\,\sin (\Thetaindex{1})  \right ) \nonumber \\
C_{32-2} &=&
\frac{1}{36}
\,\left ( -2\,{\Aindex{7}}^2 + 6\,J^2 - 9\,\Aindex{7}\,t - 33\,J\,t 
 + 
 54\,t^2 + 6\,\Aindex{7}\,U + 6\,J\,U \right .\nonumber \\
&& \hspace{1cm} \left . - 6\,t\,U + 30\,\Aindex{7}\,W - 6\,J\,W + 87\,t\,W - 54\,U\,W
 -108\,W^2 - 6\,{\sqrt{\Aindex{2}}}\,J\,\cos (\Thetaindex{1}) 
\right .  \nonumber \\
&& \hspace{1cm} 
 \left .
- 3\,{\sqrt{\Aindex{2}}}\,t\,\cos (\Thetaindex{1}) + 6\,{\sqrt{3}}\,{\sqrt{\Aindex{2}}}\,J\,\sin (\Thetaindex{1}) + 3\,{\sqrt{3}}\,{\sqrt{\Aindex{2}}}\,t\,\sin (\Thetaindex{1})
\right )  \nonumber \\
C_{32-3} &=&
\frac{1}{36}
\,\left ( 2\,{\Aindex{7}}^2 - 6\,J^2 + 9\,\Aindex{7}\,t + 21\,J\,t - 6\,\Aindex{7}\,U - 6\,J\,U - 6\,t\,U - 30\,\Aindex{7}\,W 
\right . 
\nonumber \\
&& \hspace{1cm} 
\left .
+ 6\,J\,W - 75\,t\,W + 54\,U\,W 108\,W^2 + 6\,{\sqrt{\Aindex{2}}}\,J\,\cos (\Thetaindex{1})
\right . 
\nonumber \\
&& \hspace{1cm} 
\left .
- 3\,{\sqrt{\Aindex{2}}}\,t\,\cos (\Thetaindex{1})
-6\,{\sqrt{3}}\,{\sqrt{\Aindex{2}}}\,J\,\sin (\Thetaindex{1}) + 3\,{\sqrt{3}}\,{\sqrt{\Aindex{2}}}\,t\,\sin (\Thetaindex{1})
\right )  \nonumber \\
C_{32-4} &=&
\frac{-t}{2}
\, \left ( -J + 3\,t - U + W + {\sqrt{\Aindex{2}}}\,\cos (\Thetaindex{1}) - {\sqrt{3}}\,{\sqrt{\Aindex{2}}}\,\sin (\Thetaindex{1})  \right ) \nonumber \\
 N_{32} &=& {\sqrt{2}}\,{\sqrt{{\Cindex{32}{1}}^2 + {\Cindex{32}{2}}^2 + {\Cindex{32}{3}}^2 + {\Cindex{32}{4}}^2}} \nonumber \eeq 
{\subsection*{\boldmath Unnormalized eigenvectors for ${\rm  N_e}=3$ and   ${\rm m_s}$= $ {1 \over 2} $.}
\beq
\ket{\Psi}_{33}& = &\ket{3, {1 \over 2} , {3 \over 4} ,\Gamma_1} \nonumber \\ 
&=& \frac{1}{{\sqrt{6}}}
 \left ( \ket{02u} + \ket{0u2} + \ket{20u} + \ket{2u0} + \ket{u02} + \ket{u20}\right) \nonumber 
\eeq
\beq
\ket{\Psi}_{34}& = &\ket{3, {1 \over 2} , {3 \over 4} ,\Gamma_2} \nonumber \\ 
&=& \frac{1}{{\sqrt{6}}}
 \left ( \ket{02u} - \ket{0u2} - \ket{20u} + \ket{2u0} + \ket{u02} - \ket{u20}\right) \nonumber 
\eeq
\beq
\ket{\Psi}_{35}& = &\ket{3, {1 \over 2} , {15 \over 4} ,\Gamma_2} \nonumber \\ 
&=& \frac{1}{{\sqrt{3}}}
 \left ( \ket{duu} + \ket{udu} + \ket{uud}\right) \nonumber 
\eeq
\beq
\ket{\Psi}_{36}& = &\ket{3, {1 \over 2} , {3 \over 4} ,\Gamma_{3,1}} \nonumber \\ 
& = &
\Cindex{36}{1} \left ( 
\ket{02u} - \ket{0u2}\right) 
 \nonumber \\
& + &
\Cindex{36}{2} \left ( 
\ket{20u} - \ket{2u0}\right) 
 \nonumber \\
& + &
\Cindex{36}{3} \left ( 
\ket{duu}\right) 
 \nonumber \\
& + &
\Cindex{36}{4} \left ( 
\ket{u02} - \ket{u20}\right) 
 \nonumber \\
& + &
\Cindex{36}{5} \left ( 
\ket{udu} + \ket{uud}\right) 
\nonumber \eeq
\beq
C_{36-1} &=&
\frac{-t}{6}
\, \left ( -2\,J + 9\,t - 2\,U + 2\,W + 2\,{\sqrt{\Aindex{2}}}\,\cos (\Thetaindex{1})  \right ) \nonumber \\
C_{36-2} &=&
\frac{1}{36}
\,\left ( -2\,{\Aindex{6}}^2 + 6\,\Aindex{6}\,J - 9\,\Aindex{6}\,t + 33\,J\,t\right . \nonumber \\
&& \hspace{1cm} 
 + 
 \left . 54\,t^2 - 6\,\Aindex{6}\,U + 18\,J\,U + 6\,t\,U - 30\,\Aindex{6}\,W + 36\,J\,W - 87\,t\,W - 54\,U\,W
\right .  \nonumber \\
&& \hspace{1cm} 
 \left . -108\,W^2 - 6\,{\sqrt{\Aindex{2}}}\,t\,\cos (\Thetaindex{1})
\right )  \nonumber \\
C_{36-3} &=&
\frac{-t}{3}
\, \left ( J + 9\,t + U - W + 2\,{\sqrt{\Aindex{2}}}\,\cos (\Thetaindex{1})  \right ) \nonumber \\
C_{36-4} &=&
\frac{1}{36}
\,\left ( -2\,{\Aindex{6}}^2 + 6\,\Aindex{6}\,J - 9\,\Aindex{6}\,t + 21\,J\,t\right . \nonumber \\
&& \hspace{1cm} 
 + 
 \left . 108\,t^2 - 6\,\Aindex{6}\,U + 18\,J\,U - 6\,t\,U - 30\,\Aindex{6}\,W + 36\,J\,W - 75\,t\,W - 54\,U\,W
\right .  \nonumber \\
&& \hspace{1cm} 
 \left . -108\,W^2 + 6\,{\sqrt{\Aindex{2}}}\,t\,\cos (\Thetaindex{1})
\right )  \nonumber \\
C_{36-5} &=&
\frac{t}{6}
\, \left ( J + 9\,t + U - W + 2\,{\sqrt{\Aindex{2}}}\,\cos (\Thetaindex{1})  \right ) \nonumber \\
 N_{36} &=& {\sqrt{2\,{\Cindex{36}{1}}^2 + 2\,{\Cindex{36}{2}}^2 + {\Cindex{36}{3}}^2 + 2\,{\Cindex{36}{4}}^2 + 2\,{\Cindex{36}{5}}^2}} \nonumber \eeq 
\beq
\ket{\Psi}_{37}& = &\ket{3, {1 \over 2} , {3 \over 4} ,\Gamma_{3,1}} \nonumber \\ 
& = &
\Cindex{37}{1} \left ( 
\ket{02u} - \ket{0u2}\right) 
 \nonumber \\
& + &
\Cindex{37}{2} \left ( 
\ket{20u} - \ket{2u0}\right) 
 \nonumber \\
& + &
\Cindex{37}{3} \left ( 
\ket{duu}\right) 
 \nonumber \\
& + &
\Cindex{37}{4} \left ( 
\ket{u02} - \ket{u20}\right) 
 \nonumber \\
& + &
\Cindex{37}{5} \left ( 
\ket{udu} + \ket{uud}\right) 
\nonumber \eeq
\beq
C_{37-1} &=&
\frac{t}{6}
\, \left ( 2\,J - 9\,t + 2\,U - 2\,W + {\sqrt{\Aindex{2}}}\,\cos (\Thetaindex{1}) + {\sqrt{3}}\,{\sqrt{\Aindex{2}}}\,\sin (\Thetaindex{1})  \right ) \nonumber \\
C_{37-2} &=&
\frac{1}{36}
\,\left ( -2\,{\Aindex{8}}^2 - 6\,\Aindex{8}\,J + 9\,\Aindex{8}\,t + 33\,J\,t\right . \nonumber \\
&& \hspace{1cm} 
 + 
 \left . 54\,t^2 + 6\,\Aindex{8}\,U + 18\,J\,U + 6\,t\,U + 30\,\Aindex{8}\,W + 36\,J\,W - 87\,t\,W - 54\,U\,W
\right .  \nonumber \\
&& \hspace{1cm} 
 \left . -108\,W^2 + 3\,{\sqrt{\Aindex{2}}}\,t\,\cos (\Thetaindex{1}) + 3\,{\sqrt{3}}\,{\sqrt{\Aindex{2}}}\,t\,\sin (\Thetaindex{1})
\right )  \nonumber \\
C_{37-3} &=&
\frac{t}{3}
\, \left ( -J - 9\,t - U + W + {\sqrt{\Aindex{2}}}\,\cos (\Thetaindex{1}) + {\sqrt{3}}\,{\sqrt{\Aindex{2}}}\,\sin (\Thetaindex{1})  \right ) \nonumber \\
C_{37-4} &=&
\frac{1}{36}
\,\left ( -2\,{\Aindex{8}}^2 - 6\,\Aindex{8}\,J + 9\,\Aindex{8}\,t + 21\,J\,t\right . \nonumber \\
&& \hspace{1cm} 
 + 
 \left . 108\,t^2 + 6\,\Aindex{8}\,U + 18\,J\,U - 6\,t\,U + 30\,\Aindex{8}\,W + 36\,J\,W - 75\,t\,W - 54\,U\,W
\right .  \nonumber \\
&& \hspace{1cm} 
 \left . -108\,W^2 - 3\,{\sqrt{\Aindex{2}}}\,t\,\cos (\Thetaindex{1}) - 3\,{\sqrt{3}}\,{\sqrt{\Aindex{2}}}\,t\,\sin (\Thetaindex{1})
\right )  \nonumber \\
C_{37-5} &=&
\frac{-t}{6}
\, \left ( -J - 9\,t - U + W + {\sqrt{\Aindex{2}}}\,\cos (\Thetaindex{1}) + {\sqrt{3}}\,{\sqrt{\Aindex{2}}}\,\sin (\Thetaindex{1})  \right ) \nonumber \\
 N_{37} &=& {\sqrt{2\,{\Cindex{37}{1}}^2 + 2\,{\Cindex{37}{2}}^2 + {\Cindex{37}{3}}^2 + 2\,{\Cindex{37}{4}}^2 + 2\,{\Cindex{37}{5}}^2}} \nonumber \eeq 
\beq
\ket{\Psi}_{38}& = &\ket{3, {1 \over 2} , {3 \over 4} ,\Gamma_{3,1}} \nonumber \\ 
& = &
\Cindex{38}{1} \left ( 
\ket{02u} - \ket{0u2}\right) 
 \nonumber \\
& + &
\Cindex{38}{2} \left ( 
\ket{20u} - \ket{2u0}\right) 
 \nonumber \\
& + &
\Cindex{38}{3} \left ( 
\ket{duu}\right) 
 \nonumber \\
& + &
\Cindex{38}{4} \left ( 
\ket{u02} - \ket{u20}\right) 
 \nonumber \\
& + &
\Cindex{38}{5} \left ( 
\ket{udu} + \ket{uud}\right) 
\nonumber \eeq
\beq
C_{38-1} &=&
\frac{t}{6}
\, \left ( 2\,J - 9\,t + 2\,U - 2\,W + {\sqrt{\Aindex{2}}}\,\cos (\Thetaindex{1}) - {\sqrt{3}}\,{\sqrt{\Aindex{2}}}\,\sin (\Thetaindex{1})  \right ) \nonumber \\
C_{38-2} &=&
\frac{1}{36}
\,\left ( -2\,{\Aindex{7}}^2 - 6\,\Aindex{7}\,J + 9\,\Aindex{7}\,t + 33\,J\,t\right . \nonumber \\
&& \hspace{1cm} 
 + 
 \left . 54\,t^2 + 6\,\Aindex{7}\,U + 18\,J\,U + 6\,t\,U + 30\,\Aindex{7}\,W + 36\,J\,W - 87\,t\,W - 54\,U\,W
\right .  \nonumber \\
&& \hspace{1cm} 
 \left . -108\,W^2 + 3\,{\sqrt{\Aindex{2}}}\,t\,\cos (\Thetaindex{1}) - 3\,{\sqrt{3}}\,{\sqrt{\Aindex{2}}}\,t\,\sin (\Thetaindex{1})
\right )  \nonumber \\
C_{38-3} &=&
\frac{t}{3}
\, \left ( -J - 9\,t - U + W + {\sqrt{\Aindex{2}}}\,\cos (\Thetaindex{1}) - {\sqrt{3}}\,{\sqrt{\Aindex{2}}}\,\sin (\Thetaindex{1})  \right ) \nonumber \\
C_{38-4} &=&
\frac{1}{36}
\,\left ( -2\,{\Aindex{7}}^2 - 6\,\Aindex{7}\,J + 9\,\Aindex{7}\,t + 21\,J\,t\right . \nonumber \\
&& \hspace{1cm} 
 + 
 \left . 108\,t^2 + 6\,\Aindex{7}\,U + 18\,J\,U - 6\,t\,U + 30\,\Aindex{7}\,W + 36\,J\,W - 75\,t\,W - 54\,U\,W
\right .  \nonumber \\
&& \hspace{1cm} 
 \left . -108\,W^2 - 3\,{\sqrt{\Aindex{2}}}\,t\,\cos (\Thetaindex{1}) + 3\,{\sqrt{3}}\,{\sqrt{\Aindex{2}}}\,t\,\sin (\Thetaindex{1})
\right )  \nonumber \\
C_{38-5} &=&
\frac{-t}{6}
\, \left ( -J - 9\,t - U + W + {\sqrt{\Aindex{2}}}\,\cos (\Thetaindex{1}) - {\sqrt{3}}\,{\sqrt{\Aindex{2}}}\,\sin (\Thetaindex{1})  \right ) \nonumber \\
 N_{38} &=& {\sqrt{2\,{\Cindex{38}{1}}^2 + 2\,{\Cindex{38}{2}}^2 + {\Cindex{38}{3}}^2 + 2\,{\Cindex{38}{4}}^2 + 2\,{\Cindex{38}{5}}^2}} \nonumber \eeq 
\beq
\ket{\Psi}_{39}& = &\ket{3, {1 \over 2} , {3 \over 4} ,\Gamma_{3,2}} \nonumber \\ 
& = &
\Cindex{39}{1} \left ( 
\ket{02u} + \ket{0u2}\right) 
 \nonumber \\
& + &
\Cindex{39}{2} \left ( 
\ket{20u} + \ket{2u0}\right) 
 \nonumber \\
& + &
\Cindex{39}{3} \left ( 
\ket{u02} + \ket{u20}\right) 
 \nonumber \\
& + &
\Cindex{39}{4} \left ( 
\ket{udu} - \ket{uud}\right) 
\nonumber \eeq
\beq
C_{39-1} &=&
\frac{-t}{6}
\, \left ( -2\,J + 9\,t - 2\,U + 2\,W + 2\,{\sqrt{\Aindex{2}}}\,\cos (\Thetaindex{1})  \right ) \nonumber \\
C_{39-2} &=&
\frac{1}{18}
\,\left ( -{\Aindex{6}}^2 + 3\,J^2 - 12\,J\,t + 27\,t^2 - 12\,t\,U + 6\,U^2 - 18\,J\,W + 12\,t\,W + 24\,U\,W\right . \nonumber \\
&& \hspace{1cm} 
 + 
 \left . 51\,W^2 + 6\,{\sqrt{\Aindex{2}}}\,J\,\cos (\Thetaindex{1}) + 12\,{\sqrt{\Aindex{2}}}\,t\,\cos (\Thetaindex{1})
\right .  \nonumber \\
&& \hspace{1cm} 
 \left . -6\,{\sqrt{\Aindex{2}}}\,U\,\cos (\Thetaindex{1}) - 30\,{\sqrt{\Aindex{2}}}\,W\,\cos (\Thetaindex{1})
\right )  \nonumber \\
C_{39-3} &=&
\frac{1}{18}
\,\left ( {\Aindex{6}}^2 - 3\,J^2 + 6\,J\,t + 6\,t\,U - 6\,U^2 + 18\,J\,W - 6\,t\,W - 24\,U\,W\right . \nonumber \\
&& \hspace{1cm} 
 \left . -51\,W^2 - 6\,{\sqrt{\Aindex{2}}}\,J\,\cos (\Thetaindex{1}) - 6\,{\sqrt{\Aindex{2}}}\,t\,\cos (\Thetaindex{1})
\right .  \nonumber \\
&& \hspace{1cm} 
 + 
 \left . 6\,{\sqrt{\Aindex{2}}}\,U\,\cos (\Thetaindex{1}) + 30\,{\sqrt{\Aindex{2}}}\,W\,\cos (\Thetaindex{1})
\right )  \nonumber \\
C_{39-4} &=&
\frac{-t}{2}
\, \left ( -J + 3\,t - U + W - 2\,{\sqrt{\Aindex{2}}}\,\cos (\Thetaindex{1})  \right ) \nonumber \\
 N_{39} &=& {\sqrt{2}}\,{\sqrt{{\Cindex{39}{1}}^2 + {\Cindex{39}{2}}^2 + {\Cindex{39}{3}}^2 + {\Cindex{39}{4}}^2}} \nonumber \eeq 
\beq
\ket{\Psi}_{40}& = &\ket{3, {1 \over 2} , {3 \over 4} ,\Gamma_{3,2}} \nonumber \\ 
& = &
\Cindex{40}{1} \left ( 
\ket{02u} + \ket{0u2}\right) 
 \nonumber \\
& + &
\Cindex{40}{2} \left ( 
\ket{20u} + \ket{2u0}\right) 
 \nonumber \\
& + &
\Cindex{40}{3} \left ( 
\ket{u02} + \ket{u20}\right) 
 \nonumber \\
& + &
\Cindex{40}{4} \left ( 
\ket{udu} - \ket{uud}\right) 
\nonumber \eeq
\beq
C_{40-1} &=&
\frac{t}{6}
\, \left ( 2\,J - 9\,t + 2\,U - 2\,W + {\sqrt{\Aindex{2}}}\,\cos (\Thetaindex{1}) + {\sqrt{3}}\,{\sqrt{\Aindex{2}}}\,\sin (\Thetaindex{1})  \right ) \nonumber \\
C_{40-2} &=&
\frac{1}{36}
\,\left ( -2\,{\Aindex{8}}^2 + 6\,J^2 - 9\,\Aindex{8}\,t - 33\,J\,t + 54\,t^2 + 6\,\Aindex{8}\,U + 6\,J\,U 
\right . \nonumber \\
&& \hspace{1cm} 
\left .
- 6\,t\,U + 30\,\Aindex{8}\,W - 6\,J\,W + 87\,t\,W - 54\,U\,W
 -108\,W^2 - 6\,{\sqrt{\Aindex{2}}}\,J\,\cos (\Thetaindex{1}) 
\right .  
\nonumber \\
&& \hspace{1cm} 
\left .
- 3\,{\sqrt{\Aindex{2}}}\,t\,\cos (\Thetaindex{1}) - 6\,{\sqrt{3}}\,{\sqrt{\Aindex{2}}}\,J\,\sin (\Thetaindex{1}) - 3\,{\sqrt{3}}\,{\sqrt{\Aindex{2}}}\,t\,\sin (\Thetaindex{1})
\right )  \nonumber \\
C_{40-3} &=&
\frac{1}{36}
\,\left ( 2\,{\Aindex{8}}^2 - 6\,J^2 + 9\,\Aindex{8}\,t + 21\,J\,t - 6\,\Aindex{8}\,U - 6\,J\,U - 6\,t\,U - 30\,\Aindex{8}\,W 
\right . \nonumber \\
&& \hspace{1cm} 
 \left .
+ 6\,J\,W - 75\,t\,W + 54\,U\,W + 108\,W^2 + 6\,{\sqrt{\Aindex{2}}}\,J\,\cos (\Thetaindex{1}) 
\right .  \nonumber \\
&& \hspace{1cm} 
\left .
- 3\,{\sqrt{\Aindex{2}}}\,t\,\cos (\Thetaindex{1})
+ 6\,{\sqrt{3}}\,{\sqrt{\Aindex{2}}}\,J\,\sin (\Thetaindex{1})
- 3\,{\sqrt{3}}\,{\sqrt{\Aindex{2}}}\,t\,\sin (\Thetaindex{1})
\right )  \nonumber \\
C_{40-4} &=&
\frac{-t}{2}
\, \left ( -J + 3\,t - U + W + {\sqrt{\Aindex{2}}}\,\cos (\Thetaindex{1}) + {\sqrt{3}}\,{\sqrt{\Aindex{2}}}\,\sin (\Thetaindex{1})  \right ) \nonumber \\
 N_{40} &=& {\sqrt{2}}\,{\sqrt{{\Cindex{40}{1}}^2 + {\Cindex{40}{2}}^2 + {\Cindex{40}{3}}^2 + {\Cindex{40}{4}}^2}} \nonumber \eeq 
\beq
\ket{\Psi}_{41}& = &\ket{3, {1 \over 2} , {3 \over 4} ,\Gamma_{3,2}} \nonumber \\ 
& = &
\Cindex{41}{1} \left ( 
\ket{02u} + \ket{0u2}\right) 
 \nonumber \\
& + &
\Cindex{41}{2} \left ( 
\ket{20u} + \ket{2u0}\right) 
 \nonumber \\
& + &
\Cindex{41}{3} \left ( 
\ket{u02} + \ket{u20}\right) 
 \nonumber \\
& + &
\Cindex{41}{4} \left ( 
\ket{udu} - \ket{uud}\right) 
\nonumber \eeq
\beq
C_{41-1} &=&
\frac{t}{6}
\, \left ( 2\,J - 9\,t + 2\,U - 2\,W + {\sqrt{\Aindex{2}}}\,\cos (\Thetaindex{1}) - {\sqrt{3}}\,{\sqrt{\Aindex{2}}}\,\sin (\Thetaindex{1})  \right ) \nonumber \\
C_{41-2} &=&
\frac{1}{36}
\,\left ( -2\,{\Aindex{7}}^2 + 6\,J^2 - 9\,\Aindex{7}\,t - 33\,J\,t + 54\,t^2 + 6\,\Aindex{7}\,U + 6\,J\,U - 6\,t\,U + 30\,\Aindex{7}\,W 
\right . \nonumber \\
&& \hspace{1cm} 
\left .
- 6\,J\,W + 87\,t\,W - 54\,U\,W
 -108\,W^2 - 6\,{\sqrt{\Aindex{2}}}\,J\,\cos (\Thetaindex{1})
\right .  \nonumber \\
&& \hspace{1cm} 
 \left .
 - 3\,{\sqrt{\Aindex{2}}}\,t\,\cos (\Thetaindex{1}) +
 6\,{\sqrt{3}}\,{\sqrt{\Aindex{2}}}\,J\,\sin (\Thetaindex{1}) 
+ 3\,{\sqrt{3}}\,{\sqrt{\Aindex{2}}}\,t\,\sin (\Thetaindex{1})
\right )  \nonumber \\
C_{41-3} &=&
\frac{1}{36}
\,\left ( 2\,{\Aindex{7}}^2 - 6\,J^2 + 9\,\Aindex{7}\,t + 21\,J\,t - 6\,\Aindex{7}\,U - 6\,J\,U - 6\,t\,U - 30\,\Aindex{7}\,W + 6\,J\,W 
\right . \nonumber \\
&& \hspace{1cm} 
 \left .
- 75\,t\,W + 54\,U\,W +108\,W^2 + 6\,{\sqrt{\Aindex{2}}}\,J\,\cos (\Thetaindex{1}) 
\right .  \nonumber \\
&& \hspace{1cm} 
 \left .
- 3\,{\sqrt{\Aindex{2}}}\,t\,\cos (\Thetaindex{1})
 -6\,{\sqrt{3}}\,{\sqrt{\Aindex{2}}}\,J\,\sin (\Thetaindex{1}) 
+ 3\,{\sqrt{3}}\,{\sqrt{\Aindex{2}}}\,t\,\sin (\Thetaindex{1})
\right )  \nonumber \\
C_{41-4} &=&
\frac{-t}{2}
\, \left ( -J + 3\,t - U + W + {\sqrt{\Aindex{2}}}\,\cos (\Thetaindex{1}) - {\sqrt{3}}\,{\sqrt{\Aindex{2}}}\,\sin (\Thetaindex{1})  \right ) \nonumber \\
 N_{41} &=& {\sqrt{2}}\,{\sqrt{{\Cindex{41}{1}}^2 + {\Cindex{41}{2}}^2 + {\Cindex{41}{3}}^2 + {\Cindex{41}{4}}^2}} \nonumber \eeq 
{\subsection*{\boldmath Unnormalized eigenvectors for ${\rm  N_e}=3$ and   ${\rm m_s}$= $ {3 \over 2} $.}
\beq
\ket{\Psi}_{42}& = &\ket{3, {3 \over 2} , {15 \over 4} ,\Gamma_2} \nonumber \\ 
&=& 1
 \left ( \ket{uuu}\right) \nonumber 
\eeq
{\subsection*{\boldmath Unnormalized eigenvectors for ${\rm  N_e}=4$ and   ${\rm m_s}$= $-1$.}
\beq
\ket{\Psi}_{43}& = &\ket{4,-1,2,\Gamma_2} \nonumber \\ 
&=& \frac{1}{{\sqrt{3}}}
 \left ( \ket{2dd} - \ket{d2d} + \ket{dd2}\right) \nonumber 
\eeq
\beq
\ket{\Psi}_{44}& = &\ket{4,-1,2,\Gamma_{3,1}} \nonumber \\ 
& = &
\Cindex{44}{1} \left ( 
\ket{2dd}\right) 
 \nonumber \\
& + &
\Cindex{44}{2} \left ( 
\ket{d2d} - \ket{dd2}\right) 
\nonumber \eeq
\beq
C_{44-1} &=&
{\sqrt{\frac{2}{3}}} \nonumber \\
C_{44-2} &=&
\frac{1}{{\sqrt{6}}} \nonumber \\
 N_{44} &=& {\sqrt{{\Cindex{44}{1}}^2 + 2\,{\Cindex{44}{2}}^2}} \nonumber \eeq 
\beq
\ket{\Psi}_{45}& = &\ket{4,-1,2,\Gamma_{3,2}} \nonumber \\ 
&=& \frac{1}{{\sqrt{2}}}
 \left ( \ket{d2d} + \ket{dd2}\right) \nonumber 
\eeq
{\subsection*{\boldmath Unnormalized eigenvectors for ${\rm  N_e}=4$ and   ${\rm m_s}$= $0$.}
\beq
\ket{\Psi}_{46}& = &\ket{4,0,0,\Gamma_1} \nonumber \\ 
& = &
\Cindex{46}{1} \left ( 
\ket{022} + \ket{202} + \ket{220}\right) 
 \nonumber \\
& + &
\Cindex{46}{2} \left ( 
\ket{2du} - \ket{2ud} + \ket{d2u} + \ket{du2} - \ket{u2d} - \ket{ud2}\right) 
\nonumber \eeq
\beq
C_{46-1} &=&
2\,{\sqrt{\frac{2}{3}}}\,t \nonumber \\
C_{46-2} &=&
\frac{1}{2\,{\sqrt{6}}}
\, \left ( {\sqrt{\Aindex{3}}} + J + 2\,t + U - W  \right ) \nonumber \\
 N_{46} &=& {\sqrt{3\,{\Cindex{46}{1}}^2 + 6\,{\Cindex{46}{2}}^2}} \nonumber \eeq 
\beq
\ket{\Psi}_{47}& = &\ket{4,0,0,\Gamma_1} \nonumber \\ 
& = &
\Cindex{47}{1} \left ( 
\ket{022} + \ket{202} + \ket{220}\right) 
 \nonumber \\
& + &
\Cindex{47}{2} \left ( 
\ket{2du} - \ket{2ud} + \ket{d2u} + \ket{du2} - \ket{u2d} - \ket{ud2}\right) 
\nonumber \eeq
\beq
C_{47-1} &=&
2\,{\sqrt{\frac{2}{3}}}\,t \nonumber \\
C_{47-2} &=&
\frac{-1}{2\,{\sqrt{6}}}
\, \left ( {\sqrt{\Aindex{3}}} - J - 2\,t - U + W  \right ) \nonumber \\
 N_{47} &=& {\sqrt{3\,{\Cindex{47}{1}}^2 + 6\,{\Cindex{47}{2}}^2}} \nonumber \eeq 
\beq
\ket{\Psi}_{48}& = &\ket{4,0,2,\Gamma_2} \nonumber \\ 
&=& \frac{1}{{\sqrt{6}}}
 \left ( \ket{2du} + \ket{2ud} - \ket{d2u} + \ket{du2} - \ket{u2d} + \ket{ud2}\right) \nonumber 
\eeq
\beq
\ket{\Psi}_{49}& = &\ket{4,0,0,\Gamma_{3,1}} \nonumber \\ 
& = &
\Cindex{49}{1} \left ( 
\ket{202} - \ket{220}\right) 
 \nonumber \\
& + &
\Cindex{49}{2} \left ( 
\ket{d2u} - \ket{du2} - \ket{u2d} + \ket{ud2}\right) 
\nonumber \eeq
\beq
C_{49-1} &=&
\frac{1}{2\,{\sqrt{2}}}
\, \left ( {\sqrt{\Aindex{4}}} - J + t - U + W  \right ) \nonumber \\
C_{49-2} &=&
-\left( \frac{t}{{\sqrt{2}}} \right)  \nonumber \\
 N_{49} &=& {\sqrt{2\,{\Cindex{49}{1}}^2 + 4\,{\Cindex{49}{2}}^2}} \nonumber \eeq 
\beq
\ket{\Psi}_{50}& = &\ket{4,0,0,\Gamma_{3,1}} \nonumber \\ 
& = &
\Cindex{50}{1} \left ( 
\ket{202} - \ket{220}\right) 
 \nonumber \\
& + &
\Cindex{50}{2} \left ( 
\ket{d2u} - \ket{du2} - \ket{u2d} + \ket{ud2}\right) 
\nonumber \eeq
\beq
C_{50-1} &=&
\frac{-1}{2\,{\sqrt{2}}}
\, \left ( {\sqrt{\Aindex{4}}} + J - t + U - W  \right ) \nonumber \\
C_{50-2} &=&
-\left( \frac{t}{{\sqrt{2}}} \right)  \nonumber \\
 N_{50} &=& {\sqrt{2\,{\Cindex{50}{1}}^2 + 4\,{\Cindex{50}{2}}^2}} \nonumber \eeq 
\beq
\ket{\Psi}_{51}& = &\ket{4,0,2,\Gamma_{3,1}} \nonumber \\ 
& = &
\Cindex{51}{1} \left ( 
\ket{2du} + \ket{2ud}\right) 
 \nonumber \\
& + &
\Cindex{51}{2} \left ( 
\ket{d2u} - \ket{du2} + \ket{u2d} - \ket{ud2}\right) 
\nonumber \eeq
\beq
C_{51-1} &=&
\frac{1}{{\sqrt{3}}} \nonumber \\
C_{51-2} &=&
\frac{1}{2\,{\sqrt{3}}} \nonumber \\
 N_{51} &=& {\sqrt{2\,{\Cindex{51}{1}}^2 + 4\,{\Cindex{51}{2}}^2}} \nonumber \eeq 
\beq
\ket{\Psi}_{52}& = &\ket{4,0,0,\Gamma_{3,2}} \nonumber \\ 
& = &
\Cindex{52}{1} \left ( 
\ket{022}\right) 
 \nonumber \\
& + &
\Cindex{52}{2} \left ( 
\ket{202} + \ket{220}\right) 
 \nonumber \\
& + &
\Cindex{52}{3} \left ( 
\ket{2du} - \ket{2ud}\right) 
 \nonumber \\
& + &
\Cindex{52}{4} \left ( 
\ket{d2u} + \ket{du2} - \ket{u2d} - \ket{ud2}\right) 
\nonumber \eeq
\beq
C_{52-1} &=&
\frac{2\,t}{{\sqrt{3}}} \nonumber \\
C_{52-2} &=&
-\left( \frac{t}{{\sqrt{3}}} \right)  \nonumber \\
C_{52-3} &=&
\frac{-1}{2\,{\sqrt{3}}}
\, \left ( {\sqrt{\Aindex{4}}} + J - t + U - W  \right ) \nonumber \\
C_{52-4} &=&
\frac{1}{4\,{\sqrt{3}}}
\, \left ( {\sqrt{\Aindex{4}}} + J - t + U - W  \right ) \nonumber \\
 N_{52} &=& {\sqrt{{\Cindex{52}{1}}^2 + 2\,\left( {\Cindex{52}{2}}^2 + {\Cindex{52}{3}}^2 + 2\,{\Cindex{52}{4}}^2 \right) }} \nonumber \eeq 
\beq
\ket{\Psi}_{53}& = &\ket{4,0,0,\Gamma_{3,2}} \nonumber \\ 
& = &
\Cindex{53}{1} \left ( 
\ket{022}\right) 
 \nonumber \\
& + &
\Cindex{53}{2} \left ( 
\ket{202} + \ket{220}\right) 
 \nonumber \\
& + &
\Cindex{53}{3} \left ( 
\ket{2du} - \ket{2ud}\right) 
 \nonumber \\
& + &
\Cindex{53}{4} \left ( 
\ket{d2u} + \ket{du2} - \ket{u2d} - \ket{ud2}\right) 
\nonumber \eeq
\beq
C_{53-1} &=&
\frac{2\,t}{{\sqrt{3}}} \nonumber \\
C_{53-2} &=&
-\left( \frac{t}{{\sqrt{3}}} \right)  \nonumber \\
C_{53-3} &=&
\frac{1}{2\,{\sqrt{3}}}
\, \left ( {\sqrt{\Aindex{4}}} - J + t - U + W  \right ) \nonumber \\
C_{53-4} &=&
\frac{-1}{4\,{\sqrt{3}}}
\, \left ( {\sqrt{\Aindex{4}}} - J + t - U + W  \right ) \nonumber \\
 N_{53} &=& {\sqrt{{\Cindex{53}{1}}^2 + 2\,\left( {\Cindex{53}{2}}^2 + {\Cindex{53}{3}}^2 + 2\,{\Cindex{53}{4}}^2 \right) }} \nonumber \eeq 
\beq
\ket{\Psi}_{54}& = &\ket{4,0,2,\Gamma_{3,2}} \nonumber \\ 
&=& \frac{1}{2}
 \left ( \ket{d2u} + \ket{du2} + \ket{u2d} + \ket{ud2}\right) \nonumber 
\eeq
{\subsection*{\boldmath Unnormalized eigenvectors for ${\rm  N_e}=4$ and   ${\rm m_s}$= $1$.}
\beq
\ket{\Psi}_{55}& = &\ket{4,1,2,\Gamma_2} \nonumber \\ 
&=& \frac{1}{{\sqrt{3}}}
 \left ( \ket{2uu} - \ket{u2u} + \ket{uu2}\right) \nonumber 
\eeq
\beq
\ket{\Psi}_{56}& = &\ket{4,1,2,\Gamma_{3,1}} \nonumber \\ 
& = &
\Cindex{56}{1} \left ( 
\ket{2uu}\right) 
 \nonumber \\
& + &
\Cindex{56}{2} \left ( 
\ket{u2u} - \ket{uu2}\right) 
\nonumber \eeq
\beq
C_{56-1} &=&
{\sqrt{\frac{2}{3}}} \nonumber \\
C_{56-2} &=&
\frac{1}{{\sqrt{6}}} \nonumber \\
 N_{56} &=& {\sqrt{{\Cindex{56}{1}}^2 + 2\,{\Cindex{56}{2}}^2}} \nonumber \eeq 
\beq
\ket{\Psi}_{57}& = &\ket{4,1,2,\Gamma_{3,2}} \nonumber \\ 
&=& \frac{1}{{\sqrt{2}}}
 \left ( \ket{u2u} + \ket{uu2}\right) \nonumber 
\eeq
{\subsection*{\boldmath Unnormalized eigenvectors for ${\rm  N_e}=5$ and   ${\rm m_s}$= $- {1 \over 2} $.}
\beq
\ket{\Psi}_{58}& = &\ket{5,- {1 \over 2} , {3 \over 4} ,\Gamma_1} \nonumber \\ 
&=& \frac{1}{{\sqrt{3}}}
 \left ( \ket{22d} + \ket{2d2} + \ket{d22}\right) \nonumber 
\eeq
\beq
\ket{\Psi}_{59}& = &\ket{5,- {1 \over 2} , {3 \over 4} ,\Gamma_{3,1}} \nonumber \\ 
&=& \frac{1}{{\sqrt{2}}}
 \left ( \ket{22d} - \ket{2d2}\right) \nonumber 
\eeq
\beq
\ket{\Psi}_{60}& = &\ket{5,- {1 \over 2} , {3 \over 4} ,\Gamma_{3,2}} \nonumber \\ 
& = &
\Cindex{60}{1} \left ( 
\ket{22d} + \ket{2d2}\right) 
 \nonumber \\
& + &
\Cindex{60}{2} \left ( 
\ket{d22}\right) 
\nonumber \eeq
\beq
C_{60-1} &=&
\frac{1}{{\sqrt{6}}} \nonumber \\
C_{60-2} &=&
-{\sqrt{\frac{2}{3}}} \nonumber \\
 N_{60} &=& {\sqrt{2\,{\Cindex{60}{1}}^2 + {\Cindex{60}{2}}^2}} \nonumber \eeq 
{\subsection*{\boldmath Unnormalized eigenvectors for ${\rm  N_e}=5$ and   ${\rm m_s}$= $ {1 \over 2} $.}
\beq
\ket{\Psi}_{61}& = &\ket{5, {1 \over 2} , {3 \over 4} ,\Gamma_1} \nonumber \\ 
&=& \frac{1}{{\sqrt{3}}}
 \left ( \ket{22u} + \ket{2u2} + \ket{u22}\right) \nonumber 
\eeq
\beq
\ket{\Psi}_{62}& = &\ket{5, {1 \over 2} , {3 \over 4} ,\Gamma_{3,1}} \nonumber \\ 
&=& \frac{1}{{\sqrt{2}}}
 \left ( \ket{22u} - \ket{2u2}\right) \nonumber 
\eeq
\beq
\ket{\Psi}_{63}& = &\ket{5, {1 \over 2} , {3 \over 4} ,\Gamma_{3,2}} \nonumber \\ 
& = &
\Cindex{63}{1} \left ( 
\ket{22u} + \ket{2u2}\right) 
 \nonumber \\
& + &
\Cindex{63}{2} \left ( 
\ket{u22}\right) 
\nonumber \eeq
\beq
C_{63-1} &=&
\frac{1}{{\sqrt{6}}} \nonumber \\
C_{63-2} &=&
-{\sqrt{\frac{2}{3}}} \nonumber \\
 N_{63} &=& {\sqrt{2\,{\Cindex{63}{1}}^2 + {\Cindex{63}{2}}^2}} \nonumber \eeq 
{\subsection*{\boldmath Unnormalized eigenvectors for ${\rm  N_e}=6$ and   ${\rm m_s}$= $0$.}
\beq
\ket{\Psi}_{64}& = &\ket{6,0,0,\Gamma_1} \nonumber \\ 
&=& 1
 \left ( \ket{222}\right) \nonumber 
\eeq

%% file: Parts/appendix2a.tex
\section{Eigensystem of the tetrahedron}
\label{appendix2a}
In the following tables the first column gives the number of the state for referencing
in the text. The second column labels the eigenstate by the eigenvalues of the U-independent
symmetry operators, i.e. the electron occupation number  $N_e$, the spin-projection 
$\opS_z$ in z-direction $m_s$, the eigenvalues $s(s+1)$ of $\opS^2$ and the spatial
symmetry is indicated by $\Gamma_{i,j}$, where the first index labels the irreducible
representation of the tetrahedral group and the second numbers the partner.
The notation is based on Ref. \cite{CornwellBook}.
The third column gives the enery eigenvalues in abbreviated form. The abbreviations
are listed subsequent to the tables. In the last column a numerical value is given for example.
For comparison with we have chosen the same parameters as in Ref. \cite{Schumann02}
corresponding to a pure Hubbard-model with $U=5t$, $W=J=0t$. If the grand-canonical 
energy levels in an applied magnetic field are needed one has to substract $\mu N_e+h*m_s$.
\subsection{\bf The eigenvalues}
\parindent0cm
\begin{tabular}[t]{|r|l|c|c|}
\multicolumn{4}{c}{\large \bf \boldmath Eigenkets and eigenvalues for ${\rm  N_e}$=0 and   ${\rm m_s}$= $0$. } \\[0.5ex] \hline
\parbox[c]{1cm}{No}  & \parbox[c]{2.5cm}{\begin{center} Eigenstate \end{center}}  & \parbox[c]{7.5cm}{ \begin{center}   Energy \end{center}}  & \parbox[c]{2cm}{ \begin{center} Example \end{center}}  \\ \hline 
\hline 
1 & $\ket{0,0,0,\Gamma_1}$  & $0$   & 0. \\[0.5ex] 
\hline
\end{tabular} \\[2ex]
\begin{tabular}[t]{|r|l|c|c|}
\multicolumn{4}{c}{\large \bf \boldmath Eigenkets and eigenvalues for ${\rm  N_e}$=1 and   ${\rm m_s}$= $- {1 \over 2} $. } \\[0.5ex] \hline
\parbox[c]{1cm}{No}  & \parbox[c]{2.5cm}{\begin{center} Eigenstate \end{center}}  & \parbox[c]{7.5cm}{ \begin{center}   Energy \end{center}}  & \parbox[c]{2cm}{ \begin{center} Example \end{center}}  \\ \hline 
\hline 
2 & $\ket{1,- {1 \over 2} , {3 \over 4} ,\Gamma_1}$  & $3\,t$   & 3. \\[0.5ex] 
3 & $\ket{1,- {1 \over 2} , {3 \over 4} ,\Gamma_{4,1}}$  & $-t$   & -1. \\[0.5ex] 
4 & $\ket{1,- {1 \over 2} , {3 \over 4} ,\Gamma_{4,2}}$  & $-t$   & -1. \\[0.5ex] 
5 & $\ket{1,- {1 \over 2} , {3 \over 4} ,\Gamma_{4,3}}$  & $-t$   & -1. \\[0.5ex] 
\hline
\end{tabular} \\[2ex]
\begin{tabular}[t]{|r|l|c|c|}
\multicolumn{4}{c}{\large \bf \boldmath Eigenkets and eigenvalues for ${\rm  N_e}$=1 and   ${\rm m_s}$= $ {1 \over 2} $. } \\[0.5ex] \hline
\parbox[c]{1cm}{No}  & \parbox[c]{2.5cm}{\begin{center} Eigenstate \end{center}}  & \parbox[c]{7.5cm}{ \begin{center}   Energy \end{center}}  & \parbox[c]{2cm}{ \begin{center} Example \end{center}}  \\ \hline 
\hline 
6 & $\ket{1, {1 \over 2} , {3 \over 4} ,\Gamma_1}$  & $3\,t$   & 3. \\[0.5ex] 
7 & $\ket{1, {1 \over 2} , {3 \over 4} ,\Gamma_{4,1}}$  & $-t$   & -1. \\[0.5ex] 
8 & $\ket{1, {1 \over 2} , {3 \over 4} ,\Gamma_{4,2}}$  & $-t$   & -1. \\[0.5ex] 
9 & $\ket{1, {1 \over 2} , {3 \over 4} ,\Gamma_{4,3}}$  & $-t$   & -1. \\[0.5ex] 
\hline
\end{tabular} \\[2ex]
\begin{tabular}[t]{|r|l|c|c|}
\multicolumn{4}{c}{\large \bf \boldmath Eigenkets and eigenvalues for ${\rm  N_e}$=2 and   ${\rm m_s}$= $-1$. } \\[0.5ex] \hline
\parbox[c]{1cm}{No}  & \parbox[c]{2.5cm}{\begin{center} Eigenstate \end{center}}  & \parbox[c]{7.5cm}{ \begin{center}   Energy \end{center}}  & \parbox[c]{2cm}{ \begin{center} Example \end{center}}  \\ \hline 
\hline 
10 & $\ket{2,-1,2,\Gamma_{4,1}}$  & $\frac{J}{2} + 2\,t + 2\,W$   & 2. \\[0.5ex] 
11 & $\ket{2,-1,2,\Gamma_{4,2}}$  & $\frac{J}{2} + 2\,t + 2\,W$   & 2. \\[0.5ex] 
12 & $\ket{2,-1,2,\Gamma_{4,3}}$  & $\frac{J}{2} + 2\,t + 2\,W$   & 2. \\[0.5ex] 
13 & $\ket{2,-1,2,\Gamma_{5,1}}$  & $\frac{J}{2} - 2\,t + 2\,W$   & -2. \\[0.5ex] 
14 & $\ket{2,-1,2,\Gamma_{5,2}}$  & $\frac{J}{2} - 2\,t + 2\,W$   & -2. \\[0.5ex] 
15 & $\ket{2,-1,2,\Gamma_{5,3}}$  & $\frac{J}{2} - 2\,t + 2\,W$   & -2. \\[0.5ex] 
\hline
\end{tabular} \\[2ex]
\begin{tabular}[t]{|r|l|c|c|}
\multicolumn{4}{c}{\large \bf \boldmath Eigenkets and eigenvalues for ${\rm  N_e}$=2 and   ${\rm m_s}$= $0$. } \\[0.5ex] \hline
\parbox[c]{1cm}{No}  & \parbox[c]{2.5cm}{\begin{center} Eigenstate \end{center}}  & \parbox[c]{7.5cm}{ \begin{center}   Energy \end{center}}  & \parbox[c]{2cm}{ \begin{center} Example \end{center}}  \\ \hline 
\hline 
16 & $\ket{2,0,0,\Gamma_1}$  & $\frac{-{\sqrt{\Aindex{9}}}}{2} - \frac{J}{2} + 2\,t + \frac{U}{2} + W$   & 1. \\[0.5ex] 
17 & $\ket{2,0,0,\Gamma_1}$  & $\frac{{\sqrt{\Aindex{9}}}}{2} - \frac{J}{2} + 2\,t + \frac{U}{2} + W$   & 8. \\[0.5ex] 
18 & $\ket{2,0,0,\Gamma_{3,1}}$  & $-J - 2\,t + 2\,W$   & -2. \\[0.5ex] 
19 & $\ket{2,0,0,\Gamma_{3,2}}$  & $-J - 2\,t + 2\,W$   & -2. \\[0.5ex] 
20 & $\ket{2,0,0,\Gamma_{4,1}}$  & $\frac{-{\sqrt{\Aindex{1}}}}{2} - \frac{J}{2} + \frac{U}{2} + W$   & -0.701562 \\[0.5ex] 
21 & $\ket{2,0,0,\Gamma_{4,1}}$  & $\frac{{\sqrt{\Aindex{1}}}}{2} - \frac{J}{2} + \frac{U}{2} + W$   & 5.70156 \\[0.5ex] 
22 & $\ket{2,0,2,\Gamma_{4,1}}$  & $2\,t + 2\,W$   & 2. \\[0.5ex] 
23 & $\ket{2,0,0,\Gamma_{4,2}}$  & $\frac{-{\sqrt{\Aindex{1}}}}{2} - \frac{J}{2} + \frac{U}{2} + W$   & -0.701562 \\[0.5ex] 
24 & $\ket{2,0,0,\Gamma_{4,2}}$  & $\frac{{\sqrt{\Aindex{1}}}}{2} - \frac{J}{2} + \frac{U}{2} + W$   & 5.70156 \\[0.5ex] 
25 & $\ket{2,0,2,\Gamma_{4,2}}$  & $2\,t + 2\,W$   & 2. \\[0.5ex] 
26 & $\ket{2,0,0,\Gamma_{4,3}}$  & $\frac{-{\sqrt{\Aindex{1}}}}{2} - \frac{J}{2} + \frac{U}{2} + W$   & -0.701562 \\[0.5ex] 
27 & $\ket{2,0,0,\Gamma_{4,3}}$  & $\frac{{\sqrt{\Aindex{1}}}}{2} - \frac{J}{2} + \frac{U}{2} + W$   & 5.70156 \\[0.5ex] 
28 & $\ket{2,0,2,\Gamma_{4,3}}$  & $2\,t + 2\,W$   & 2. \\[0.5ex] 
29 & $\ket{2,0,2,\Gamma_{5,1}}$  & $-2\,t + 2\,W$   & -2. \\[0.5ex] 
30 & $\ket{2,0,2,\Gamma_{5,2}}$  & $-2\,t + 2\,W$   & -2. \\[0.5ex] 
31 & $\ket{2,0,2,\Gamma_{5,3}}$  & $-2\,t + 2\,W$   & -2. \\[0.5ex] 
\hline
\end{tabular} \\[2ex]
\begin{tabular}[t]{|r|l|c|c|}
\multicolumn{4}{c}{\large \bf \boldmath Eigenkets and eigenvalues for ${\rm  N_e}$=2 and   ${\rm m_s}$= $1$. } \\[0.5ex] \hline
\parbox[c]{1cm}{No}  & \parbox[c]{2.5cm}{\begin{center} Eigenstate \end{center}}  & \parbox[c]{7.5cm}{ \begin{center}   Energy \end{center}}  & \parbox[c]{2cm}{ \begin{center} Example \end{center}}  \\ \hline 
\hline 
32 & $\ket{2,1,2,\Gamma_{4,1}}$  & $\frac{J}{2} + 2\,t + 2\,W$   & 2. \\[0.5ex] 
33 & $\ket{2,1,2,\Gamma_{4,2}}$  & $\frac{J}{2} + 2\,t + 2\,W$   & 2. \\[0.5ex] 
34 & $\ket{2,1,2,\Gamma_{4,3}}$  & $\frac{J}{2} + 2\,t + 2\,W$   & 2. \\[0.5ex] 
35 & $\ket{2,1,2,\Gamma_{5,1}}$  & $\frac{J}{2} - 2\,t + 2\,W$   & -2. \\[0.5ex] 
36 & $\ket{2,1,2,\Gamma_{5,2}}$  & $\frac{J}{2} - 2\,t + 2\,W$   & -2. \\[0.5ex] 
37 & $\ket{2,1,2,\Gamma_{5,3}}$  & $\frac{J}{2} - 2\,t + 2\,W$   & -2. \\[0.5ex] 
\hline
\end{tabular} \\[2ex]
\begin{tabular}[t]{|r|l|c|c|}
\multicolumn{4}{c}{\large \bf \boldmath Eigenkets and eigenvalues for ${\rm  N_e}$=3 and   ${\rm m_s}$= $- {3 \over 2} $. } \\[0.5ex] \hline
\parbox[c]{1cm}{No}  & \parbox[c]{2.5cm}{\begin{center} Eigenstate \end{center}}  & \parbox[c]{7.5cm}{ \begin{center}   Energy \end{center}}  & \parbox[c]{2cm}{ \begin{center} Example \end{center}}  \\ \hline 
\hline 
38 & $\ket{3,- {3 \over 2} , {15 \over 4} ,\Gamma_2}$  & $\frac{3\,J}{2} - 3\,t + 6\,W$   & -3. \\[0.5ex] 
39 & $\ket{3,- {3 \over 2} , {15 \over 4} ,\Gamma_{5,1}}$  & $\frac{3\,J}{2} + t + 6\,W$   & 1. \\[0.5ex] 
40 & $\ket{3,- {3 \over 2} , {15 \over 4} ,\Gamma_{5,2}}$  & $\frac{3\,J}{2} + t + 6\,W$   & 1. \\[0.5ex] 
41 & $\ket{3,- {3 \over 2} , {15 \over 4} ,\Gamma_{5,3}}$  & $\frac{3\,J}{2} + t + 6\,W$   & 1. \\[0.5ex] 
\hline
\end{tabular} \\[2ex]
\begin{tabular}[t]{|r|l|c|c|}
\multicolumn{4}{c}{\large \bf \boldmath Eigenkets and eigenvalues for ${\rm  N_e}$=3 and   ${\rm m_s}$= $- {1 \over 2} $. } \\[0.5ex] \hline
\parbox[c]{1cm}{No}  & \parbox[c]{2.5cm}{\begin{center} Eigenstate \end{center}}  & \parbox[c]{7.5cm}{ \begin{center}   Energy \end{center}}  & \parbox[c]{2cm}{ \begin{center} Example \end{center}}  \\ \hline 
\hline 
42 & $\ket{3,- {1 \over 2} , {3 \over 4} ,\Gamma_1}$  & $t + U + 4\,W$   & 6. \\[0.5ex] 
43 & $\ket{3,- {1 \over 2} , {15 \over 4} ,\Gamma_2}$  & $\frac{J}{2} - 3\,t + 6\,W$   & -3. \\[0.5ex] 
44 & $\ket{3,- {1 \over 2} , {3 \over 4} ,\Gamma_{3,1}}$  & $\frac{-{\sqrt{\Aindex{4}}}}{2} - \frac{J}{2} - t + \frac{U}{2} + 5\,W$   & -0.791288 \\[0.5ex] 
45 & $\ket{3,- {1 \over 2} , {3 \over 4} ,\Gamma_{3,1}}$  & $\frac{{\sqrt{\Aindex{4}}}}{2} - \frac{J}{2} - t + \frac{U}{2} + 5\,W$   & 3.79129 \\[0.5ex] 
46 & $\ket{3,- {1 \over 2} , {3 \over 4} ,\Gamma_{3,2}}$  & $\frac{-{\sqrt{\Aindex{4}}}}{2} - \frac{J}{2} - t + \frac{U}{2} + 5\,W$   & -0.791288 \\[0.5ex] 
47 & $\ket{3,- {1 \over 2} , {3 \over 4} ,\Gamma_{3,2}}$  & $\frac{{\sqrt{\Aindex{4}}}}{2} - \frac{J}{2} - t + \frac{U}{2} + 5\,W$   & 3.79129 \\[0.5ex] 
48 & $\ket{3,- {1 \over 2} , {3 \over 4} ,\Gamma_{4,1}}$  & $\frac{-J}{3} + t + \frac{2\,U}{3} + \frac{14\,W}{3} - \frac{2\,{\sqrt{\Aindex{5}}}\,\cos (\Thetaindex{4})}{3}$   & -0.248211 \\[0.5ex] 
49 & $\ket{3,- {1 \over 2} , {3 \over 4} ,\Gamma_{4,1}}$  & $\frac{\Aindex{21}}{3}$   & 8.51895 \\[0.5ex] 
50 & $\ket{3,- {1 \over 2} , {3 \over 4} ,\Gamma_{4,1}}$  & $\frac{\Aindex{20}}{3}$   & 4.72926 \\[0.5ex] 
51 & $\ket{3,- {1 \over 2} , {3 \over 4} ,\Gamma_{4,2}}$  & $\frac{-J}{3} + t + \frac{2\,U}{3} + \frac{14\,W}{3} - \frac{2\,{\sqrt{\Aindex{5}}}\,\cos (\Thetaindex{4})}{3}$   & -0.248211 \\[0.5ex] 
52 & $\ket{3,- {1 \over 2} , {3 \over 4} ,\Gamma_{4,2}}$  & $\frac{\Aindex{21}}{3}$   & 8.51895 \\[0.5ex] 
53 & $\ket{3,- {1 \over 2} , {3 \over 4} ,\Gamma_{4,2}}$  & $\frac{\Aindex{20}}{3}$   & 4.72926 \\[0.5ex] 
54 & $\ket{3,- {1 \over 2} , {3 \over 4} ,\Gamma_{4,3}}$  & $\frac{-J}{3} + t + \frac{2\,U}{3} + \frac{14\,W}{3} - \frac{2\,{\sqrt{\Aindex{5}}}\,\cos (\Thetaindex{4})}{3}$   & -0.248211 \\[0.5ex] 
55 & $\ket{3,- {1 \over 2} , {3 \over 4} ,\Gamma_{4,3}}$  & $\frac{\Aindex{21}}{3}$   & 8.51895 \\[0.5ex] 
56 & $\ket{3,- {1 \over 2} , {3 \over 4} ,\Gamma_{4,3}}$  & $\frac{\Aindex{20}}{3}$   & 4.72926 \\[0.5ex] 
57 & $\ket{3,- {1 \over 2} , {3 \over 4} ,\Gamma_{5,1}}$  & $\frac{-{\sqrt{\Aindex{7}}}}{2} - \frac{J}{2} - t + \frac{U}{2} + 5\,W$   & -2.40512 \\[0.5ex] 
58 & $\ket{3,- {1 \over 2} , {3 \over 4} ,\Gamma_{5,1}}$  & $\frac{{\sqrt{\Aindex{7}}}}{2} - \frac{J}{2} - t + \frac{U}{2} + 5\,W$   & 5.40512 \\[0.5ex] 
59 & $\ket{3,- {1 \over 2} , {15 \over 4} ,\Gamma_{5,1}}$  & $\frac{J}{2} + t + 6\,W$   & 1. \\[0.5ex] 
60 & $\ket{3,- {1 \over 2} , {3 \over 4} ,\Gamma_{5,2}}$  & $\frac{-{\sqrt{\Aindex{7}}}}{2} - \frac{J}{2} - t + \frac{U}{2} + 5\,W$   & -2.40512 \\[0.5ex] 
61 & $\ket{3,- {1 \over 2} , {3 \over 4} ,\Gamma_{5,2}}$  & $\frac{{\sqrt{\Aindex{7}}}}{2} - \frac{J}{2} - t + \frac{U}{2} + 5\,W$   & 5.40512 \\[0.5ex] 
62 & $\ket{3,- {1 \over 2} , {15 \over 4} ,\Gamma_{5,2}}$  & $\frac{J}{2} + t + 6\,W$   & 1. \\[0.5ex] 
63 & $\ket{3,- {1 \over 2} , {3 \over 4} ,\Gamma_{5,3}}$  & $\frac{-{\sqrt{\Aindex{7}}}}{2} - \frac{J}{2} - t + \frac{U}{2} + 5\,W$   & -2.40512 \\[0.5ex] 
64 & $\ket{3,- {1 \over 2} , {3 \over 4} ,\Gamma_{5,3}}$  & $\frac{{\sqrt{\Aindex{7}}}}{2} - \frac{J}{2} - t + \frac{U}{2} + 5\,W$   & 5.40512 \\[0.5ex] 
65 & $\ket{3,- {1 \over 2} , {15 \over 4} ,\Gamma_{5,3}}$  & $\frac{J}{2} + t + 6\,W$   & 1. \\[0.5ex] 
\hline
\end{tabular} \\[2ex]
\begin{tabular}[t]{|r|l|c|c|}
\multicolumn{4}{c}{\large \bf \boldmath Eigenkets and eigenvalues for ${\rm  N_e}$=3 and   ${\rm m_s}$= $ {1 \over 2} $. } \\[0.5ex] \hline
\parbox[c]{1cm}{No}  & \parbox[c]{2.5cm}{\begin{center} Eigenstate \end{center}}  & \parbox[c]{7.5cm}{ \begin{center}   Energy \end{center}}  & \parbox[c]{2cm}{ \begin{center} Example \end{center}}  \\ \hline 
\hline 
66 & $\ket{3, {1 \over 2} , {3 \over 4} ,\Gamma_1}$  & $t + U + 4\,W$   & 6. \\[0.5ex] 
67 & $\ket{3, {1 \over 2} , {15 \over 4} ,\Gamma_2}$  & $\frac{J}{2} - 3\,t + 6\,W$   & -3. \\[0.5ex] 
68 & $\ket{3, {1 \over 2} , {3 \over 4} ,\Gamma_{3,1}}$  & $\frac{-{\sqrt{\Aindex{4}}}}{2} - \frac{J}{2} - t + \frac{U}{2} + 5\,W$   & -0.791288 \\[0.5ex] 
69 & $\ket{3, {1 \over 2} , {3 \over 4} ,\Gamma_{3,1}}$  & $\frac{{\sqrt{\Aindex{4}}}}{2} - \frac{J}{2} - t + \frac{U}{2} + 5\,W$   & 3.79129 \\[0.5ex] 
70 & $\ket{3, {1 \over 2} , {3 \over 4} ,\Gamma_{3,2}}$  & $\frac{-{\sqrt{\Aindex{4}}}}{2} - \frac{J}{2} - t + \frac{U}{2} + 5\,W$   & -0.791288 \\[0.5ex] 
71 & $\ket{3, {1 \over 2} , {3 \over 4} ,\Gamma_{3,2}}$  & $\frac{{\sqrt{\Aindex{4}}}}{2} - \frac{J}{2} - t + \frac{U}{2} + 5\,W$   & 3.79129 \\[0.5ex] 
72 & $\ket{3, {1 \over 2} , {3 \over 4} ,\Gamma_{4,1}}$  & $\frac{-J}{3} + t + \frac{2\,U}{3} + \frac{14\,W}{3} - \frac{2\,{\sqrt{\Aindex{5}}}\,\cos (\Thetaindex{4})}{3}$   & -0.248211 \\[0.5ex] 
73 & $\ket{3, {1 \over 2} , {3 \over 4} ,\Gamma_{4,1}}$  & $\frac{\Aindex{21}}{3}$   & 8.51895 \\[0.5ex] 
74 & $\ket{3, {1 \over 2} , {3 \over 4} ,\Gamma_{4,1}}$  & $\frac{\Aindex{20}}{3}$   & 4.72926 \\[0.5ex] 
75 & $\ket{3, {1 \over 2} , {3 \over 4} ,\Gamma_{4,2}}$  & $\frac{-J}{3} + t + \frac{2\,U}{3} + \frac{14\,W}{3} - \frac{2\,{\sqrt{\Aindex{5}}}\,\cos (\Thetaindex{4})}{3}$   & -0.248211 \\[0.5ex] 
76 & $\ket{3, {1 \over 2} , {3 \over 4} ,\Gamma_{4,2}}$  & $\frac{\Aindex{21}}{3}$   & 8.51895 \\[0.5ex] 
77 & $\ket{3, {1 \over 2} , {3 \over 4} ,\Gamma_{4,2}}$  & $\frac{\Aindex{20}}{3}$   & 4.72926 \\[0.5ex] 
78 & $\ket{3, {1 \over 2} , {3 \over 4} ,\Gamma_{4,3}}$  & $\frac{-J}{3} + t + \frac{2\,U}{3} + \frac{14\,W}{3} - \frac{2\,{\sqrt{\Aindex{5}}}\,\cos (\Thetaindex{4})}{3}$   & -0.248211 \\[0.5ex] 
79 & $\ket{3, {1 \over 2} , {3 \over 4} ,\Gamma_{4,3}}$  & $\frac{\Aindex{21}}{3}$   & 8.51895 \\[0.5ex] 
80 & $\ket{3, {1 \over 2} , {3 \over 4} ,\Gamma_{4,3}}$  & $\frac{\Aindex{20}}{3}$   & 4.72926 \\[0.5ex] 
81 & $\ket{3, {1 \over 2} , {3 \over 4} ,\Gamma_{5,1}}$  & $\frac{-{\sqrt{\Aindex{7}}}}{2} - \frac{J}{2} - t + \frac{U}{2} + 5\,W$   & -2.40512 \\[0.5ex] 
82 & $\ket{3, {1 \over 2} , {3 \over 4} ,\Gamma_{5,1}}$  & $\frac{{\sqrt{\Aindex{7}}}}{2} - \frac{J}{2} - t + \frac{U}{2} + 5\,W$   & 5.40512 \\[0.5ex] 
83 & $\ket{3, {1 \over 2} , {15 \over 4} ,\Gamma_{5,1}}$  & $\frac{J}{2} + t + 6\,W$   & 1. \\[0.5ex] 
84 & $\ket{3, {1 \over 2} , {3 \over 4} ,\Gamma_{5,2}}$  & $\frac{-{\sqrt{\Aindex{7}}}}{2} - \frac{J}{2} - t + \frac{U}{2} + 5\,W$   & -2.40512 \\[0.5ex] 
85 & $\ket{3, {1 \over 2} , {3 \over 4} ,\Gamma_{5,2}}$  & $\frac{{\sqrt{\Aindex{7}}}}{2} - \frac{J}{2} - t + \frac{U}{2} + 5\,W$   & 5.40512 \\[0.5ex] 
86 & $\ket{3, {1 \over 2} , {15 \over 4} ,\Gamma_{5,2}}$  & $\frac{J}{2} + t + 6\,W$   & 1. \\[0.5ex] 
87 & $\ket{3, {1 \over 2} , {3 \over 4} ,\Gamma_{5,3}}$  & $\frac{-{\sqrt{\Aindex{7}}}}{2} - \frac{J}{2} - t + \frac{U}{2} + 5\,W$   & -2.40512 \\[0.5ex] 
88 & $\ket{3, {1 \over 2} , {3 \over 4} ,\Gamma_{5,3}}$  & $\frac{{\sqrt{\Aindex{7}}}}{2} - \frac{J}{2} - t + \frac{U}{2} + 5\,W$   & 5.40512 \\[0.5ex] 
89 & $\ket{3, {1 \over 2} , {15 \over 4} ,\Gamma_{5,3}}$  & $\frac{J}{2} + t + 6\,W$   & 1. \\[0.5ex] 
\hline
\end{tabular} \\[2ex]
\begin{tabular}[t]{|r|l|c|c|}
\multicolumn{4}{c}{\large \bf \boldmath Eigenkets and eigenvalues for ${\rm  N_e}$=3 and   ${\rm m_s}$= $ {3 \over 2} $. } \\[0.5ex] \hline
\parbox[c]{1cm}{No}  & \parbox[c]{2.5cm}{\begin{center} Eigenstate \end{center}}  & \parbox[c]{7.5cm}{ \begin{center}   Energy \end{center}}  & \parbox[c]{2cm}{ \begin{center} Example \end{center}}  \\ \hline 
\hline 
90 & $\ket{3, {3 \over 2} , {15 \over 4} ,\Gamma_2}$  & $\frac{3\,J}{2} - 3\,t + 6\,W$   & -3. \\[0.5ex] 
91 & $\ket{3, {3 \over 2} , {15 \over 4} ,\Gamma_{5,1}}$  & $\frac{3\,J}{2} + t + 6\,W$   & 1. \\[0.5ex] 
92 & $\ket{3, {3 \over 2} , {15 \over 4} ,\Gamma_{5,2}}$  & $\frac{3\,J}{2} + t + 6\,W$   & 1. \\[0.5ex] 
93 & $\ket{3, {3 \over 2} , {15 \over 4} ,\Gamma_{5,3}}$  & $\frac{3\,J}{2} + t + 6\,W$   & 1. \\[0.5ex] 
\hline
\end{tabular} \\[2ex]
\begin{tabular}[t]{|r|l|c|c|}
\multicolumn{4}{c}{\large \bf \boldmath Eigenkets and eigenvalues for ${\rm  N_e}$=4 and   ${\rm m_s}$= $-2$. } \\[0.5ex] \hline
\parbox[c]{1cm}{No}  & \parbox[c]{2.5cm}{\begin{center} Eigenstate \end{center}}  & \parbox[c]{7.5cm}{ \begin{center}   Energy \end{center}}  & \parbox[c]{2cm}{ \begin{center} Example \end{center}}  \\ \hline 
\hline 
94 & $\ket{4,-2,6,\Gamma_2}$  & $3\,J + 12\,W$   & 0. \\[0.5ex] 
\hline
\end{tabular} \\[2ex]
\begin{tabular}[t]{|r|l|c|c|}
\multicolumn{4}{c}{\large \bf \boldmath Eigenkets and eigenvalues for ${\rm  N_e}$=4 and   ${\rm m_s}$= $-1$. } \\[0.5ex] \hline
\parbox[c]{1cm}{No}  & \parbox[c]{2.5cm}{\begin{center} Eigenstate \end{center}}  & \parbox[c]{7.5cm}{ \begin{center}   Energy \end{center}}  & \parbox[c]{2cm}{ \begin{center} Example \end{center}}  \\ \hline 
\hline 
95 & $\ket{4,-1,2,\Gamma_2}$  & $\frac{J}{2} + U + 10\,W$   & 5. \\[0.5ex] 
96 & $\ket{4,-1,6,\Gamma_2}$  & $\frac{3\,J}{2} + 12\,W$   & 0. \\[0.5ex] 
97 & $\ket{4,-1,2,\Gamma_{3,1}}$  & $\frac{J}{2} + U + 10\,W$   & 5. \\[0.5ex] 
98 & $\ket{4,-1,2,\Gamma_{3,2}}$  & $\frac{J}{2} + U + 10\,W$   & 5. \\[0.5ex] 
99 & $\ket{4,-1,2,\Gamma_{4,1}}$  & $\frac{J}{2} + U + 10\,W$   & 5. \\[0.5ex] 
100 & $\ket{4,-1,2,\Gamma_{4,2}}$  & $\frac{J}{2} + U + 10\,W$   & 5. \\[0.5ex] 
101 & $\ket{4,-1,2,\Gamma_{4,3}}$  & $\frac{J}{2} + U + 10\,W$   & 5. \\[0.5ex] 
102 & $\ket{4,-1,2,\Gamma_{5,1}}$  & $\frac{\Aindex{11}}{6}$   & -1.5136 \\[0.5ex] 
103 & $\ket{4,-1,2,\Gamma_{5,1}}$  & $\frac{\Aindex{18}}{3}$   & 8.34789 \\[0.5ex] 
104 & $\ket{4,-1,2,\Gamma_{5,1}}$  & $\frac{\Aindex{16}}{3}$   & 3.16571 \\[0.5ex] 
105 & $\ket{4,-1,2,\Gamma_{5,2}}$  & $\frac{\Aindex{11}}{6}$   & -1.5136 \\[0.5ex] 
106 & $\ket{4,-1,2,\Gamma_{5,2}}$  & $\frac{\Aindex{18}}{3}$   & 8.34789 \\[0.5ex] 
107 & $\ket{4,-1,2,\Gamma_{5,2}}$  & $\frac{\Aindex{16}}{3}$   & 3.16571 \\[0.5ex] 
108 & $\ket{4,-1,2,\Gamma_{5,3}}$  & $\frac{\Aindex{11}}{6}$   & -1.5136 \\[0.5ex] 
109 & $\ket{4,-1,2,\Gamma_{5,3}}$  & $\frac{\Aindex{18}}{3}$   & 8.34789 \\[0.5ex] 
110 & $\ket{4,-1,2,\Gamma_{5,3}}$  & $\frac{\Aindex{16}}{3}$   & 3.16571 \\[0.5ex] 
\hline
\end{tabular} \\[2ex]
\begin{tabular}[t]{|r|l|c|c|}
\multicolumn{4}{c}{\large \bf \boldmath Eigenkets and eigenvalues for ${\rm  N_e}$=4 and   ${\rm m_s}$= $0$. } \\[0.5ex] \hline
\parbox[c]{1cm}{No}  & \parbox[c]{2.5cm}{\begin{center} Eigenstate \end{center}}  & \parbox[c]{7.5cm}{ \begin{center}   Energy \end{center}}  & \parbox[c]{2cm}{ \begin{center} Example \end{center}}  \\ \hline 
\hline 
111 & $\ket{4,0,0,\Gamma_1}$  & $\frac{-{\sqrt{\Aindex{3}}}}{2} - \frac{J}{2} + \frac{3\,U}{2} + 9\,W$   & 2.78301 \\[0.5ex] 
112 & $\ket{4,0,0,\Gamma_1}$  & $\frac{{\sqrt{\Aindex{3}}}}{2} - \frac{J}{2} + \frac{3\,U}{2} + 9\,W$   & 12.217 \\[0.5ex] 
113 & $\ket{4,0,2,\Gamma_2}$  & $U + 10\,W$   & 5. \\[0.5ex] 
114 & $\ket{4,0,6,\Gamma_2}$  & $J + 12\,W$   & 0. \\[0.5ex] 
115 & $\ket{4,0,0,\Gamma_{3,1}}$  & $-J + U + 10\,W - \frac{2\,{\sqrt{\Aindex{1}}}\,\cos (\Thetaindex{1})}{{\sqrt{3}}}$   & -1.84429 \\[0.5ex] 
116 & $\ket{4,0,0,\Gamma_{3,1}}$  & $-J + U + 10\,W + \frac{{\sqrt{\Aindex{1}}}\,\cos (\Thetaindex{1})}{{\sqrt{3}}} + {\sqrt{\Aindex{1}}}\,\sin (\Thetaindex{1})$   & 10.8443 \\[0.5ex] 
117 & $\ket{4,0,0,\Gamma_{3,1}}$  & $-J + U + 10\,W + \frac{{\sqrt{\Aindex{1}}}\,\cos (\Thetaindex{1})}{{\sqrt{3}}} - {\sqrt{\Aindex{1}}}\,\sin (\Thetaindex{1})$   & 6. \\[0.5ex] 
118 & $\ket{4,0,2,\Gamma_{3,1}}$  & $U + 10\,W$   & 5. \\[0.5ex] 
119 & $\ket{4,0,0,\Gamma_{3,2}}$  & $-J + U + 10\,W - \frac{2\,{\sqrt{\Aindex{1}}}\,\cos (\Thetaindex{1})}{{\sqrt{3}}}$   & -1.84429 \\[0.5ex] 
120 & $\ket{4,0,0,\Gamma_{3,2}}$  & $-J + U + 10\,W + \frac{{\sqrt{\Aindex{1}}}\,\cos (\Thetaindex{1})}{{\sqrt{3}}} + {\sqrt{\Aindex{1}}}\,\sin (\Thetaindex{1})$   & 10.8443 \\[0.5ex] 
121 & $\ket{4,0,0,\Gamma_{3,2}}$  & $-J + U + 10\,W + \frac{{\sqrt{\Aindex{1}}}\,\cos (\Thetaindex{1})}{{\sqrt{3}}} - {\sqrt{\Aindex{1}}}\,\sin (\Thetaindex{1})$   & 6. \\[0.5ex] 
122 & $\ket{4,0,2,\Gamma_{3,2}}$  & $U + 10\,W$   & 5. \\[0.5ex] 
123 & $\ket{4,0,0,\Gamma_{4,1}}$  & $\frac{-2\,\Aindex{10}}{3}$   & 1.65211 \\[0.5ex] 
124 & $\ket{4,0,0,\Gamma_{4,1}}$  & $\frac{\Aindex{14}}{3}$   & 11.5136 \\[0.5ex] 
125 & $\ket{4,0,0,\Gamma_{4,1}}$  & $\frac{\Aindex{13}}{3}$   & 6.83429 \\[0.5ex] 
126 & $\ket{4,0,2,\Gamma_{4,1}}$  & $U + 10\,W$   & 5. \\[0.5ex] 
127 & $\ket{4,0,0,\Gamma_{4,2}}$  & $\frac{-2\,\Aindex{10}}{3}$   & 1.65211 \\[0.5ex] 
128 & $\ket{4,0,0,\Gamma_{4,2}}$  & $\frac{\Aindex{14}}{3}$   & 11.5136 \\[0.5ex] 
129 & $\ket{4,0,0,\Gamma_{4,2}}$  & $\frac{\Aindex{13}}{3}$   & 6.83429 \\[0.5ex] 
130 & $\ket{4,0,2,\Gamma_{4,2}}$  & $U + 10\,W$   & 5. \\[0.5ex] 
131 & $\ket{4,0,0,\Gamma_{4,3}}$  & $\frac{-2\,\Aindex{10}}{3}$   & 1.65211 \\[0.5ex] 
132 & $\ket{4,0,0,\Gamma_{4,3}}$  & $\frac{\Aindex{14}}{3}$   & 11.5136 \\[0.5ex] 
133 & $\ket{4,0,0,\Gamma_{4,3}}$  & $\frac{\Aindex{13}}{3}$   & 6.83429 \\[0.5ex] 
134 & $\ket{4,0,2,\Gamma_{4,3}}$  & $U + 10\,W$   & 5. \\[0.5ex] 
135 & $\ket{4,0,0,\Gamma_{5,1}}$  & $-J + U + 10\,W$   & 5. \\[0.5ex] 
136 & $\ket{4,0,2,\Gamma_{5,1}}$  & $\frac{-J}{3} + \frac{2\,U}{3} + \frac{32\,W}{3} - \frac{2\,{\sqrt{\Aindex{2}}}\,\cos (\Thetaindex{3})}{3}$   & -1.5136 \\[0.5ex] 
137 & $\ket{4,0,2,\Gamma_{5,1}}$  & $\frac{\Aindex{17}}{3}$   & 8.34789 \\[0.5ex] 
138 & $\ket{4,0,2,\Gamma_{5,1}}$  & $\frac{\Aindex{15}}{3}$   & 3.16571 \\[0.5ex] 
139 & $\ket{4,0,0,\Gamma_{5,2}}$  & $-J + U + 10\,W$   & 5. \\[0.5ex] 
140 & $\ket{4,0,2,\Gamma_{5,2}}$  & $\frac{-J}{3} + \frac{2\,U}{3} + \frac{32\,W}{3} - \frac{2\,{\sqrt{\Aindex{2}}}\,\cos (\Thetaindex{3})}{3}$   & -1.5136 \\[0.5ex] 
141 & $\ket{4,0,2,\Gamma_{5,2}}$  & $\frac{\Aindex{17}}{3}$   & 8.34789 \\[0.5ex] 
142 & $\ket{4,0,2,\Gamma_{5,2}}$  & $\frac{\Aindex{15}}{3}$   & 3.16571 \\[0.5ex] 
143 & $\ket{4,0,0,\Gamma_{5,3}}$  & $-J + U + 10\,W$   & 5. \\[0.5ex] 
144 & $\ket{4,0,2,\Gamma_{5,3}}$  & $\frac{-J}{3} + \frac{2\,U}{3} + \frac{32\,W}{3} - \frac{2\,{\sqrt{\Aindex{2}}}\,\cos (\Thetaindex{3})}{3}$   & -1.5136 \\[0.5ex] 
145 & $\ket{4,0,2,\Gamma_{5,3}}$  & $\frac{\Aindex{17}}{3}$   & 8.34789 \\[0.5ex] 
146 & $\ket{4,0,2,\Gamma_{5,3}}$  & $\frac{\Aindex{15}}{3}$   & 3.16571 \\[0.5ex] 
\hline
\end{tabular} \\[2ex]
\begin{tabular}[t]{|r|l|c|c|}
\multicolumn{4}{c}{\large \bf \boldmath Eigenkets and eigenvalues for ${\rm  N_e}$=4 and   ${\rm m_s}$= $1$. } \\[0.5ex] \hline
\parbox[c]{1cm}{No}  & \parbox[c]{2.5cm}{\begin{center} Eigenstate \end{center}}  & \parbox[c]{7.5cm}{ \begin{center}   Energy \end{center}}  & \parbox[c]{2cm}{ \begin{center} Example \end{center}}  \\ \hline 
\hline 
147 & $\ket{4,1,2,\Gamma_2}$  & $\frac{J}{2} + U + 10\,W$   & 5. \\[0.5ex] 
148 & $\ket{4,1,6,\Gamma_2}$  & $\frac{3\,J}{2} + 12\,W$   & 0. \\[0.5ex] 
149 & $\ket{4,1,2,\Gamma_{3,1}}$  & $\frac{J}{2} + U + 10\,W$   & 5. \\[0.5ex] 
150 & $\ket{4,1,2,\Gamma_{3,2}}$  & $\frac{J}{2} + U + 10\,W$   & 5. \\[0.5ex] 
151 & $\ket{4,1,2,\Gamma_{4,1}}$  & $\frac{J}{2} + U + 10\,W$   & 5. \\[0.5ex] 
152 & $\ket{4,1,2,\Gamma_{4,2}}$  & $\frac{J}{2} + U + 10\,W$   & 5. \\[0.5ex] 
153 & $\ket{4,1,2,\Gamma_{4,3}}$  & $\frac{J}{2} + U + 10\,W$   & 5. \\[0.5ex] 
154 & $\ket{4,1,2,\Gamma_{5,1}}$  & $\frac{\Aindex{11}}{6}$   & -1.5136 \\[0.5ex] 
155 & $\ket{4,1,2,\Gamma_{5,1}}$  & $\frac{\Aindex{18}}{3}$   & 8.34789 \\[0.5ex] 
156 & $\ket{4,1,2,\Gamma_{5,1}}$  & $\frac{\Aindex{16}}{3}$   & 3.16571 \\[0.5ex] 
157 & $\ket{4,1,2,\Gamma_{5,2}}$  & $\frac{\Aindex{11}}{6}$   & -1.5136 \\[0.5ex] 
158 & $\ket{4,1,2,\Gamma_{5,2}}$  & $\frac{\Aindex{18}}{3}$   & 8.34789 \\[0.5ex] 
159 & $\ket{4,1,2,\Gamma_{5,2}}$  & $\frac{\Aindex{16}}{3}$   & 3.16571 \\[0.5ex] 
160 & $\ket{4,1,2,\Gamma_{5,3}}$  & $\frac{\Aindex{11}}{6}$   & -1.5136 \\[0.5ex] 
161 & $\ket{4,1,2,\Gamma_{5,3}}$  & $\frac{\Aindex{18}}{3}$   & 8.34789 \\[0.5ex] 
162 & $\ket{4,1,2,\Gamma_{5,3}}$  & $\frac{\Aindex{16}}{3}$   & 3.16571 \\[0.5ex] 
\hline
\end{tabular} \\[2ex]
\begin{tabular}[t]{|r|l|c|c|}
\multicolumn{4}{c}{\large \bf \boldmath Eigenkets and eigenvalues for ${\rm  N_e}$=4 and   ${\rm m_s}$= $2$. } \\[0.5ex] \hline
\parbox[c]{1cm}{No}  & \parbox[c]{2.5cm}{\begin{center} Eigenstate \end{center}}  & \parbox[c]{7.5cm}{ \begin{center}   Energy \end{center}}  & \parbox[c]{2cm}{ \begin{center} Example \end{center}}  \\ \hline 
\hline 
163 & $\ket{4,2,6,\Gamma_2}$  & $3\,J + 12\,W$   & 0. \\[0.5ex] 
\hline
\end{tabular} \\[2ex]
\begin{tabular}[t]{|r|l|c|c|}
\multicolumn{4}{c}{\large \bf \boldmath Eigenkets and eigenvalues for ${\rm  N_e}$=5 and   ${\rm m_s}$= $- {3 \over 2} $. } \\[0.5ex] \hline
\parbox[c]{1cm}{No}  & \parbox[c]{2.5cm}{\begin{center} Eigenstate \end{center}}  & \parbox[c]{7.5cm}{ \begin{center}   Energy \end{center}}  & \parbox[c]{2cm}{ \begin{center} Example \end{center}}  \\ \hline 
\hline 
164 & $\ket{5,- {3 \over 2} , {15 \over 4} ,\Gamma_2}$  & $\frac{3\,J}{2} + 3\,t + U + 18\,W$   & 8. \\[0.5ex] 
165 & $\ket{5,- {3 \over 2} , {15 \over 4} ,\Gamma_{5,1}}$  & $\frac{3\,J}{2} - t + U + 18\,W$   & 4. \\[0.5ex] 
166 & $\ket{5,- {3 \over 2} , {15 \over 4} ,\Gamma_{5,2}}$  & $\frac{3\,J}{2} - t + U + 18\,W$   & 4. \\[0.5ex] 
167 & $\ket{5,- {3 \over 2} , {15 \over 4} ,\Gamma_{5,3}}$  & $\frac{3\,J}{2} - t + U + 18\,W$   & 4. \\[0.5ex] 
\hline
\end{tabular} \\[2ex]
\begin{tabular}[t]{|r|l|c|c|}
\multicolumn{4}{c}{\large \bf \boldmath Eigenkets and eigenvalues for ${\rm  N_e}$=5 and   ${\rm m_s}$= $- {1 \over 2} $. } \\[0.5ex] \hline
\parbox[c]{1cm}{No}  & \parbox[c]{2.5cm}{\begin{center} Eigenstate \end{center}}  & \parbox[c]{7.5cm}{ \begin{center}   Energy \end{center}}  & \parbox[c]{2cm}{ \begin{center} Example \end{center}}  \\ \hline 
\hline 
168 & $\ket{5,- {1 \over 2} , {3 \over 4} ,\Gamma_1}$  & $-t + 2\,U + 16\,W$   & 9. \\[0.5ex] 
169 & $\ket{5,- {1 \over 2} , {15 \over 4} ,\Gamma_2}$  & $\frac{J}{2} + 3\,t + U + 18\,W$   & 8. \\[0.5ex] 
170 & $\ket{5,- {1 \over 2} , {3 \over 4} ,\Gamma_{3,1}}$  & $\frac{-{\sqrt{\Aindex{7}}}}{2} - \frac{J}{2} + t + \frac{3\,U}{2} + 17\,W$   & 4.59488 \\[0.5ex] 
171 & $\ket{5,- {1 \over 2} , {3 \over 4} ,\Gamma_{3,1}}$  & $\frac{{\sqrt{\Aindex{7}}}}{2} - \frac{J}{2} + t + \frac{3\,U}{2} + 17\,W$   & 12.4051 \\[0.5ex] 
172 & $\ket{5,- {1 \over 2} , {3 \over 4} ,\Gamma_{3,2}}$  & $\frac{-{\sqrt{\Aindex{7}}}}{2} - \frac{J}{2} + t + \frac{3\,U}{2} + 17\,W$   & 4.59488 \\[0.5ex] 
173 & $\ket{5,- {1 \over 2} , {3 \over 4} ,\Gamma_{3,2}}$  & $\frac{{\sqrt{\Aindex{7}}}}{2} - \frac{J}{2} + t + \frac{3\,U}{2} + 17\,W$   & 12.4051 \\[0.5ex] 
174 & $\ket{5,- {1 \over 2} , {3 \over 4} ,\Gamma_{4,1}}$  & $\frac{-J}{3} - t + \frac{5\,U}{3} + \frac{50\,W}{3} - \frac{2\,{\sqrt{\Aindex{6}}}\,\cos (\Thetaindex{5})}{3}$   & 1.54341 \\[0.5ex] 
175 & $\ket{5,- {1 \over 2} , {3 \over 4} ,\Gamma_{4,1}}$  & $\frac{-J}{3} - t + \frac{5\,U}{3} + \frac{50\,W}{3} + \frac{{\sqrt{\Aindex{6}}}\,\cos (\Thetaindex{5})}{3} + \frac{{\sqrt{\Aindex{6}}}\,\sin (\Thetaindex{5})}{{\sqrt{3}}}$   & 12.2755 \\[0.5ex] 
176 & $\ket{5,- {1 \over 2} , {3 \over 4} ,\Gamma_{4,1}}$  & $\frac{-J}{3} - t + \frac{5\,U}{3} + \frac{50\,W}{3} + \frac{{\sqrt{\Aindex{6}}}\,\cos (\Thetaindex{5})}{3} - \frac{{\sqrt{\Aindex{6}}}\,\sin (\Thetaindex{5})}{{\sqrt{3}}}$   & 8.18113 \\[0.5ex] 
177 & $\ket{5,- {1 \over 2} , {3 \over 4} ,\Gamma_{4,2}}$  & $\frac{-J}{3} - t + \frac{5\,U}{3} + \frac{50\,W}{3} - \frac{2\,{\sqrt{\Aindex{6}}}\,\cos (\Thetaindex{5})}{3}$   & 1.54341 \\[0.5ex] 
178 & $\ket{5,- {1 \over 2} , {3 \over 4} ,\Gamma_{4,2}}$  & $\frac{-J}{3} - t + \frac{5\,U}{3} + \frac{50\,W}{3} + \frac{{\sqrt{\Aindex{6}}}\,\cos (\Thetaindex{5})}{3} + \frac{{\sqrt{\Aindex{6}}}\,\sin (\Thetaindex{5})}{{\sqrt{3}}}$   & 12.2755 \\[0.5ex] 
179 & $\ket{5,- {1 \over 2} , {3 \over 4} ,\Gamma_{4,2}}$  & $\frac{-J}{3} - t + \frac{5\,U}{3} + \frac{50\,W}{3} + \frac{{\sqrt{\Aindex{6}}}\,\cos (\Thetaindex{5})}{3} - \frac{{\sqrt{\Aindex{6}}}\,\sin (\Thetaindex{5})}{{\sqrt{3}}}$   & 8.18113 \\[0.5ex] 
180 & $\ket{5,- {1 \over 2} , {3 \over 4} ,\Gamma_{4,3}}$  & $\frac{-J}{3} - t + \frac{5\,U}{3} + \frac{50\,W}{3} - \frac{2\,{\sqrt{\Aindex{6}}}\,\cos (\Thetaindex{5})}{3}$   & 1.54341 \\[0.5ex] 
181 & $\ket{5,- {1 \over 2} , {3 \over 4} ,\Gamma_{4,3}}$  & $\frac{-J}{3} - t + \frac{5\,U}{3} + \frac{50\,W}{3} + \frac{{\sqrt{\Aindex{6}}}\,\cos (\Thetaindex{5})}{3} + \frac{{\sqrt{\Aindex{6}}}\,\sin (\Thetaindex{5})}{{\sqrt{3}}}$   & 12.2755 \\[0.5ex] 
182 & $\ket{5,- {1 \over 2} , {3 \over 4} ,\Gamma_{4,3}}$  & $\frac{-J}{3} - t + \frac{5\,U}{3} + \frac{50\,W}{3} + \frac{{\sqrt{\Aindex{6}}}\,\cos (\Thetaindex{5})}{3} - \frac{{\sqrt{\Aindex{6}}}\,\sin (\Thetaindex{5})}{{\sqrt{3}}}$   & 8.18113 \\[0.5ex] 
183 & $\ket{5,- {1 \over 2} , {3 \over 4} ,\Gamma_{5,1}}$  & $\frac{-{\sqrt{\Aindex{4}}}}{2} - \frac{J}{2} + t + \frac{3\,U}{2} + 17\,W$   & 6.20871 \\[0.5ex] 
184 & $\ket{5,- {1 \over 2} , {3 \over 4} ,\Gamma_{5,1}}$  & $\frac{{\sqrt{\Aindex{4}}}}{2} - \frac{J}{2} + t + \frac{3\,U}{2} + 17\,W$   & 10.7913 \\[0.5ex] 
185 & $\ket{5,- {1 \over 2} , {15 \over 4} ,\Gamma_{5,1}}$  & $\frac{J}{2} - t + U + 18\,W$   & 4. \\[0.5ex] 
186 & $\ket{5,- {1 \over 2} , {3 \over 4} ,\Gamma_{5,2}}$  & $\frac{-{\sqrt{\Aindex{4}}}}{2} - \frac{J}{2} + t + \frac{3\,U}{2} + 17\,W$   & 6.20871 \\[0.5ex] 
187 & $\ket{5,- {1 \over 2} , {3 \over 4} ,\Gamma_{5,2}}$  & $\frac{{\sqrt{\Aindex{4}}}}{2} - \frac{J}{2} + t + \frac{3\,U}{2} + 17\,W$   & 10.7913 \\[0.5ex] 
188 & $\ket{5,- {1 \over 2} , {15 \over 4} ,\Gamma_{5,2}}$  & $\frac{J}{2} - t + U + 18\,W$   & 4. \\[0.5ex] 
189 & $\ket{5,- {1 \over 2} , {3 \over 4} ,\Gamma_{5,3}}$  & $\frac{-{\sqrt{\Aindex{4}}}}{2} - \frac{J}{2} + t + \frac{3\,U}{2} + 17\,W$   & 6.20871 \\[0.5ex] 
190 & $\ket{5,- {1 \over 2} , {3 \over 4} ,\Gamma_{5,3}}$  & $\frac{{\sqrt{\Aindex{4}}}}{2} - \frac{J}{2} + t + \frac{3\,U}{2} + 17\,W$   & 10.7913 \\[0.5ex] 
191 & $\ket{5,- {1 \over 2} , {15 \over 4} ,\Gamma_{5,3}}$  & $\frac{J}{2} - t + U + 18\,W$   & 4. \\[0.5ex] 
\hline
\end{tabular} \\[2ex]
\begin{tabular}[t]{|r|l|c|c|}
\multicolumn{4}{c}{\large \bf \boldmath Eigenkets and eigenvalues for ${\rm  N_e}$=5 and   ${\rm m_s}$= $ {1 \over 2} $. } \\[0.5ex] \hline
\parbox[c]{1cm}{No}  & \parbox[c]{2.5cm}{\begin{center} Eigenstate \end{center}}  & \parbox[c]{7.5cm}{ \begin{center}   Energy \end{center}}  & \parbox[c]{2cm}{ \begin{center} Example \end{center}}  \\ \hline 
\hline 
192 & $\ket{5, {1 \over 2} , {3 \over 4} ,\Gamma_1}$  & $-t + 2\,U + 16\,W$   & 9. \\[0.5ex] 
193 & $\ket{5, {1 \over 2} , {15 \over 4} ,\Gamma_2}$  & $\frac{J}{2} + 3\,t + U + 18\,W$   & 8. \\[0.5ex] 
194 & $\ket{5, {1 \over 2} , {3 \over 4} ,\Gamma_{3,1}}$  & $\frac{-{\sqrt{\Aindex{7}}}}{2} - \frac{J}{2} + t + \frac{3\,U}{2} + 17\,W$   & 4.59488 \\[0.5ex] 
195 & $\ket{5, {1 \over 2} , {3 \over 4} ,\Gamma_{3,1}}$  & $\frac{{\sqrt{\Aindex{7}}}}{2} - \frac{J}{2} + t + \frac{3\,U}{2} + 17\,W$   & 12.4051 \\[0.5ex] 
196 & $\ket{5, {1 \over 2} , {3 \over 4} ,\Gamma_{3,2}}$  & $\frac{-{\sqrt{\Aindex{7}}}}{2} - \frac{J}{2} + t + \frac{3\,U}{2} + 17\,W$   & 4.59488 \\[0.5ex] 
197 & $\ket{5, {1 \over 2} , {3 \over 4} ,\Gamma_{3,2}}$  & $\frac{{\sqrt{\Aindex{7}}}}{2} - \frac{J}{2} + t + \frac{3\,U}{2} + 17\,W$   & 12.4051 \\[0.5ex] 
198 & $\ket{5, {1 \over 2} , {3 \over 4} ,\Gamma_{4,1}}$  & $\frac{-J}{3} - t + \frac{5\,U}{3} + \frac{50\,W}{3} - \frac{2\,{\sqrt{\Aindex{6}}}\,\cos (\Thetaindex{5})}{3}$   & 1.54341 \\[0.5ex] 
199 & $\ket{5, {1 \over 2} , {3 \over 4} ,\Gamma_{4,1}}$  & $\frac{-J}{3} - t + \frac{5\,U}{3} + \frac{50\,W}{3} + \frac{{\sqrt{\Aindex{6}}}\,\cos (\Thetaindex{5})}{3} + \frac{{\sqrt{\Aindex{6}}}\,\sin (\Thetaindex{5})}{{\sqrt{3}}}$   & 12.2755 \\[0.5ex] 
200 & $\ket{5, {1 \over 2} , {3 \over 4} ,\Gamma_{4,1}}$  & $\frac{-J}{3} - t + \frac{5\,U}{3} + \frac{50\,W}{3} + \frac{{\sqrt{\Aindex{6}}}\,\cos (\Thetaindex{5})}{3} - \frac{{\sqrt{\Aindex{6}}}\,\sin (\Thetaindex{5})}{{\sqrt{3}}}$   & 8.18113 \\[0.5ex] 
201 & $\ket{5, {1 \over 2} , {3 \over 4} ,\Gamma_{4,2}}$  & $\frac{-J}{3} - t + \frac{5\,U}{3} + \frac{50\,W}{3} - \frac{2\,{\sqrt{\Aindex{6}}}\,\cos (\Thetaindex{5})}{3}$   & 1.54341 \\[0.5ex] 
202 & $\ket{5, {1 \over 2} , {3 \over 4} ,\Gamma_{4,2}}$  & $\frac{-J}{3} - t + \frac{5\,U}{3} + \frac{50\,W}{3} + \frac{{\sqrt{\Aindex{6}}}\,\cos (\Thetaindex{5})}{3} + \frac{{\sqrt{\Aindex{6}}}\,\sin (\Thetaindex{5})}{{\sqrt{3}}}$   & 12.2755 \\[0.5ex] 
203 & $\ket{5, {1 \over 2} , {3 \over 4} ,\Gamma_{4,2}}$  & $\frac{-J}{3} - t + \frac{5\,U}{3} + \frac{50\,W}{3} + \frac{{\sqrt{\Aindex{6}}}\,\cos (\Thetaindex{5})}{3} - \frac{{\sqrt{\Aindex{6}}}\,\sin (\Thetaindex{5})}{{\sqrt{3}}}$   & 8.18113 \\[0.5ex] 
204 & $\ket{5, {1 \over 2} , {3 \over 4} ,\Gamma_{4,3}}$  & $\frac{-J}{3} - t + \frac{5\,U}{3} + \frac{50\,W}{3} - \frac{2\,{\sqrt{\Aindex{6}}}\,\cos (\Thetaindex{5})}{3}$   & 1.54341 \\[0.5ex] 
205 & $\ket{5, {1 \over 2} , {3 \over 4} ,\Gamma_{4,3}}$  & $\frac{-J}{3} - t + \frac{5\,U}{3} + \frac{50\,W}{3} + \frac{{\sqrt{\Aindex{6}}}\,\cos (\Thetaindex{5})}{3} + \frac{{\sqrt{\Aindex{6}}}\,\sin (\Thetaindex{5})}{{\sqrt{3}}}$   & 12.2755 \\[0.5ex] 
206 & $\ket{5, {1 \over 2} , {3 \over 4} ,\Gamma_{4,3}}$  & $\frac{-J}{3} - t + \frac{5\,U}{3} + \frac{50\,W}{3} + \frac{{\sqrt{\Aindex{6}}}\,\cos (\Thetaindex{5})}{3} - \frac{{\sqrt{\Aindex{6}}}\,\sin (\Thetaindex{5})}{{\sqrt{3}}}$   & 8.18113 \\[0.5ex] 
207 & $\ket{5, {1 \over 2} , {3 \over 4} ,\Gamma_{5,1}}$  & $\frac{-{\sqrt{\Aindex{4}}}}{2} - \frac{J}{2} + t + \frac{3\,U}{2} + 17\,W$   & 6.20871 \\[0.5ex] 
208 & $\ket{5, {1 \over 2} , {3 \over 4} ,\Gamma_{5,1}}$  & $\frac{{\sqrt{\Aindex{4}}}}{2} - \frac{J}{2} + t + \frac{3\,U}{2} + 17\,W$   & 10.7913 \\[0.5ex] 
209 & $\ket{5, {1 \over 2} , {15 \over 4} ,\Gamma_{5,1}}$  & $\frac{J}{2} - t + U + 18\,W$   & 4. \\[0.5ex] 
210 & $\ket{5, {1 \over 2} , {3 \over 4} ,\Gamma_{5,2}}$  & $\frac{-{\sqrt{\Aindex{4}}}}{2} - \frac{J}{2} + t + \frac{3\,U}{2} + 17\,W$   & 6.20871 \\[0.5ex] 
211 & $\ket{5, {1 \over 2} , {3 \over 4} ,\Gamma_{5,2}}$  & $\frac{{\sqrt{\Aindex{4}}}}{2} - \frac{J}{2} + t + \frac{3\,U}{2} + 17\,W$   & 10.7913 \\[0.5ex] 
212 & $\ket{5, {1 \over 2} , {15 \over 4} ,\Gamma_{5,2}}$  & $\frac{J}{2} - t + U + 18\,W$   & 4. \\[0.5ex] 
213 & $\ket{5, {1 \over 2} , {3 \over 4} ,\Gamma_{5,3}}$  & $\frac{-{\sqrt{\Aindex{4}}}}{2} - \frac{J}{2} + t + \frac{3\,U}{2} + 17\,W$   & 6.20871 \\[0.5ex] 
214 & $\ket{5, {1 \over 2} , {3 \over 4} ,\Gamma_{5,3}}$  & $\frac{{\sqrt{\Aindex{4}}}}{2} - \frac{J}{2} + t + \frac{3\,U}{2} + 17\,W$   & 10.7913 \\[0.5ex] 
215 & $\ket{5, {1 \over 2} , {15 \over 4} ,\Gamma_{5,3}}$  & $\frac{J}{2} - t + U + 18\,W$   & 4. \\[0.5ex] 
\hline
\end{tabular} \\[2ex]
\begin{tabular}[t]{|r|l|c|c|}
\multicolumn{4}{c}{\large \bf \boldmath Eigenkets and eigenvalues for ${\rm  N_e}$=5 and   ${\rm m_s}$= $ {3 \over 2} $. } \\[0.5ex] \hline
\parbox[c]{1cm}{No}  & \parbox[c]{2.5cm}{\begin{center} Eigenstate \end{center}}  & \parbox[c]{7.5cm}{ \begin{center}   Energy \end{center}}  & \parbox[c]{2cm}{ \begin{center} Example \end{center}}  \\ \hline 
\hline 
216 & $\ket{5, {3 \over 2} , {15 \over 4} ,\Gamma_2}$  & $\frac{3\,J}{2} + 3\,t + U + 18\,W$   & 8. \\[0.5ex] 
217 & $\ket{5, {3 \over 2} , {15 \over 4} ,\Gamma_{5,1}}$  & $\frac{3\,J}{2} - t + U + 18\,W$   & 4. \\[0.5ex] 
218 & $\ket{5, {3 \over 2} , {15 \over 4} ,\Gamma_{5,2}}$  & $\frac{3\,J}{2} - t + U + 18\,W$   & 4. \\[0.5ex] 
219 & $\ket{5, {3 \over 2} , {15 \over 4} ,\Gamma_{5,3}}$  & $\frac{3\,J}{2} - t + U + 18\,W$   & 4. \\[0.5ex] 
\hline
\end{tabular} \\[2ex]
\begin{tabular}[t]{|r|l|c|c|}
\multicolumn{4}{c}{\large \bf \boldmath Eigenkets and eigenvalues for ${\rm  N_e}$=6 and   ${\rm m_s}$= $-1$. } \\[0.5ex] \hline
\parbox[c]{1cm}{No}  & \parbox[c]{2.5cm}{\begin{center} Eigenstate \end{center}}  & \parbox[c]{7.5cm}{ \begin{center}   Energy \end{center}}  & \parbox[c]{2cm}{ \begin{center} Example \end{center}}  \\ \hline 
\hline 
220 & $\ket{6,-1,2,\Gamma_{4,1}}$  & $\frac{J}{2} - 2\,t + 2\,U + 26\,W$   & 8. \\[0.5ex] 
221 & $\ket{6,-1,2,\Gamma_{4,2}}$  & $\frac{J}{2} - 2\,t + 2\,U + 26\,W$   & 8. \\[0.5ex] 
222 & $\ket{6,-1,2,\Gamma_{4,3}}$  & $\frac{J}{2} - 2\,t + 2\,U + 26\,W$   & 8. \\[0.5ex] 
223 & $\ket{6,-1,2,\Gamma_{5,1}}$  & $\frac{J}{2} + 2\,t + 2\,U + 26\,W$   & 12. \\[0.5ex] 
224 & $\ket{6,-1,2,\Gamma_{5,2}}$  & $\frac{J}{2} + 2\,t + 2\,U + 26\,W$   & 12. \\[0.5ex] 
225 & $\ket{6,-1,2,\Gamma_{5,3}}$  & $\frac{J}{2} + 2\,t + 2\,U + 26\,W$   & 12. \\[0.5ex] 
\hline
\end{tabular} \\[2ex]
\begin{tabular}[t]{|r|l|c|c|}
\multicolumn{4}{c}{\large \bf \boldmath Eigenkets and eigenvalues for ${\rm  N_e}$=6 and   ${\rm m_s}$= $0$. } \\[0.5ex] \hline
\parbox[c]{1cm}{No}  & \parbox[c]{2.5cm}{\begin{center} Eigenstate \end{center}}  & \parbox[c]{7.5cm}{ \begin{center}   Energy \end{center}}  & \parbox[c]{2cm}{ \begin{center} Example \end{center}}  \\ \hline 
\hline 
226 & $\ket{6,0,0,\Gamma_1}$  & $\frac{-{\sqrt{\Aindex{8}}}}{2} - \frac{J}{2} - 2\,t + \frac{5\,U}{2} + 25\,W$   & 4.82109 \\[0.5ex] 
227 & $\ket{6,0,0,\Gamma_1}$  & $\frac{{\sqrt{\Aindex{8}}}}{2} - \frac{J}{2} - 2\,t + \frac{5\,U}{2} + 25\,W$   & 16.1789 \\[0.5ex] 
228 & $\ket{6,0,0,\Gamma_{3,1}}$  & $-J + 2\,t + 2\,U + 26\,W$   & 12. \\[0.5ex] 
229 & $\ket{6,0,0,\Gamma_{3,2}}$  & $-J + 2\,t + 2\,U + 26\,W$   & 12. \\[0.5ex] 
230 & $\ket{6,0,0,\Gamma_{4,1}}$  & $\frac{-{\sqrt{\Aindex{1}}}}{2} - \frac{J}{2} + \frac{5\,U}{2} + 25\,W$   & 9.29844 \\[0.5ex] 
231 & $\ket{6,0,0,\Gamma_{4,1}}$  & $\frac{{\sqrt{\Aindex{1}}}}{2} - \frac{J}{2} + \frac{5\,U}{2} + 25\,W$   & 15.7016 \\[0.5ex] 
232 & $\ket{6,0,2,\Gamma_{4,1}}$  & $-2\,t + 2\,U + 26\,W$   & 8. \\[0.5ex] 
233 & $\ket{6,0,0,\Gamma_{4,2}}$  & $\frac{-{\sqrt{\Aindex{1}}}}{2} - \frac{J}{2} + \frac{5\,U}{2} + 25\,W$   & 9.29844 \\[0.5ex] 
234 & $\ket{6,0,0,\Gamma_{4,2}}$  & $\frac{{\sqrt{\Aindex{1}}}}{2} - \frac{J}{2} + \frac{5\,U}{2} + 25\,W$   & 15.7016 \\[0.5ex] 
235 & $\ket{6,0,2,\Gamma_{4,2}}$  & $-2\,t + 2\,U + 26\,W$   & 8. \\[0.5ex] 
236 & $\ket{6,0,0,\Gamma_{4,3}}$  & $\frac{-{\sqrt{\Aindex{1}}}}{2} - \frac{J}{2} + \frac{5\,U}{2} + 25\,W$   & 9.29844 \\[0.5ex] 
237 & $\ket{6,0,0,\Gamma_{4,3}}$  & $\frac{{\sqrt{\Aindex{1}}}}{2} - \frac{J}{2} + \frac{5\,U}{2} + 25\,W$   & 15.7016 \\[0.5ex] 
238 & $\ket{6,0,2,\Gamma_{4,3}}$  & $-2\,t + 2\,U + 26\,W$   & 8. \\[0.5ex] 
239 & $\ket{6,0,2,\Gamma_{5,1}}$  & $2\,t + 2\,U + 26\,W$   & 12. \\[0.5ex] 
240 & $\ket{6,0,2,\Gamma_{5,2}}$  & $2\,t + 2\,U + 26\,W$   & 12. \\[0.5ex] 
241 & $\ket{6,0,2,\Gamma_{5,3}}$  & $2\,t + 2\,U + 26\,W$   & 12. \\[0.5ex] 
\hline
\end{tabular} \\[2ex]
\begin{tabular}[t]{|r|l|c|c|}
\multicolumn{4}{c}{\large \bf \boldmath Eigenkets and eigenvalues for ${\rm  N_e}$=6 and   ${\rm m_s}$= $1$. } \\[0.5ex] \hline
\parbox[c]{1cm}{No}  & \parbox[c]{2.5cm}{\begin{center} Eigenstate \end{center}}  & \parbox[c]{7.5cm}{ \begin{center}   Energy \end{center}}  & \parbox[c]{2cm}{ \begin{center} Example \end{center}}  \\ \hline 
\hline 
242 & $\ket{6,1,2,\Gamma_{4,1}}$  & $\frac{J}{2} - 2\,t + 2\,U + 26\,W$   & 8. \\[0.5ex] 
243 & $\ket{6,1,2,\Gamma_{4,2}}$  & $\frac{J}{2} - 2\,t + 2\,U + 26\,W$   & 8. \\[0.5ex] 
244 & $\ket{6,1,2,\Gamma_{4,3}}$  & $\frac{J}{2} - 2\,t + 2\,U + 26\,W$   & 8. \\[0.5ex] 
245 & $\ket{6,1,2,\Gamma_{5,1}}$  & $\frac{J}{2} + 2\,t + 2\,U + 26\,W$   & 12. \\[0.5ex] 
246 & $\ket{6,1,2,\Gamma_{5,2}}$  & $\frac{J}{2} + 2\,t + 2\,U + 26\,W$   & 12. \\[0.5ex] 
247 & $\ket{6,1,2,\Gamma_{5,3}}$  & $\frac{J}{2} + 2\,t + 2\,U + 26\,W$   & 12. \\[0.5ex] 
\hline
\end{tabular} \\[2ex]
\begin{tabular}[t]{|r|l|c|c|}
\multicolumn{4}{c}{\large \bf \boldmath Eigenkets and eigenvalues for ${\rm  N_e}$=7 and   ${\rm m_s}$= $- {1 \over 2} $. } \\[0.5ex] \hline
\parbox[c]{1cm}{No}  & \parbox[c]{2.5cm}{\begin{center} Eigenstate \end{center}}  & \parbox[c]{7.5cm}{ \begin{center}   Energy \end{center}}  & \parbox[c]{2cm}{ \begin{center} Example \end{center}}  \\ \hline 
\hline 
248 & $\ket{7,- {1 \over 2} , {3 \over 4} ,\Gamma_1}$  & $-3\,t + 3\,U + 36\,W$   & 12. \\[0.5ex] 
249 & $\ket{7,- {1 \over 2} , {3 \over 4} ,\Gamma_{4,1}}$  & $t + 3\,U + 36\,W$   & 16. \\[0.5ex] 
250 & $\ket{7,- {1 \over 2} , {3 \over 4} ,\Gamma_{4,2}}$  & $t + 3\,U + 36\,W$   & 16. \\[0.5ex] 
251 & $\ket{7,- {1 \over 2} , {3 \over 4} ,\Gamma_{4,3}}$  & $t + 3\,U + 36\,W$   & 16. \\[0.5ex] 
\hline
\end{tabular} \\[2ex]
\begin{tabular}[t]{|r|l|c|c|}
\multicolumn{4}{c}{\large \bf \boldmath Eigenkets and eigenvalues for ${\rm  N_e}$=7 and   ${\rm m_s}$= $ {1 \over 2} $. } \\[0.5ex] \hline
\parbox[c]{1cm}{No}  & \parbox[c]{2.5cm}{\begin{center} Eigenstate \end{center}}  & \parbox[c]{7.5cm}{ \begin{center}   Energy \end{center}}  & \parbox[c]{2cm}{ \begin{center} Example \end{center}}  \\ \hline 
\hline 
252 & $\ket{7, {1 \over 2} , {3 \over 4} ,\Gamma_1}$  & $-3\,t + 3\,U + 36\,W$   & 12. \\[0.5ex] 
253 & $\ket{7, {1 \over 2} , {3 \over 4} ,\Gamma_{4,1}}$  & $t + 3\,U + 36\,W$   & 16. \\[0.5ex] 
254 & $\ket{7, {1 \over 2} , {3 \over 4} ,\Gamma_{4,2}}$  & $t + 3\,U + 36\,W$   & 16. \\[0.5ex] 
255 & $\ket{7, {1 \over 2} , {3 \over 4} ,\Gamma_{4,3}}$  & $t + 3\,U + 36\,W$   & 16. \\[0.5ex] 
\hline
\end{tabular} \\[2ex]
\begin{tabular}[t]{|r|l|c|c|}
\multicolumn{4}{c}{\large \bf \boldmath Eigenkets and eigenvalues for ${\rm  N_e}$=8 and   ${\rm m_s}$= $0$. } \\[0.5ex] \hline
\parbox[c]{1cm}{No}  & \parbox[c]{2.5cm}{\begin{center} Eigenstate \end{center}}  & \parbox[c]{7.5cm}{ \begin{center}   Energy \end{center}}  & \parbox[c]{2cm}{ \begin{center} Example \end{center}}  \\ \hline 
\hline 
256 & $\ket{8,0,0,\Gamma_1}$  & $4\,U + 48\,W$   & 20. \\[0.5ex] 
\hline
\end{tabular}

%% file: Parts/appendix2b.tex
\parindent0cm
\subsection{\bf List of abbreviations }
\beq 
\Aindex{1} &=& J^2 + 16\,t^2 + 2\,J\,\left( U - 2\,W \right)  + {\left( U - 2\,W \right) }^2
 \nonumber \\
\Aindex{2} &=& J^2 + 48\,t^2 + 2\,J\,\left( U - 2\,W \right)  + {\left( U - 2\,W \right) }^2
 \nonumber \\
\Aindex{3} &=& J^2 + 64\,t^2 + 2\,J\,\left( U - 2\,W \right)  + {\left( U - 2\,W \right) }^2
 \nonumber \\
\Aindex{4} &=& J^2 + 16\,t^2 + J\,\left( -4\,t + 2\,U - 4\,W \right)  - 4\,t\,\left( U - 2\,W \right)  + {\left( U - 2\,W \right) }^2
 \nonumber \\
\Aindex{5} &=& J^2 + 48\,t^2 + J\,\left( -3\,t + 2\,U - 4\,W \right)  - 3\,t\,\left( U - 2\,W \right)  + {\left( U - 2\,W \right) }^2
 \nonumber \\
\Aindex{6} &=& J^2 + 48\,t^2 + J\,\left( 3\,t + 2\,U - 4\,W \right)  + 3\,t\,\left( U - 2\,W \right)  + {\left( U - 2\,W \right) }^2
 \nonumber \\
\Aindex{7} &=& J^2 + 16\,t^2 + 4\,t\,\left( U - 2\,W \right)  + {\left( U - 2\,W \right) }^2 + 2\,J\,\left( 2\,t + U - 2\,W \right) 
 \nonumber \\
\Aindex{8} &=& J^2 + 64\,t^2 + 8\,t\,\left( U - 2\,W \right)  + {\left( U - 2\,W \right) }^2 + 2\,J\,\left( 4\,t + U - 2\,W \right) 
 \nonumber \\
\Aindex{9} &=& 4\,\left( 12\,t^2 - 4\,t\,U + U\,\left( J - 2\,W \right)  \right)  + {\left( -J + 4\,t + U + 2\,W \right) }^2
 \nonumber \\
\Aindex{10} &=& J - 2\,U - 14\,W + {\sqrt{\Aindex{2}}}\,\cos (\Thetaindex{2})
 \nonumber \\
\Aindex{11} &=& J + 4\,U + 64\,W - 4\,{\sqrt{\Aindex{2}}}\,\cos (\Thetaindex{3})
 \nonumber \\
\Aindex{12} &=& J - 2\,U - 32\,W + 2\,{\sqrt{\Aindex{2}}}\,\cos (\Thetaindex{3})
 \nonumber \\
\Aindex{13} &=& -2\,J + 4\,U + 28\,W + {\sqrt{\Aindex{2}}}\,\left( \cos (\Thetaindex{2}) - {\sqrt{3}}\,\sin (\Thetaindex{2}) \right) 
 \nonumber \\
\Aindex{14} &=& -2\,J + 4\,U + 28\,W + {\sqrt{\Aindex{2}}}\,\left( \cos (\Thetaindex{2}) + {\sqrt{3}}\,\sin (\Thetaindex{2}) \right) 
 \nonumber \\
\Aindex{15} &=& -J + 2\,U + 32\,W + {\sqrt{\Aindex{2}}}\,\left( \cos (\Thetaindex{3}) - {\sqrt{3}}\,\sin (\Thetaindex{3}) \right) 
 \nonumber \\
\Aindex{16} &=& \frac{J}{2} + 2\,U + 32\,W + {\sqrt{\Aindex{2}}}\,\left( \cos (\Thetaindex{3}) - {\sqrt{3}}\,\sin (\Thetaindex{3}) \right) 
 \nonumber \\
\Aindex{17} &=& -J + 2\,U + 32\,W + {\sqrt{\Aindex{2}}}\,\left( \cos (\Thetaindex{3}) + {\sqrt{3}}\,\sin (\Thetaindex{3}) \right) 
 \nonumber \\
\Aindex{18} &=& \frac{J}{2} + 2\,U + 32\,W + {\sqrt{\Aindex{2}}}\,\left( \cos (\Thetaindex{3}) + {\sqrt{3}}\,\sin (\Thetaindex{3}) \right) 
 \nonumber \\
\Aindex{19} &=& J - 3\,t - 2\,U - 14\,W + 2\,{\sqrt{\Aindex{5}}}\,\cos (\Thetaindex{4})
 \nonumber \\
\Aindex{20} &=& -J + 3\,t + 2\,U + 14\,W + {\sqrt{\Aindex{5}}}\,\left( \cos (\Thetaindex{4}) - {\sqrt{3}}\,\sin (\Thetaindex{4}) \right) 
 \nonumber \\
\Aindex{21} &=& -J + 3\,t + 2\,U + 14\,W + {\sqrt{\Aindex{5}}}\,\left( \cos (\Thetaindex{4}) + {\sqrt{3}}\,\sin (\Thetaindex{4}) \right) 
 \nonumber \\
\Aindex{22} &=& J + 3\,t - 5\,U - 50\,W + 2\,{\sqrt{\Aindex{6}}}\,\cos (\Thetaindex{5})
 \nonumber \\
\Aindex{23} &=& J + 3\,t - 5\,U - 50\,W + {\sqrt{\Aindex{6}}}\,\left( -\cos (\Thetaindex{5}) + {\sqrt{3}}\,\sin (\Thetaindex{5}) \right) 
 \nonumber \\
\Aindex{24} &=& J + 3\,t - 5\,U - 50\,W - {\sqrt{\Aindex{6}}}\,\left( \cos (\Thetaindex{5}) + {\sqrt{3}}\,\sin (\Thetaindex{5}) \right) 
 \nonumber \\
\eeq \newpage \beq
\Thetaindex{1} &=& \frac{\arccos (\frac{12\,{\sqrt{3}}\,t^2\,\left( J + U - 2\,W \right) }{{\Aindex{1}}^{\frac{3}{2}}})}{3}
 \nonumber \\[1ex]
\Thetaindex{2} &=& \frac{\arccos (-\left( \frac{\left( J^2 - 36\,t^2 + 2\,J\,\left( U - 2\,W \right)  + {\left( U - 2\,W \right) }^2 \right) \,\left( J + U - 2\,W \right) }{{\Aindex{2}}^{\frac{3}{2}}} \right) )}{3}
 \nonumber \\[1ex]
\Thetaindex{3} &=& \frac{\arccos (\frac{\left( J^2 - 36\,t^2 + 2\,J\,\left( U - 2\,W \right)  + {\left( U - 2\,W \right) }^2 \right) \,\left( J + U - 2\,W \right) }{{\Aindex{2}}^{\frac{3}{2}}})}{3}
 \nonumber \\[1ex]
\Thetaindex{4} &=& \frac{\arccos (\frac{\left( 2\,J^2 + 36\,t^2 + J\,\left( -9\,t + 4\,U - 8\,W \right)  - 9\,t\,\left( U - 2\,W \right)  + 2\,{\left( U - 2\,W \right) }^2 \right) \,\left( J + U - 2\,W \right) }{2\,{\Aindex{5}}^{\frac{3}{2}}})}{3}
 \nonumber \\[1ex]
\Thetaindex{5} &=& \frac{\arccos (\frac{\left( 2\,J^2 + 36\,t^2 + J\,\left( 9\,t + 4\,U - 8\,W \right)  + 9\,t\,\left( U - 2\,W \right)  + 2\,{\left( U - 2\,W \right) }^2 \right) \,\left( J + U - 2\,W \right) }{2\,{\Aindex{6}}^{\frac{3}{2}}})}{3}
\nonumber \eeq

%% file: Parts/appendix2c.tex
\subsection{\bf The eigenvectors of the tetrahedron}
In the following the unnormalised eigenvectors are given in abbreviated form, whereby
the indices give the number of the state in accordance with the numbering of
the energy levels given above.
For getting the normalised eigenvectors the coefficients have to be divided by the
normalisation constant $N_i$, which is given for the vectors  containing more than one coefficient.
\mathindent0cm\mathindent0cm
{\subsection*{\boldmath Unnormalized eigenvectors for ${\rm  N_e}=0$ and   ${\rm m_s}$= $0$.}
\beq
\ket{\Psi}_{1}& = &\ket{0,0,0,\Gamma_1} \nonumber \\ 
&=& 1
 \left ( \ket{0000}\right) \nonumber 
\eeq
{\subsection*{\boldmath Unnormalized eigenvectors for ${\rm  N_e}=1$ and   ${\rm m_s}$= $- {1 \over 2} $.}
\beq
\ket{\Psi}_{2}& = &\ket{1,- {1 \over 2} , {3 \over 4} ,\Gamma_1} \nonumber \\ 
&=& \frac{1}{2}
 \left ( \ket{000d} + \ket{00d0} + \ket{0d00} + \ket{d000}\right) \nonumber 
\eeq
\beq
\ket{\Psi}_{3}& = &\ket{1,- {1 \over 2} , {3 \over 4} ,\Gamma_{4,1}} \nonumber \\ 
&=& \frac{1}{2}
 \left ( \ket{000d} + \ket{00d0} - \ket{0d00} - \ket{d000}\right) \nonumber 
\eeq
\beq
\ket{\Psi}_{4}& = &\ket{1,- {1 \over 2} , {3 \over 4} ,\Gamma_{4,2}} \nonumber \\ 
&=& \frac{1}{2}
 \left ( \ket{000d} - \ket{00d0} - \ket{0d00} + \ket{d000}\right) \nonumber 
\eeq
\beq
\ket{\Psi}_{5}& = &\ket{1,- {1 \over 2} , {3 \over 4} ,\Gamma_{4,3}} \nonumber \\ 
&=& \frac{1}{2}
 \left ( \ket{000d} - \ket{00d0} + \ket{0d00} - \ket{d000}\right) \nonumber 
\eeq
{\subsection*{\boldmath Unnormalized eigenvectors for ${\rm  N_e}=1$ and   ${\rm m_s}$= $ {1 \over 2} $.}
\beq
\ket{\Psi}_{6}& = &\ket{1, {1 \over 2} , {3 \over 4} ,\Gamma_1} \nonumber \\ 
&=& \frac{1}{2}
 \left ( \ket{000u} + \ket{00u0} + \ket{0u00} + \ket{u000}\right) \nonumber 
\eeq
\beq
\ket{\Psi}_{7}& = &\ket{1, {1 \over 2} , {3 \over 4} ,\Gamma_{4,1}} \nonumber \\ 
&=& \frac{1}{2}
 \left ( \ket{000u} + \ket{00u0} - \ket{0u00} - \ket{u000}\right) \nonumber 
\eeq
\beq
\ket{\Psi}_{8}& = &\ket{1, {1 \over 2} , {3 \over 4} ,\Gamma_{4,2}} \nonumber \\ 
&=& \frac{1}{2}
 \left ( \ket{000u} - \ket{00u0} - \ket{0u00} + \ket{u000}\right) \nonumber 
\eeq
\beq
\ket{\Psi}_{9}& = &\ket{1, {1 \over 2} , {3 \over 4} ,\Gamma_{4,3}} \nonumber \\ 
&=& \frac{1}{2}
 \left ( \ket{000u} - \ket{00u0} + \ket{0u00} - \ket{u000}\right) \nonumber 
\eeq
{\subsection*{\boldmath Unnormalized eigenvectors for ${\rm  N_e}=2$ and   ${\rm m_s}$= $-1$.}
\beq
\ket{\Psi}_{10}& = &\ket{2,-1,2,\Gamma_{4,1}} \nonumber \\ 
&=& \frac{1}{2}
 \left ( \ket{0d0d} + \ket{0dd0} + \ket{d00d} + \ket{d0d0}\right) \nonumber 
\eeq
\beq
\ket{\Psi}_{11}& = &\ket{2,-1,2,\Gamma_{4,2}} \nonumber \\ 
&=& \frac{1}{2}
 \left ( \ket{00dd} + \ket{0d0d} - \ket{d0d0} - \ket{dd00}\right) \nonumber 
\eeq
\beq
\ket{\Psi}_{12}& = &\ket{2,-1,2,\Gamma_{4,3}} \nonumber \\ 
&=& \frac{1}{2}
 \left ( \ket{00dd} - \ket{0dd0} + \ket{d00d} + \ket{dd00}\right) \nonumber 
\eeq
\beq
\ket{\Psi}_{13}& = &\ket{2,-1,2,\Gamma_{5,1}} \nonumber \\ 
&=& \frac{1}{2}
 \left ( \ket{00dd} + \ket{0dd0} - \ket{d00d} + \ket{dd00}\right) \nonumber 
\eeq
\beq
\ket{\Psi}_{14}& = &\ket{2,-1,2,\Gamma_{5,2}} \nonumber \\ 
&=& \frac{1}{2}
 \left ( \ket{0d0d} - \ket{0dd0} - \ket{d00d} + \ket{d0d0}\right) \nonumber 
\eeq
\beq
\ket{\Psi}_{15}& = &\ket{2,-1,2,\Gamma_{5,3}} \nonumber \\ 
&=& \frac{1}{2}
 \left ( \ket{00dd} - \ket{0d0d} + \ket{d0d0} - \ket{dd00}\right) \nonumber 
\eeq
{\subsection*{\boldmath Unnormalized eigenvectors for ${\rm  N_e}=2$ and   ${\rm m_s}$= $0$.}
\beq
\ket{\Psi}_{16}& = &\ket{2,0,0,\Gamma_1} \nonumber \\ 
& = &
\Cindex{16}{1} \left ( 
\ket{0002} + \ket{0020} + \ket{0200} + \ket{2000}\right) 
 \nonumber \\
& + &
\Cindex{16}{2} \left ( 
\ket{00du} - \ket{00ud} + \ket{0d0u} + \ket{0du0} - \ket{0u0d} - \ket{0ud0}\right . \nonumber \\
&& \hspace{3em} 
 + 
\left . \ket{d00u} + \ket{d0u0} + \ket{du00} - \ket{u00d} - \ket{u0d0} - \ket{ud00}\right ) 
\nonumber \eeq
\beq
C_{16-1} &=&
{\sqrt{3}}\,t \nonumber \\
C_{16-2} &=&
\frac{1}{4\,{\sqrt{3}}}
\, \left ( {\sqrt{\Aindex{9}}} + J - 4\,t + U - 2\,W  \right ) \nonumber \\
 N_{16} &=& 2\,{\sqrt{{\Cindex{16}{1}}^2 + 3\,{\Cindex{16}{2}}^2}} \nonumber \eeq 
\beq
\ket{\Psi}_{17}& = &\ket{2,0,0,\Gamma_1} \nonumber \\ 
& = &
\Cindex{17}{1} \left ( 
\ket{0002} + \ket{0020} + \ket{0200} + \ket{2000}\right) 
 \nonumber \\
& + &
\Cindex{17}{2} \left ( 
\ket{00du} - \ket{00ud} + \ket{0d0u} + \ket{0du0} - \ket{0u0d} - \ket{0ud0}\right . \nonumber \\
&& \hspace{3em} 
 + 
\left . \ket{d00u} + \ket{d0u0} + \ket{du00} - \ket{u00d} - \ket{u0d0} - \ket{ud00}\right ) 
\nonumber \eeq
\beq
C_{17-1} &=&
{\sqrt{3}}\,t \nonumber \\
C_{17-2} &=&
\frac{-1}{4\,{\sqrt{3}}}
\, \left ( {\sqrt{\Aindex{9}}} - J + 4\,t - U + 2\,W  \right ) \nonumber \\
 N_{17} &=& 2\,{\sqrt{{\Cindex{17}{1}}^2 + 3\,{\Cindex{17}{2}}^2}} \nonumber \eeq 
\beq
\ket{\Psi}_{18}& = &\ket{2,0,0,\Gamma_{3,1}} \nonumber \\ 
& = &
\Cindex{18}{1} \left ( 
\ket{00du} - \ket{00ud} + \ket{0du0} - \ket{0ud0} + \ket{d00u} + \ket{du00} - \ket{u00d} - \ket{ud00}\right) 
 \nonumber \\
& + &
\Cindex{18}{2} \left ( 
\ket{0d0u} - \ket{0u0d} + \ket{d0u0} - \ket{u0d0}\right) 
\nonumber \eeq
\beq
C_{18-1} &=&
\frac{-1}{2\,{\sqrt{6}}} \nonumber \\
C_{18-2} &=&
\frac{1}{{\sqrt{6}}} \nonumber \\
 N_{18} &=& 2\,{\sqrt{2\,{\Cindex{18}{1}}^2 + {\Cindex{18}{2}}^2}} \nonumber \eeq 
\beq
\ket{\Psi}_{19}& = &\ket{2,0,0,\Gamma_{3,2}} \nonumber \\ 
&=& \frac{1}{2\,{\sqrt{2}}}
 \left ( \ket{00du} - \ket{00ud} - \ket{0du0} + \ket{0ud0} - \ket{d00u} + \ket{du00} + \ket{u00d} - \ket{ud00}\right) \nonumber 
\eeq
\beq
\ket{\Psi}_{20}& = &\ket{2,0,0,\Gamma_{4,1}} \nonumber \\ 
& = &
\Cindex{20}{1} \left ( 
\ket{0002} + \ket{0020} - \ket{0200} - \ket{2000}\right) 
 \nonumber \\
& + &
\Cindex{20}{2} \left ( 
\ket{00du} - \ket{00ud} - \ket{du00} + \ket{ud00}\right) 
\nonumber \eeq
\beq
C_{20-1} &=&
-t \nonumber \\
C_{20-2} &=&
\frac{1}{4}
\, \left ( -{\sqrt{\Aindex{1}}} - J - U + 2\,W  \right ) \nonumber \\
 N_{20} &=& 2\,{\sqrt{{\Cindex{20}{1}}^2 + {\Cindex{20}{2}}^2}} \nonumber \eeq 
\beq
\ket{\Psi}_{21}& = &\ket{2,0,0,\Gamma_{4,1}} \nonumber \\ 
& = &
\Cindex{21}{1} \left ( 
\ket{0002} + \ket{0020} - \ket{0200} - \ket{2000}\right) 
 \nonumber \\
& + &
\Cindex{21}{2} \left ( 
\ket{00du} - \ket{00ud} - \ket{du00} + \ket{ud00}\right) 
\nonumber \eeq
\beq
C_{21-1} &=&
-t \nonumber \\
C_{21-2} &=&
\frac{1}{4}
\, \left ( {\sqrt{\Aindex{1}}} - J - U + 2\,W  \right ) \nonumber \\
 N_{21} &=& 2\,{\sqrt{{\Cindex{21}{1}}^2 + {\Cindex{21}{2}}^2}} \nonumber \eeq 
\beq
\ket{\Psi}_{22}& = &\ket{2,0,2,\Gamma_{4,1}} \nonumber \\ 
&=& \frac{1}{2\,{\sqrt{2}}}
 \left ( \ket{0d0u} + \ket{0du0} + \ket{0u0d} + \ket{0ud0} + \ket{d00u} + \ket{d0u0} + \ket{u00d} + \ket{u0d0}\right) \nonumber 
\eeq
\beq
\ket{\Psi}_{23}& = &\ket{2,0,0,\Gamma_{4,2}} \nonumber \\ 
& = &
\Cindex{23}{1} \left ( 
\ket{0002} - \ket{0020} - \ket{0200} + \ket{2000}\right) 
 \nonumber \\
& + &
\Cindex{23}{2} \left ( 
\ket{0du0} - \ket{0ud0} - \ket{d00u} + \ket{u00d}\right) 
\nonumber \eeq
\beq
C_{23-1} &=&
t \nonumber \\
C_{23-2} &=&
\frac{1}{4}
\, \left ( -{\sqrt{\Aindex{1}}} - J - U + 2\,W  \right ) \nonumber \\
 N_{23} &=& 2\,{\sqrt{{\Cindex{23}{1}}^2 + {\Cindex{23}{2}}^2}} \nonumber \eeq 
\beq
\ket{\Psi}_{24}& = &\ket{2,0,0,\Gamma_{4,2}} \nonumber \\ 
& = &
\Cindex{24}{1} \left ( 
\ket{0002} - \ket{0020} - \ket{0200} + \ket{2000}\right) 
 \nonumber \\
& + &
\Cindex{24}{2} \left ( 
\ket{0du0} - \ket{0ud0} - \ket{d00u} + \ket{u00d}\right) 
\nonumber \eeq
\beq
C_{24-1} &=&
t \nonumber \\
C_{24-2} &=&
\frac{1}{4}
\, \left ( {\sqrt{\Aindex{1}}} - J - U + 2\,W  \right ) \nonumber \\
 N_{24} &=& 2\,{\sqrt{{\Cindex{24}{1}}^2 + {\Cindex{24}{2}}^2}} \nonumber \eeq 
\beq
\ket{\Psi}_{25}& = &\ket{2,0,2,\Gamma_{4,2}} \nonumber \\ 
&=& \frac{1}{2\,{\sqrt{2}}}
 \left ( \ket{00du} + \ket{00ud} + \ket{0d0u} + \ket{0u0d} - \ket{d0u0} - \ket{du00} - \ket{u0d0} - \ket{ud00}\right) \nonumber 
\eeq
\beq
\ket{\Psi}_{26}& = &\ket{2,0,0,\Gamma_{4,3}} \nonumber \\ 
& = &
\Cindex{26}{1} \left ( 
\ket{0002} - \ket{0020} + \ket{0200} - \ket{2000}\right) 
 \nonumber \\
& + &
\Cindex{26}{2} \left ( 
\ket{0d0u} - \ket{0u0d} - \ket{d0u0} + \ket{u0d0}\right) 
\nonumber \eeq
\beq
C_{26-1} &=&
-t \nonumber \\
C_{26-2} &=&
\frac{1}{4}
\, \left ( -{\sqrt{\Aindex{1}}} - J - U + 2\,W  \right ) \nonumber \\
 N_{26} &=& 2\,{\sqrt{{\Cindex{26}{1}}^2 + {\Cindex{26}{2}}^2}} \nonumber \eeq 
\beq
\ket{\Psi}_{27}& = &\ket{2,0,0,\Gamma_{4,3}} \nonumber \\ 
& = &
\Cindex{27}{1} \left ( 
\ket{0002} - \ket{0020} + \ket{0200} - \ket{2000}\right) 
 \nonumber \\
& + &
\Cindex{27}{2} \left ( 
\ket{0d0u} - \ket{0u0d} - \ket{d0u0} + \ket{u0d0}\right) 
\nonumber \eeq
\beq
C_{27-1} &=&
-t \nonumber \\
C_{27-2} &=&
\frac{1}{4}
\, \left ( {\sqrt{\Aindex{1}}} - J - U + 2\,W  \right ) \nonumber \\
 N_{27} &=& 2\,{\sqrt{{\Cindex{27}{1}}^2 + {\Cindex{27}{2}}^2}} \nonumber \eeq 
\beq
\ket{\Psi}_{28}& = &\ket{2,0,2,\Gamma_{4,3}} \nonumber \\ 
&=& \frac{1}{2\,{\sqrt{2}}}
 \left ( \ket{00du} + \ket{00ud} - \ket{0du0} - \ket{0ud0} + \ket{d00u} + \ket{du00} + \ket{u00d} + \ket{ud00}\right) \nonumber 
\eeq
\beq
\ket{\Psi}_{29}& = &\ket{2,0,2,\Gamma_{5,1}} \nonumber \\ 
&=& \frac{1}{2\,{\sqrt{2}}}
 \left ( \ket{00du} + \ket{00ud} + \ket{0du0} + \ket{0ud0} - \ket{d00u} + \ket{du00} - \ket{u00d} + \ket{ud00}\right) \nonumber 
\eeq
\beq
\ket{\Psi}_{30}& = &\ket{2,0,2,\Gamma_{5,2}} \nonumber \\ 
&=& \frac{1}{2\,{\sqrt{2}}}
 \left ( \ket{0d0u} - \ket{0du0} + \ket{0u0d} - \ket{0ud0} - \ket{d00u} + \ket{d0u0} - \ket{u00d} + \ket{u0d0}\right) \nonumber 
\eeq
\beq
\ket{\Psi}_{31}& = &\ket{2,0,2,\Gamma_{5,3}} \nonumber \\ 
&=& \frac{1}{2\,{\sqrt{2}}}
 \left ( \ket{00du} + \ket{00ud} - \ket{0d0u} - \ket{0u0d} + \ket{d0u0} - \ket{du00} + \ket{u0d0} - \ket{ud00}\right) \nonumber 
\eeq
{\subsection*{\boldmath Unnormalized eigenvectors for ${\rm  N_e}=2$ and   ${\rm m_s}$= $1$.}
\beq
\ket{\Psi}_{32}& = &\ket{2,1,2,\Gamma_{4,1}} \nonumber \\ 
&=& \frac{1}{2}
 \left ( \ket{0u0u} + \ket{0uu0} + \ket{u00u} + \ket{u0u0}\right) \nonumber 
\eeq
\beq
\ket{\Psi}_{33}& = &\ket{2,1,2,\Gamma_{4,2}} \nonumber \\ 
&=& \frac{1}{2}
 \left ( \ket{00uu} + \ket{0u0u} - \ket{u0u0} - \ket{uu00}\right) \nonumber 
\eeq
\beq
\ket{\Psi}_{34}& = &\ket{2,1,2,\Gamma_{4,3}} \nonumber \\ 
&=& \frac{1}{2}
 \left ( \ket{00uu} - \ket{0uu0} + \ket{u00u} + \ket{uu00}\right) \nonumber 
\eeq
\beq
\ket{\Psi}_{35}& = &\ket{2,1,2,\Gamma_{5,1}} \nonumber \\ 
&=& \frac{1}{2}
 \left ( \ket{00uu} + \ket{0uu0} - \ket{u00u} + \ket{uu00}\right) \nonumber 
\eeq
\beq
\ket{\Psi}_{36}& = &\ket{2,1,2,\Gamma_{5,2}} \nonumber \\ 
&=& \frac{1}{2}
 \left ( \ket{0u0u} - \ket{0uu0} - \ket{u00u} + \ket{u0u0}\right) \nonumber 
\eeq
\beq
\ket{\Psi}_{37}& = &\ket{2,1,2,\Gamma_{5,3}} \nonumber \\ 
&=& \frac{1}{2}
 \left ( \ket{00uu} - \ket{0u0u} + \ket{u0u0} - \ket{uu00}\right) \nonumber 
\eeq
{\subsection*{\boldmath Unnormalized eigenvectors for ${\rm  N_e}=3$ and   ${\rm m_s}$= $- {3 \over 2} $.}
\beq
\ket{\Psi}_{38}& = &\ket{3,- {3 \over 2} , {15 \over 4} ,\Gamma_2} \nonumber \\ 
&=& \frac{1}{2}
 \left ( \ket{0ddd} - \ket{d0dd} + \ket{dd0d} - \ket{ddd0}\right) \nonumber 
\eeq
\beq
\ket{\Psi}_{39}& = &\ket{3,- {3 \over 2} , {15 \over 4} ,\Gamma_{5,1}} \nonumber \\ 
&=& \frac{1}{2}
 \left ( \ket{0ddd} + \ket{d0dd} + \ket{dd0d} + \ket{ddd0}\right) \nonumber 
\eeq
\beq
\ket{\Psi}_{40}& = &\ket{3,- {3 \over 2} , {15 \over 4} ,\Gamma_{5,2}} \nonumber \\ 
&=& \frac{1}{2}
 \left ( \ket{0ddd} - \ket{d0dd} - \ket{dd0d} + \ket{ddd0}\right) \nonumber 
\eeq
\beq
\ket{\Psi}_{41}& = &\ket{3,- {3 \over 2} , {15 \over 4} ,\Gamma_{5,3}} \nonumber \\ 
&=& \frac{1}{2}
 \left ( \ket{0ddd} + \ket{d0dd} - \ket{dd0d} - \ket{ddd0}\right) \nonumber 
\eeq
{\subsection*{\boldmath Unnormalized eigenvectors for ${\rm  N_e}=3$ and   ${\rm m_s}$= $- {1 \over 2} $.}
\beq
\ket{\Psi}_{42}& = &\ket{3,- {1 \over 2} , {3 \over 4} ,\Gamma_1} \nonumber \\ 
&=& \frac{1}{2\,{\sqrt{3}}}
 \left ( \ket{002d} + \ket{00d2} + \ket{020d} + \ket{02d0} + \ket{0d02} + \ket{0d20}\right . \nonumber \\
&& \hspace{1em} 
 + 
\left . \ket{200d} + \ket{20d0} + \ket{2d00} + \ket{d002} + \ket{d020} + \ket{d200}\right ) \nonumber  
\eeq
\beq
\ket{\Psi}_{43}& = &\ket{3,- {1 \over 2} , {15 \over 4} ,\Gamma_2} \nonumber \\ 
&=& \frac{1}{2\,{\sqrt{3}}}
 \left ( \ket{0ddu} + \ket{0dud} + \ket{0udd} - \ket{d0du} - \ket{d0ud} + \ket{dd0u}\right . \nonumber \\
&& \hspace{1em} 
\left . -\ket{ddu0} + \ket{du0d} - \ket{dud0} - \ket{u0dd} + \ket{ud0d} - \ket{udd0}\right ) \nonumber  
\eeq
\beq
\ket{\Psi}_{44}& = &\ket{3,- {1 \over 2} , {3 \over 4} ,\Gamma_{3,1}} \nonumber \\ 
& = &
\Cindex{44}{1} \left ( 
\ket{002d} + \ket{00d2} + \ket{02d0} + \ket{0d20} + \ket{200d} + \ket{2d00} + \ket{d002} + \ket{d200}\right) 
 \nonumber \\
& + &
\Cindex{44}{2} \left ( 
\ket{020d} + \ket{0d02} + \ket{20d0} + \ket{d020}\right) 
 \nonumber \\
& + &
\Cindex{44}{3} \left ( 
\ket{0ddu} - \ket{0udd} - \ket{d0ud} - \ket{dd0u} + \ket{ddu0} + \ket{du0d} + \ket{u0dd} - \ket{udd0}\right) 
\nonumber \eeq
\beq
C_{44-1} &=&
\frac{-t}{2\,{\sqrt{2}}} \nonumber \\
C_{44-2} &=&
\frac{t}{{\sqrt{2}}} \nonumber \\
C_{44-3} &=&
\frac{1}{4\,{\sqrt{2}}}
\, \left ( {\sqrt{\Aindex{4}}} + J - 2\,t + U - 2\,W  \right ) \nonumber \\
 N_{44} &=& 2\,{\sqrt{2\,{\Cindex{44}{1}}^2 + {\Cindex{44}{2}}^2 + 2\,{\Cindex{44}{3}}^2}} \nonumber \eeq 
\beq
\ket{\Psi}_{45}& = &\ket{3,- {1 \over 2} , {3 \over 4} ,\Gamma_{3,1}} \nonumber \\ 
& = &
\Cindex{45}{1} \left ( 
\ket{002d} + \ket{00d2} + \ket{02d0} + \ket{0d20} + \ket{200d} + \ket{2d00} + \ket{d002} + \ket{d200}\right) 
 \nonumber \\
& + &
\Cindex{45}{2} \left ( 
\ket{020d} + \ket{0d02} + \ket{20d0} + \ket{d020}\right) 
 \nonumber \\
& + &
\Cindex{45}{3} \left ( 
\ket{0ddu} - \ket{0udd} - \ket{d0ud} - \ket{dd0u} + \ket{ddu0} + \ket{du0d} + \ket{u0dd} - \ket{udd0}\right) 
\nonumber \eeq
\beq
C_{45-1} &=&
\frac{-t}{2\,{\sqrt{2}}} \nonumber \\
C_{45-2} &=&
\frac{t}{{\sqrt{2}}} \nonumber \\
C_{45-3} &=&
\frac{-1}{4\,{\sqrt{2}}}
\, \left ( {\sqrt{\Aindex{4}}} - J + 2\,t - U + 2\,W  \right ) \nonumber \\
 N_{45} &=& 2\,{\sqrt{2\,{\Cindex{45}{1}}^2 + {\Cindex{45}{2}}^2 + 2\,{\Cindex{45}{3}}^2}} \nonumber \eeq 
\beq
\ket{\Psi}_{46}& = &\ket{3,- {1 \over 2} , {3 \over 4} ,\Gamma_{3,2}} \nonumber \\ 
& = &
\Cindex{46}{1} \left ( 
\ket{002d} + \ket{00d2} - \ket{02d0} - \ket{0d20} - \ket{200d} + \ket{2d00} - \ket{d002} + \ket{d200}\right) 
 \nonumber \\
& + &
\Cindex{46}{2} \left ( 
\ket{0ddu} + \ket{0udd} - \ket{d0ud} + \ket{dd0u} - \ket{ddu0} + \ket{du0d} - \ket{u0dd} - \ket{udd0}\right) 
 \nonumber \\
& + &
\Cindex{46}{3} \left ( 
\ket{0dud} - \ket{d0du} - \ket{dud0} + \ket{ud0d}\right) 
\nonumber \eeq
\beq
C_{46-1} &=&
\frac{-\left( {\sqrt{\frac{3}{2}}}\,t \right) }{2} \nonumber \\
C_{46-2} &=&
\frac{1}{4\,{\sqrt{6}}}
\, \left ( {\sqrt{\Aindex{4}}} + J - 2\,t + U - 2\,W  \right ) \nonumber \\
C_{46-3} &=&
\frac{-1}{2\,{\sqrt{6}}}
\, \left ( {\sqrt{\Aindex{4}}} + J - 2\,t + U - 2\,W  \right ) \nonumber \\
 N_{46} &=& 2\,{\sqrt{2\,{\Cindex{46}{1}}^2 + 2\,{\Cindex{46}{2}}^2 + {\Cindex{46}{3}}^2}} \nonumber \eeq 
\beq
\ket{\Psi}_{47}& = &\ket{3,- {1 \over 2} , {3 \over 4} ,\Gamma_{3,2}} \nonumber \\ 
& = &
\Cindex{47}{1} \left ( 
\ket{002d} + \ket{00d2} - \ket{02d0} - \ket{0d20} - \ket{200d} + \ket{2d00} - \ket{d002} + \ket{d200}\right) 
 \nonumber \\
& + &
\Cindex{47}{2} \left ( 
\ket{0ddu} + \ket{0udd} - \ket{d0ud} + \ket{dd0u} - \ket{ddu0} + \ket{du0d} - \ket{u0dd} - \ket{udd0}\right) 
 \nonumber \\
& + &
\Cindex{47}{3} \left ( 
\ket{0dud} - \ket{d0du} - \ket{dud0} + \ket{ud0d}\right) 
\nonumber \eeq
\beq
C_{47-1} &=&
\frac{-\left( {\sqrt{\frac{3}{2}}}\,t \right) }{2} \nonumber \\
C_{47-2} &=&
\frac{-1}{4\,{\sqrt{6}}}
\, \left ( {\sqrt{\Aindex{4}}} - J + 2\,t - U + 2\,W  \right ) \nonumber \\
C_{47-3} &=&
\frac{1}{2\,{\sqrt{6}}}
\, \left ( {\sqrt{\Aindex{4}}} - J + 2\,t - U + 2\,W  \right ) \nonumber \\
 N_{47} &=& 2\,{\sqrt{2\,{\Cindex{47}{1}}^2 + 2\,{\Cindex{47}{2}}^2 + {\Cindex{47}{3}}^2}} \nonumber \eeq 
\beq
\ket{\Psi}_{48}& = &\ket{3,- {1 \over 2} , {3 \over 4} ,\Gamma_{4,1}} \nonumber \\ 
& = &
\Cindex{48}{1} \left ( 
\ket{002d} + \ket{00d2} - \ket{2d00} - \ket{d200}\right) 
 \nonumber \\
& + &
\Cindex{48}{2} \left ( 
\ket{020d} + \ket{02d0} - \ket{0d02} - \ket{0d20} + \ket{200d} + \ket{20d0} - \ket{d002} - \ket{d020}\right) 
 \nonumber \\
& + &
\Cindex{48}{3} \left ( 
\ket{0ddu} - \ket{0dud} + \ket{d0du} - \ket{d0ud} - \ket{du0d} - \ket{dud0} + \ket{ud0d} + \ket{udd0}\right) 
\nonumber \eeq
\beq
C_{48-1} &=&
\frac{-t}{3\,{\sqrt{2}}}
\, \left ( J + 12\,t + U - 2\,W + 2\,{\sqrt{\Aindex{5}}}\,\cos (\Thetaindex{4})  \right ) \nonumber \\
C_{48-2} &=&
\frac{-t}{2\,{\sqrt{2}}}
\, \left ( J - 4\,t + U - 2\,W + 2\,{\sqrt{\Aindex{5}}}\,\cos (\Thetaindex{4})  \right ) \nonumber \\
C_{48-3} &=&
\frac{1}{18\,{\sqrt{2}}}
\,\left ( {\Aindex{19}}^2 + 3\,J\,t - 45\,t^2 + 6\,J\,U - 15\,t\,U - 3\,U^2 + 24\,J\,W - 78\,t\,W - 60\,U\,W\right . \nonumber \\
&& \hspace{1cm} 
 \left . -192\,W^2 + 6\,{\sqrt{\Aindex{5}}}\,t\,\cos (\Thetaindex{4})
\right .  \nonumber \\
&& \hspace{1cm} 
 + 
 \left . 12\,{\sqrt{\Aindex{5}}}\,U\,\cos (\Thetaindex{4}) + 48\,{\sqrt{\Aindex{5}}}\,W\,\cos (\Thetaindex{4})
\right )  \nonumber \\
 N_{48} &=& 2\,{\sqrt{{\Cindex{48}{1}}^2 + 2\,\left( {\Cindex{48}{2}}^2 + {\Cindex{48}{3}}^2 \right) }} \nonumber \eeq 
\beq
\ket{\Psi}_{49}& = &\ket{3,- {1 \over 2} , {3 \over 4} ,\Gamma_{4,1}} \nonumber \\ 
& = &
\Cindex{49}{1} \left ( 
\ket{002d} + \ket{00d2} - \ket{2d00} - \ket{d200}\right) 
 \nonumber \\
& + &
\Cindex{49}{2} \left ( 
\ket{020d} + \ket{02d0} - \ket{0d02} - \ket{0d20} + \ket{200d} + \ket{20d0} - \ket{d002} - \ket{d020}\right) 
 \nonumber \\
& + &
\Cindex{49}{3} \left ( 
\ket{0ddu} - \ket{0dud} + \ket{d0du} - \ket{d0ud} - \ket{du0d} - \ket{dud0} + \ket{ud0d} + \ket{udd0}\right) 
\nonumber \eeq
\beq
C_{49-1} &=&
\frac{-t}{3\,{\sqrt{2}}}
\, \left ( J + 12\,t + U - 2\,W - {\sqrt{\Aindex{5}}}\,\cos (\Thetaindex{4}) - {\sqrt{3}}\,{\sqrt{\Aindex{5}}}\,\sin (\Thetaindex{4})  \right ) \nonumber \\
C_{49-2} &=&
\frac{-t}{2\,{\sqrt{2}}}
\, \left ( J - 4\,t + U - 2\,W - {\sqrt{\Aindex{5}}}\,\cos (\Thetaindex{4}) - {\sqrt{3}}\,{\sqrt{\Aindex{5}}}\,\sin (\Thetaindex{4})  \right ) \nonumber \\
C_{49-3} &=&
\frac{1}{18\,{\sqrt{2}}}
\,\left ( {\Aindex{21}}^2 - 3\,\Aindex{21}\,t - 36\,t^2 - 6\,\Aindex{21}\,U\right . \nonumber \\
&& \hspace{1cm} 
 + 
 \left . 9\,t\,U + 9\,U^2 - 24\,\Aindex{21}\,W + 36\,t\,W + 72\,U\,W + 144\,W^2
\right  )  \nonumber \\
 N_{49} &=& 2\,{\sqrt{{\Cindex{49}{1}}^2 + 2\,\left( {\Cindex{49}{2}}^2 + {\Cindex{49}{3}}^2 \right) }} \nonumber \eeq 
\beq
\ket{\Psi}_{50}& = &\ket{3,- {1 \over 2} , {3 \over 4} ,\Gamma_{4,1}} \nonumber \\ 
& = &
\Cindex{50}{1} \left ( 
\ket{002d} + \ket{00d2} - \ket{2d00} - \ket{d200}\right) 
 \nonumber \\
& + &
\Cindex{50}{2} \left ( 
\ket{020d} + \ket{02d0} - \ket{0d02} - \ket{0d20} + \ket{200d} + \ket{20d0} - \ket{d002} - \ket{d020}\right) 
 \nonumber \\
& + &
\Cindex{50}{3} \left ( 
\ket{0ddu} - \ket{0dud} + \ket{d0du} - \ket{d0ud} - \ket{du0d} - \ket{dud0} + \ket{ud0d} + \ket{udd0}\right) 
\nonumber \eeq
\beq
C_{50-1} &=&
\frac{-t}{3\,{\sqrt{2}}}
\, \left ( J + 12\,t + U - 2\,W - {\sqrt{\Aindex{5}}}\,\cos (\Thetaindex{4}) + {\sqrt{3}}\,{\sqrt{\Aindex{5}}}\,\sin (\Thetaindex{4})  \right ) \nonumber \\
C_{50-2} &=&
\frac{-t}{2\,{\sqrt{2}}}
\, \left ( J - 4\,t + U - 2\,W - {\sqrt{\Aindex{5}}}\,\cos (\Thetaindex{4}) + {\sqrt{3}}\,{\sqrt{\Aindex{5}}}\,\sin (\Thetaindex{4})  \right ) \nonumber \\
C_{50-3} &=&
\frac{1}{18\,{\sqrt{2}}}
\,\left ( {\Aindex{20}}^2 - 3\,\Aindex{20}\,t - 36\,t^2 - 6\,\Aindex{20}\,U\right . \nonumber \\
&& \hspace{1cm} 
 + 
 \left . 9\,t\,U + 9\,U^2 - 24\,\Aindex{20}\,W + 36\,t\,W + 72\,U\,W + 144\,W^2
\right  )  \nonumber \\
 N_{50} &=& 2\,{\sqrt{{\Cindex{50}{1}}^2 + 2\,\left( {\Cindex{50}{2}}^2 + {\Cindex{50}{3}}^2 \right) }} \nonumber \eeq 
\beq
\ket{\Psi}_{51}& = &\ket{3,- {1 \over 2} , {3 \over 4} ,\Gamma_{4,2}} \nonumber \\ 
& = &
\Cindex{51}{1} \left ( 
\ket{002d} - \ket{00d2} + \ket{020d} - \ket{0d02} - \ket{20d0} - \ket{2d00} + \ket{d020} + \ket{d200}\right) 
 \nonumber \\
& + &
\Cindex{51}{2} \left ( 
\ket{02d0} + \ket{0d20} - \ket{200d} - \ket{d002}\right) 
 \nonumber \\
& + &
\Cindex{51}{3} \left ( 
\ket{0dud} - \ket{0udd} + \ket{d0du} + \ket{dd0u} + \ket{ddu0} - \ket{dud0} - \ket{u0dd} - \ket{ud0d}\right) 
\nonumber \eeq
\beq
C_{51-1} &=&
\frac{-t}{2\,{\sqrt{2}}}
\, \left ( J - 4\,t + U - 2\,W + 2\,{\sqrt{\Aindex{5}}}\,\cos (\Thetaindex{4})  \right ) \nonumber \\
C_{51-2} &=&
\frac{t}{3\,{\sqrt{2}}}
\, \left ( J + 12\,t + U - 2\,W + 2\,{\sqrt{\Aindex{5}}}\,\cos (\Thetaindex{4})  \right ) \nonumber \\
C_{51-3} &=&
\frac{-1}{18\,{\sqrt{2}}}
\,\left ( {\Aindex{19}}^2 + 3\,J\,t - 45\,t^2 + 6\,J\,U - 15\,t\,U - 3\,U^2 + 24\,J\,W - 78\,t\,W - 60\,U\,W\right . \nonumber \\
&& \hspace{1cm} 
 \left . -192\,W^2 + 6\,{\sqrt{\Aindex{5}}}\,t\,\cos (\Thetaindex{4})
\right .  \nonumber \\
&& \hspace{1cm} 
 + 
 \left . 12\,{\sqrt{\Aindex{5}}}\,U\,\cos (\Thetaindex{4}) + 48\,{\sqrt{\Aindex{5}}}\,W\,\cos (\Thetaindex{4})
\right )  \nonumber \\
 N_{51} &=& 2\,{\sqrt{2\,{\Cindex{51}{1}}^2 + {\Cindex{51}{2}}^2 + 2\,{\Cindex{51}{3}}^2}} \nonumber \eeq 
\beq
\ket{\Psi}_{52}& = &\ket{3,- {1 \over 2} , {3 \over 4} ,\Gamma_{4,2}} \nonumber \\ 
& = &
\Cindex{52}{1} \left ( 
\ket{002d} - \ket{00d2} + \ket{020d} - \ket{0d02} - \ket{20d0} - \ket{2d00} + \ket{d020} + \ket{d200}\right) 
 \nonumber \\
& + &
\Cindex{52}{2} \left ( 
\ket{02d0} + \ket{0d20} - \ket{200d} - \ket{d002}\right) 
 \nonumber \\
& + &
\Cindex{52}{3} \left ( 
\ket{0dud} - \ket{0udd} + \ket{d0du} + \ket{dd0u} + \ket{ddu0} - \ket{dud0} - \ket{u0dd} - \ket{ud0d}\right) 
\nonumber \eeq
\beq
C_{52-1} &=&
\frac{-t}{2\,{\sqrt{2}}}
\, \left ( J - 4\,t + U - 2\,W - {\sqrt{\Aindex{5}}}\,\cos (\Thetaindex{4}) - {\sqrt{3}}\,{\sqrt{\Aindex{5}}}\,\sin (\Thetaindex{4})  \right ) \nonumber \\
C_{52-2} &=&
\frac{t}{3\,{\sqrt{2}}}
\, \left ( J + 12\,t + U - 2\,W - {\sqrt{\Aindex{5}}}\,\cos (\Thetaindex{4}) - {\sqrt{3}}\,{\sqrt{\Aindex{5}}}\,\sin (\Thetaindex{4})  \right ) \nonumber \\
C_{52-3} &=&
\frac{-1}{18\,{\sqrt{2}}}
\,\left ( {\Aindex{21}}^2 - 3\,\Aindex{21}\,t - 36\,t^2 - 6\,\Aindex{21}\,U\right . \nonumber \\
&& \hspace{1cm} 
 + 
 \left . 9\,t\,U + 9\,U^2 - 24\,\Aindex{21}\,W + 36\,t\,W + 72\,U\,W + 144\,W^2
\right  )  \nonumber \\
 N_{52} &=& 2\,{\sqrt{2\,{\Cindex{52}{1}}^2 + {\Cindex{52}{2}}^2 + 2\,{\Cindex{52}{3}}^2}} \nonumber \eeq 
\beq
\ket{\Psi}_{53}& = &\ket{3,- {1 \over 2} , {3 \over 4} ,\Gamma_{4,2}} \nonumber \\ 
& = &
\Cindex{53}{1} \left ( 
\ket{002d} - \ket{00d2} + \ket{020d} - \ket{0d02} - \ket{20d0} - \ket{2d00} + \ket{d020} + \ket{d200}\right) 
 \nonumber \\
& + &
\Cindex{53}{2} \left ( 
\ket{02d0} + \ket{0d20} - \ket{200d} - \ket{d002}\right) 
 \nonumber \\
& + &
\Cindex{53}{3} \left ( 
\ket{0dud} - \ket{0udd} + \ket{d0du} + \ket{dd0u} + \ket{ddu0} - \ket{dud0} - \ket{u0dd} - \ket{ud0d}\right) 
\nonumber \eeq
\beq
C_{53-1} &=&
\frac{-t}{2\,{\sqrt{2}}}
\, \left ( J - 4\,t + U - 2\,W - {\sqrt{\Aindex{5}}}\,\cos (\Thetaindex{4}) + {\sqrt{3}}\,{\sqrt{\Aindex{5}}}\,\sin (\Thetaindex{4})  \right ) \nonumber \\
C_{53-2} &=&
\frac{t}{3\,{\sqrt{2}}}
\, \left ( J + 12\,t + U - 2\,W - {\sqrt{\Aindex{5}}}\,\cos (\Thetaindex{4}) + {\sqrt{3}}\,{\sqrt{\Aindex{5}}}\,\sin (\Thetaindex{4})  \right ) \nonumber \\
C_{53-3} &=&
\frac{-1}{18\,{\sqrt{2}}}
\,\left ( {\Aindex{20}}^2 - 3\,\Aindex{20}\,t - 36\,t^2 - 6\,\Aindex{20}\,U\right . \nonumber \\
&& \hspace{1cm} 
 + 
 \left . 9\,t\,U + 9\,U^2 - 24\,\Aindex{20}\,W + 36\,t\,W + 72\,U\,W + 144\,W^2
\right  )  \nonumber \\
 N_{53} &=& 2\,{\sqrt{2\,{\Cindex{53}{1}}^2 + {\Cindex{53}{2}}^2 + 2\,{\Cindex{53}{3}}^2}} \nonumber \eeq 
\beq
\ket{\Psi}_{54}& = &\ket{3,- {1 \over 2} , {3 \over 4} ,\Gamma_{4,3}} \nonumber \\ 
& = &
\Cindex{54}{1} \left ( 
\ket{002d} - \ket{00d2} - \ket{02d0} + \ket{0d20} + \ket{200d} + \ket{2d00} - \ket{d002} - \ket{d200}\right) 
 \nonumber \\
& + &
\Cindex{54}{2} \left ( 
\ket{020d} + \ket{0d02} - \ket{20d0} - \ket{d020}\right) 
 \nonumber \\
& + &
\Cindex{54}{3} \left ( 
\ket{0ddu} - \ket{0udd} + \ket{d0ud} - \ket{dd0u} - \ket{ddu0} + \ket{du0d} - \ket{u0dd} + \ket{udd0}\right) 
\nonumber \eeq
\beq
C_{54-1} &=&
\frac{t}{2\,{\sqrt{2}}}
\, \left ( J - 4\,t + U - 2\,W + 2\,{\sqrt{\Aindex{5}}}\,\cos (\Thetaindex{4})  \right ) \nonumber \\
C_{54-2} &=&
\frac{t}{3\,{\sqrt{2}}}
\, \left ( J + 12\,t + U - 2\,W + 2\,{\sqrt{\Aindex{5}}}\,\cos (\Thetaindex{4})  \right ) \nonumber \\
C_{54-3} &=&
\frac{1}{18\,{\sqrt{2}}}
\,\left ( {\Aindex{19}}^2 + 3\,J\,t - 45\,t^2 + 6\,J\,U - 15\,t\,U - 3\,U^2 + 24\,J\,W - 78\,t\,W - 60\,U\,W\right . \nonumber \\
&& \hspace{1cm} 
 \left . -192\,W^2 + 6\,{\sqrt{\Aindex{5}}}\,t\,\cos (\Thetaindex{4})
\right .  \nonumber \\
&& \hspace{1cm} 
 + 
 \left . 12\,{\sqrt{\Aindex{5}}}\,U\,\cos (\Thetaindex{4}) + 48\,{\sqrt{\Aindex{5}}}\,W\,\cos (\Thetaindex{4})
\right )  \nonumber \\
 N_{54} &=& 2\,{\sqrt{2\,{\Cindex{54}{1}}^2 + {\Cindex{54}{2}}^2 + 2\,{\Cindex{54}{3}}^2}} \nonumber \eeq 
\beq
\ket{\Psi}_{55}& = &\ket{3,- {1 \over 2} , {3 \over 4} ,\Gamma_{4,3}} \nonumber \\ 
& = &
\Cindex{55}{1} \left ( 
\ket{002d} - \ket{00d2} - \ket{02d0} + \ket{0d20} + \ket{200d} + \ket{2d00} - \ket{d002} - \ket{d200}\right) 
 \nonumber \\
& + &
\Cindex{55}{2} \left ( 
\ket{020d} + \ket{0d02} - \ket{20d0} - \ket{d020}\right) 
 \nonumber \\
& + &
\Cindex{55}{3} \left ( 
\ket{0ddu} - \ket{0udd} + \ket{d0ud} - \ket{dd0u} - \ket{ddu0} + \ket{du0d} - \ket{u0dd} + \ket{udd0}\right) 
\nonumber \eeq
\beq
C_{55-1} &=&
\frac{t}{2\,{\sqrt{2}}}
\, \left ( J - 4\,t + U - 2\,W - {\sqrt{\Aindex{5}}}\,\cos (\Thetaindex{4}) - {\sqrt{3}}\,{\sqrt{\Aindex{5}}}\,\sin (\Thetaindex{4})  \right ) \nonumber \\
C_{55-2} &=&
\frac{t}{3\,{\sqrt{2}}}
\, \left ( J + 12\,t + U - 2\,W - {\sqrt{\Aindex{5}}}\,\cos (\Thetaindex{4}) - {\sqrt{3}}\,{\sqrt{\Aindex{5}}}\,\sin (\Thetaindex{4})  \right ) \nonumber \\
C_{55-3} &=&
\frac{1}{18\,{\sqrt{2}}}
\,\left ( {\Aindex{21}}^2 - 3\,\Aindex{21}\,t - 36\,t^2 - 6\,\Aindex{21}\,U\right . \nonumber \\
&& \hspace{1cm} 
 + 
 \left . 9\,t\,U + 9\,U^2 - 24\,\Aindex{21}\,W + 36\,t\,W + 72\,U\,W + 144\,W^2
\right  )  \nonumber \\
 N_{55} &=& 2\,{\sqrt{2\,{\Cindex{55}{1}}^2 + {\Cindex{55}{2}}^2 + 2\,{\Cindex{55}{3}}^2}} \nonumber \eeq 
\beq
\ket{\Psi}_{56}& = &\ket{3,- {1 \over 2} , {3 \over 4} ,\Gamma_{4,3}} \nonumber \\ 
& = &
\Cindex{56}{1} \left ( 
\ket{002d} - \ket{00d2} - \ket{02d0} + \ket{0d20} + \ket{200d} + \ket{2d00} - \ket{d002} - \ket{d200}\right) 
 \nonumber \\
& + &
\Cindex{56}{2} \left ( 
\ket{020d} + \ket{0d02} - \ket{20d0} - \ket{d020}\right) 
 \nonumber \\
& + &
\Cindex{56}{3} \left ( 
\ket{0ddu} - \ket{0udd} + \ket{d0ud} - \ket{dd0u} - \ket{ddu0} + \ket{du0d} - \ket{u0dd} + \ket{udd0}\right) 
\nonumber \eeq
\beq
C_{56-1} &=&
\frac{t}{2\,{\sqrt{2}}}
\, \left ( J - 4\,t + U - 2\,W - {\sqrt{\Aindex{5}}}\,\cos (\Thetaindex{4}) + {\sqrt{3}}\,{\sqrt{\Aindex{5}}}\,\sin (\Thetaindex{4})  \right ) \nonumber \\
C_{56-2} &=&
\frac{t}{3\,{\sqrt{2}}}
\, \left ( J + 12\,t + U - 2\,W - {\sqrt{\Aindex{5}}}\,\cos (\Thetaindex{4}) + {\sqrt{3}}\,{\sqrt{\Aindex{5}}}\,\sin (\Thetaindex{4})  \right ) \nonumber \\
C_{56-3} &=&
\frac{1}{18\,{\sqrt{2}}}
\,\left ( {\Aindex{20}}^2 - 3\,\Aindex{20}\,t - 36\,t^2 - 6\,\Aindex{20}\,U\right . \nonumber \\
&& \hspace{1cm} 
 + 
 \left . 9\,t\,U + 9\,U^2 - 24\,\Aindex{20}\,W + 36\,t\,W + 72\,U\,W + 144\,W^2
\right  )  \nonumber \\
 N_{56} &=& 2\,{\sqrt{2\,{\Cindex{56}{1}}^2 + {\Cindex{56}{2}}^2 + 2\,{\Cindex{56}{3}}^2}} \nonumber \eeq 
\beq
\ket{\Psi}_{57}& = &\ket{3,- {1 \over 2} , {3 \over 4} ,\Gamma_{5,1}} \nonumber \\ 
& = &
\Cindex{57}{1} \left ( 
\ket{002d} - \ket{00d2} + \ket{02d0} - \ket{0d20} - \ket{200d} + \ket{2d00} + \ket{d002} - \ket{d200}\right) 
 \nonumber \\
& + &
\Cindex{57}{2} \left ( 
\ket{0ddu} + \ket{0udd} + \ket{d0ud} + \ket{dd0u} + \ket{ddu0} + \ket{du0d} + \ket{u0dd} + \ket{udd0}\right) 
 \nonumber \\
& + &
\Cindex{57}{3} \left ( 
\ket{0dud} + \ket{d0du} + \ket{dud0} + \ket{ud0d}\right) 
\nonumber \eeq
\beq
C_{57-1} &=&
\frac{-\left( {\sqrt{\frac{3}{2}}}\,t \right) }{2} \nonumber \\
C_{57-2} &=&
\frac{1}{4\,{\sqrt{6}}}
\, \left ( {\sqrt{\Aindex{7}}} + J + 2\,t + U - 2\,W  \right ) \nonumber \\
C_{57-3} &=&
\frac{-1}{2\,{\sqrt{6}}}
\, \left ( {\sqrt{\Aindex{7}}} + J + 2\,t + U - 2\,W  \right ) \nonumber \\
 N_{57} &=& 2\,{\sqrt{2\,{\Cindex{57}{1}}^2 + 2\,{\Cindex{57}{2}}^2 + {\Cindex{57}{3}}^2}} \nonumber \eeq 
\beq
\ket{\Psi}_{58}& = &\ket{3,- {1 \over 2} , {3 \over 4} ,\Gamma_{5,1}} \nonumber \\ 
& = &
\Cindex{58}{1} \left ( 
\ket{002d} - \ket{00d2} + \ket{02d0} - \ket{0d20} - \ket{200d} + \ket{2d00} + \ket{d002} - \ket{d200}\right) 
 \nonumber \\
& + &
\Cindex{58}{2} \left ( 
\ket{0ddu} + \ket{0udd} + \ket{d0ud} + \ket{dd0u} + \ket{ddu0} + \ket{du0d} + \ket{u0dd} + \ket{udd0}\right) 
 \nonumber \\
& + &
\Cindex{58}{3} \left ( 
\ket{0dud} + \ket{d0du} + \ket{dud0} + \ket{ud0d}\right) 
\nonumber \eeq
\beq
C_{58-1} &=&
\frac{-\left( {\sqrt{\frac{3}{2}}}\,t \right) }{2} \nonumber \\
C_{58-2} &=&
\frac{-1}{4\,{\sqrt{6}}}
\, \left ( {\sqrt{\Aindex{7}}} - J - 2\,t - U + 2\,W  \right ) \nonumber \\
C_{58-3} &=&
\frac{1}{2\,{\sqrt{6}}}
\, \left ( {\sqrt{\Aindex{7}}} - J - 2\,t - U + 2\,W  \right ) \nonumber \\
 N_{58} &=& 2\,{\sqrt{2\,{\Cindex{58}{1}}^2 + 2\,{\Cindex{58}{2}}^2 + {\Cindex{58}{3}}^2}} \nonumber \eeq 
\beq
\ket{\Psi}_{59}& = &\ket{3,- {1 \over 2} , {15 \over 4} ,\Gamma_{5,1}} \nonumber \\ 
&=& \frac{1}{2\,{\sqrt{3}}}
 \left ( \ket{0ddu} + \ket{0dud} + \ket{0udd} + \ket{d0du} + \ket{d0ud} + \ket{dd0u}\right . \nonumber \\
&& \hspace{1em} 
 + 
\left . \ket{ddu0} + \ket{du0d} + \ket{dud0} + \ket{u0dd} + \ket{ud0d} + \ket{udd0}\right ) \nonumber  
\eeq
\beq
\ket{\Psi}_{60}& = &\ket{3,- {1 \over 2} , {3 \over 4} ,\Gamma_{5,2}} \nonumber \\ 
& = &
\Cindex{60}{1} \left ( 
\ket{020d} - \ket{02d0} - \ket{0d02} + \ket{0d20} - \ket{200d} + \ket{20d0} + \ket{d002} - \ket{d020}\right) 
 \nonumber \\
& + &
\Cindex{60}{2} \left ( 
\ket{0ddu} + \ket{0dud} - \ket{d0du} - \ket{d0ud} - \ket{du0d} + \ket{dud0} - \ket{ud0d} + \ket{udd0}\right) 
 \nonumber \\
& + &
\Cindex{60}{3} \left ( 
\ket{0udd} - \ket{dd0u} + \ket{ddu0} - \ket{u0dd}\right) 
\nonumber \eeq
\beq
C_{60-1} &=&
\frac{-\left( {\sqrt{\frac{3}{2}}}\,t \right) }{2} \nonumber \\
C_{60-2} &=&
\frac{-1}{4\,{\sqrt{6}}}
\, \left ( {\sqrt{\Aindex{7}}} + J + 2\,t + U - 2\,W  \right ) \nonumber \\
C_{60-3} &=&
\frac{1}{2\,{\sqrt{6}}}
\, \left ( {\sqrt{\Aindex{7}}} + J + 2\,t + U - 2\,W  \right ) \nonumber \\
 N_{60} &=& 2\,{\sqrt{2\,{\Cindex{60}{1}}^2 + 2\,{\Cindex{60}{2}}^2 + {\Cindex{60}{3}}^2}} \nonumber \eeq 
\beq
\ket{\Psi}_{61}& = &\ket{3,- {1 \over 2} , {3 \over 4} ,\Gamma_{5,2}} \nonumber \\ 
& = &
\Cindex{61}{1} \left ( 
\ket{020d} - \ket{02d0} - \ket{0d02} + \ket{0d20} - \ket{200d} + \ket{20d0} + \ket{d002} - \ket{d020}\right) 
 \nonumber \\
& + &
\Cindex{61}{2} \left ( 
\ket{0ddu} + \ket{0dud} - \ket{d0du} - \ket{d0ud} - \ket{du0d} + \ket{dud0} - \ket{ud0d} + \ket{udd0}\right) 
 \nonumber \\
& + &
\Cindex{61}{3} \left ( 
\ket{0udd} - \ket{dd0u} + \ket{ddu0} - \ket{u0dd}\right) 
\nonumber \eeq
\beq
C_{61-1} &=&
\frac{-\left( {\sqrt{\frac{3}{2}}}\,t \right) }{2} \nonumber \\
C_{61-2} &=&
\frac{1}{4\,{\sqrt{6}}}
\, \left ( {\sqrt{\Aindex{7}}} - J - 2\,t - U + 2\,W  \right ) \nonumber \\
C_{61-3} &=&
\frac{-1}{2\,{\sqrt{6}}}
\, \left ( {\sqrt{\Aindex{7}}} - J - 2\,t - U + 2\,W  \right ) \nonumber \\
 N_{61} &=& 2\,{\sqrt{2\,{\Cindex{61}{1}}^2 + 2\,{\Cindex{61}{2}}^2 + {\Cindex{61}{3}}^2}} \nonumber \eeq 
\beq
\ket{\Psi}_{62}& = &\ket{3,- {1 \over 2} , {15 \over 4} ,\Gamma_{5,2}} \nonumber \\ 
&=& \frac{1}{2\,{\sqrt{3}}}
 \left ( \ket{0ddu} + \ket{0dud} + \ket{0udd} - \ket{d0du} - \ket{d0ud} - \ket{dd0u}\right . \nonumber \\
&& \hspace{1em} 
 + 
\left . \ket{ddu0} - \ket{du0d} + \ket{dud0} - \ket{u0dd} - \ket{ud0d} + \ket{udd0}\right ) \nonumber  
\eeq
\beq
\ket{\Psi}_{63}& = &\ket{3,- {1 \over 2} , {3 \over 4} ,\Gamma_{5,3}} \nonumber \\ 
& = &
\Cindex{63}{1} \left ( 
\ket{002d} - \ket{00d2} - \ket{020d} + \ket{0d02} + \ket{20d0} - \ket{2d00} - \ket{d020} + \ket{d200}\right) 
 \nonumber \\
& + &
\Cindex{63}{2} \left ( 
\ket{0ddu} + \ket{d0ud} - \ket{du0d} - \ket{udd0}\right) 
 \nonumber \\
& + &
\Cindex{63}{3} \left ( 
\ket{0dud} + \ket{0udd} + \ket{d0du} - \ket{dd0u} - \ket{ddu0} - \ket{dud0} + \ket{u0dd} - \ket{ud0d}\right) 
\nonumber \eeq
\beq
C_{63-1} &=&
\frac{-\left( {\sqrt{\frac{3}{2}}}\,t \right) }{2} \nonumber \\
C_{63-2} &=&
\frac{-1}{2\,{\sqrt{6}}}
\, \left ( {\sqrt{\Aindex{7}}} + J + 2\,t + U - 2\,W  \right ) \nonumber \\
C_{63-3} &=&
\frac{1}{4\,{\sqrt{6}}}
\, \left ( {\sqrt{\Aindex{7}}} + J + 2\,t + U - 2\,W  \right ) \nonumber \\
 N_{63} &=& 2\,{\sqrt{2\,{\Cindex{63}{1}}^2 + {\Cindex{63}{2}}^2 + 2\,{\Cindex{63}{3}}^2}} \nonumber \eeq 
\beq
\ket{\Psi}_{64}& = &\ket{3,- {1 \over 2} , {3 \over 4} ,\Gamma_{5,3}} \nonumber \\ 
& = &
\Cindex{64}{1} \left ( 
\ket{002d} - \ket{00d2} - \ket{020d} + \ket{0d02} + \ket{20d0} - \ket{2d00} - \ket{d020} + \ket{d200}\right) 
 \nonumber \\
& + &
\Cindex{64}{2} \left ( 
\ket{0ddu} + \ket{d0ud} - \ket{du0d} - \ket{udd0}\right) 
 \nonumber \\
& + &
\Cindex{64}{3} \left ( 
\ket{0dud} + \ket{0udd} + \ket{d0du} - \ket{dd0u} - \ket{ddu0} - \ket{dud0} + \ket{u0dd} - \ket{ud0d}\right) 
\nonumber \eeq
\beq
C_{64-1} &=&
\frac{-\left( {\sqrt{\frac{3}{2}}}\,t \right) }{2} \nonumber \\
C_{64-2} &=&
\frac{1}{2\,{\sqrt{6}}}
\, \left ( {\sqrt{\Aindex{7}}} - J - 2\,t - U + 2\,W  \right ) \nonumber \\
C_{64-3} &=&
\frac{-1}{4\,{\sqrt{6}}}
\, \left ( {\sqrt{\Aindex{7}}} - J - 2\,t - U + 2\,W  \right ) \nonumber \\
 N_{64} &=& 2\,{\sqrt{2\,{\Cindex{64}{1}}^2 + {\Cindex{64}{2}}^2 + 2\,{\Cindex{64}{3}}^2}} \nonumber \eeq 
\beq
\ket{\Psi}_{65}& = &\ket{3,- {1 \over 2} , {15 \over 4} ,\Gamma_{5,3}} \nonumber \\ 
&=& \frac{1}{2\,{\sqrt{3}}}
 \left ( \ket{0ddu} + \ket{0dud} + \ket{0udd} + \ket{d0du} + \ket{d0ud} - \ket{dd0u}\right . \nonumber \\
&& \hspace{1em} 
\left . -\ket{ddu0} - \ket{du0d} - \ket{dud0} + \ket{u0dd} - \ket{ud0d} - \ket{udd0}\right ) \nonumber  
\eeq
{\subsection*{\boldmath Unnormalized eigenvectors for ${\rm  N_e}=3$ and   ${\rm m_s}$= $ {1 \over 2} $.}
\beq
\ket{\Psi}_{66}& = &\ket{3, {1 \over 2} , {3 \over 4} ,\Gamma_1} \nonumber \\ 
&=& \frac{1}{2\,{\sqrt{3}}}
 \left ( \ket{002u} + \ket{00u2} + \ket{020u} + \ket{02u0} + \ket{0u02} + \ket{0u20}\right . \nonumber \\
&& \hspace{1em} 
 + 
\left . \ket{200u} + \ket{20u0} + \ket{2u00} + \ket{u002} + \ket{u020} + \ket{u200}\right ) \nonumber  
\eeq
\beq
\ket{\Psi}_{67}& = &\ket{3, {1 \over 2} , {15 \over 4} ,\Gamma_2} \nonumber \\ 
&=& \frac{1}{2\,{\sqrt{3}}}
 \left ( \ket{0duu} + \ket{0udu} + \ket{0uud} - \ket{d0uu} + \ket{du0u} - \ket{duu0}\right . \nonumber \\
&& \hspace{1em} 
\left . -\ket{u0du} - \ket{u0ud} + \ket{ud0u} - \ket{udu0} + \ket{uu0d} - \ket{uud0}\right ) \nonumber  
\eeq
\beq
\ket{\Psi}_{68}& = &\ket{3, {1 \over 2} , {3 \over 4} ,\Gamma_{3,1}} \nonumber \\ 
& = &
\Cindex{68}{1} \left ( 
\ket{002u} + \ket{00u2} + \ket{02u0} + \ket{0u20} + \ket{200u} + \ket{2u00} + \ket{u002} + \ket{u200}\right) 
 \nonumber \\
& + &
\Cindex{68}{2} \left ( 
\ket{020u} + \ket{0u02} + \ket{20u0} + \ket{u020}\right) 
 \nonumber \\
& + &
\Cindex{68}{3} \left ( 
\ket{0duu} - \ket{0uud} - \ket{d0uu} + \ket{duu0} + \ket{u0du} - \ket{ud0u} + \ket{uu0d} - \ket{uud0}\right) 
\nonumber \eeq
\beq
C_{68-1} &=&
\frac{-t}{2\,{\sqrt{2}}} \nonumber \\
C_{68-2} &=&
\frac{t}{{\sqrt{2}}} \nonumber \\
C_{68-3} &=&
\frac{1}{4\,{\sqrt{2}}}
\, \left ( {\sqrt{\Aindex{4}}} + J - 2\,t + U - 2\,W  \right ) \nonumber \\
 N_{68} &=& 2\,{\sqrt{2\,{\Cindex{68}{1}}^2 + {\Cindex{68}{2}}^2 + 2\,{\Cindex{68}{3}}^2}} \nonumber \eeq 
\beq
\ket{\Psi}_{69}& = &\ket{3, {1 \over 2} , {3 \over 4} ,\Gamma_{3,1}} \nonumber \\ 
& = &
\Cindex{69}{1} \left ( 
\ket{002u} + \ket{00u2} + \ket{02u0} + \ket{0u20} + \ket{200u} + \ket{2u00} + \ket{u002} + \ket{u200}\right) 
 \nonumber \\
& + &
\Cindex{69}{2} \left ( 
\ket{020u} + \ket{0u02} + \ket{20u0} + \ket{u020}\right) 
 \nonumber \\
& + &
\Cindex{69}{3} \left ( 
\ket{0duu} - \ket{0uud} - \ket{d0uu} + \ket{duu0} + \ket{u0du} - \ket{ud0u} + \ket{uu0d} - \ket{uud0}\right) 
\nonumber \eeq
\beq
C_{69-1} &=&
\frac{-t}{2\,{\sqrt{2}}} \nonumber \\
C_{69-2} &=&
\frac{t}{{\sqrt{2}}} \nonumber \\
C_{69-3} &=&
\frac{-1}{4\,{\sqrt{2}}}
\, \left ( {\sqrt{\Aindex{4}}} - J + 2\,t - U + 2\,W  \right ) \nonumber \\
 N_{69} &=& 2\,{\sqrt{2\,{\Cindex{69}{1}}^2 + {\Cindex{69}{2}}^2 + 2\,{\Cindex{69}{3}}^2}} \nonumber \eeq 
\beq
\ket{\Psi}_{70}& = &\ket{3, {1 \over 2} , {3 \over 4} ,\Gamma_{3,2}} \nonumber \\ 
& = &
\Cindex{70}{1} \left ( 
\ket{002u} + \ket{00u2} - \ket{02u0} - \ket{0u20} - \ket{200u} + \ket{2u00} - \ket{u002} + \ket{u200}\right) 
 \nonumber \\
& + &
\Cindex{70}{2} \left ( 
\ket{0duu} + \ket{0uud} - \ket{d0uu} - \ket{duu0} - \ket{u0du} + \ket{ud0u} + \ket{uu0d} - \ket{uud0}\right) 
 \nonumber \\
& + &
\Cindex{70}{3} \left ( 
\ket{0udu} + \ket{du0u} - \ket{u0ud} - \ket{udu0}\right) 
\nonumber \eeq
\beq
C_{70-1} &=&
\frac{{\sqrt{\frac{3}{2}}}\,t}{2} \nonumber \\
C_{70-2} &=&
\frac{1}{4\,{\sqrt{6}}}
\, \left ( {\sqrt{\Aindex{4}}} + J - 2\,t + U - 2\,W  \right ) \nonumber \\
C_{70-3} &=&
\frac{-1}{2\,{\sqrt{6}}}
\, \left ( {\sqrt{\Aindex{4}}} + J - 2\,t + U - 2\,W  \right ) \nonumber \\
 N_{70} &=& 2\,{\sqrt{2\,{\Cindex{70}{1}}^2 + 2\,{\Cindex{70}{2}}^2 + {\Cindex{70}{3}}^2}} \nonumber \eeq 
\beq
\ket{\Psi}_{71}& = &\ket{3, {1 \over 2} , {3 \over 4} ,\Gamma_{3,2}} \nonumber \\ 
& = &
\Cindex{71}{1} \left ( 
\ket{002u} + \ket{00u2} - \ket{02u0} - \ket{0u20} - \ket{200u} + \ket{2u00} - \ket{u002} + \ket{u200}\right) 
 \nonumber \\
& + &
\Cindex{71}{2} \left ( 
\ket{0duu} + \ket{0uud} - \ket{d0uu} - \ket{duu0} - \ket{u0du} + \ket{ud0u} + \ket{uu0d} - \ket{uud0}\right) 
 \nonumber \\
& + &
\Cindex{71}{3} \left ( 
\ket{0udu} + \ket{du0u} - \ket{u0ud} - \ket{udu0}\right) 
\nonumber \eeq
\beq
C_{71-1} &=&
\frac{{\sqrt{\frac{3}{2}}}\,t}{2} \nonumber \\
C_{71-2} &=&
\frac{-1}{4\,{\sqrt{6}}}
\, \left ( {\sqrt{\Aindex{4}}} - J + 2\,t - U + 2\,W  \right ) \nonumber \\
C_{71-3} &=&
\frac{1}{2\,{\sqrt{6}}}
\, \left ( {\sqrt{\Aindex{4}}} - J + 2\,t - U + 2\,W  \right ) \nonumber \\
 N_{71} &=& 2\,{\sqrt{2\,{\Cindex{71}{1}}^2 + 2\,{\Cindex{71}{2}}^2 + {\Cindex{71}{3}}^2}} \nonumber \eeq 
\beq
\ket{\Psi}_{72}& = &\ket{3, {1 \over 2} , {3 \over 4} ,\Gamma_{4,1}} \nonumber \\ 
& = &
\Cindex{72}{1} \left ( 
\ket{002u} + \ket{00u2} - \ket{2u00} - \ket{u200}\right) 
 \nonumber \\
& + &
\Cindex{72}{2} \left ( 
\ket{020u} + \ket{02u0} - \ket{0u02} - \ket{0u20} + \ket{200u} + \ket{20u0} - \ket{u002} - \ket{u020}\right) 
 \nonumber \\
& + &
\Cindex{72}{3} \left ( 
\ket{0udu} - \ket{0uud} - \ket{du0u} - \ket{duu0} + \ket{u0du} - \ket{u0ud} + \ket{ud0u} + \ket{udu0}\right) 
\nonumber \eeq
\beq
C_{72-1} &=&
\frac{-t}{3\,{\sqrt{2}}}
\, \left ( J + 12\,t + U - 2\,W + 2\,{\sqrt{\Aindex{5}}}\,\cos (\Thetaindex{4})  \right ) \nonumber \\
C_{72-2} &=&
\frac{-t}{2\,{\sqrt{2}}}
\, \left ( J - 4\,t + U - 2\,W + 2\,{\sqrt{\Aindex{5}}}\,\cos (\Thetaindex{4})  \right ) \nonumber \\
C_{72-3} &=&
\frac{1}{18\,{\sqrt{2}}}
\,\left ( {\Aindex{19}}^2 + 3\,J\,t - 45\,t^2 + 6\,J\,U - 15\,t\,U - 3\,U^2 + 24\,J\,W - 78\,t\,W - 60\,U\,W\right . \nonumber \\
&& \hspace{1cm} 
 \left . -192\,W^2 + 6\,{\sqrt{\Aindex{5}}}\,t\,\cos (\Thetaindex{4})
\right .  \nonumber \\
&& \hspace{1cm} 
 + 
 \left . 12\,{\sqrt{\Aindex{5}}}\,U\,\cos (\Thetaindex{4}) + 48\,{\sqrt{\Aindex{5}}}\,W\,\cos (\Thetaindex{4})
\right )  \nonumber \\
 N_{72} &=& 2\,{\sqrt{{\Cindex{72}{1}}^2 + 2\,\left( {\Cindex{72}{2}}^2 + {\Cindex{72}{3}}^2 \right) }} \nonumber \eeq 
\beq
\ket{\Psi}_{73}& = &\ket{3, {1 \over 2} , {3 \over 4} ,\Gamma_{4,1}} \nonumber \\ 
& = &
\Cindex{73}{1} \left ( 
\ket{002u} + \ket{00u2} - \ket{2u00} - \ket{u200}\right) 
 \nonumber \\
& + &
\Cindex{73}{2} \left ( 
\ket{020u} + \ket{02u0} - \ket{0u02} - \ket{0u20} + \ket{200u} + \ket{20u0} - \ket{u002} - \ket{u020}\right) 
 \nonumber \\
& + &
\Cindex{73}{3} \left ( 
\ket{0udu} - \ket{0uud} - \ket{du0u} - \ket{duu0} + \ket{u0du} - \ket{u0ud} + \ket{ud0u} + \ket{udu0}\right) 
\nonumber \eeq
\beq
C_{73-1} &=&
\frac{-t}{3\,{\sqrt{2}}}
\, \left ( J + 12\,t + U - 2\,W - {\sqrt{\Aindex{5}}}\,\cos (\Thetaindex{4}) - {\sqrt{3}}\,{\sqrt{\Aindex{5}}}\,\sin (\Thetaindex{4})  \right ) \nonumber \\
C_{73-2} &=&
\frac{-t}{2\,{\sqrt{2}}}
\, \left ( J - 4\,t + U - 2\,W - {\sqrt{\Aindex{5}}}\,\cos (\Thetaindex{4}) - {\sqrt{3}}\,{\sqrt{\Aindex{5}}}\,\sin (\Thetaindex{4})  \right ) \nonumber \\
C_{73-3} &=&
\frac{1}{18\,{\sqrt{2}}}
\,\left ( {\Aindex{21}}^2 - 3\,\Aindex{21}\,t - 36\,t^2 - 6\,\Aindex{21}\,U\right . \nonumber \\
&& \hspace{1cm} 
 + 
 \left . 9\,t\,U + 9\,U^2 - 24\,\Aindex{21}\,W + 36\,t\,W + 72\,U\,W + 144\,W^2
\right  )  \nonumber \\
 N_{73} &=& 2\,{\sqrt{{\Cindex{73}{1}}^2 + 2\,\left( {\Cindex{73}{2}}^2 + {\Cindex{73}{3}}^2 \right) }} \nonumber \eeq 
\beq
\ket{\Psi}_{74}& = &\ket{3, {1 \over 2} , {3 \over 4} ,\Gamma_{4,1}} \nonumber \\ 
& = &
\Cindex{74}{1} \left ( 
\ket{002u} + \ket{00u2} - \ket{2u00} - \ket{u200}\right) 
 \nonumber \\
& + &
\Cindex{74}{2} \left ( 
\ket{020u} + \ket{02u0} - \ket{0u02} - \ket{0u20} + \ket{200u} + \ket{20u0} - \ket{u002} - \ket{u020}\right) 
 \nonumber \\
& + &
\Cindex{74}{3} \left ( 
\ket{0udu} - \ket{0uud} - \ket{du0u} - \ket{duu0} + \ket{u0du} - \ket{u0ud} + \ket{ud0u} + \ket{udu0}\right) 
\nonumber \eeq
\beq
C_{74-1} &=&
\frac{-t}{3\,{\sqrt{2}}}
\, \left ( J + 12\,t + U - 2\,W - {\sqrt{\Aindex{5}}}\,\cos (\Thetaindex{4}) + {\sqrt{3}}\,{\sqrt{\Aindex{5}}}\,\sin (\Thetaindex{4})  \right ) \nonumber \\
C_{74-2} &=&
\frac{-t}{2\,{\sqrt{2}}}
\, \left ( J - 4\,t + U - 2\,W - {\sqrt{\Aindex{5}}}\,\cos (\Thetaindex{4}) + {\sqrt{3}}\,{\sqrt{\Aindex{5}}}\,\sin (\Thetaindex{4})  \right ) \nonumber \\
C_{74-3} &=&
\frac{1}{18\,{\sqrt{2}}}
\,\left ( {\Aindex{20}}^2 - 3\,\Aindex{20}\,t - 36\,t^2 - 6\,\Aindex{20}\,U\right . \nonumber \\
&& \hspace{1cm} 
 + 
 \left . 9\,t\,U + 9\,U^2 - 24\,\Aindex{20}\,W + 36\,t\,W + 72\,U\,W + 144\,W^2
\right  )  \nonumber \\
 N_{74} &=& 2\,{\sqrt{{\Cindex{74}{1}}^2 + 2\,\left( {\Cindex{74}{2}}^2 + {\Cindex{74}{3}}^2 \right) }} \nonumber \eeq 
\beq
\ket{\Psi}_{75}& = &\ket{3, {1 \over 2} , {3 \over 4} ,\Gamma_{4,2}} \nonumber \\ 
& = &
\Cindex{75}{1} \left ( 
\ket{002u} - \ket{00u2} + \ket{020u} - \ket{0u02} - \ket{20u0} - \ket{2u00} + \ket{u020} + \ket{u200}\right) 
 \nonumber \\
& + &
\Cindex{75}{2} \left ( 
\ket{02u0} + \ket{0u20} - \ket{200u} - \ket{u002}\right) 
 \nonumber \\
& + &
\Cindex{75}{3} \left ( 
\ket{0duu} - \ket{0udu} + \ket{d0uu} + \ket{du0u} - \ket{u0ud} + \ket{udu0} - \ket{uu0d} - \ket{uud0}\right) 
\nonumber \eeq
\beq
C_{75-1} &=&
\frac{-t}{2\,{\sqrt{2}}}
\, \left ( J - 4\,t + U - 2\,W + 2\,{\sqrt{\Aindex{5}}}\,\cos (\Thetaindex{4})  \right ) \nonumber \\
C_{75-2} &=&
\frac{t}{3\,{\sqrt{2}}}
\, \left ( J + 12\,t + U - 2\,W + 2\,{\sqrt{\Aindex{5}}}\,\cos (\Thetaindex{4})  \right ) \nonumber \\
C_{75-3} &=&
\frac{-1}{18\,{\sqrt{2}}}
\,\left ( {\Aindex{19}}^2 + 3\,J\,t - 45\,t^2 + 6\,J\,U - 15\,t\,U - 3\,U^2 + 24\,J\,W - 78\,t\,W - 60\,U\,W\right . \nonumber \\
&& \hspace{1cm} 
 \left . -192\,W^2 + 6\,{\sqrt{\Aindex{5}}}\,t\,\cos (\Thetaindex{4})
\right .  \nonumber \\
&& \hspace{1cm} 
 + 
 \left . 12\,{\sqrt{\Aindex{5}}}\,U\,\cos (\Thetaindex{4}) + 48\,{\sqrt{\Aindex{5}}}\,W\,\cos (\Thetaindex{4})
\right )  \nonumber \\
 N_{75} &=& 2\,{\sqrt{2\,{\Cindex{75}{1}}^2 + {\Cindex{75}{2}}^2 + 2\,{\Cindex{75}{3}}^2}} \nonumber \eeq 
\beq
\ket{\Psi}_{76}& = &\ket{3, {1 \over 2} , {3 \over 4} ,\Gamma_{4,2}} \nonumber \\ 
& = &
\Cindex{76}{1} \left ( 
\ket{002u} - \ket{00u2} + \ket{020u} - \ket{0u02} - \ket{20u0} - \ket{2u00} + \ket{u020} + \ket{u200}\right) 
 \nonumber \\
& + &
\Cindex{76}{2} \left ( 
\ket{02u0} + \ket{0u20} - \ket{200u} - \ket{u002}\right) 
 \nonumber \\
& + &
\Cindex{76}{3} \left ( 
\ket{0duu} - \ket{0udu} + \ket{d0uu} + \ket{du0u} - \ket{u0ud} + \ket{udu0} - \ket{uu0d} - \ket{uud0}\right) 
\nonumber \eeq
\beq
C_{76-1} &=&
\frac{-t}{2\,{\sqrt{2}}}
\, \left ( J - 4\,t + U - 2\,W - {\sqrt{\Aindex{5}}}\,\cos (\Thetaindex{4}) - {\sqrt{3}}\,{\sqrt{\Aindex{5}}}\,\sin (\Thetaindex{4})  \right ) \nonumber \\
C_{76-2} &=&
\frac{t}{3\,{\sqrt{2}}}
\, \left ( J + 12\,t + U - 2\,W - {\sqrt{\Aindex{5}}}\,\cos (\Thetaindex{4}) - {\sqrt{3}}\,{\sqrt{\Aindex{5}}}\,\sin (\Thetaindex{4})  \right ) \nonumber \\
C_{76-3} &=&
\frac{-1}{18\,{\sqrt{2}}}
\,\left ( {\Aindex{21}}^2 - 3\,\Aindex{21}\,t - 36\,t^2 - 6\,\Aindex{21}\,U\right . \nonumber \\
&& \hspace{1cm} 
 + 
 \left . 9\,t\,U + 9\,U^2 - 24\,\Aindex{21}\,W + 36\,t\,W + 72\,U\,W + 144\,W^2
\right  )  \nonumber \\
 N_{76} &=& 2\,{\sqrt{2\,{\Cindex{76}{1}}^2 + {\Cindex{76}{2}}^2 + 2\,{\Cindex{76}{3}}^2}} \nonumber \eeq 
\beq
\ket{\Psi}_{77}& = &\ket{3, {1 \over 2} , {3 \over 4} ,\Gamma_{4,2}} \nonumber \\ 
& = &
\Cindex{77}{1} \left ( 
\ket{002u} - \ket{00u2} + \ket{020u} - \ket{0u02} - \ket{20u0} - \ket{2u00} + \ket{u020} + \ket{u200}\right) 
 \nonumber \\
& + &
\Cindex{77}{2} \left ( 
\ket{02u0} + \ket{0u20} - \ket{200u} - \ket{u002}\right) 
 \nonumber \\
& + &
\Cindex{77}{3} \left ( 
\ket{0duu} - \ket{0udu} + \ket{d0uu} + \ket{du0u} - \ket{u0ud} + \ket{udu0} - \ket{uu0d} - \ket{uud0}\right) 
\nonumber \eeq
\beq
C_{77-1} &=&
\frac{-t}{2\,{\sqrt{2}}}
\, \left ( J - 4\,t + U - 2\,W - {\sqrt{\Aindex{5}}}\,\cos (\Thetaindex{4}) + {\sqrt{3}}\,{\sqrt{\Aindex{5}}}\,\sin (\Thetaindex{4})  \right ) \nonumber \\
C_{77-2} &=&
\frac{t}{3\,{\sqrt{2}}}
\, \left ( J + 12\,t + U - 2\,W - {\sqrt{\Aindex{5}}}\,\cos (\Thetaindex{4}) + {\sqrt{3}}\,{\sqrt{\Aindex{5}}}\,\sin (\Thetaindex{4})  \right ) \nonumber \\
C_{77-3} &=&
\frac{-1}{18\,{\sqrt{2}}}
\,\left ( {\Aindex{20}}^2 - 3\,\Aindex{20}\,t - 36\,t^2 - 6\,\Aindex{20}\,U\right . \nonumber \\
&& \hspace{1cm} 
 + 
 \left . 9\,t\,U + 9\,U^2 - 24\,\Aindex{20}\,W + 36\,t\,W + 72\,U\,W + 144\,W^2
\right  )  \nonumber \\
 N_{77} &=& 2\,{\sqrt{2\,{\Cindex{77}{1}}^2 + {\Cindex{77}{2}}^2 + 2\,{\Cindex{77}{3}}^2}} \nonumber \eeq 
\beq
\ket{\Psi}_{78}& = &\ket{3, {1 \over 2} , {3 \over 4} ,\Gamma_{4,3}} \nonumber \\ 
& = &
\Cindex{78}{1} \left ( 
\ket{002u} - \ket{00u2} - \ket{02u0} + \ket{0u20} + \ket{200u} + \ket{2u00} - \ket{u002} - \ket{u200}\right) 
 \nonumber \\
& + &
\Cindex{78}{2} \left ( 
\ket{020u} + \ket{0u02} - \ket{20u0} - \ket{u020}\right) 
 \nonumber \\
& + &
\Cindex{78}{3} \left ( 
\ket{0duu} - \ket{0uud} + \ket{d0uu} - \ket{duu0} - \ket{u0du} - \ket{ud0u} + \ket{uu0d} + \ket{uud0}\right) 
\nonumber \eeq
\beq
C_{78-1} &=&
\frac{t}{2\,{\sqrt{2}}}
\, \left ( J - 4\,t + U - 2\,W + 2\,{\sqrt{\Aindex{5}}}\,\cos (\Thetaindex{4})  \right ) \nonumber \\
C_{78-2} &=&
\frac{t}{3\,{\sqrt{2}}}
\, \left ( J + 12\,t + U - 2\,W + 2\,{\sqrt{\Aindex{5}}}\,\cos (\Thetaindex{4})  \right ) \nonumber \\
C_{78-3} &=&
\frac{1}{18\,{\sqrt{2}}}
\,\left ( {\Aindex{19}}^2 + 3\,J\,t - 45\,t^2 + 6\,J\,U - 15\,t\,U - 3\,U^2 + 24\,J\,W - 78\,t\,W - 60\,U\,W\right . \nonumber \\
&& \hspace{1cm} 
 \left . -192\,W^2 + 6\,{\sqrt{\Aindex{5}}}\,t\,\cos (\Thetaindex{4})
\right .  \nonumber \\
&& \hspace{1cm} 
 + 
 \left . 12\,{\sqrt{\Aindex{5}}}\,U\,\cos (\Thetaindex{4}) + 48\,{\sqrt{\Aindex{5}}}\,W\,\cos (\Thetaindex{4})
\right )  \nonumber \\
 N_{78} &=& 2\,{\sqrt{2\,{\Cindex{78}{1}}^2 + {\Cindex{78}{2}}^2 + 2\,{\Cindex{78}{3}}^2}} \nonumber \eeq 
\beq
\ket{\Psi}_{79}& = &\ket{3, {1 \over 2} , {3 \over 4} ,\Gamma_{4,3}} \nonumber \\ 
& = &
\Cindex{79}{1} \left ( 
\ket{002u} - \ket{00u2} - \ket{02u0} + \ket{0u20} + \ket{200u} + \ket{2u00} - \ket{u002} - \ket{u200}\right) 
 \nonumber \\
& + &
\Cindex{79}{2} \left ( 
\ket{020u} + \ket{0u02} - \ket{20u0} - \ket{u020}\right) 
 \nonumber \\
& + &
\Cindex{79}{3} \left ( 
\ket{0duu} - \ket{0uud} + \ket{d0uu} - \ket{duu0} - \ket{u0du} - \ket{ud0u} + \ket{uu0d} + \ket{uud0}\right) 
\nonumber \eeq
\beq
C_{79-1} &=&
\frac{t}{2\,{\sqrt{2}}}
\, \left ( J - 4\,t + U - 2\,W - {\sqrt{\Aindex{5}}}\,\cos (\Thetaindex{4}) - {\sqrt{3}}\,{\sqrt{\Aindex{5}}}\,\sin (\Thetaindex{4})  \right ) \nonumber \\
C_{79-2} &=&
\frac{t}{3\,{\sqrt{2}}}
\, \left ( J + 12\,t + U - 2\,W - {\sqrt{\Aindex{5}}}\,\cos (\Thetaindex{4}) - {\sqrt{3}}\,{\sqrt{\Aindex{5}}}\,\sin (\Thetaindex{4})  \right ) \nonumber \\
C_{79-3} &=&
\frac{1}{18\,{\sqrt{2}}}
\,\left ( {\Aindex{21}}^2 - 3\,\Aindex{21}\,t - 36\,t^2 - 6\,\Aindex{21}\,U\right . \nonumber \\
&& \hspace{1cm} 
 + 
 \left . 9\,t\,U + 9\,U^2 - 24\,\Aindex{21}\,W + 36\,t\,W + 72\,U\,W + 144\,W^2
\right  )  \nonumber \\
 N_{79} &=& 2\,{\sqrt{2\,{\Cindex{79}{1}}^2 + {\Cindex{79}{2}}^2 + 2\,{\Cindex{79}{3}}^2}} \nonumber \eeq 
\beq
\ket{\Psi}_{80}& = &\ket{3, {1 \over 2} , {3 \over 4} ,\Gamma_{4,3}} \nonumber \\ 
& = &
\Cindex{80}{1} \left ( 
\ket{002u} - \ket{00u2} - \ket{02u0} + \ket{0u20} + \ket{200u} + \ket{2u00} - \ket{u002} - \ket{u200}\right) 
 \nonumber \\
& + &
\Cindex{80}{2} \left ( 
\ket{020u} + \ket{0u02} - \ket{20u0} - \ket{u020}\right) 
 \nonumber \\
& + &
\Cindex{80}{3} \left ( 
\ket{0duu} - \ket{0uud} + \ket{d0uu} - \ket{duu0} - \ket{u0du} - \ket{ud0u} + \ket{uu0d} + \ket{uud0}\right) 
\nonumber \eeq
\beq
C_{80-1} &=&
\frac{t}{2\,{\sqrt{2}}}
\, \left ( J - 4\,t + U - 2\,W - {\sqrt{\Aindex{5}}}\,\cos (\Thetaindex{4}) + {\sqrt{3}}\,{\sqrt{\Aindex{5}}}\,\sin (\Thetaindex{4})  \right ) \nonumber \\
C_{80-2} &=&
\frac{t}{3\,{\sqrt{2}}}
\, \left ( J + 12\,t + U - 2\,W - {\sqrt{\Aindex{5}}}\,\cos (\Thetaindex{4}) + {\sqrt{3}}\,{\sqrt{\Aindex{5}}}\,\sin (\Thetaindex{4})  \right ) \nonumber \\
C_{80-3} &=&
\frac{1}{18\,{\sqrt{2}}}
\,\left ( {\Aindex{20}}^2 - 3\,\Aindex{20}\,t - 36\,t^2 - 6\,\Aindex{20}\,U\right . \nonumber \\
&& \hspace{1cm} 
 + 
 \left . 9\,t\,U + 9\,U^2 - 24\,\Aindex{20}\,W + 36\,t\,W + 72\,U\,W + 144\,W^2
\right  )  \nonumber \\
 N_{80} &=& 2\,{\sqrt{2\,{\Cindex{80}{1}}^2 + {\Cindex{80}{2}}^2 + 2\,{\Cindex{80}{3}}^2}} \nonumber \eeq 
\beq
\ket{\Psi}_{81}& = &\ket{3, {1 \over 2} , {3 \over 4} ,\Gamma_{5,1}} \nonumber \\ 
& = &
\Cindex{81}{1} \left ( 
\ket{002u} - \ket{00u2} + \ket{02u0} - \ket{0u20} - \ket{200u} + \ket{2u00} + \ket{u002} - \ket{u200}\right) 
 \nonumber \\
& + &
\Cindex{81}{2} \left ( 
\ket{0duu} + \ket{0uud} + \ket{d0uu} + \ket{duu0} + \ket{u0du} + \ket{ud0u} + \ket{uu0d} + \ket{uud0}\right) 
 \nonumber \\
& + &
\Cindex{81}{3} \left ( 
\ket{0udu} + \ket{du0u} + \ket{u0ud} + \ket{udu0}\right) 
\nonumber \eeq
\beq
C_{81-1} &=&
\frac{{\sqrt{\frac{3}{2}}}\,t}{2} \nonumber \\
C_{81-2} &=&
\frac{1}{4\,{\sqrt{6}}}
\, \left ( {\sqrt{\Aindex{7}}} + J + 2\,t + U - 2\,W  \right ) \nonumber \\
C_{81-3} &=&
\frac{-1}{2\,{\sqrt{6}}}
\, \left ( {\sqrt{\Aindex{7}}} + J + 2\,t + U - 2\,W  \right ) \nonumber \\
 N_{81} &=& 2\,{\sqrt{2\,{\Cindex{81}{1}}^2 + 2\,{\Cindex{81}{2}}^2 + {\Cindex{81}{3}}^2}} \nonumber \eeq 
\beq
\ket{\Psi}_{82}& = &\ket{3, {1 \over 2} , {3 \over 4} ,\Gamma_{5,1}} \nonumber \\ 
& = &
\Cindex{82}{1} \left ( 
\ket{002u} - \ket{00u2} + \ket{02u0} - \ket{0u20} - \ket{200u} + \ket{2u00} + \ket{u002} - \ket{u200}\right) 
 \nonumber \\
& + &
\Cindex{82}{2} \left ( 
\ket{0duu} + \ket{0uud} + \ket{d0uu} + \ket{duu0} + \ket{u0du} + \ket{ud0u} + \ket{uu0d} + \ket{uud0}\right) 
 \nonumber \\
& + &
\Cindex{82}{3} \left ( 
\ket{0udu} + \ket{du0u} + \ket{u0ud} + \ket{udu0}\right) 
\nonumber \eeq
\beq
C_{82-1} &=&
\frac{{\sqrt{\frac{3}{2}}}\,t}{2} \nonumber \\
C_{82-2} &=&
\frac{-1}{4\,{\sqrt{6}}}
\, \left ( {\sqrt{\Aindex{7}}} - J - 2\,t - U + 2\,W  \right ) \nonumber \\
C_{82-3} &=&
\frac{1}{2\,{\sqrt{6}}}
\, \left ( {\sqrt{\Aindex{7}}} - J - 2\,t - U + 2\,W  \right ) \nonumber \\
 N_{82} &=& 2\,{\sqrt{2\,{\Cindex{82}{1}}^2 + 2\,{\Cindex{82}{2}}^2 + {\Cindex{82}{3}}^2}} \nonumber \eeq 
\beq
\ket{\Psi}_{83}& = &\ket{3, {1 \over 2} , {15 \over 4} ,\Gamma_{5,1}} \nonumber \\ 
&=& \frac{1}{2\,{\sqrt{3}}}
 \left ( \ket{0duu} + \ket{0udu} + \ket{0uud} + \ket{d0uu} + \ket{du0u} + \ket{duu0}\right . \nonumber \\
&& \hspace{1em} 
 + 
\left . \ket{u0du} + \ket{u0ud} + \ket{ud0u} + \ket{udu0} + \ket{uu0d} + \ket{uud0}\right ) \nonumber  
\eeq
\beq
\ket{\Psi}_{84}& = &\ket{3, {1 \over 2} , {3 \over 4} ,\Gamma_{5,2}} \nonumber \\ 
& = &
\Cindex{84}{1} \left ( 
\ket{020u} - \ket{02u0} - \ket{0u02} + \ket{0u20} - \ket{200u} + \ket{20u0} + \ket{u002} - \ket{u020}\right) 
 \nonumber \\
& + &
\Cindex{84}{2} \left ( 
\ket{0duu} - \ket{d0uu} - \ket{uu0d} + \ket{uud0}\right) 
 \nonumber \\
& + &
\Cindex{84}{3} \left ( 
\ket{0udu} + \ket{0uud} - \ket{du0u} + \ket{duu0} - \ket{u0du} - \ket{u0ud} - \ket{ud0u} + \ket{udu0}\right) 
\nonumber \eeq
\beq
C_{84-1} &=&
\frac{{\sqrt{\frac{3}{2}}}\,t}{2} \nonumber \\
C_{84-2} &=&
\frac{1}{2\,{\sqrt{6}}}
\, \left ( {\sqrt{\Aindex{7}}} + J + 2\,t + U - 2\,W  \right ) \nonumber \\
C_{84-3} &=&
\frac{-1}{4\,{\sqrt{6}}}
\, \left ( {\sqrt{\Aindex{7}}} + J + 2\,t + U - 2\,W  \right ) \nonumber \\
 N_{84} &=& 2\,{\sqrt{2\,{\Cindex{84}{1}}^2 + {\Cindex{84}{2}}^2 + 2\,{\Cindex{84}{3}}^2}} \nonumber \eeq 
\beq
\ket{\Psi}_{85}& = &\ket{3, {1 \over 2} , {3 \over 4} ,\Gamma_{5,2}} \nonumber \\ 
& = &
\Cindex{85}{1} \left ( 
\ket{020u} - \ket{02u0} - \ket{0u02} + \ket{0u20} - \ket{200u} + \ket{20u0} + \ket{u002} - \ket{u020}\right) 
 \nonumber \\
& + &
\Cindex{85}{2} \left ( 
\ket{0duu} - \ket{d0uu} - \ket{uu0d} + \ket{uud0}\right) 
 \nonumber \\
& + &
\Cindex{85}{3} \left ( 
\ket{0udu} + \ket{0uud} - \ket{du0u} + \ket{duu0} - \ket{u0du} - \ket{u0ud} - \ket{ud0u} + \ket{udu0}\right) 
\nonumber \eeq
\beq
C_{85-1} &=&
\frac{{\sqrt{\frac{3}{2}}}\,t}{2} \nonumber \\
C_{85-2} &=&
\frac{-1}{2\,{\sqrt{6}}}
\, \left ( {\sqrt{\Aindex{7}}} - J - 2\,t - U + 2\,W  \right ) \nonumber \\
C_{85-3} &=&
\frac{1}{4\,{\sqrt{6}}}
\, \left ( {\sqrt{\Aindex{7}}} - J - 2\,t - U + 2\,W  \right ) \nonumber \\
 N_{85} &=& 2\,{\sqrt{2\,{\Cindex{85}{1}}^2 + {\Cindex{85}{2}}^2 + 2\,{\Cindex{85}{3}}^2}} \nonumber \eeq 
\beq
\ket{\Psi}_{86}& = &\ket{3, {1 \over 2} , {15 \over 4} ,\Gamma_{5,2}} \nonumber \\ 
&=& \frac{1}{2\,{\sqrt{3}}}
 \left ( \ket{0duu} + \ket{0udu} + \ket{0uud} - \ket{d0uu} - \ket{du0u} + \ket{duu0}\right . \nonumber \\
&& \hspace{1em} 
\left . -\ket{u0du} - \ket{u0ud} - \ket{ud0u} + \ket{udu0} - \ket{uu0d} + \ket{uud0}\right ) \nonumber  
\eeq
\beq
\ket{\Psi}_{87}& = &\ket{3, {1 \over 2} , {3 \over 4} ,\Gamma_{5,3}} \nonumber \\ 
& = &
\Cindex{87}{1} \left ( 
\ket{002u} - \ket{00u2} - \ket{020u} + \ket{0u02} + \ket{20u0} - \ket{2u00} - \ket{u020} + \ket{u200}\right) 
 \nonumber \\
& + &
\Cindex{87}{2} \left ( 
\ket{0duu} + \ket{0udu} + \ket{d0uu} - \ket{du0u} + \ket{u0ud} - \ket{udu0} - \ket{uu0d} - \ket{uud0}\right) 
 \nonumber \\
& + &
\Cindex{87}{3} \left ( 
\ket{0uud} - \ket{duu0} + \ket{u0du} - \ket{ud0u}\right) 
\nonumber \eeq
\beq
C_{87-1} &=&
\frac{-\left( {\sqrt{\frac{3}{2}}}\,t \right) }{2} \nonumber \\
C_{87-2} &=&
\frac{-1}{4\,{\sqrt{6}}}
\, \left ( {\sqrt{\Aindex{7}}} + J + 2\,t + U - 2\,W  \right ) \nonumber \\
C_{87-3} &=&
\frac{1}{2\,{\sqrt{6}}}
\, \left ( {\sqrt{\Aindex{7}}} + J + 2\,t + U - 2\,W  \right ) \nonumber \\
 N_{87} &=& 2\,{\sqrt{2\,{\Cindex{87}{1}}^2 + 2\,{\Cindex{87}{2}}^2 + {\Cindex{87}{3}}^2}} \nonumber \eeq 
\beq
\ket{\Psi}_{88}& = &\ket{3, {1 \over 2} , {3 \over 4} ,\Gamma_{5,3}} \nonumber \\ 
& = &
\Cindex{88}{1} \left ( 
\ket{002u} - \ket{00u2} - \ket{020u} + \ket{0u02} + \ket{20u0} - \ket{2u00} - \ket{u020} + \ket{u200}\right) 
 \nonumber \\
& + &
\Cindex{88}{2} \left ( 
\ket{0duu} + \ket{0udu} + \ket{d0uu} - \ket{du0u} + \ket{u0ud} - \ket{udu0} - \ket{uu0d} - \ket{uud0}\right) 
 \nonumber \\
& + &
\Cindex{88}{3} \left ( 
\ket{0uud} - \ket{duu0} + \ket{u0du} - \ket{ud0u}\right) 
\nonumber \eeq
\beq
C_{88-1} &=&
\frac{-\left( {\sqrt{\frac{3}{2}}}\,t \right) }{2} \nonumber \\
C_{88-2} &=&
\frac{1}{4\,{\sqrt{6}}}
\, \left ( {\sqrt{\Aindex{7}}} - J - 2\,t - U + 2\,W  \right ) \nonumber \\
C_{88-3} &=&
\frac{-1}{2\,{\sqrt{6}}}
\, \left ( {\sqrt{\Aindex{7}}} - J - 2\,t - U + 2\,W  \right ) \nonumber \\
 N_{88} &=& 2\,{\sqrt{2\,{\Cindex{88}{1}}^2 + 2\,{\Cindex{88}{2}}^2 + {\Cindex{88}{3}}^2}} \nonumber \eeq 
\beq
\ket{\Psi}_{89}& = &\ket{3, {1 \over 2} , {15 \over 4} ,\Gamma_{5,3}} \nonumber \\ 
&=& \frac{1}{2\,{\sqrt{3}}}
 \left ( \ket{0duu} + \ket{0udu} + \ket{0uud} + \ket{d0uu} - \ket{du0u} - \ket{duu0}\right . \nonumber \\
&& \hspace{1em} 
 + 
\left . \ket{u0du} + \ket{u0ud} - \ket{ud0u} - \ket{udu0} - \ket{uu0d} - \ket{uud0}\right ) \nonumber  
\eeq
{\subsection*{\boldmath Unnormalized eigenvectors for ${\rm  N_e}=3$ and   ${\rm m_s}$= $ {3 \over 2} $.}
\beq
\ket{\Psi}_{90}& = &\ket{3, {3 \over 2} , {15 \over 4} ,\Gamma_2} \nonumber \\ 
&=& \frac{1}{2}
 \left ( \ket{0uuu} - \ket{u0uu} + \ket{uu0u} - \ket{uuu0}\right) \nonumber 
\eeq
\beq
\ket{\Psi}_{91}& = &\ket{3, {3 \over 2} , {15 \over 4} ,\Gamma_{5,1}} \nonumber \\ 
&=& \frac{1}{2}
 \left ( \ket{0uuu} + \ket{u0uu} + \ket{uu0u} + \ket{uuu0}\right) \nonumber 
\eeq
\beq
\ket{\Psi}_{92}& = &\ket{3, {3 \over 2} , {15 \over 4} ,\Gamma_{5,2}} \nonumber \\ 
&=& \frac{1}{2}
 \left ( \ket{0uuu} - \ket{u0uu} - \ket{uu0u} + \ket{uuu0}\right) \nonumber 
\eeq
\beq
\ket{\Psi}_{93}& = &\ket{3, {3 \over 2} , {15 \over 4} ,\Gamma_{5,3}} \nonumber \\ 
&=& \frac{1}{2}
 \left ( \ket{0uuu} + \ket{u0uu} - \ket{uu0u} - \ket{uuu0}\right) \nonumber 
\eeq
{\subsection*{\boldmath Unnormalized eigenvectors for ${\rm  N_e}=4$ and   ${\rm m_s}$= $-2$.}
\beq
\ket{\Psi}_{94}& = &\ket{4,-2,6,\Gamma_2} \nonumber \\ 
&=& 1
 \left ( \ket{dddd}\right) \nonumber 
\eeq
{\subsection*{\boldmath Unnormalized eigenvectors for ${\rm  N_e}=4$ and   ${\rm m_s}$= $-1$.}
\beq
\ket{\Psi}_{95}& = &\ket{4,-1,2,\Gamma_2} \nonumber \\ 
&=& \frac{1}{2\,{\sqrt{3}}}
 \left ( \ket{02dd} - \ket{0d2d} + \ket{0dd2} - \ket{20dd} + \ket{2d0d} - \ket{2dd0}\right . \nonumber \\
&& \hspace{1em} 
 + 
\left . \ket{d02d} - \ket{d0d2} - \ket{d20d} + \ket{d2d0} + \ket{dd02} - \ket{dd20}\right ) \nonumber  
\eeq
\beq
\ket{\Psi}_{96}& = &\ket{4,-1,6,\Gamma_2} \nonumber \\ 
&=& \frac{1}{2}
 \left ( \ket{dddu} + \ket{ddud} + \ket{dudd} + \ket{uddd}\right) \nonumber 
\eeq
\beq
\ket{\Psi}_{97}& = &\ket{4,-1,2,\Gamma_{3,1}} \nonumber \\ 
&=& \frac{1}{2\,{\sqrt{2}}}
 \left ( \ket{02dd} - \ket{0dd2} - \ket{20dd} + \ket{2dd0} - \ket{d02d} + \ket{d20d} + \ket{dd02} - \ket{dd20}\right) \nonumber 
\eeq
\beq
\ket{\Psi}_{98}& = &\ket{4,-1,2,\Gamma_{3,2}} \nonumber \\ 
& = &
\Cindex{98}{1} \left ( 
\ket{02dd} + \ket{0dd2} - \ket{20dd} - \ket{2dd0} + \ket{d02d} - \ket{d20d} + \ket{dd02} - \ket{dd20}\right) 
 \nonumber \\
& + &
\Cindex{98}{2} \left ( 
\ket{0d2d} - \ket{2d0d} + \ket{d0d2} - \ket{d2d0}\right) 
\nonumber \eeq
\beq
C_{98-1} &=&
\frac{-1}{2\,{\sqrt{6}}} \nonumber \\
C_{98-2} &=&
-\left( \frac{1}{{\sqrt{6}}} \right)  \nonumber \\
 N_{98} &=& 2\,{\sqrt{2\,{\Cindex{98}{1}}^2 + {\Cindex{98}{2}}^2}} \nonumber \eeq 
\beq
\ket{\Psi}_{99}& = &\ket{4,-1,2,\Gamma_{4,1}} \nonumber \\ 
&=& \frac{1}{2\,{\sqrt{2}}}
 \left ( \ket{0d2d} + \ket{0dd2} + \ket{2d0d} + \ket{2dd0} + \ket{d02d} + \ket{d0d2} + \ket{d20d} + \ket{d2d0}\right) \nonumber 
\eeq
\beq
\ket{\Psi}_{100}& = &\ket{4,-1,2,\Gamma_{4,2}} \nonumber \\ 
&=& \frac{1}{2\,{\sqrt{2}}}
 \left ( \ket{02dd} + \ket{0d2d} + \ket{20dd} + \ket{2d0d} - \ket{d0d2} - \ket{d2d0} - \ket{dd02} - \ket{dd20}\right) \nonumber 
\eeq
\beq
\ket{\Psi}_{101}& = &\ket{4,-1,2,\Gamma_{4,3}} \nonumber \\ 
&=& \frac{1}{2\,{\sqrt{2}}}
 \left ( \ket{02dd} - \ket{0dd2} + \ket{20dd} - \ket{2dd0} + \ket{d02d} + \ket{d20d} + \ket{dd02} + \ket{dd20}\right) \nonumber 
\eeq
\beq
\ket{\Psi}_{102}& = &\ket{4,-1,2,\Gamma_{5,1}} \nonumber \\ 
& = &
\Cindex{102}{1} \left ( 
\ket{02dd} + \ket{0dd2} + \ket{20dd} + \ket{2dd0} - \ket{d02d} - \ket{d20d} + \ket{dd02} + \ket{dd20}\right) 
 \nonumber \\
& + &
\Cindex{102}{2} \left ( 
\ket{0d2d} - \ket{2d0d} - \ket{d0d2} + \ket{d2d0}\right) 
 \nonumber \\
& + &
\Cindex{102}{3} \left ( 
\ket{dddu} - \ket{ddud} + \ket{dudd} - \ket{uddd}\right) 
\nonumber \eeq
\beq
C_{102-1} &=&
\frac{-t}{3}
\, \left ( J + U - 2\,W + 2\,{\sqrt{\Aindex{2}}}\,\cos (\Thetaindex{3})  \right ) \nonumber \\
C_{102-2} &=&
-4\,t^2 \nonumber \\
C_{102-3} &=&
\frac{1}{72}
\,\left ( -{\Aindex{11}}^2 + 6\,\Aindex{11}\,J\right . \nonumber \\
&& \hspace{1cm} 
 \left . -9\,J^2 + 288\,t^2 + 12\,\Aindex{11}\,U - 36\,J\,U
\right .  \nonumber \\
&& \hspace{1cm} 
 \left . -36\,U^2 + 120\,\Aindex{11}\,W - 360\,J\,W - 720\,U\,W - 3600\,W^2
\right )  \nonumber \\
 N_{102} &=& 2\,{\sqrt{2\,{\Cindex{102}{1}}^2 + {\Cindex{102}{2}}^2 + {\Cindex{102}{3}}^2}} \nonumber \eeq 
\beq
\ket{\Psi}_{103}& = &\ket{4,-1,2,\Gamma_{5,1}} \nonumber \\ 
& = &
\Cindex{103}{1} \left ( 
\ket{02dd} + \ket{0dd2} + \ket{20dd} + \ket{2dd0} - \ket{d02d} - \ket{d20d} + \ket{dd02} + \ket{dd20}\right) 
 \nonumber \\
& + &
\Cindex{103}{2} \left ( 
\ket{0d2d} - \ket{2d0d} - \ket{d0d2} + \ket{d2d0}\right) 
 \nonumber \\
& + &
\Cindex{103}{3} \left ( 
\ket{dddu} - \ket{ddud} + \ket{dudd} - \ket{uddd}\right) 
\nonumber \eeq
\beq
C_{103-1} &=&
\frac{t}{6}
\, \left ( 2\,\Aindex{18} - 3\,J - 6\,U - 60\,W  \right ) \nonumber \\
C_{103-2} &=&
-4\,t^2 \nonumber \\
C_{103-3} &=&
\frac{1}{72}
\,\left ( -4\,{\Aindex{18}}^2 + 12\,\Aindex{18}\,J\right . \nonumber \\
&& \hspace{1cm} 
 \left . -9\,J^2 + 288\,t^2 + 24\,\Aindex{18}\,U - 36\,J\,U
\right .  \nonumber \\
&& \hspace{1cm} 
 \left . -36\,U^2 + 240\,\Aindex{18}\,W - 360\,J\,W - 720\,U\,W - 3600\,W^2
\right )  \nonumber \\
 N_{103} &=& 2\,{\sqrt{2\,{\Cindex{103}{1}}^2 + {\Cindex{103}{2}}^2 + {\Cindex{103}{3}}^2}} \nonumber \eeq 
\beq
\ket{\Psi}_{104}& = &\ket{4,-1,2,\Gamma_{5,1}} \nonumber \\ 
& = &
\Cindex{104}{1} \left ( 
\ket{02dd} + \ket{0dd2} + \ket{20dd} + \ket{2dd0} - \ket{d02d} - \ket{d20d} + \ket{dd02} + \ket{dd20}\right) 
 \nonumber \\
& + &
\Cindex{104}{2} \left ( 
\ket{0d2d} - \ket{2d0d} - \ket{d0d2} + \ket{d2d0}\right) 
 \nonumber \\
& + &
\Cindex{104}{3} \left ( 
\ket{dddu} - \ket{ddud} + \ket{dudd} - \ket{uddd}\right) 
\nonumber \eeq
\beq
C_{104-1} &=&
\frac{t}{6}
\, \left ( 2\,\Aindex{16} - 3\,J - 6\,U - 60\,W  \right ) \nonumber \\
C_{104-2} &=&
-4\,t^2 \nonumber \\
C_{104-3} &=&
\frac{1}{72}
\,\left ( -4\,{\Aindex{16}}^2 + 12\,\Aindex{16}\,J\right . \nonumber \\
&& \hspace{1cm} 
 \left . -9\,J^2 + 288\,t^2 + 24\,\Aindex{16}\,U - 36\,J\,U
\right .  \nonumber \\
&& \hspace{1cm} 
 \left . -36\,U^2 + 240\,\Aindex{16}\,W - 360\,J\,W - 720\,U\,W - 3600\,W^2
\right )  \nonumber \\
 N_{104} &=& 2\,{\sqrt{2\,{\Cindex{104}{1}}^2 + {\Cindex{104}{2}}^2 + {\Cindex{104}{3}}^2}} \nonumber \eeq 
\beq
\ket{\Psi}_{105}& = &\ket{4,-1,2,\Gamma_{5,2}} \nonumber \\ 
& = &
\Cindex{105}{1} \left ( 
\ket{02dd} - \ket{20dd} - \ket{dd02} + \ket{dd20}\right) 
 \nonumber \\
& + &
\Cindex{105}{2} \left ( 
\ket{0d2d} - \ket{0dd2} + \ket{2d0d} - \ket{2dd0} - \ket{d02d} + \ket{d0d2} - \ket{d20d} + \ket{d2d0}\right) 
 \nonumber \\
& + &
\Cindex{105}{3} \left ( 
\ket{dddu} + \ket{ddud} - \ket{dudd} - \ket{uddd}\right) 
\nonumber \eeq
\beq
C_{105-1} &=&
4\,t^2 \nonumber \\
C_{105-2} &=&
\frac{t}{3}
\, \left ( J + U - 2\,W + 2\,{\sqrt{\Aindex{2}}}\,\cos (\Thetaindex{3})  \right ) \nonumber \\
C_{105-3} &=&
\frac{1}{72}
\,\left ( -{\Aindex{11}}^2 + 6\,\Aindex{11}\,J\right . \nonumber \\
&& \hspace{1cm} 
 \left . -9\,J^2 + 288\,t^2 + 12\,\Aindex{11}\,U - 36\,J\,U
\right .  \nonumber \\
&& \hspace{1cm} 
 \left . -36\,U^2 + 120\,\Aindex{11}\,W - 360\,J\,W - 720\,U\,W - 3600\,W^2
\right )  \nonumber \\
 N_{105} &=& 2\,{\sqrt{{\Cindex{105}{1}}^2 + 2\,{\Cindex{105}{2}}^2 + {\Cindex{105}{3}}^2}} \nonumber \eeq 
\beq
\ket{\Psi}_{106}& = &\ket{4,-1,2,\Gamma_{5,2}} \nonumber \\ 
& = &
\Cindex{106}{1} \left ( 
\ket{02dd} - \ket{20dd} - \ket{dd02} + \ket{dd20}\right) 
 \nonumber \\
& + &
\Cindex{106}{2} \left ( 
\ket{0d2d} - \ket{0dd2} + \ket{2d0d} - \ket{2dd0} - \ket{d02d} + \ket{d0d2} - \ket{d20d} + \ket{d2d0}\right) 
 \nonumber \\
& + &
\Cindex{106}{3} \left ( 
\ket{dddu} + \ket{ddud} - \ket{dudd} - \ket{uddd}\right) 
\nonumber \eeq
\beq
C_{106-1} &=&
4\,t^2 \nonumber \\
C_{106-2} &=&
\frac{-t}{6}
\, \left ( 2\,\Aindex{18} - 3\,J - 6\,U - 60\,W  \right ) \nonumber \\
C_{106-3} &=&
\frac{1}{72}
\,\left ( -4\,{\Aindex{18}}^2 + 12\,\Aindex{18}\,J\right . \nonumber \\
&& \hspace{1cm} 
 \left . -9\,J^2 + 288\,t^2 + 24\,\Aindex{18}\,U - 36\,J\,U
\right .  \nonumber \\
&& \hspace{1cm} 
 \left . -36\,U^2 + 240\,\Aindex{18}\,W - 360\,J\,W - 720\,U\,W - 3600\,W^2
\right )  \nonumber \\
 N_{106} &=& 2\,{\sqrt{{\Cindex{106}{1}}^2 + 2\,{\Cindex{106}{2}}^2 + {\Cindex{106}{3}}^2}} \nonumber \eeq 
\beq
\ket{\Psi}_{107}& = &\ket{4,-1,2,\Gamma_{5,2}} \nonumber \\ 
& = &
\Cindex{107}{1} \left ( 
\ket{02dd} - \ket{20dd} - \ket{dd02} + \ket{dd20}\right) 
 \nonumber \\
& + &
\Cindex{107}{2} \left ( 
\ket{0d2d} - \ket{0dd2} + \ket{2d0d} - \ket{2dd0} - \ket{d02d} + \ket{d0d2} - \ket{d20d} + \ket{d2d0}\right) 
 \nonumber \\
& + &
\Cindex{107}{3} \left ( 
\ket{dddu} + \ket{ddud} - \ket{dudd} - \ket{uddd}\right) 
\nonumber \eeq
\beq
C_{107-1} &=&
4\,t^2 \nonumber \\
C_{107-2} &=&
\frac{-t}{6}
\, \left ( 2\,\Aindex{16} - 3\,J - 6\,U - 60\,W  \right ) \nonumber \\
C_{107-3} &=&
\frac{1}{72}
\,\left ( -4\,{\Aindex{16}}^2 + 12\,\Aindex{16}\,J\right . \nonumber \\
&& \hspace{1cm} 
 \left . -9\,J^2 + 288\,t^2 + 24\,\Aindex{16}\,U - 36\,J\,U
\right .  \nonumber \\
&& \hspace{1cm} 
 \left . -36\,U^2 + 240\,\Aindex{16}\,W - 360\,J\,W - 720\,U\,W - 3600\,W^2
\right )  \nonumber \\
 N_{107} &=& 2\,{\sqrt{{\Cindex{107}{1}}^2 + 2\,{\Cindex{107}{2}}^2 + {\Cindex{107}{3}}^2}} \nonumber \eeq 
\beq
\ket{\Psi}_{108}& = &\ket{4,-1,2,\Gamma_{5,3}} \nonumber \\ 
& = &
\Cindex{108}{1} \left ( 
\ket{02dd} - \ket{0d2d} + \ket{20dd} - \ket{2d0d} + \ket{d0d2} + \ket{d2d0} - \ket{dd02} - \ket{dd20}\right) 
 \nonumber \\
& + &
\Cindex{108}{2} \left ( 
\ket{0dd2} - \ket{2dd0} - \ket{d02d} + \ket{d20d}\right) 
 \nonumber \\
& + &
\Cindex{108}{3} \left ( 
\ket{dddu} - \ket{ddud} - \ket{dudd} + \ket{uddd}\right) 
\nonumber \eeq
\beq
C_{108-1} &=&
\frac{-t}{3}
\, \left ( J + U - 2\,W + 2\,{\sqrt{\Aindex{2}}}\,\cos (\Thetaindex{3})  \right ) \nonumber \\
C_{108-2} &=&
4\,t^2 \nonumber \\
C_{108-3} &=&
\frac{1}{72}
\,\left ( {\Aindex{11}}^2 - 6\,\Aindex{11}\,J\right . \nonumber \\
&& \hspace{1cm} 
 + 
 \left . 9\,J^2 - 288\,t^2 - 12\,\Aindex{11}\,U + 36\,J\,U
\right .  \nonumber \\
&& \hspace{1cm} 
 + 
 \left . 36\,U^2 - 120\,\Aindex{11}\,W + 360\,J\,W + 720\,U\,W + 3600\,W^2
\right )  \nonumber \\
 N_{108} &=& 2\,{\sqrt{2\,{\Cindex{108}{1}}^2 + {\Cindex{108}{2}}^2 + {\Cindex{108}{3}}^2}} \nonumber \eeq 
\beq
\ket{\Psi}_{109}& = &\ket{4,-1,2,\Gamma_{5,3}} \nonumber \\ 
& = &
\Cindex{109}{1} \left ( 
\ket{02dd} - \ket{0d2d} + \ket{20dd} - \ket{2d0d} + \ket{d0d2} + \ket{d2d0} - \ket{dd02} - \ket{dd20}\right) 
 \nonumber \\
& + &
\Cindex{109}{2} \left ( 
\ket{0dd2} - \ket{2dd0} - \ket{d02d} + \ket{d20d}\right) 
 \nonumber \\
& + &
\Cindex{109}{3} \left ( 
\ket{dddu} - \ket{ddud} - \ket{dudd} + \ket{uddd}\right) 
\nonumber \eeq
\beq
C_{109-1} &=&
\frac{t}{6}
\, \left ( 2\,\Aindex{18} - 3\,J - 6\,U - 60\,W  \right ) \nonumber \\
C_{109-2} &=&
4\,t^2 \nonumber \\
C_{109-3} &=&
\frac{1}{72}
\,\left ( 4\,{\Aindex{18}}^2 - 12\,\Aindex{18}\,J\right . \nonumber \\
&& \hspace{1cm} 
 + 
 \left . 9\,J^2 - 288\,t^2 - 24\,\Aindex{18}\,U + 36\,J\,U
\right .  \nonumber \\
&& \hspace{1cm} 
 + 
 \left . 36\,U^2 - 240\,\Aindex{18}\,W + 360\,J\,W + 720\,U\,W + 3600\,W^2
\right )  \nonumber \\
 N_{109} &=& 2\,{\sqrt{2\,{\Cindex{109}{1}}^2 + {\Cindex{109}{2}}^2 + {\Cindex{109}{3}}^2}} \nonumber \eeq 
\beq
\ket{\Psi}_{110}& = &\ket{4,-1,2,\Gamma_{5,3}} \nonumber \\ 
& = &
\Cindex{110}{1} \left ( 
\ket{02dd} - \ket{0d2d} + \ket{20dd} - \ket{2d0d} + \ket{d0d2} + \ket{d2d0} - \ket{dd02} - \ket{dd20}\right) 
 \nonumber \\
& + &
\Cindex{110}{2} \left ( 
\ket{0dd2} - \ket{2dd0} - \ket{d02d} + \ket{d20d}\right) 
 \nonumber \\
& + &
\Cindex{110}{3} \left ( 
\ket{dddu} - \ket{ddud} - \ket{dudd} + \ket{uddd}\right) 
\nonumber \eeq
\beq
C_{110-1} &=&
\frac{t}{6}
\, \left ( 2\,\Aindex{16} - 3\,J - 6\,U - 60\,W  \right ) \nonumber \\
C_{110-2} &=&
4\,t^2 \nonumber \\
C_{110-3} &=&
\frac{1}{72}
\,\left ( 4\,{\Aindex{16}}^2 - 12\,\Aindex{16}\,J\right . \nonumber \\
&& \hspace{1cm} 
 + 
 \left . 9\,J^2 - 288\,t^2 - 24\,\Aindex{16}\,U + 36\,J\,U
\right .  \nonumber \\
&& \hspace{1cm} 
 + 
 \left . 36\,U^2 - 240\,\Aindex{16}\,W + 360\,J\,W + 720\,U\,W + 3600\,W^2
\right )  \nonumber \\
 N_{110} &=& 2\,{\sqrt{2\,{\Cindex{110}{1}}^2 + {\Cindex{110}{2}}^2 + {\Cindex{110}{3}}^2}} \nonumber \eeq 
{\subsection*{\boldmath Unnormalized eigenvectors for ${\rm  N_e}=4$ and   ${\rm m_s}$= $0$.}
\beq
\ket{\Psi}_{111}& = &\ket{4,0,0,\Gamma_1} \nonumber \\ 
& = &
\Cindex{111}{1} \left ( 
\ket{0022} + \ket{0202} + \ket{0220} + \ket{2002} + \ket{2020} + \ket{2200}\right) 
 \nonumber \\
& + &
\Cindex{111}{2} \left ( 
	\ket{02du} - \ket{02ud} + \ket{0d2u} + \ket{0du2} 
	- \ket{0u2d} - \ket{0ud2} \right . \nonumber \\
&& \hspace{3em}  
\left . + \ket{20du} - \ket{20ud} + \ket{2d0u} + \ket{2du0} - \ket{2u0d} - \ket{2ud0}\right . \nonumber \\
&& \hspace{3em}  
\left . + \ket{d02u} + \ket{d0u2} + \ket{d20u} + \ket{d2u0} + \ket{du02} + \ket{du20}
\right . 
\nonumber \\
&& \hspace{3em}
\left . 
- \ket{u02d} - \ket{u0d2} - \ket{u20d} - \ket{u2d0} - \ket{ud02} - \ket{ud20}\right ) 
\nonumber \eeq
\beq
C_{111-1} &=&
2\,{\sqrt{\frac{2}{3}}}\,t \nonumber \\
C_{111-2} &=&
\frac{1}{4\,{\sqrt{6}}}
\, \left ( {\sqrt{\Aindex{3}}} + J + U - 2\,W  \right ) \nonumber \\
 N_{111} &=& {\sqrt{6\,{\Cindex{111}{1}}^2 + 24\,{\Cindex{111}{2}}^2}} \nonumber \eeq 
\beq
\ket{\Psi}_{112}& = &\ket{4,0,0,\Gamma_1} \nonumber \\ 
& = &
\Cindex{112}{1} \left ( 
\ket{0022} + \ket{0202} + \ket{0220} + \ket{2002} + \ket{2020} + \ket{2200}\right) 
 \nonumber \\
& + &
\Cindex{112}{2} \left ( 
\ket{02du} - \ket{02ud} + \ket{0d2u} + \ket{0du2} - \ket{0u2d} - \ket{0ud2} 
\right .
\nonumber 
\\
&& \hspace{3em} 
\left . 
+ \ket{20du} - \ket{20ud} + \ket{2d0u} + \ket{2du0} - \ket{2u0d} - \ket{2ud0}\right . \nonumber \\
&& \hspace{3em} 
\left . + \ket{d02u} + \ket{d0u2} + \ket{d20u} + \ket{d2u0} + \ket{du02} + \ket{du20} 
\right .
\nonumber 
\\
&& \hspace{3em} 
\left . 
- \ket{u02d} - \ket{u0d2} - \ket{u20d} - \ket{u2d0} - \ket{ud02} - \ket{ud20}\right ) 
\nonumber \eeq
\beq
C_{112-1} &=&
2\,{\sqrt{\frac{2}{3}}}\,t \nonumber \\
C_{112-2} &=&
\frac{-1}{4\,{\sqrt{6}}}
\, \left ( {\sqrt{\Aindex{3}}} - J - U + 2\,W  \right ) \nonumber \\
 N_{112} &=& {\sqrt{6\,{\Cindex{112}{1}}^2 + 24\,{\Cindex{112}{2}}^2}} \nonumber \eeq 
\beq
\ket{\Psi}_{113}& = &\ket{4,0,2,\Gamma_2} \nonumber \\ 
&=& \frac{1}{2\,{\sqrt{6}}}
 \left ( \ket{02du} + \ket{02ud} - \ket{0d2u} + \ket{0du2} - \ket{0u2d} + \ket{0ud2} 
\right .
\nonumber 
\\
&& \hspace{3em} 
\left . 
- \ket{20du} - \ket{20ud} + \ket{2d0u} - \ket{2du0} + \ket{2u0d} - \ket{2ud0}
\right . 
\nonumber 
\\
&& \hspace{3em} 
\left . 
+ \ket{d02u} - \ket{d0u2} - \ket{d20u} + \ket{d2u0} + \ket{du02} - \ket{du20} 
\right .
\nonumber 
\\
&& \hspace{3em} 
\left . + \ket{u02d} - \ket{u0d2} - \ket{u20d} + \ket{u2d0} + \ket{ud02} - \ket{ud20}\right ) \nonumber  
\eeq
\beq
\ket{\Psi}_{114}& = &\ket{4,0,6,\Gamma_2} \nonumber \\ 
&=& \frac{1}{{\sqrt{6}}}
 \left ( \ket{dduu} + \ket{dudu} + \ket{duud} + \ket{uddu} + \ket{udud} + \ket{uudd}\right) \nonumber 
\eeq
\beq
\ket{\Psi}_{115}& = &\ket{4,0,0,\Gamma_{3,1}} \nonumber \\ 
& = &
\Cindex{115}{1} \left ( 
\ket{0022} + \ket{0220} + \ket{2002} + \ket{2200}\right) 
 \nonumber \\
& + &
\Cindex{115}{2} \left ( 
\ket{0202} + \ket{2020}\right) 
 \nonumber \\
& + &
\Cindex{115}{3} \left ( 
\ket{02du} - \ket{02ud} + \ket{0du2} - \ket{0ud2} + \ket{20du} - \ket{20ud} + \ket{2du0} - \ket{2ud0}\right . \nonumber \\
&& \hspace{3em} 
 + 
\left . \ket{d02u} + \ket{d20u} + \ket{du02} + \ket{du20} - \ket{u02d} - \ket{u20d} - \ket{ud02} - \ket{ud20}\right ) 
\nonumber \\
& + &
\Cindex{115}{4} \left ( 
\ket{0d2u} - \ket{0u2d} + \ket{2d0u} - \ket{2u0d} + \ket{d0u2} + \ket{d2u0} - \ket{u0d2} - \ket{u2d0}\right) 
 \nonumber \\
& + &
\Cindex{115}{5} \left ( 
\ket{dduu} - \ket{duud} - \ket{uddu} + \ket{uudd}\right) 
\nonumber \eeq
\beq
C_{115-1} &=&
\frac{t}{3}
\, \left ( {\sqrt{3}}\,J + {\sqrt{3}}\,U - 2\,{\sqrt{3}}\,W - 2\,{\sqrt{\Aindex{1}}}\,\cos (\Thetaindex{1})  \right ) \nonumber \\
C_{115-2} &=&
\frac{-2\,t}{3}
\, \left ( {\sqrt{3}}\,J + {\sqrt{3}}\,U - 2\,{\sqrt{3}}\,W - 2\,{\sqrt{\Aindex{1}}}\,\cos (\Thetaindex{1})  \right ) \nonumber \\
C_{115-3} &=&
\frac{-1}{12\,{\sqrt{3}}}
\,\left ( 3\,J^2 + 6\,J\,U + 3\,U^2 - 12\,J\,W - 12\,U\,W\right . \nonumber \\
&& \hspace{1cm} 
 + 
 \left . 12\,W^2 - 4\,\Aindex{1}\,{\cos (\Thetaindex{1})}^2
\right  )  \nonumber \\
C_{115-4} &=&
\frac{1}{6\,{\sqrt{3}}}
\,\left ( 3\,J^2 + 6\,J\,U + 3\,U^2 - 12\,J\,W - 12\,U\,W\right . \nonumber \\
&& \hspace{1cm} 
 + 
 \left . 12\,W^2 - 4\,\Aindex{1}\,{\cos (\Thetaindex{1})}^2
\right  )  \nonumber \\
C_{115-5} &=&
t
\, \left ( {\sqrt{3}}\,J + {\sqrt{3}}\,U - 2\,{\sqrt{3}}\,W + 2\,{\sqrt{\Aindex{1}}}\,\cos (\Thetaindex{1})  \right ) \nonumber \\
 N_{115} &=& {\sqrt{4\,{\Cindex{115}{1}}^2 + 2\,{\Cindex{115}{2}}^2 + 16\,{\Cindex{115}{3}}^2 + 8\,{\Cindex{115}{4}}^2 + 4\,{\Cindex{115}{5}}^2}} \nonumber \eeq 
\beq
\ket{\Psi}_{116}& = &\ket{4,0,0,\Gamma_{3,1}} \nonumber \\ 
& = &
\Cindex{116}{1} \left ( 
\ket{0022} + \ket{0220} + \ket{2002} + \ket{2200}\right) 
 \nonumber \\
& + &
\Cindex{116}{2} \left ( 
\ket{0202} + \ket{2020}\right) 
 \nonumber \\
& + &
\Cindex{116}{3} \left ( 
\ket{02du} - \ket{02ud} + \ket{0du2} - \ket{0ud2} + \ket{20du} - \ket{20ud} + \ket{2du0} - \ket{2ud0}\right . \nonumber \\
&& \hspace{3em} 
 + 
\left . \ket{d02u} + \ket{d20u} + \ket{du02} + \ket{du20} - \ket{u02d} - \ket{u20d} - \ket{ud02} - \ket{ud20}\right ) 
\nonumber \\
& + &
\Cindex{116}{4} \left ( 
\ket{0d2u} - \ket{0u2d} + \ket{2d0u} - \ket{2u0d} + \ket{d0u2} + \ket{d2u0} - \ket{u0d2} - \ket{u2d0}\right) 
 \nonumber \\
& + &
\Cindex{116}{5} \left ( 
\ket{dduu} - \ket{duud} - \ket{uddu} + \ket{uudd}\right) 
\nonumber \eeq
\beq
C_{116-1} &=&
\frac{t}{3}
\, \left ( {\sqrt{3}}\,J + {\sqrt{3}}\,U - 2\,{\sqrt{3}}\,W + {\sqrt{\Aindex{1}}}\,\cos (\Thetaindex{1}) + {\sqrt{3}}\,{\sqrt{\Aindex{1}}}\,\sin (\Thetaindex{1})  \right ) \nonumber \\
C_{116-2} &=&
\frac{-2\,t}{3}
\, \left ( {\sqrt{3}}\,J + {\sqrt{3}}\,U - 2\,{\sqrt{3}}\,W + {\sqrt{\Aindex{1}}}\,\cos (\Thetaindex{1}) + {\sqrt{3}}\,{\sqrt{\Aindex{1}}}\,\sin (\Thetaindex{1})  \right ) \nonumber \\
C_{116-3} &=&
\frac{1}{36}
\,\left ( 2\,{\sqrt{3}}\,\Aindex{1} - 3\,{\sqrt{3}}\,J^2 - 6\,{\sqrt{3}}\,J\,U - 3\,{\sqrt{3}}\,U^2 + 12\,{\sqrt{3}}\,J\,W\right . \nonumber \\
&& \hspace{1cm} 
 + 
 \left . 12\,{\sqrt{3}}\,U\,W - 12\,{\sqrt{3}}\,W^2 - {\sqrt{3}}\,\Aindex{1}\,\cos (2\,\Thetaindex{1}) + 3\,\Aindex{1}\,\sin (2\,\Thetaindex{1})
\right  )  \nonumber \\
C_{116-4} &=&
\frac{1}{18}
\,\left ( -2\,{\sqrt{3}}\,\Aindex{1} + 3\,{\sqrt{3}}\,J^2 + 6\,{\sqrt{3}}\,J\,U + 3\,{\sqrt{3}}\,U^2 - 12\,{\sqrt{3}}\,J\,W - 12\,{\sqrt{3}}\,U\,W\right . \nonumber \\
&& \hspace{1cm} 
 + 
 \left . 12\,{\sqrt{3}}\,W^2 + {\sqrt{3}}\,\Aindex{1}\,\cos (2\,\Thetaindex{1}) - 3\,\Aindex{1}\,\sin (2\,\Thetaindex{1})
\right  )  \nonumber \\
C_{116-5} &=&
t
\, \left ( {\sqrt{3}}\,J + {\sqrt{3}}\,U - 2\,{\sqrt{3}}\,W - {\sqrt{\Aindex{1}}}\,\cos (\Thetaindex{1}) - {\sqrt{3}}\,{\sqrt{\Aindex{1}}}\,\sin (\Thetaindex{1})  \right ) \nonumber \\
 N_{116} &=& {\sqrt{4\,{\Cindex{116}{1}}^2 + 2\,{\Cindex{116}{2}}^2 + 16\,{\Cindex{116}{3}}^2 + 8\,{\Cindex{116}{4}}^2 + 4\,{\Cindex{116}{5}}^2}} \nonumber \eeq 
\beq
\ket{\Psi}_{117}& = &\ket{4,0,0,\Gamma_{3,1}} \nonumber \\ 
& = &
\Cindex{117}{1} \left ( 
\ket{0022} + \ket{0220} + \ket{2002} + \ket{2200}\right) 
 \nonumber \\
& + &
\Cindex{117}{2} \left ( 
\ket{0202} + \ket{2020}\right) 
 \nonumber \\
& + &
\Cindex{117}{3} \left ( 
\ket{02du} - \ket{02ud} + \ket{0du2} - \ket{0ud2} + \ket{20du} - \ket{20ud} + \ket{2du0} - \ket{2ud0}\right . \nonumber \\
&& \hspace{3em} 
 + 
\left . \ket{d02u} + \ket{d20u} + \ket{du02} + \ket{du20} - \ket{u02d} - \ket{u20d} - \ket{ud02} - \ket{ud20}\right ) 
\nonumber \\
& + &
\Cindex{117}{4} \left ( 
\ket{0d2u} - \ket{0u2d} + \ket{2d0u} - \ket{2u0d} + \ket{d0u2} + \ket{d2u0} - \ket{u0d2} - \ket{u2d0}\right) 
 \nonumber \\
& + &
\Cindex{117}{5} \left ( 
\ket{dduu} - \ket{duud} - \ket{uddu} + \ket{uudd}\right) 
\nonumber \eeq
\beq
C_{117-1} &=&
\frac{t}{3}
\, \left ( {\sqrt{3}}\,J + {\sqrt{3}}\,U - 2\,{\sqrt{3}}\,W + {\sqrt{\Aindex{1}}}\,\cos (\Thetaindex{1}) - {\sqrt{3}}\,{\sqrt{\Aindex{1}}}\,\sin (\Thetaindex{1})  \right ) \nonumber \\
C_{117-2} &=&
\frac{-2\,t}{3}
\, \left ( {\sqrt{3}}\,J + {\sqrt{3}}\,U - 2\,{\sqrt{3}}\,W + {\sqrt{\Aindex{1}}}\,\cos (\Thetaindex{1}) - {\sqrt{3}}\,{\sqrt{\Aindex{1}}}\,\sin (\Thetaindex{1})  \right ) \nonumber \\
C_{117-3} &=&
\frac{1}{36}
\,\left ( 2\,{\sqrt{3}}\,\Aindex{1} - 3\,{\sqrt{3}}\,J^2 - 6\,{\sqrt{3}}\,J\,U - 3\,{\sqrt{3}}\,U^2 + 12\,{\sqrt{3}}\,J\,W\right . \nonumber \\
&& \hspace{1cm} 
 + 
 \left . 12\,{\sqrt{3}}\,U\,W - 12\,{\sqrt{3}}\,W^2 - {\sqrt{3}}\,\Aindex{1}\,\cos (2\,\Thetaindex{1}) - 3\,\Aindex{1}\,\sin (2\,\Thetaindex{1})
\right  )  \nonumber \\
C_{117-4} &=&
\frac{1}{18}
\,\left ( -2\,{\sqrt{3}}\,\Aindex{1} + 3\,{\sqrt{3}}\,J^2 + 6\,{\sqrt{3}}\,J\,U + 3\,{\sqrt{3}}\,U^2 - 12\,{\sqrt{3}}\,J\,W - 12\,{\sqrt{3}}\,U\,W\right . \nonumber \\
&& \hspace{1cm} 
 + 
 \left . 12\,{\sqrt{3}}\,W^2 + {\sqrt{3}}\,\Aindex{1}\,\cos (2\,\Thetaindex{1}) + 3\,\Aindex{1}\,\sin (2\,\Thetaindex{1})
\right  )  \nonumber \\
C_{117-5} &=&
t
\, \left ( {\sqrt{3}}\,J + {\sqrt{3}}\,U - 2\,{\sqrt{3}}\,W - {\sqrt{\Aindex{1}}}\,\cos (\Thetaindex{1}) + {\sqrt{3}}\,{\sqrt{\Aindex{1}}}\,\sin (\Thetaindex{1})  \right ) \nonumber \\
 N_{117} &=& {\sqrt{4\,{\Cindex{117}{1}}^2 + 2\,{\Cindex{117}{2}}^2 + 16\,{\Cindex{117}{3}}^2 + 8\,{\Cindex{117}{4}}^2 + 4\,{\Cindex{117}{5}}^2}} \nonumber \eeq 
\beq
\ket{\Psi}_{118}& = &\ket{4,0,2,\Gamma_{3,1}} \nonumber \\ 
&=& \frac{1}{4}
 \left ( \ket{02du} + \ket{02ud} - \ket{0du2} - \ket{0ud2} - \ket{20du} - \ket{20ud} + \ket{2du0} + \ket{2ud0}\right . \nonumber \\
&& \hspace{1em} 
\left . -\ket{d02u} + \ket{d20u} + \ket{du02} - \ket{du20} - \ket{u02d} + \ket{u20d} + \ket{ud02} - \ket{ud20}\right ) \nonumber  
\eeq
\beq
\ket{\Psi}_{119}& = &\ket{4,0,0,\Gamma_{3,2}} \nonumber \\ 
& = &
\Cindex{119}{1} \left ( 
\ket{0022} - \ket{0220} - \ket{2002} + \ket{2200}\right) 
 \nonumber \\
& + &
\Cindex{119}{2} \left ( 
\ket{02du} - \ket{02ud} - \ket{0du2} + \ket{0ud2} + \ket{20du} - \ket{20ud} - \ket{2du0} + \ket{2ud0}\right . \nonumber \\
&& \hspace{3em} 
\left . -\ket{d02u} - \ket{d20u} + \ket{du02} + \ket{du20} + \ket{u02d} + \ket{u20d} - \ket{ud02} - \ket{ud20}\right ) 
\nonumber \\
& + &
\Cindex{119}{3} \left ( 
\ket{dduu} + \ket{duud} + \ket{uddu} + \ket{uudd}\right) 
 \nonumber \\
& + &
\Cindex{119}{4} \left ( 
\ket{dudu} + \ket{udud}\right) 
\nonumber \eeq
\beq
C_{119-1} &=&
\frac{-t}{3}
\, \left ( -3\,J - 3\,U + 6\,W + 2\,{\sqrt{3}}\,{\sqrt{\Aindex{1}}}\,\cos (\Thetaindex{1})  \right ) \nonumber \\
C_{119-2} &=&
\frac{1}{12}
\,\left ( -3\,J^2 - 6\,J\,U - 3\,U^2 + 12\,J\,W + 12\,U\,W\right . \nonumber \\
&& \hspace{1cm} 
 \left . -12\,W^2 + 4\,\Aindex{1}\,{\cos (\Thetaindex{1})}^2
\right  )  \nonumber \\
C_{119-3} &=&
\frac{t}{3}
\, \left ( -3\,J - 3\,U + 6\,W - 2\,{\sqrt{3}}\,{\sqrt{\Aindex{1}}}\,\cos (\Thetaindex{1})  \right ) \nonumber \\
C_{119-4} &=&
\frac{-2\,t}{3}
\, \left ( -3\,J - 3\,U + 6\,W - 2\,{\sqrt{3}}\,{\sqrt{\Aindex{1}}}\,\cos (\Thetaindex{1})  \right ) \nonumber \\
 N_{119} &=& {\sqrt{4\,{\Cindex{119}{1}}^2 + 16\,{\Cindex{119}{2}}^2 + 4\,{\Cindex{119}{3}}^2 + 2\,{\Cindex{119}{4}}^2}} \nonumber \eeq 
\beq
\ket{\Psi}_{120}& = &\ket{4,0,0,\Gamma_{3,2}} \nonumber \\ 
& = &
\Cindex{120}{1} \left ( 
\ket{0022} - \ket{0220} - \ket{2002} + \ket{2200}\right) 
 \nonumber \\
& + &
\Cindex{120}{2} \left ( 
\ket{02du} - \ket{02ud} - \ket{0du2} + \ket{0ud2} + \ket{20du} - \ket{20ud} - \ket{2du0} + \ket{2ud0}\right . \nonumber \\
&& \hspace{3em} 
\left . -\ket{d02u} - \ket{d20u} + \ket{du02} + \ket{du20} + \ket{u02d} + \ket{u20d} - \ket{ud02} - \ket{ud20}\right ) 
\nonumber \\
& + &
\Cindex{120}{3} \left ( 
\ket{dduu} + \ket{duud} + \ket{uddu} + \ket{uudd}\right) 
 \nonumber \\
& + &
\Cindex{120}{4} \left ( 
\ket{dudu} + \ket{udud}\right) 
\nonumber \eeq
\beq
C_{120-1} &=&
\frac{-t}{3}
\, \left ( -3\,J - 3\,U + 6\,W - {\sqrt{3}}\,{\sqrt{\Aindex{1}}}\,\cos (\Thetaindex{1}) - 3\,{\sqrt{\Aindex{1}}}\,\sin (\Thetaindex{1})  \right ) \nonumber \\
C_{120-2} &=&
\frac{1}{12}
\,\left ( 2\,\Aindex{1} - 3\,J^2 - 6\,J\,U - 3\,U^2 + 12\,J\,W + 12\,U\,W\right . \nonumber \\
&& \hspace{1cm} 
 \left . -12\,W^2 - \Aindex{1}\,\cos (2\,\Thetaindex{1}) + {\sqrt{3}}\,\Aindex{1}\,\sin (2\,\Thetaindex{1})
\right  )  \nonumber \\
C_{120-3} &=&
\frac{t}{3}
\, \left ( -3\,J - 3\,U + 6\,W + {\sqrt{3}}\,{\sqrt{\Aindex{1}}}\,\cos (\Thetaindex{1}) + 3\,{\sqrt{\Aindex{1}}}\,\sin (\Thetaindex{1})  \right ) \nonumber \\
C_{120-4} &=&
\frac{-2\,t}{3}
\, \left ( -3\,J - 3\,U + 6\,W + {\sqrt{3}}\,{\sqrt{\Aindex{1}}}\,\cos (\Thetaindex{1}) + 3\,{\sqrt{\Aindex{1}}}\,\sin (\Thetaindex{1})  \right ) \nonumber \\
 N_{120} &=& {\sqrt{4\,{\Cindex{120}{1}}^2 + 16\,{\Cindex{120}{2}}^2 + 4\,{\Cindex{120}{3}}^2 + 2\,{\Cindex{120}{4}}^2}} \nonumber \eeq 
\beq
\ket{\Psi}_{121}& = &\ket{4,0,0,\Gamma_{3,2}} \nonumber \\ 
& = &
\Cindex{121}{1} \left ( 
\ket{0022} - \ket{0220} - \ket{2002} + \ket{2200}\right) 
 \nonumber \\
& + &
\Cindex{121}{2} \left ( 
\ket{02du} - \ket{02ud} - \ket{0du2} + \ket{0ud2} + \ket{20du} - \ket{20ud} - \ket{2du0} + \ket{2ud0}\right . \nonumber \\
&& \hspace{3em} 
\left . -\ket{d02u} - \ket{d20u} + \ket{du02} + \ket{du20} + \ket{u02d} + \ket{u20d} - \ket{ud02} - \ket{ud20}\right ) 
\nonumber \\
& + &
\Cindex{121}{3} \left ( 
\ket{dduu} + \ket{duud} + \ket{uddu} + \ket{uudd}\right) 
 \nonumber \\
& + &
\Cindex{121}{4} \left ( 
\ket{dudu} + \ket{udud}\right) 
\nonumber \eeq
\beq
C_{121-1} &=&
\frac{-t}{3}
\, \left ( -3\,J - 3\,U + 6\,W - {\sqrt{3}}\,{\sqrt{\Aindex{1}}}\,\cos (\Thetaindex{1}) + 3\,{\sqrt{\Aindex{1}}}\,\sin (\Thetaindex{1})  \right ) \nonumber \\
C_{121-2} &=&
\frac{1}{12}
\,\left ( 2\,\Aindex{1} - 3\,J^2 - 6\,J\,U - 3\,U^2 + 12\,J\,W + 12\,U\,W\right . \nonumber \\
&& \hspace{1cm} 
 \left . -12\,W^2 - \Aindex{1}\,\cos (2\,\Thetaindex{1}) - {\sqrt{3}}\,\Aindex{1}\,\sin (2\,\Thetaindex{1})
\right  )  \nonumber \\
C_{121-3} &=&
\frac{t}{3}
\, \left ( -3\,J - 3\,U + 6\,W + {\sqrt{3}}\,{\sqrt{\Aindex{1}}}\,\cos (\Thetaindex{1}) - 3\,{\sqrt{\Aindex{1}}}\,\sin (\Thetaindex{1})  \right ) \nonumber \\
C_{121-4} &=&
\frac{-2\,t}{3}
\, \left ( -3\,J - 3\,U + 6\,W + {\sqrt{3}}\,{\sqrt{\Aindex{1}}}\,\cos (\Thetaindex{1}) - 3\,{\sqrt{\Aindex{1}}}\,\sin (\Thetaindex{1})  \right ) \nonumber \\
 N_{121} &=& {\sqrt{4\,{\Cindex{121}{1}}^2 + 16\,{\Cindex{121}{2}}^2 + 4\,{\Cindex{121}{3}}^2 + 2\,{\Cindex{121}{4}}^2}} \nonumber \eeq 
\beq
\ket{\Psi}_{122}& = &\ket{4,0,2,\Gamma_{3,2}} \nonumber \\ 
& = &
\Cindex{122}{1} \left ( 
\ket{02du} + \ket{02ud} + \ket{0du2} + \ket{0ud2} - \ket{20du} - \ket{20ud} - \ket{2du0} - \ket{2ud0}\right . \nonumber \\
&& \hspace{3em} 
 + 
\left . \ket{d02u} - \ket{d20u} + \ket{du02} - \ket{du20} + \ket{u02d} - \ket{u20d} + \ket{ud02} - \ket{ud20}\right ) 
\nonumber \\
& + &
\Cindex{122}{2} \left ( 
\ket{0d2u} + \ket{0u2d} - \ket{2d0u} - \ket{2u0d} + \ket{d0u2} - \ket{d2u0} + \ket{u0d2} - \ket{u2d0}\right) 
\nonumber \eeq
\beq
C_{122-1} &=&
\frac{-1}{4\,{\sqrt{3}}} \nonumber \\
C_{122-2} &=&
\frac{-1}{2\,{\sqrt{3}}} \nonumber \\
 N_{122} &=& {\sqrt{16\,{\Cindex{122}{1}}^2 + 8\,{\Cindex{122}{2}}^2}} \nonumber \eeq 
\beq
\ket{\Psi}_{123}& = &\ket{4,0,0,\Gamma_{4,1}} \nonumber \\ 
& = &
\Cindex{123}{1} \left ( 
\ket{0022} - \ket{2200}\right) 
 \nonumber \\
& + &
\Cindex{123}{2} \left ( 
\ket{02du} - \ket{02ud} + \ket{20du} - \ket{20ud} - \ket{du02} - \ket{du20} + \ket{ud02} + \ket{ud20}\right) 
 \nonumber \\
& + &
\Cindex{123}{3} \left ( 
\ket{0d2u} + \ket{0du2} - \ket{0u2d} - \ket{0ud2} - \ket{2d0u} - \ket{2du0} + \ket{2u0d} + \ket{2ud0}\right . \nonumber \\
&& \hspace{3em} 
 + 
\left . \ket{d02u} + \ket{d0u2} - \ket{d20u} - \ket{d2u0} - \ket{u02d} - \ket{u0d2} + \ket{u20d} + \ket{u2d0}\right ) 
\nonumber \eeq
\beq
C_{123-1} &=&
-4\,{\sqrt{2}}\,t^2 \nonumber \\
C_{123-2} &=&
\frac{-1}{9\,{\sqrt{2}}}
\,\left ( 2\,{\Aindex{10}}^2 - 3\,\Aindex{10}\,J - 36\,t^2 + 9\,\Aindex{10}\,U - 9\,J\,U\right . \nonumber \\
&& \hspace{1cm} 
 + 
 \left . 9\,U^2 + 54\,\Aindex{10}\,W - 36\,J\,W + 126\,U\,W + 360\,W^2
\right  )  \nonumber \\
C_{123-3} &=&
\frac{-\left( {\sqrt{2}}\,t \right) }{3}
\, \left ( J + U - 2\,W + {\sqrt{\Aindex{2}}}\,\cos (\Thetaindex{2})  \right ) \nonumber \\
 N_{123} &=& {\sqrt{2\,{\Cindex{123}{1}}^2 + 8\,{\Cindex{123}{2}}^2 + 16\,{\Cindex{123}{3}}^2}} \nonumber \eeq 
\beq
\ket{\Psi}_{124}& = &\ket{4,0,0,\Gamma_{4,1}} \nonumber \\ 
& = &
\Cindex{124}{1} \left ( 
\ket{0022} - \ket{2200}\right) 
 \nonumber \\
& + &
\Cindex{124}{2} \left ( 
\ket{02du} - \ket{02ud} + \ket{20du} - \ket{20ud} - \ket{du02} - \ket{du20} + \ket{ud02} + \ket{ud20}\right) 
 \nonumber \\
& + &
\Cindex{124}{3} \left ( 
\ket{0d2u} + \ket{0du2} - \ket{0u2d} - \ket{0ud2} - \ket{2d0u} - \ket{2du0} + \ket{2u0d} + \ket{2ud0}\right . \nonumber \\
&& \hspace{3em} 
 + 
\left . \ket{d02u} + \ket{d0u2} - \ket{d20u} - \ket{d2u0} - \ket{u02d} - \ket{u0d2} + \ket{u20d} + \ket{u2d0}\right ) 
\nonumber \eeq
\beq
C_{124-1} &=&
-4\,{\sqrt{2}}\,t^2 \nonumber \\
C_{124-2} &=&
\frac{-1}{18\,{\sqrt{2}}}
\,\left ( {\Aindex{14}}^2 + 3\,\Aindex{14}\,J - 72\,t^2 - 9\,\Aindex{14}\,U - 18\,J\,U\right . \nonumber \\
&& \hspace{1cm} 
 + 
 \left . 18\,U^2 - 54\,\Aindex{14}\,W - 72\,J\,W + 252\,U\,W + 720\,W^2
\right  )  \nonumber \\
C_{124-3} &=&
\frac{-t}{3\,{\sqrt{2}}}
\, \left ( 2\,J + 2\,U - 4\,W - {\sqrt{\Aindex{2}}}\,\cos (\Thetaindex{2}) - {\sqrt{3}}\,{\sqrt{\Aindex{2}}}\,\sin (\Thetaindex{2})  \right ) \nonumber \\
 N_{124} &=& {\sqrt{2\,{\Cindex{124}{1}}^2 + 8\,{\Cindex{124}{2}}^2 + 16\,{\Cindex{124}{3}}^2}} \nonumber \eeq 
\beq
\ket{\Psi}_{125}& = &\ket{4,0,0,\Gamma_{4,1}} \nonumber \\ 
& = &
\Cindex{125}{1} \left ( 
\ket{0022} - \ket{2200}\right) 
 \nonumber \\
& + &
\Cindex{125}{2} \left ( 
\ket{02du} - \ket{02ud} + \ket{20du} - \ket{20ud} - \ket{du02} - \ket{du20} + \ket{ud02} + \ket{ud20}\right) 
 \nonumber \\
& + &
\Cindex{125}{3} \left ( 
\ket{0d2u} + \ket{0du2} - \ket{0u2d} - \ket{0ud2} - \ket{2d0u} - \ket{2du0} + \ket{2u0d} + \ket{2ud0}\right . \nonumber \\
&& \hspace{3em} 
 + 
\left . \ket{d02u} + \ket{d0u2} - \ket{d20u} - \ket{d2u0} - \ket{u02d} - \ket{u0d2} + \ket{u20d} + \ket{u2d0}\right ) 
\nonumber \eeq
\beq
C_{125-1} &=&
-4\,{\sqrt{2}}\,t^2 \nonumber \\
C_{125-2} &=&
\frac{-1}{18\,{\sqrt{2}}}
\,\left ( {\Aindex{13}}^2 + 3\,\Aindex{13}\,J - 72\,t^2 - 9\,\Aindex{13}\,U - 18\,J\,U\right . \nonumber \\
&& \hspace{1cm} 
 + 
 \left . 18\,U^2 - 54\,\Aindex{13}\,W - 72\,J\,W + 252\,U\,W + 720\,W^2
\right  )  \nonumber \\
C_{125-3} &=&
\frac{-t}{3\,{\sqrt{2}}}
\, \left ( 2\,J + 2\,U - 4\,W - {\sqrt{\Aindex{2}}}\,\cos (\Thetaindex{2}) + {\sqrt{3}}\,{\sqrt{\Aindex{2}}}\,\sin (\Thetaindex{2})  \right ) \nonumber \\
 N_{125} &=& {\sqrt{2\,{\Cindex{125}{1}}^2 + 8\,{\Cindex{125}{2}}^2 + 16\,{\Cindex{125}{3}}^2}} \nonumber \eeq 
\beq
\ket{\Psi}_{126}& = &\ket{4,0,2,\Gamma_{4,1}} \nonumber \\ 
&=& \frac{1}{4}
 \left ( \ket{0d2u} + \ket{0du2} + \ket{0u2d} + \ket{0ud2} + \ket{2d0u} + \ket{2du0} + \ket{2u0d} + \ket{2ud0}\right . \nonumber \\
&& \hspace{1em} 
 + 
\left . \ket{d02u} + \ket{d0u2} + \ket{d20u} + \ket{d2u0} + \ket{u02d} + \ket{u0d2} + \ket{u20d} + \ket{u2d0}\right ) \nonumber  
\eeq
\beq
\ket{\Psi}_{127}& = &\ket{4,0,0,\Gamma_{4,2}} \nonumber \\ 
& = &
\Cindex{127}{1} \left ( 
\ket{0220} - \ket{2002}\right) 
 \nonumber \\
& + &
\Cindex{127}{2} \left ( 
\ket{02du} - \ket{02ud} + \ket{0d2u} - \ket{0u2d} - \ket{20du} + \ket{20ud} - \ket{2d0u} + \ket{2u0d}\right . \nonumber \\
&& \hspace{3em} 
\left . -\ket{d0u2} + \ket{d2u0} - \ket{du02} + \ket{du20} + \ket{u0d2} - \ket{u2d0} + \ket{ud02} - \ket{ud20}\right ) 
\nonumber \\
& + &
\Cindex{127}{3} \left ( 
\ket{0du2} - \ket{0ud2} + \ket{2du0} - \ket{2ud0} - \ket{d02u} - \ket{d20u} + \ket{u02d} + \ket{u20d}\right) 
\nonumber \eeq
\beq
C_{127-1} &=&
\frac{-1}{9\,{\sqrt{2}}}
\,\left ( 4\,{\Aindex{10}}^2 - 12\,\Aindex{10}\,J\right . \nonumber \\
&& \hspace{1cm} 
 + 
 \left . 9\,J^2 - 72\,t^2 + 12\,\Aindex{10}\,U - 18\,J\,U
\right .  \nonumber \\
&& \hspace{1cm} 
 + 
 \left . 9\,U^2 + 120\,\Aindex{10}\,W - 180\,J\,W + 180\,U\,W + 900\,W^2
\right )  \nonumber \\
C_{127-2} &=&
\frac{t}{3\,{\sqrt{2}}}
\, \left ( J + U - 2\,W - 2\,{\sqrt{\Aindex{2}}}\,\cos (\Thetaindex{2})  \right ) \nonumber \\
C_{127-3} &=&
-2\,{\sqrt{2}}\,t^2 \nonumber \\
 N_{127} &=& {\sqrt{2\,{\Cindex{127}{1}}^2 + 16\,{\Cindex{127}{2}}^2 + 8\,{\Cindex{127}{3}}^2}} \nonumber \eeq 
\beq
\ket{\Psi}_{128}& = &\ket{4,0,0,\Gamma_{4,2}} \nonumber \\ 
& = &
\Cindex{128}{1} \left ( 
\ket{0220} - \ket{2002}\right) 
 \nonumber \\
& + &
\Cindex{128}{2} \left ( 
\ket{02du} - \ket{02ud} + \ket{0d2u} - \ket{0u2d} - \ket{20du} + \ket{20ud} - \ket{2d0u} + \ket{2u0d}\right . \nonumber \\
&& \hspace{3em} 
\left . -\ket{d0u2} + \ket{d2u0} - \ket{du02} + \ket{du20} + \ket{u0d2} - \ket{u2d0} + \ket{ud02} - \ket{ud20}\right ) 
\nonumber \\
& + &
\Cindex{128}{3} \left ( 
\ket{0du2} - \ket{0ud2} + \ket{2du0} - \ket{2ud0} - \ket{d02u} - \ket{d20u} + \ket{u02d} + \ket{u20d}\right) 
\nonumber \eeq
\beq
C_{128-1} &=&
\frac{-1}{9\,{\sqrt{2}}}
\,\left ( {\Aindex{14}}^2 + 6\,\Aindex{14}\,J\right . \nonumber \\
&& \hspace{1cm} 
 + 
 \left . 9\,J^2 - 72\,t^2 - 6\,\Aindex{14}\,U - 18\,J\,U
\right .  \nonumber \\
&& \hspace{1cm} 
 + 
 \left . 9\,U^2 - 60\,\Aindex{14}\,W - 180\,J\,W + 180\,U\,W + 900\,W^2
\right )  \nonumber \\
C_{128-2} &=&
\frac{t}{3\,{\sqrt{2}}}
\, \left ( J + U - 2\,W + {\sqrt{\Aindex{2}}}\,\cos (\Thetaindex{2}) + {\sqrt{3}}\,{\sqrt{\Aindex{2}}}\,\sin (\Thetaindex{2})  \right ) \nonumber \\
C_{128-3} &=&
-2\,{\sqrt{2}}\,t^2 \nonumber \\
 N_{128} &=& {\sqrt{2\,{\Cindex{128}{1}}^2 + 16\,{\Cindex{128}{2}}^2 + 8\,{\Cindex{128}{3}}^2}} \nonumber \eeq 
\beq
\ket{\Psi}_{129}& = &\ket{4,0,0,\Gamma_{4,2}} \nonumber \\ 
& = &
\Cindex{129}{1} \left ( 
\ket{0220} - \ket{2002}\right) 
 \nonumber \\
& + &
\Cindex{129}{2} \left ( 
\ket{02du} - \ket{02ud} + \ket{0d2u} - \ket{0u2d} - \ket{20du} + \ket{20ud} - \ket{2d0u} + \ket{2u0d}\right . \nonumber \\
&& \hspace{3em} 
\left . -\ket{d0u2} + \ket{d2u0} - \ket{du02} + \ket{du20} + \ket{u0d2} - \ket{u2d0} + \ket{ud02} - \ket{ud20}\right ) 
\nonumber \\
& + &
\Cindex{129}{3} \left ( 
\ket{0du2} - \ket{0ud2} + \ket{2du0} - \ket{2ud0} - \ket{d02u} - \ket{d20u} + \ket{u02d} + \ket{u20d}\right) 
\nonumber \eeq
\beq
C_{129-1} &=&
\frac{-1}{9\,{\sqrt{2}}}
\,\left ( {\Aindex{13}}^2 + 6\,\Aindex{13}\,J\right . \nonumber \\
&& \hspace{1cm} 
 + 
 \left . 9\,J^2 - 72\,t^2 - 6\,\Aindex{13}\,U - 18\,J\,U
\right .  \nonumber \\
&& \hspace{1cm} 
 + 
 \left . 9\,U^2 - 60\,\Aindex{13}\,W - 180\,J\,W + 180\,U\,W + 900\,W^2
\right )  \nonumber \\
C_{129-2} &=&
\frac{t}{3\,{\sqrt{2}}}
\, \left ( J + U - 2\,W + {\sqrt{\Aindex{2}}}\,\cos (\Thetaindex{2}) - {\sqrt{3}}\,{\sqrt{\Aindex{2}}}\,\sin (\Thetaindex{2})  \right ) \nonumber \\
C_{129-3} &=&
-2\,{\sqrt{2}}\,t^2 \nonumber \\
 N_{129} &=& {\sqrt{2\,{\Cindex{129}{1}}^2 + 16\,{\Cindex{129}{2}}^2 + 8\,{\Cindex{129}{3}}^2}} \nonumber \eeq 
\beq
\ket{\Psi}_{130}& = &\ket{4,0,2,\Gamma_{4,2}} \nonumber \\ 
&=& \frac{1}{4}
 \left ( \ket{02du} + \ket{02ud} + \ket{0d2u} + \ket{0u2d} + \ket{20du} + \ket{20ud} + \ket{2d0u} + \ket{2u0d}\right . \nonumber \\
&& \hspace{1em} 
\left . -\ket{d0u2} - \ket{d2u0} - \ket{du02} - \ket{du20} - \ket{u0d2} - \ket{u2d0} - \ket{ud02} - \ket{ud20}\right ) \nonumber  
\eeq
\beq
\ket{\Psi}_{131}& = &\ket{4,0,0,\Gamma_{4,3}} \nonumber \\ 
& = &
\Cindex{131}{1} \left ( 
\ket{0202} - \ket{2020}\right) 
 \nonumber \\
& + &
\Cindex{131}{2} \left ( 
\ket{02du} - \ket{02ud} + \ket{0du2} - \ket{0ud2} - \ket{20du} + \ket{20ud} - \ket{2du0} + \ket{2ud0}\right . \nonumber \\
&& \hspace{3em} 
\left . -\ket{d02u} + \ket{d20u} + \ket{du02} - \ket{du20} + \ket{u02d} - \ket{u20d} - \ket{ud02} + \ket{ud20}\right ) 
\nonumber \\
& + &
\Cindex{131}{3} \left ( 
\ket{0d2u} - \ket{0u2d} + \ket{2d0u} - \ket{2u0d} - \ket{d0u2} - \ket{d2u0} + \ket{u0d2} + \ket{u2d0}\right) 
\nonumber \eeq
\beq
C_{131-1} &=&
\frac{-1}{9\,{\sqrt{2}}}
\,\left ( 4\,{\Aindex{10}}^2 - 12\,\Aindex{10}\,J\right . \nonumber \\
&& \hspace{1cm} 
 + 
 \left . 9\,J^2 - 72\,t^2 + 12\,\Aindex{10}\,U - 18\,J\,U
\right .  \nonumber \\
&& \hspace{1cm} 
 + 
 \left . 9\,U^2 + 120\,\Aindex{10}\,W - 180\,J\,W + 180\,U\,W + 900\,W^2
\right )  \nonumber \\
C_{131-2} &=&
\frac{t}{3\,{\sqrt{2}}}
\, \left ( J + U - 2\,W - 2\,{\sqrt{\Aindex{2}}}\,\cos (\Thetaindex{2})  \right ) \nonumber \\
C_{131-3} &=&
-2\,{\sqrt{2}}\,t^2 \nonumber \\
 N_{131} &=& {\sqrt{2\,{\Cindex{131}{1}}^2 + 16\,{\Cindex{131}{2}}^2 + 8\,{\Cindex{131}{3}}^2}} \nonumber \eeq 
\beq
\ket{\Psi}_{132}& = &\ket{4,0,0,\Gamma_{4,3}} \nonumber \\ 
& = &
\Cindex{132}{1} \left ( 
\ket{0202} - \ket{2020}\right) 
 \nonumber \\
& + &
\Cindex{132}{2} \left ( 
\ket{02du} - \ket{02ud} + \ket{0du2} - \ket{0ud2} - \ket{20du} + \ket{20ud} - \ket{2du0} + \ket{2ud0}\right . \nonumber \\
&& \hspace{3em} 
\left . -\ket{d02u} + \ket{d20u} + \ket{du02} - \ket{du20} + \ket{u02d} - \ket{u20d} - \ket{ud02} + \ket{ud20}\right ) 
\nonumber \\
& + &
\Cindex{132}{3} \left ( 
\ket{0d2u} - \ket{0u2d} + \ket{2d0u} - \ket{2u0d} - \ket{d0u2} - \ket{d2u0} + \ket{u0d2} + \ket{u2d0}\right) 
\nonumber \eeq
\beq
C_{132-1} &=&
\frac{-1}{9\,{\sqrt{2}}}
\,\left ( {\Aindex{14}}^2 + 6\,\Aindex{14}\,J\right . \nonumber \\
&& \hspace{1cm} 
 + 
 \left . 9\,J^2 - 72\,t^2 - 6\,\Aindex{14}\,U - 18\,J\,U
\right .  \nonumber \\
&& \hspace{1cm} 
 + 
 \left . 9\,U^2 - 60\,\Aindex{14}\,W - 180\,J\,W + 180\,U\,W + 900\,W^2
\right )  \nonumber \\
C_{132-2} &=&
\frac{t}{3\,{\sqrt{2}}}
\, \left ( J + U - 2\,W + {\sqrt{\Aindex{2}}}\,\cos (\Thetaindex{2}) + {\sqrt{3}}\,{\sqrt{\Aindex{2}}}\,\sin (\Thetaindex{2})  \right ) \nonumber \\
C_{132-3} &=&
-2\,{\sqrt{2}}\,t^2 \nonumber \\
 N_{132} &=& {\sqrt{2\,{\Cindex{132}{1}}^2 + 16\,{\Cindex{132}{2}}^2 + 8\,{\Cindex{132}{3}}^2}} \nonumber \eeq 
\beq
\ket{\Psi}_{133}& = &\ket{4,0,0,\Gamma_{4,3}} \nonumber \\ 
& = &
\Cindex{133}{1} \left ( 
\ket{0202} - \ket{2020}\right) 
 \nonumber \\
& + &
\Cindex{133}{2} \left ( 
\ket{02du} - \ket{02ud} + \ket{0du2} - \ket{0ud2} - \ket{20du} + \ket{20ud} - \ket{2du0} + \ket{2ud0}\right . \nonumber \\
&& \hspace{3em} 
\left . -\ket{d02u} + \ket{d20u} + \ket{du02} - \ket{du20} + \ket{u02d} - \ket{u20d} - \ket{ud02} + \ket{ud20}\right ) 
\nonumber \\
& + &
\Cindex{133}{3} \left ( 
\ket{0d2u} - \ket{0u2d} + \ket{2d0u} - \ket{2u0d} - \ket{d0u2} - \ket{d2u0} + \ket{u0d2} + \ket{u2d0}\right) 
\nonumber \eeq
\beq
C_{133-1} &=&
\frac{-1}{9\,{\sqrt{2}}}
\,\left ( {\Aindex{13}}^2 + 6\,\Aindex{13}\,J\right . \nonumber \\
&& \hspace{1cm} 
 + 
 \left . 9\,J^2 - 72\,t^2 - 6\,\Aindex{13}\,U - 18\,J\,U
\right .  \nonumber \\
&& \hspace{1cm} 
 + 
 \left . 9\,U^2 - 60\,\Aindex{13}\,W - 180\,J\,W + 180\,U\,W + 900\,W^2
\right )  \nonumber \\
C_{133-2} &=&
\frac{t}{3\,{\sqrt{2}}}
\, \left ( J + U - 2\,W + {\sqrt{\Aindex{2}}}\,\cos (\Thetaindex{2}) - {\sqrt{3}}\,{\sqrt{\Aindex{2}}}\,\sin (\Thetaindex{2})  \right ) \nonumber \\
C_{133-3} &=&
-2\,{\sqrt{2}}\,t^2 \nonumber \\
 N_{133} &=& {\sqrt{2\,{\Cindex{133}{1}}^2 + 16\,{\Cindex{133}{2}}^2 + 8\,{\Cindex{133}{3}}^2}} \nonumber \eeq 
\beq
\ket{\Psi}_{134}& = &\ket{4,0,2,\Gamma_{4,3}} \nonumber \\ 
&=& \frac{1}{4}
 \left ( \ket{02du} + \ket{02ud} - \ket{0du2} - \ket{0ud2} + \ket{20du} + \ket{20ud} - \ket{2du0} - \ket{2ud0}\right . \nonumber \\
&& \hspace{1em} 
 + 
\left . \ket{d02u} + \ket{d20u} + \ket{du02} + \ket{du20} + \ket{u02d} + \ket{u20d} + \ket{ud02} + \ket{ud20}\right ) \nonumber  
\eeq
\beq
\ket{\Psi}_{135}& = &\ket{4,0,0,\Gamma_{5,1}} \nonumber \\ 
&=& \frac{1}{4}
 \left ( \ket{02du} - \ket{02ud} - \ket{0du2} + \ket{0ud2} - \ket{20du} + \ket{20ud} + \ket{2du0} - \ket{2ud0}\right . \nonumber \\
&& \hspace{1em} 
 + 
\left . \ket{d02u} - \ket{d20u} + \ket{du02} - \ket{du20} - \ket{u02d} + \ket{u20d} - \ket{ud02} + \ket{ud20}\right ) \nonumber  
\eeq
\beq
\ket{\Psi}_{136}& = &\ket{4,0,2,\Gamma_{5,1}} \nonumber \\ 
& = &
\Cindex{136}{1} \left ( 
\ket{02du} + \ket{02ud} + \ket{0du2} + \ket{0ud2} + \ket{20du} + \ket{20ud} + \ket{2du0} + \ket{2ud0}\right . \nonumber \\
&& \hspace{3em} 
\left . -\ket{d02u} - \ket{d20u} + \ket{du02} + \ket{du20} - \ket{u02d} - \ket{u20d} + \ket{ud02} + \ket{ud20}\right ) 
\nonumber \\
& + &
\Cindex{136}{2} \left ( 
\ket{0d2u} + \ket{0u2d} - \ket{2d0u} - \ket{2u0d} - \ket{d0u2} + \ket{d2u0} - \ket{u0d2} + \ket{u2d0}\right) 
 \nonumber \\
& + &
\Cindex{136}{3} \left ( 
\ket{dudu} - \ket{udud}\right) 
\nonumber \eeq
\beq
C_{136-1} &=&
\frac{-t}{3\,{\sqrt{2}}}
\, \left ( J + U - 2\,W + 2\,{\sqrt{\Aindex{2}}}\,\cos (\Thetaindex{3})  \right ) \nonumber \\
C_{136-2} &=&
-2\,{\sqrt{2}}\,t^2 \nonumber \\
C_{136-3} &=&
\frac{-1}{9\,{\sqrt{2}}}
\,\left ( {\Aindex{12}}^2 - 72\,t^2 + 6\,J\,U - 3\,U^2 + 60\,J\,W - 132\,U\,W - 1020\,W^2\right . \nonumber \\
&& \hspace{1cm} 
 + 
 \left . 12\,{\sqrt{\Aindex{2}}}\,U\,\cos (\Thetaindex{3}) + 120\,{\sqrt{\Aindex{2}}}\,W\,\cos (\Thetaindex{3})
\right  )  \nonumber \\
 N_{136} &=& {\sqrt{16\,{\Cindex{136}{1}}^2 + 8\,{\Cindex{136}{2}}^2 + 2\,{\Cindex{136}{3}}^2}} \nonumber \eeq 
\beq
\ket{\Psi}_{137}& = &\ket{4,0,2,\Gamma_{5,1}} \nonumber \\ 
& = &
\Cindex{137}{1} \left ( 
\ket{02du} + \ket{02ud} + \ket{0du2} + \ket{0ud2} + \ket{20du} + \ket{20ud} + \ket{2du0} + \ket{2ud0}\right . \nonumber \\
&& \hspace{3em} 
\left . -\ket{d02u} - \ket{d20u} + \ket{du02} + \ket{du20} - \ket{u02d} - \ket{u20d} + \ket{ud02} + \ket{ud20}\right ) 
\nonumber \\
& + &
\Cindex{137}{2} \left ( 
\ket{0d2u} + \ket{0u2d} - \ket{2d0u} - \ket{2u0d} - \ket{d0u2} + \ket{d2u0} - \ket{u0d2} + \ket{u2d0}\right) 
 \nonumber \\
& + &
\Cindex{137}{3} \left ( 
\ket{dudu} - \ket{udud}\right) 
\nonumber \eeq
\beq
C_{137-1} &=&
\frac{-t}{3\,{\sqrt{2}}}
\, \left ( J + U - 2\,W - {\sqrt{\Aindex{2}}}\,\cos (\Thetaindex{3}) - {\sqrt{3}}\,{\sqrt{\Aindex{2}}}\,\sin (\Thetaindex{3})  \right ) \nonumber \\
C_{137-2} &=&
-2\,{\sqrt{2}}\,t^2 \nonumber \\
C_{137-3} &=&
\frac{-1}{9\,{\sqrt{2}}}
\,\left ( {\Aindex{17}}^2 - 72\,t^2 - 6\,\Aindex{17}\,U\right . \nonumber \\
&& \hspace{1cm} 
 + 
 \left . 9\,U^2 - 60\,\Aindex{17}\,W + 180\,U\,W + 900\,W^2
\right  )  \nonumber \\
 N_{137} &=& {\sqrt{16\,{\Cindex{137}{1}}^2 + 8\,{\Cindex{137}{2}}^2 + 2\,{\Cindex{137}{3}}^2}} \nonumber \eeq 
\beq
\ket{\Psi}_{138}& = &\ket{4,0,2,\Gamma_{5,1}} \nonumber \\ 
& = &
\Cindex{138}{1} \left ( 
\ket{02du} + \ket{02ud} + \ket{0du2} + \ket{0ud2} + \ket{20du} + \ket{20ud} + \ket{2du0} + \ket{2ud0}\right . \nonumber \\
&& \hspace{3em} 
\left . -\ket{d02u} - \ket{d20u} + \ket{du02} + \ket{du20} - \ket{u02d} - \ket{u20d} + \ket{ud02} + \ket{ud20}\right ) 
\nonumber \\
& + &
\Cindex{138}{2} \left ( 
\ket{0d2u} + \ket{0u2d} - \ket{2d0u} - \ket{2u0d} - \ket{d0u2} + \ket{d2u0} - \ket{u0d2} + \ket{u2d0}\right) 
 \nonumber \\
& + &
\Cindex{138}{3} \left ( 
\ket{dudu} - \ket{udud}\right) 
\nonumber \eeq
\beq
C_{138-1} &=&
\frac{-t}{3\,{\sqrt{2}}}
\, \left ( J + U - 2\,W - {\sqrt{\Aindex{2}}}\,\cos (\Thetaindex{3}) + {\sqrt{3}}\,{\sqrt{\Aindex{2}}}\,\sin (\Thetaindex{3})  \right ) \nonumber \\
C_{138-2} &=&
-2\,{\sqrt{2}}\,t^2 \nonumber \\
C_{138-3} &=&
\frac{-1}{9\,{\sqrt{2}}}
\,\left ( {\Aindex{15}}^2 - 72\,t^2 - 6\,\Aindex{15}\,U\right . \nonumber \\
&& \hspace{1cm} 
 + 
 \left . 9\,U^2 - 60\,\Aindex{15}\,W + 180\,U\,W + 900\,W^2
\right  )  \nonumber \\
 N_{138} &=& {\sqrt{16\,{\Cindex{138}{1}}^2 + 8\,{\Cindex{138}{2}}^2 + 2\,{\Cindex{138}{3}}^2}} \nonumber \eeq 
\beq
\ket{\Psi}_{139}& = &\ket{4,0,0,\Gamma_{5,2}} \nonumber \\ 
&=& \frac{1}{4}
 \left ( \ket{0d2u} - \ket{0du2} - \ket{0u2d} + \ket{0ud2} - \ket{2d0u} + \ket{2du0} + \ket{2u0d} - \ket{2ud0}\right . \nonumber \\
&& \hspace{1em} 
\left . -\ket{d02u} + \ket{d0u2} + \ket{d20u} - \ket{d2u0} + \ket{u02d} - \ket{u0d2} - \ket{u20d} + \ket{u2d0}\right ) \nonumber  
\eeq
\beq
\ket{\Psi}_{140}& = &\ket{4,0,2,\Gamma_{5,2}} \nonumber \\ 
& = &
\Cindex{140}{1} \left ( 
\ket{02du} + \ket{02ud} - \ket{20du} - \ket{20ud} - \ket{du02} + \ket{du20} - \ket{ud02} + \ket{ud20}\right) 
 \nonumber \\
& + &
\Cindex{140}{2} \left ( 
\ket{0d2u} - \ket{0du2} + \ket{0u2d} - \ket{0ud2} + \ket{2d0u} - \ket{2du0} + \ket{2u0d} - \ket{2ud0}\right . \nonumber \\
&& \hspace{3em} 
\left . -\ket{d02u} + \ket{d0u2} - \ket{d20u} + \ket{d2u0} - \ket{u02d} + \ket{u0d2} - \ket{u20d} + \ket{u2d0}\right ) 
\nonumber \\
& + &
\Cindex{140}{3} \left ( 
\ket{dduu} - \ket{uudd}\right) 
\nonumber \eeq
\beq
C_{140-1} &=&
2\,{\sqrt{2}}\,t^2 \nonumber \\
C_{140-2} &=&
\frac{t}{3\,{\sqrt{2}}}
\, \left ( J + U - 2\,W + 2\,{\sqrt{\Aindex{2}}}\,\cos (\Thetaindex{3})  \right ) \nonumber \\
C_{140-3} &=&
\frac{-1}{9\,{\sqrt{2}}}
\,\left ( {\Aindex{12}}^2 - 72\,t^2 + 6\,J\,U - 3\,U^2 + 60\,J\,W - 132\,U\,W - 1020\,W^2\right . \nonumber \\
&& \hspace{1cm} 
 + 
 \left . 12\,{\sqrt{\Aindex{2}}}\,U\,\cos (\Thetaindex{3}) + 120\,{\sqrt{\Aindex{2}}}\,W\,\cos (\Thetaindex{3})
\right  )  \nonumber \\
 N_{140} &=& {\sqrt{8\,{\Cindex{140}{1}}^2 + 16\,{\Cindex{140}{2}}^2 + 2\,{\Cindex{140}{3}}^2}} \nonumber \eeq 
\beq
\ket{\Psi}_{141}& = &\ket{4,0,2,\Gamma_{5,2}} \nonumber \\ 
& = &
\Cindex{141}{1} \left ( 
\ket{02du} + \ket{02ud} - \ket{20du} - \ket{20ud} - \ket{du02} + \ket{du20} - \ket{ud02} + \ket{ud20}\right) 
 \nonumber \\
& + &
\Cindex{141}{2} \left ( 
\ket{0d2u} - \ket{0du2} + \ket{0u2d} - \ket{0ud2} + \ket{2d0u} - \ket{2du0} + \ket{2u0d} - \ket{2ud0}\right . \nonumber \\
&& \hspace{3em} 
\left . -\ket{d02u} + \ket{d0u2} - \ket{d20u} + \ket{d2u0} - \ket{u02d} + \ket{u0d2} - \ket{u20d} + \ket{u2d0}\right ) 
\nonumber \\
& + &
\Cindex{141}{3} \left ( 
\ket{dduu} - \ket{uudd}\right) 
\nonumber \eeq
\beq
C_{141-1} &=&
2\,{\sqrt{2}}\,t^2 \nonumber \\
C_{141-2} &=&
\frac{t}{3\,{\sqrt{2}}}
\, \left ( J + U - 2\,W - {\sqrt{\Aindex{2}}}\,\cos (\Thetaindex{3}) - {\sqrt{3}}\,{\sqrt{\Aindex{2}}}\,\sin (\Thetaindex{3})  \right ) \nonumber \\
C_{141-3} &=&
\frac{-1}{9\,{\sqrt{2}}}
\,\left ( {\Aindex{17}}^2 - 72\,t^2 - 6\,\Aindex{17}\,U\right . \nonumber \\
&& \hspace{1cm} 
 + 
 \left . 9\,U^2 - 60\,\Aindex{17}\,W + 180\,U\,W + 900\,W^2
\right  )  \nonumber \\
 N_{141} &=& {\sqrt{8\,{\Cindex{141}{1}}^2 + 16\,{\Cindex{141}{2}}^2 + 2\,{\Cindex{141}{3}}^2}} \nonumber \eeq 
\beq
\ket{\Psi}_{142}& = &\ket{4,0,2,\Gamma_{5,2}} \nonumber \\ 
& = &
\Cindex{142}{1} \left ( 
\ket{02du} + \ket{02ud} - \ket{20du} - \ket{20ud} - \ket{du02} + \ket{du20} - \ket{ud02} + \ket{ud20}\right) 
 \nonumber \\
& + &
\Cindex{142}{2} \left ( 
\ket{0d2u} - \ket{0du2} + \ket{0u2d} - \ket{0ud2} + \ket{2d0u} - \ket{2du0} + \ket{2u0d} - \ket{2ud0}\right . \nonumber \\
&& \hspace{3em} 
\left . -\ket{d02u} + \ket{d0u2} - \ket{d20u} + \ket{d2u0} - \ket{u02d} + \ket{u0d2} - \ket{u20d} + \ket{u2d0}\right ) 
\nonumber \\
& + &
\Cindex{142}{3} \left ( 
\ket{dduu} - \ket{uudd}\right) 
\nonumber \eeq
\beq
C_{142-1} &=&
2\,{\sqrt{2}}\,t^2 \nonumber \\
C_{142-2} &=&
\frac{t}{3\,{\sqrt{2}}}
\, \left ( J + U - 2\,W - {\sqrt{\Aindex{2}}}\,\cos (\Thetaindex{3}) + {\sqrt{3}}\,{\sqrt{\Aindex{2}}}\,\sin (\Thetaindex{3})  \right ) \nonumber \\
C_{142-3} &=&
\frac{-1}{9\,{\sqrt{2}}}
\,\left ( {\Aindex{15}}^2 - 72\,t^2 - 6\,\Aindex{15}\,U\right . \nonumber \\
&& \hspace{1cm} 
 + 
 \left . 9\,U^2 - 60\,\Aindex{15}\,W + 180\,U\,W + 900\,W^2
\right  )  \nonumber \\
 N_{142} &=& {\sqrt{8\,{\Cindex{142}{1}}^2 + 16\,{\Cindex{142}{2}}^2 + 2\,{\Cindex{142}{3}}^2}} \nonumber \eeq 
\beq
\ket{\Psi}_{143}& = &\ket{4,0,0,\Gamma_{5,3}} \nonumber \\ 
&=& \frac{1}{4}
 \left ( \ket{02du} - \ket{02ud} - \ket{0d2u} + \ket{0u2d} - \ket{20du} + \ket{20ud} + \ket{2d0u} - \ket{2u0d}\right . \nonumber \\
&& \hspace{1em} 
 + 
\left . \ket{d0u2} - \ket{d2u0} - \ket{du02} + \ket{du20} - \ket{u0d2} + \ket{u2d0} + \ket{ud02} - \ket{ud20}\right ) \nonumber  
\eeq
\beq
\ket{\Psi}_{144}& = &\ket{4,0,2,\Gamma_{5,3}} \nonumber \\ 
& = &
\Cindex{144}{1} \left ( 
\ket{02du} + \ket{02ud} - \ket{0d2u} - \ket{0u2d} + \ket{20du} + \ket{20ud} - \ket{2d0u} - \ket{2u0d}\right . \nonumber \\
&& \hspace{3em} 
 + 
\left . \ket{d0u2} + \ket{d2u0} - \ket{du02} - \ket{du20} + \ket{u0d2} + \ket{u2d0} - \ket{ud02} - \ket{ud20}\right ) 
\nonumber \\
& + &
\Cindex{144}{2} \left ( 
\ket{0du2} + \ket{0ud2} - \ket{2du0} - \ket{2ud0} - \ket{d02u} + \ket{d20u} - \ket{u02d} + \ket{u20d}\right) 
 \nonumber \\
& + &
\Cindex{144}{3} \left ( 
\ket{duud} - \ket{uddu}\right) 
\nonumber \eeq
\beq
C_{144-1} &=&
\frac{-t}{3\,{\sqrt{2}}}
\, \left ( J + U - 2\,W + 2\,{\sqrt{\Aindex{2}}}\,\cos (\Thetaindex{3})  \right ) \nonumber \\
C_{144-2} &=&
2\,{\sqrt{2}}\,t^2 \nonumber \\
C_{144-3} &=&
\frac{-1}{9\,{\sqrt{2}}}
\,\left ( {\Aindex{12}}^2 - 72\,t^2 + 6\,J\,U - 3\,U^2 + 60\,J\,W - 132\,U\,W - 1020\,W^2\right . \nonumber \\
&& \hspace{1cm} 
 + 
 \left . 12\,{\sqrt{\Aindex{2}}}\,U\,\cos (\Thetaindex{3}) + 120\,{\sqrt{\Aindex{2}}}\,W\,\cos (\Thetaindex{3})
\right  )  \nonumber \\
 N_{144} &=& {\sqrt{16\,{\Cindex{144}{1}}^2 + 8\,{\Cindex{144}{2}}^2 + 2\,{\Cindex{144}{3}}^2}} \nonumber \eeq 
\beq
\ket{\Psi}_{145}& = &\ket{4,0,2,\Gamma_{5,3}} \nonumber \\ 
& = &
\Cindex{145}{1} \left ( 
\ket{02du} + \ket{02ud} - \ket{0d2u} - \ket{0u2d} + \ket{20du} + \ket{20ud} - \ket{2d0u} - \ket{2u0d}\right . \nonumber \\
&& \hspace{3em} 
 + 
\left . \ket{d0u2} + \ket{d2u0} - \ket{du02} - \ket{du20} + \ket{u0d2} + \ket{u2d0} - \ket{ud02} - \ket{ud20}\right ) 
\nonumber \\
& + &
\Cindex{145}{2} \left ( 
\ket{0du2} + \ket{0ud2} - \ket{2du0} - \ket{2ud0} - \ket{d02u} + \ket{d20u} - \ket{u02d} + \ket{u20d}\right) 
 \nonumber \\
& + &
\Cindex{145}{3} \left ( 
\ket{duud} - \ket{uddu}\right) 
\nonumber \eeq
\beq
C_{145-1} &=&
\frac{-t}{3\,{\sqrt{2}}}
\, \left ( J + U - 2\,W - {\sqrt{\Aindex{2}}}\,\cos (\Thetaindex{3}) - {\sqrt{3}}\,{\sqrt{\Aindex{2}}}\,\sin (\Thetaindex{3})  \right ) \nonumber \\
C_{145-2} &=&
2\,{\sqrt{2}}\,t^2 \nonumber \\
C_{145-3} &=&
\frac{-1}{9\,{\sqrt{2}}}
\,\left ( {\Aindex{17}}^2 - 72\,t^2 - 6\,\Aindex{17}\,U\right . \nonumber \\
&& \hspace{1cm} 
 + 
 \left . 9\,U^2 - 60\,\Aindex{17}\,W + 180\,U\,W + 900\,W^2
\right  )  \nonumber \\
 N_{145} &=& {\sqrt{16\,{\Cindex{145}{1}}^2 + 8\,{\Cindex{145}{2}}^2 + 2\,{\Cindex{145}{3}}^2}} \nonumber \eeq 
\beq
\ket{\Psi}_{146}& = &\ket{4,0,2,\Gamma_{5,3}} \nonumber \\ 
& = &
\Cindex{146}{1} \left ( 
\ket{02du} + \ket{02ud} - \ket{0d2u} - \ket{0u2d} + \ket{20du} + \ket{20ud} - \ket{2d0u} - \ket{2u0d}\right . \nonumber \\
&& \hspace{3em} 
 + 
\left . \ket{d0u2} + \ket{d2u0} - \ket{du02} - \ket{du20} + \ket{u0d2} + \ket{u2d0} - \ket{ud02} - \ket{ud20}\right ) 
\nonumber \\
& + &
\Cindex{146}{2} \left ( 
\ket{0du2} + \ket{0ud2} - \ket{2du0} - \ket{2ud0} - \ket{d02u} + \ket{d20u} - \ket{u02d} + \ket{u20d}\right) 
 \nonumber \\
& + &
\Cindex{146}{3} \left ( 
\ket{duud} - \ket{uddu}\right) 
\nonumber \eeq
\beq
C_{146-1} &=&
\frac{-t}{3\,{\sqrt{2}}}
\, \left ( J + U - 2\,W - {\sqrt{\Aindex{2}}}\,\cos (\Thetaindex{3}) + {\sqrt{3}}\,{\sqrt{\Aindex{2}}}\,\sin (\Thetaindex{3})  \right ) \nonumber \\
C_{146-2} &=&
2\,{\sqrt{2}}\,t^2 \nonumber \\
C_{146-3} &=&
\frac{-1}{9\,{\sqrt{2}}}
\,\left ( {\Aindex{15}}^2 - 72\,t^2 - 6\,\Aindex{15}\,U\right . \nonumber \\
&& \hspace{1cm} 
 + 
 \left . 9\,U^2 - 60\,\Aindex{15}\,W + 180\,U\,W + 900\,W^2
\right  )  \nonumber \\
 N_{146} &=& {\sqrt{16\,{\Cindex{146}{1}}^2 + 8\,{\Cindex{146}{2}}^2 + 2\,{\Cindex{146}{3}}^2}} \nonumber \eeq 
{\subsection*{\boldmath Unnormalized eigenvectors for ${\rm  N_e}=4$ and   ${\rm m_s}$= $1$.}
\beq
\ket{\Psi}_{147}& = &\ket{4,1,2,\Gamma_2} \nonumber \\ 
&=& \frac{1}{2\,{\sqrt{3}}}
 \left ( \ket{02uu} - \ket{0u2u} + \ket{0uu2} - \ket{20uu} + \ket{2u0u} - \ket{2uu0}\right . \nonumber \\
&& \hspace{1em} 
 + 
\left . \ket{u02u} - \ket{u0u2} - \ket{u20u} + \ket{u2u0} + \ket{uu02} - \ket{uu20}\right ) \nonumber  
\eeq
\beq
\ket{\Psi}_{148}& = &\ket{4,1,6,\Gamma_2} \nonumber \\ 
&=& \frac{1}{2}
 \left ( \ket{duuu} + \ket{uduu} + \ket{uudu} + \ket{uuud}\right) \nonumber 
\eeq
\beq
\ket{\Psi}_{149}& = &\ket{4,1,2,\Gamma_{3,1}} \nonumber \\ 
&=& \frac{1}{2\,{\sqrt{2}}}
 \left ( \ket{02uu} - \ket{0uu2} - \ket{20uu} + \ket{2uu0} - \ket{u02u} + \ket{u20u} + \ket{uu02} - \ket{uu20}\right) \nonumber 
\eeq
\beq
\ket{\Psi}_{150}& = &\ket{4,1,2,\Gamma_{3,2}} \nonumber \\ 
& = &
\Cindex{150}{1} \left ( 
\ket{02uu} + \ket{0uu2} - \ket{20uu} - \ket{2uu0} + \ket{u02u} - \ket{u20u} + \ket{uu02} - \ket{uu20}\right) 
 \nonumber \\
& + &
\Cindex{150}{2} \left ( 
\ket{0u2u} - \ket{2u0u} + \ket{u0u2} - \ket{u2u0}\right) 
\nonumber \eeq
\beq
C_{150-1} &=&
\frac{-1}{2\,{\sqrt{6}}} \nonumber \\
C_{150-2} &=&
-\left( \frac{1}{{\sqrt{6}}} \right)  \nonumber \\
 N_{150} &=& 2\,{\sqrt{2\,{\Cindex{150}{1}}^2 + {\Cindex{150}{2}}^2}} \nonumber \eeq 
\beq
\ket{\Psi}_{151}& = &\ket{4,1,2,\Gamma_{4,1}} \nonumber \\ 
&=& \frac{1}{2\,{\sqrt{2}}}
 \left ( \ket{0u2u} + \ket{0uu2} + \ket{2u0u} + \ket{2uu0} + \ket{u02u} + \ket{u0u2} + \ket{u20u} + \ket{u2u0}\right) \nonumber 
\eeq
\beq
\ket{\Psi}_{152}& = &\ket{4,1,2,\Gamma_{4,2}} \nonumber \\ 
&=& \frac{1}{2\,{\sqrt{2}}}
 \left ( \ket{02uu} + \ket{0u2u} + \ket{20uu} + \ket{2u0u} - \ket{u0u2} - \ket{u2u0} - \ket{uu02} - \ket{uu20}\right) \nonumber 
\eeq
\beq
\ket{\Psi}_{153}& = &\ket{4,1,2,\Gamma_{4,3}} \nonumber \\ 
&=& \frac{1}{2\,{\sqrt{2}}}
 \left ( \ket{02uu} - \ket{0uu2} + \ket{20uu} - \ket{2uu0} + \ket{u02u} + \ket{u20u} + \ket{uu02} + \ket{uu20}\right) \nonumber 
\eeq
\beq
\ket{\Psi}_{154}& = &\ket{4,1,2,\Gamma_{5,1}} \nonumber \\ 
& = &
\Cindex{154}{1} \left ( 
\ket{02uu} + \ket{0uu2} + \ket{20uu} + \ket{2uu0} - \ket{u02u} - \ket{u20u} + \ket{uu02} + \ket{uu20}\right) 
 \nonumber \\
& + &
\Cindex{154}{2} \left ( 
\ket{0u2u} - \ket{2u0u} - \ket{u0u2} + \ket{u2u0}\right) 
 \nonumber \\
& + &
\Cindex{154}{3} \left ( 
\ket{duuu} - \ket{uduu} + \ket{uudu} - \ket{uuud}\right) 
\nonumber \eeq
\beq
C_{154-1} &=&
\frac{-t}{3}
\, \left ( J + U - 2\,W + 2\,{\sqrt{\Aindex{2}}}\,\cos (\Thetaindex{3})  \right ) \nonumber \\
C_{154-2} &=&
-4\,t^2 \nonumber \\
C_{154-3} &=&
\frac{1}{72}
\,\left ( -{\Aindex{11}}^2 + 6\,\Aindex{11}\,J\right . \nonumber \\
&& \hspace{1cm} 
 \left . -9\,J^2 + 288\,t^2 + 12\,\Aindex{11}\,U - 36\,J\,U
\right .  \nonumber \\
&& \hspace{1cm} 
 \left . -36\,U^2 + 120\,\Aindex{11}\,W - 360\,J\,W - 720\,U\,W - 3600\,W^2
\right )  \nonumber \\
 N_{154} &=& 2\,{\sqrt{2\,{\Cindex{154}{1}}^2 + {\Cindex{154}{2}}^2 + {\Cindex{154}{3}}^2}} \nonumber \eeq 
\beq
\ket{\Psi}_{155}& = &\ket{4,1,2,\Gamma_{5,1}} \nonumber \\ 
& = &
\Cindex{155}{1} \left ( 
\ket{02uu} + \ket{0uu2} + \ket{20uu} + \ket{2uu0} - \ket{u02u} - \ket{u20u} + \ket{uu02} + \ket{uu20}\right) 
 \nonumber \\
& + &
\Cindex{155}{2} \left ( 
\ket{0u2u} - \ket{2u0u} - \ket{u0u2} + \ket{u2u0}\right) 
 \nonumber \\
& + &
\Cindex{155}{3} \left ( 
\ket{duuu} - \ket{uduu} + \ket{uudu} - \ket{uuud}\right) 
\nonumber \eeq
\beq
C_{155-1} &=&
\frac{t}{6}
\, \left ( 2\,\Aindex{18} - 3\,J - 6\,U - 60\,W  \right ) \nonumber \\
C_{155-2} &=&
-4\,t^2 \nonumber \\
C_{155-3} &=&
\frac{1}{72}
\,\left ( -4\,{\Aindex{18}}^2 + 12\,\Aindex{18}\,J\right . \nonumber \\
&& \hspace{1cm} 
 \left . -9\,J^2 + 288\,t^2 + 24\,\Aindex{18}\,U - 36\,J\,U
\right .  \nonumber \\
&& \hspace{1cm} 
 \left . -36\,U^2 + 240\,\Aindex{18}\,W - 360\,J\,W - 720\,U\,W - 3600\,W^2
\right )  \nonumber \\
 N_{155} &=& 2\,{\sqrt{2\,{\Cindex{155}{1}}^2 + {\Cindex{155}{2}}^2 + {\Cindex{155}{3}}^2}} \nonumber \eeq 
\beq
\ket{\Psi}_{156}& = &\ket{4,1,2,\Gamma_{5,1}} \nonumber \\ 
& = &
\Cindex{156}{1} \left ( 
\ket{02uu} + \ket{0uu2} + \ket{20uu} + \ket{2uu0} - \ket{u02u} - \ket{u20u} + \ket{uu02} + \ket{uu20}\right) 
 \nonumber \\
& + &
\Cindex{156}{2} \left ( 
\ket{0u2u} - \ket{2u0u} - \ket{u0u2} + \ket{u2u0}\right) 
 \nonumber \\
& + &
\Cindex{156}{3} \left ( 
\ket{duuu} - \ket{uduu} + \ket{uudu} - \ket{uuud}\right) 
\nonumber \eeq
\beq
C_{156-1} &=&
\frac{t}{6}
\, \left ( 2\,\Aindex{16} - 3\,J - 6\,U - 60\,W  \right ) \nonumber \\
C_{156-2} &=&
-4\,t^2 \nonumber \\
C_{156-3} &=&
\frac{1}{72}
\,\left ( -4\,{\Aindex{16}}^2 + 12\,\Aindex{16}\,J\right . \nonumber \\
&& \hspace{1cm} 
 \left . -9\,J^2 + 288\,t^2 + 24\,\Aindex{16}\,U - 36\,J\,U
\right .  \nonumber \\
&& \hspace{1cm} 
 \left . -36\,U^2 + 240\,\Aindex{16}\,W - 360\,J\,W - 720\,U\,W - 3600\,W^2
\right )  \nonumber \\
 N_{156} &=& 2\,{\sqrt{2\,{\Cindex{156}{1}}^2 + {\Cindex{156}{2}}^2 + {\Cindex{156}{3}}^2}} \nonumber \eeq 
\beq
\ket{\Psi}_{157}& = &\ket{4,1,2,\Gamma_{5,2}} \nonumber \\ 
& = &
\Cindex{157}{1} \left ( 
\ket{02uu} - \ket{20uu} - \ket{uu02} + \ket{uu20}\right) 
 \nonumber \\
& + &
\Cindex{157}{2} \left ( 
\ket{0u2u} - \ket{0uu2} + \ket{2u0u} - \ket{2uu0} - \ket{u02u} + \ket{u0u2} - \ket{u20u} + \ket{u2u0}\right) 
 \nonumber \\
& + &
\Cindex{157}{3} \left ( 
\ket{duuu} + \ket{uduu} - \ket{uudu} - \ket{uuud}\right) 
\nonumber \eeq
\beq
C_{157-1} &=&
4\,t^2 \nonumber \\
C_{157-2} &=&
\frac{t}{3}
\, \left ( J + U - 2\,W + 2\,{\sqrt{\Aindex{2}}}\,\cos (\Thetaindex{3})  \right ) \nonumber \\
C_{157-3} &=&
\frac{1}{72}
\,\left ( -{\Aindex{11}}^2 + 6\,\Aindex{11}\,J\right . \nonumber \\
&& \hspace{1cm} 
 \left . -9\,J^2 + 288\,t^2 + 12\,\Aindex{11}\,U - 36\,J\,U
\right .  \nonumber \\
&& \hspace{1cm} 
 \left . -36\,U^2 + 120\,\Aindex{11}\,W - 360\,J\,W - 720\,U\,W - 3600\,W^2
\right )  \nonumber \\
 N_{157} &=& 2\,{\sqrt{{\Cindex{157}{1}}^2 + 2\,{\Cindex{157}{2}}^2 + {\Cindex{157}{3}}^2}} \nonumber \eeq 
\beq
\ket{\Psi}_{158}& = &\ket{4,1,2,\Gamma_{5,2}} \nonumber \\ 
& = &
\Cindex{158}{1} \left ( 
\ket{02uu} - \ket{20uu} - \ket{uu02} + \ket{uu20}\right) 
 \nonumber \\
& + &
\Cindex{158}{2} \left ( 
\ket{0u2u} - \ket{0uu2} + \ket{2u0u} - \ket{2uu0} - \ket{u02u} + \ket{u0u2} - \ket{u20u} + \ket{u2u0}\right) 
 \nonumber \\
& + &
\Cindex{158}{3} \left ( 
\ket{duuu} + \ket{uduu} - \ket{uudu} - \ket{uuud}\right) 
\nonumber \eeq
\beq
C_{158-1} &=&
4\,t^2 \nonumber \\
C_{158-2} &=&
\frac{-t}{6}
\, \left ( 2\,\Aindex{18} - 3\,J - 6\,U - 60\,W  \right ) \nonumber \\
C_{158-3} &=&
\frac{1}{72}
\,\left ( -4\,{\Aindex{18}}^2 + 12\,\Aindex{18}\,J\right . \nonumber \\
&& \hspace{1cm} 
 \left . -9\,J^2 + 288\,t^2 + 24\,\Aindex{18}\,U - 36\,J\,U
\right .  \nonumber \\
&& \hspace{1cm} 
 \left . -36\,U^2 + 240\,\Aindex{18}\,W - 360\,J\,W - 720\,U\,W - 3600\,W^2
\right )  \nonumber \\
 N_{158} &=& 2\,{\sqrt{{\Cindex{158}{1}}^2 + 2\,{\Cindex{158}{2}}^2 + {\Cindex{158}{3}}^2}} \nonumber \eeq 
\beq
\ket{\Psi}_{159}& = &\ket{4,1,2,\Gamma_{5,2}} \nonumber \\ 
& = &
\Cindex{159}{1} \left ( 
\ket{02uu} - \ket{20uu} - \ket{uu02} + \ket{uu20}\right) 
 \nonumber \\
& + &
\Cindex{159}{2} \left ( 
\ket{0u2u} - \ket{0uu2} + \ket{2u0u} - \ket{2uu0} - \ket{u02u} + \ket{u0u2} - \ket{u20u} + \ket{u2u0}\right) 
 \nonumber \\
& + &
\Cindex{159}{3} \left ( 
\ket{duuu} + \ket{uduu} - \ket{uudu} - \ket{uuud}\right) 
\nonumber \eeq
\beq
C_{159-1} &=&
4\,t^2 \nonumber \\
C_{159-2} &=&
\frac{-t}{6}
\, \left ( 2\,\Aindex{16} - 3\,J - 6\,U - 60\,W  \right ) \nonumber \\
C_{159-3} &=&
\frac{1}{72}
\,\left ( -4\,{\Aindex{16}}^2 + 12\,\Aindex{16}\,J\right . \nonumber \\
&& \hspace{1cm} 
 \left . -9\,J^2 + 288\,t^2 + 24\,\Aindex{16}\,U - 36\,J\,U
\right .  \nonumber \\
&& \hspace{1cm} 
 \left . -36\,U^2 + 240\,\Aindex{16}\,W - 360\,J\,W - 720\,U\,W - 3600\,W^2
\right )  \nonumber \\
 N_{159} &=& 2\,{\sqrt{{\Cindex{159}{1}}^2 + 2\,{\Cindex{159}{2}}^2 + {\Cindex{159}{3}}^2}} \nonumber \eeq 
\beq
\ket{\Psi}_{160}& = &\ket{4,1,2,\Gamma_{5,3}} \nonumber \\ 
& = &
\Cindex{160}{1} \left ( 
\ket{02uu} - \ket{0u2u} + \ket{20uu} - \ket{2u0u} + \ket{u0u2} + \ket{u2u0} - \ket{uu02} - \ket{uu20}\right) 
 \nonumber \\
& + &
\Cindex{160}{2} \left ( 
\ket{0uu2} - \ket{2uu0} - \ket{u02u} + \ket{u20u}\right) 
 \nonumber \\
& + &
\Cindex{160}{3} \left ( 
\ket{duuu} - \ket{uduu} - \ket{uudu} + \ket{uuud}\right) 
\nonumber \eeq
\beq
C_{160-1} &=&
\frac{t}{3}
\, \left ( J + U - 2\,W + 2\,{\sqrt{\Aindex{2}}}\,\cos (\Thetaindex{3})  \right ) \nonumber \\
C_{160-2} &=&
-4\,t^2 \nonumber \\
C_{160-3} &=&
\frac{1}{72}
\,\left ( {\Aindex{11}}^2 - 6\,\Aindex{11}\,J\right . \nonumber \\
&& \hspace{1cm} 
 + 
 \left . 9\,J^2 - 288\,t^2 - 12\,\Aindex{11}\,U + 36\,J\,U
\right .  \nonumber \\
&& \hspace{1cm} 
 + 
 \left . 36\,U^2 - 120\,\Aindex{11}\,W + 360\,J\,W + 720\,U\,W + 3600\,W^2
\right )  \nonumber \\
 N_{160} &=& 2\,{\sqrt{2\,{\Cindex{160}{1}}^2 + {\Cindex{160}{2}}^2 + {\Cindex{160}{3}}^2}} \nonumber \eeq 
\beq
\ket{\Psi}_{161}& = &\ket{4,1,2,\Gamma_{5,3}} \nonumber \\ 
& = &
\Cindex{161}{1} \left ( 
\ket{02uu} - \ket{0u2u} + \ket{20uu} - \ket{2u0u} + \ket{u0u2} + \ket{u2u0} - \ket{uu02} - \ket{uu20}\right) 
 \nonumber \\
& + &
\Cindex{161}{2} \left ( 
\ket{0uu2} - \ket{2uu0} - \ket{u02u} + \ket{u20u}\right) 
 \nonumber \\
& + &
\Cindex{161}{3} \left ( 
\ket{duuu} - \ket{uduu} - \ket{uudu} + \ket{uuud}\right) 
\nonumber \eeq
\beq
C_{161-1} &=&
\frac{-t}{6}
\, \left ( 2\,\Aindex{18} - 3\,J - 6\,U - 60\,W  \right ) \nonumber \\
C_{161-2} &=&
-4\,t^2 \nonumber \\
C_{161-3} &=&
\frac{1}{72}
\,\left ( 4\,{\Aindex{18}}^2 - 12\,\Aindex{18}\,J\right . \nonumber \\
&& \hspace{1cm} 
 + 
 \left . 9\,J^2 - 288\,t^2 - 24\,\Aindex{18}\,U + 36\,J\,U
\right .  \nonumber \\
&& \hspace{1cm} 
 + 
 \left . 36\,U^2 - 240\,\Aindex{18}\,W + 360\,J\,W + 720\,U\,W + 3600\,W^2
\right )  \nonumber \\
 N_{161} &=& 2\,{\sqrt{2\,{\Cindex{161}{1}}^2 + {\Cindex{161}{2}}^2 + {\Cindex{161}{3}}^2}} \nonumber \eeq 
\beq
\ket{\Psi}_{162}& = &\ket{4,1,2,\Gamma_{5,3}} \nonumber \\ 
& = &
\Cindex{162}{1} \left ( 
\ket{02uu} - \ket{0u2u} + \ket{20uu} - \ket{2u0u} + \ket{u0u2} + \ket{u2u0} - \ket{uu02} - \ket{uu20}\right) 
 \nonumber \\
& + &
\Cindex{162}{2} \left ( 
\ket{0uu2} - \ket{2uu0} - \ket{u02u} + \ket{u20u}\right) 
 \nonumber \\
& + &
\Cindex{162}{3} \left ( 
\ket{duuu} - \ket{uduu} - \ket{uudu} + \ket{uuud}\right) 
\nonumber \eeq
\beq
C_{162-1} &=&
\frac{-t}{6}
\, \left ( 2\,\Aindex{16} - 3\,J - 6\,U - 60\,W  \right ) \nonumber \\
C_{162-2} &=&
-4\,t^2 \nonumber \\
C_{162-3} &=&
\frac{1}{72}
\,\left ( 4\,{\Aindex{16}}^2 - 12\,\Aindex{16}\,J\right . \nonumber \\
&& \hspace{1cm} 
 + 
 \left . 9\,J^2 - 288\,t^2 - 24\,\Aindex{16}\,U + 36\,J\,U
\right .  \nonumber \\
&& \hspace{1cm} 
 + 
 \left . 36\,U^2 - 240\,\Aindex{16}\,W + 360\,J\,W + 720\,U\,W + 3600\,W^2
\right )  \nonumber \\
 N_{162} &=& 2\,{\sqrt{2\,{\Cindex{162}{1}}^2 + {\Cindex{162}{2}}^2 + {\Cindex{162}{3}}^2}} \nonumber \eeq 
{\subsection*{\boldmath Unnormalized eigenvectors for ${\rm  N_e}=4$ and   ${\rm m_s}$= $2$.}
\beq
\ket{\Psi}_{163}& = &\ket{4,2,6,\Gamma_2} \nonumber \\ 
&=& 1
 \left ( \ket{uuuu}\right) \nonumber 
\eeq
{\subsection*{\boldmath Unnormalized eigenvectors for ${\rm  N_e}=5$ and   ${\rm m_s}$= $- {3 \over 2} $.}
\beq
\ket{\Psi}_{164}& = &\ket{5,- {3 \over 2} , {15 \over 4} ,\Gamma_2} \nonumber \\ 
&=& \frac{1}{2}
 \left ( \ket{2ddd} - \ket{d2dd} + \ket{dd2d} - \ket{ddd2}\right) \nonumber 
\eeq
\beq
\ket{\Psi}_{165}& = &\ket{5,- {3 \over 2} , {15 \over 4} ,\Gamma_{5,1}} \nonumber \\ 
&=& \frac{1}{2}
 \left ( \ket{2ddd} + \ket{d2dd} + \ket{dd2d} + \ket{ddd2}\right) \nonumber 
\eeq
\beq
\ket{\Psi}_{166}& = &\ket{5,- {3 \over 2} , {15 \over 4} ,\Gamma_{5,2}} \nonumber \\ 
&=& \frac{1}{2}
 \left ( \ket{2ddd} - \ket{d2dd} - \ket{dd2d} + \ket{ddd2}\right) \nonumber 
\eeq
\beq
\ket{\Psi}_{167}& = &\ket{5,- {3 \over 2} , {15 \over 4} ,\Gamma_{5,3}} \nonumber \\ 
&=& \frac{1}{2}
 \left ( \ket{2ddd} + \ket{d2dd} - \ket{dd2d} - \ket{ddd2}\right) \nonumber 
\eeq
{\subsection*{\boldmath Unnormalized eigenvectors for ${\rm  N_e}=5$ and   ${\rm m_s}$= $- {1 \over 2} $.}
\beq
\ket{\Psi}_{168}& = &\ket{5,- {1 \over 2} , {3 \over 4} ,\Gamma_1} \nonumber \\ 
&=& \frac{1}{2\,{\sqrt{3}}}
 \left ( \ket{022d} + \ket{02d2} + \ket{0d22} + \ket{202d} + \ket{20d2} + \ket{220d}\right . \nonumber \\
&& \hspace{1em} 
 + 
\left . \ket{22d0} + \ket{2d02} + \ket{2d20} + \ket{d022} + \ket{d202} + \ket{d220}\right ) \nonumber  
\eeq
\beq
\ket{\Psi}_{169}& = &\ket{5,- {1 \over 2} , {15 \over 4} ,\Gamma_2} \nonumber \\ 
&=& \frac{1}{2\,{\sqrt{3}}}
 \left ( \ket{2ddu} + \ket{2dud} + \ket{2udd} - \ket{d2du} - \ket{d2ud} + \ket{dd2u}\right . \nonumber \\
&& \hspace{1em} 
\left . -\ket{ddu2} + \ket{du2d} - \ket{dud2} - \ket{u2dd} + \ket{ud2d} - \ket{udd2}\right ) \nonumber  
\eeq
\beq
\ket{\Psi}_{170}& = &\ket{5,- {1 \over 2} , {3 \over 4} ,\Gamma_{3,1}} \nonumber \\ 
& = &
\Cindex{170}{1} \left ( 
\ket{022d} + \ket{0d22} + \ket{20d2} + \ket{220d} + \ket{22d0} + \ket{2d02} + \ket{d022} + \ket{d220}\right) 
 \nonumber \\
& + &
\Cindex{170}{2} \left ( 
\ket{02d2} + \ket{202d} + \ket{2d20} + \ket{d202}\right) 
 \nonumber \\
& + &
\Cindex{170}{3} \left ( 
\ket{2ddu} - \ket{2udd} - \ket{d2ud} - \ket{dd2u} + \ket{ddu2} + \ket{du2d} + \ket{u2dd} - \ket{udd2}\right) 
\nonumber \eeq
\beq
C_{170-1} &=&
\frac{-t}{2\,{\sqrt{2}}} \nonumber \\
C_{170-2} &=&
\frac{t}{{\sqrt{2}}} \nonumber \\
C_{170-3} &=&
\frac{1}{4\,{\sqrt{2}}}
\, \left ( {\sqrt{\Aindex{7}}} + J + 2\,t + U - 2\,W  \right ) \nonumber \\
 N_{170} &=& 2\,{\sqrt{2\,{\Cindex{170}{1}}^2 + {\Cindex{170}{2}}^2 + 2\,{\Cindex{170}{3}}^2}} \nonumber \eeq 
\beq
\ket{\Psi}_{171}& = &\ket{5,- {1 \over 2} , {3 \over 4} ,\Gamma_{3,1}} \nonumber \\ 
& = &
\Cindex{171}{1} \left ( 
\ket{022d} + \ket{0d22} + \ket{20d2} + \ket{220d} + \ket{22d0} + \ket{2d02} + \ket{d022} + \ket{d220}\right) 
 \nonumber \\
& + &
\Cindex{171}{2} \left ( 
\ket{02d2} + \ket{202d} + \ket{2d20} + \ket{d202}\right) 
 \nonumber \\
& + &
\Cindex{171}{3} \left ( 
\ket{2ddu} - \ket{2udd} - \ket{d2ud} - \ket{dd2u} + \ket{ddu2} + \ket{du2d} + \ket{u2dd} - \ket{udd2}\right) 
\nonumber \eeq
\beq
C_{171-1} &=&
\frac{-t}{2\,{\sqrt{2}}} \nonumber \\
C_{171-2} &=&
\frac{t}{{\sqrt{2}}} \nonumber \\
C_{171-3} &=&
\frac{-1}{4\,{\sqrt{2}}}
\, \left ( {\sqrt{\Aindex{7}}} - J - 2\,t - U + 2\,W  \right ) \nonumber \\
 N_{171} &=& 2\,{\sqrt{2\,{\Cindex{171}{1}}^2 + {\Cindex{171}{2}}^2 + 2\,{\Cindex{171}{3}}^2}} \nonumber \eeq 
\beq
\ket{\Psi}_{172}& = &\ket{5,- {1 \over 2} , {3 \over 4} ,\Gamma_{3,2}} \nonumber \\ 
& = &
\Cindex{172}{1} \left ( 
\ket{022d} - \ket{0d22} + \ket{20d2} - \ket{220d} - \ket{22d0} + \ket{2d02} - \ket{d022} + \ket{d220}\right) 
 \nonumber \\
& + &
\Cindex{172}{2} \left ( 
\ket{2ddu} + \ket{2udd} - \ket{d2ud} + \ket{dd2u} - \ket{ddu2} + \ket{du2d} - \ket{u2dd} - \ket{udd2}\right) 
 \nonumber \\
& + &
\Cindex{172}{3} \left ( 
\ket{2dud} - \ket{d2du} - \ket{dud2} + \ket{ud2d}\right) 
\nonumber \eeq
\beq
C_{172-1} &=&
\frac{-\left( {\sqrt{\frac{3}{2}}}\,t \right) }{2} \nonumber \\
C_{172-2} &=&
\frac{-1}{4\,{\sqrt{6}}}
\, \left ( {\sqrt{\Aindex{7}}} + J + 2\,t + U - 2\,W  \right ) \nonumber \\
C_{172-3} &=&
\frac{1}{2\,{\sqrt{6}}}
\, \left ( {\sqrt{\Aindex{7}}} + J + 2\,t + U - 2\,W  \right ) \nonumber \\
 N_{172} &=& 2\,{\sqrt{2\,{\Cindex{172}{1}}^2 + 2\,{\Cindex{172}{2}}^2 + {\Cindex{172}{3}}^2}} \nonumber \eeq 
\beq
\ket{\Psi}_{173}& = &\ket{5,- {1 \over 2} , {3 \over 4} ,\Gamma_{3,2}} \nonumber \\ 
& = &
\Cindex{173}{1} \left ( 
\ket{022d} - \ket{0d22} + \ket{20d2} - \ket{220d} - \ket{22d0} + \ket{2d02} - \ket{d022} + \ket{d220}\right) 
 \nonumber \\
& + &
\Cindex{173}{2} \left ( 
\ket{2ddu} + \ket{2udd} - \ket{d2ud} + \ket{dd2u} - \ket{ddu2} + \ket{du2d} - \ket{u2dd} - \ket{udd2}\right) 
 \nonumber \\
& + &
\Cindex{173}{3} \left ( 
\ket{2dud} - \ket{d2du} - \ket{dud2} + \ket{ud2d}\right) 
\nonumber \eeq
\beq
C_{173-1} &=&
\frac{-\left( {\sqrt{\frac{3}{2}}}\,t \right) }{2} \nonumber \\
C_{173-2} &=&
\frac{1}{4\,{\sqrt{6}}}
\, \left ( {\sqrt{\Aindex{7}}} - J - 2\,t - U + 2\,W  \right ) \nonumber \\
C_{173-3} &=&
\frac{-1}{2\,{\sqrt{6}}}
\, \left ( {\sqrt{\Aindex{7}}} - J - 2\,t - U + 2\,W  \right ) \nonumber \\
 N_{173} &=& 2\,{\sqrt{2\,{\Cindex{173}{1}}^2 + 2\,{\Cindex{173}{2}}^2 + {\Cindex{173}{3}}^2}} \nonumber \eeq 
\beq
\ket{\Psi}_{174}& = &\ket{5,- {1 \over 2} , {3 \over 4} ,\Gamma_{4,1}} \nonumber \\ 
& = &
\Cindex{174}{1} \left ( 
\ket{022d} + \ket{02d2} + \ket{202d} + \ket{20d2} - \ket{2d02} - \ket{2d20} - \ket{d202} - \ket{d220}\right) 
 \nonumber \\
& + &
\Cindex{174}{2} \left ( 
\ket{0d22} - \ket{220d} - \ket{22d0} + \ket{d022}\right) 
 \nonumber \\
& + &
\Cindex{174}{3} \left ( 
\ket{2ddu} - \ket{2dud} + \ket{d2du} - \ket{d2ud} - \ket{du2d} - \ket{dud2} + \ket{ud2d} + \ket{udd2}\right) 
\nonumber \eeq
\beq
C_{174-1} &=&
\frac{t}{2\,{\sqrt{2}}}
\, \left ( J + 4\,t + U - 2\,W + 2\,{\sqrt{\Aindex{6}}}\,\cos (\Thetaindex{5})  \right ) \nonumber \\
C_{174-2} &=&
\frac{-t}{3\,{\sqrt{2}}}
\, \left ( J - 12\,t + U - 2\,W + 2\,{\sqrt{\Aindex{6}}}\,\cos (\Thetaindex{5})  \right ) \nonumber \\
C_{174-3} &=&
\frac{-1}{18\,{\sqrt{2}}}
\,\left ( {\Aindex{22}}^2 - 3\,\Aindex{22}\,t - 36\,t^2 + 12\,\Aindex{22}\,U - 18\,t\,U\right . \nonumber \\
&& \hspace{1cm} 
 + 
 \left . 36\,U^2 + 96\,\Aindex{22}\,W - 144\,t\,W + 576\,U\,W + 2304\,W^2
\right  )  \nonumber \\
 N_{174} &=& 2\,{\sqrt{2\,{\Cindex{174}{1}}^2 + {\Cindex{174}{2}}^2 + 2\,{\Cindex{174}{3}}^2}} \nonumber \eeq 
\beq
\ket{\Psi}_{175}& = &\ket{5,- {1 \over 2} , {3 \over 4} ,\Gamma_{4,1}} \nonumber \\ 
& = &
\Cindex{175}{1} \left ( 
\ket{022d} + \ket{02d2} + \ket{202d} + \ket{20d2} - \ket{2d02} - \ket{2d20} - \ket{d202} - \ket{d220}\right) 
 \nonumber \\
& + &
\Cindex{175}{2} \left ( 
\ket{0d22} - \ket{220d} - \ket{22d0} + \ket{d022}\right) 
 \nonumber \\
& + &
\Cindex{175}{3} \left ( 
\ket{2ddu} - \ket{2dud} + \ket{d2du} - \ket{d2ud} - \ket{du2d} - \ket{dud2} + \ket{ud2d} + \ket{udd2}\right) 
\nonumber \eeq
\beq
C_{175-1} &=&
\frac{t}{2\,{\sqrt{2}}}
\, \left ( J + 4\,t + U - 2\,W - {\sqrt{\Aindex{6}}}\,\cos (\Thetaindex{5}) - {\sqrt{3}}\,{\sqrt{\Aindex{6}}}\,\sin (\Thetaindex{5})  \right ) \nonumber \\
C_{175-2} &=&
\frac{-t}{3\,{\sqrt{2}}}
\, \left ( J - 12\,t + U - 2\,W - {\sqrt{\Aindex{6}}}\,\cos (\Thetaindex{5}) - {\sqrt{3}}\,{\sqrt{\Aindex{6}}}\,\sin (\Thetaindex{5})  \right ) \nonumber \\
C_{175-3} &=&
\frac{-1}{18\,{\sqrt{2}}}
\,\left ( {\Aindex{24}}^2 - 3\,J\,t - 45\,t^2 + 12\,J\,U + 33\,t\,U - 24\,U^2 + 96\,J\,W + 294\,t\,W - 504\,U\,W\right . \nonumber \\
&& \hspace{1cm} 
 \left . -2496\,W^2 + 3\,{\sqrt{\Aindex{6}}}\,t\,\cos (\Thetaindex{5}) - 12\,{\sqrt{\Aindex{6}}}\,U\,\cos (\Thetaindex{5}) - 96\,{\sqrt{\Aindex{6}}}\,W\,\cos (\Thetaindex{5})
\right .  \nonumber \\
&& \hspace{1cm} 
 + 
 \left . 3\,{\sqrt{3}}\,{\sqrt{\Aindex{6}}}\,t\,\sin (\Thetaindex{5}) - 12\,{\sqrt{3}}\,{\sqrt{\Aindex{6}}}\,U\,\sin (\Thetaindex{5}) - 96\,{\sqrt{3}}\,{\sqrt{\Aindex{6}}}\,W\,\sin (\Thetaindex{5})
\right )  \nonumber \\
 N_{175} &=& 2\,{\sqrt{2\,{\Cindex{175}{1}}^2 + {\Cindex{175}{2}}^2 + 2\,{\Cindex{175}{3}}^2}} \nonumber \eeq 
\beq
\ket{\Psi}_{176}& = &\ket{5,- {1 \over 2} , {3 \over 4} ,\Gamma_{4,1}} \nonumber \\ 
& = &
\Cindex{176}{1} \left ( 
\ket{022d} + \ket{02d2} + \ket{202d} + \ket{20d2} - \ket{2d02} - \ket{2d20} - \ket{d202} - \ket{d220}\right) 
 \nonumber \\
& + &
\Cindex{176}{2} \left ( 
\ket{0d22} - \ket{220d} - \ket{22d0} + \ket{d022}\right) 
 \nonumber \\
& + &
\Cindex{176}{3} \left ( 
\ket{2ddu} - \ket{2dud} + \ket{d2du} - \ket{d2ud} - \ket{du2d} - \ket{dud2} + \ket{ud2d} + \ket{udd2}\right) 
\nonumber \eeq
\beq
C_{176-1} &=&
\frac{t}{2\,{\sqrt{2}}}
\, \left ( J + 4\,t + U - 2\,W - {\sqrt{\Aindex{6}}}\,\cos (\Thetaindex{5}) + {\sqrt{3}}\,{\sqrt{\Aindex{6}}}\,\sin (\Thetaindex{5})  \right ) \nonumber \\
C_{176-2} &=&
\frac{-t}{3\,{\sqrt{2}}}
\, \left ( J - 12\,t + U - 2\,W - {\sqrt{\Aindex{6}}}\,\cos (\Thetaindex{5}) + {\sqrt{3}}\,{\sqrt{\Aindex{6}}}\,\sin (\Thetaindex{5})  \right ) \nonumber \\
C_{176-3} &=&
\frac{-1}{18\,{\sqrt{2}}}
\,\left ( {\Aindex{23}}^2 - 3\,J\,t - 45\,t^2 + 12\,J\,U + 33\,t\,U - 24\,U^2 + 96\,J\,W + 294\,t\,W - 504\,U\,W\right . \nonumber \\
&& \hspace{1cm} 
 \left . -2496\,W^2 + 3\,{\sqrt{\Aindex{6}}}\,t\,\cos (\Thetaindex{5}) - 12\,{\sqrt{\Aindex{6}}}\,U\,\cos (\Thetaindex{5}) - 96\,{\sqrt{\Aindex{6}}}\,W\,\cos (\Thetaindex{5})
\right .  \nonumber \\
&& \hspace{1cm} 
 \left . -3\,{\sqrt{3}}\,{\sqrt{\Aindex{6}}}\,t\,\sin (\Thetaindex{5}) + 12\,{\sqrt{3}}\,{\sqrt{\Aindex{6}}}\,U\,\sin (\Thetaindex{5}) + 96\,{\sqrt{3}}\,{\sqrt{\Aindex{6}}}\,W\,\sin (\Thetaindex{5})
\right )  \nonumber \\
 N_{176} &=& 2\,{\sqrt{2\,{\Cindex{176}{1}}^2 + {\Cindex{176}{2}}^2 + 2\,{\Cindex{176}{3}}^2}} \nonumber \eeq 
\beq
\ket{\Psi}_{177}& = &\ket{5,- {1 \over 2} , {3 \over 4} ,\Gamma_{4,2}} \nonumber \\ 
& = &
\Cindex{177}{1} \left ( 
\ket{022d} - \ket{20d2} - \ket{2d02} + \ket{d220}\right) 
 \nonumber \\
& + &
\Cindex{177}{2} \left ( 
\ket{02d2} + \ket{0d22} - \ket{202d} - \ket{220d} + \ket{22d0} + \ket{2d20} - \ket{d022} - \ket{d202}\right) 
 \nonumber \\
& + &
\Cindex{177}{3} \left ( 
\ket{2dud} - \ket{2udd} + \ket{d2du} + \ket{dd2u} + \ket{ddu2} - \ket{dud2} - \ket{u2dd} - \ket{ud2d}\right) 
\nonumber \eeq
\beq
C_{177-1} &=&
\frac{1}{18}
\,\left ( -{\Aindex{22}}^2 + 3\,\Aindex{22}\,J + 12\,\Aindex{22}\,t - 18\,J\,t\right . \nonumber \\
&& \hspace{1cm} 
 + 
 \left . 45\,t^2 - 9\,\Aindex{22}\,U + 18\,J\,U + 54\,t\,U - 18\,U^2 - 102\,\Aindex{22}\,W + 144\,J\,W
\right .  \nonumber \\
&& \hspace{1cm} 
 + 
 \left . 612\,t\,W - 468\,U\,W - 2592\,W^2
\right )  \nonumber \\
C_{177-2} &=&
\frac{t}{3}
\, \left ( J + 6\,t + U - 2\,W - {\sqrt{\Aindex{6}}}\,\cos (\Thetaindex{5})  \right ) \nonumber \\
C_{177-3} &=&
\frac{t}{6}
\, \left ( -J + 12\,t - U + 2\,W - 2\,{\sqrt{\Aindex{6}}}\,\cos (\Thetaindex{5})  \right ) \nonumber \\
 N_{177} &=& 2\,{\sqrt{{\Cindex{177}{1}}^2 + 2\,\left( {\Cindex{177}{2}}^2 + {\Cindex{177}{3}}^2 \right) }} \nonumber \eeq 
\beq
\ket{\Psi}_{178}& = &\ket{5,- {1 \over 2} , {3 \over 4} ,\Gamma_{4,2}} \nonumber \\ 
& = &
\Cindex{178}{1} \left ( 
\ket{022d} - \ket{20d2} - \ket{2d02} + \ket{d220}\right) 
 \nonumber \\
& + &
\Cindex{178}{2} \left ( 
\ket{02d2} + \ket{0d22} - \ket{202d} - \ket{220d} + \ket{22d0} + \ket{2d20} - \ket{d022} - \ket{d202}\right) 
 \nonumber \\
& + &
\Cindex{178}{3} \left ( 
\ket{2dud} - \ket{2udd} + \ket{d2du} + \ket{dd2u} + \ket{ddu2} - \ket{dud2} - \ket{u2dd} - \ket{ud2d}\right) 
\nonumber \eeq
\beq
C_{178-1} &=&
\frac{1}{18}
\,\left ( -{\Aindex{24}}^2 + 3\,J^2 + 3\,J\,t + 81\,t^2 - 6\,J\,U - 33\,t\,U 
\right . 
\nonumber \\
&& \hspace{1cm}
\left . 
+ 27\,U^2 - 108\,J\,W - 294\,t\,W + 492\,U\,W + 2508\,W^2\right . \nonumber \\
&& \hspace{1cm} 
  \left .
 - 3\,{\sqrt{\Aindex{6}}}\,J\,\cos (\Thetaindex{5}) -
 12\,{\sqrt{\Aindex{6}}}\,t\,\cos (\Thetaindex{5}) 
+ 9\,{\sqrt{\Aindex{6}}}\,U\,\cos (\Thetaindex{5})
\right . 
\nonumber \\
&& \hspace{3em}
\left . 
+ 102\,{\sqrt{\Aindex{6}}}\,W\,\cos (\Thetaindex{5})
 -3\,{\sqrt{3}}\,{\sqrt{\Aindex{6}}}\,J\,\sin (\Thetaindex{5})
 - 12\,{\sqrt{3}}\,{\sqrt{\Aindex{6}}}\,t\,\sin (\Thetaindex{5}) 
\right .  \nonumber \\
&& \hspace{1cm} 
 \left .
+ 9\,{\sqrt{3}}\,{\sqrt{\Aindex{6}}}\,U\,\sin (\Thetaindex{5}) 
+ 102\,{\sqrt{3}}\,{\sqrt{\Aindex{6}}}\,W\,\sin (\Thetaindex{5})
\right )  \nonumber \\
C_{178-2} &=&
\frac{t}{6}
\, \left ( 2\,J + 12\,t + 2\,U - 4\,W + {\sqrt{\Aindex{6}}}\,\cos (\Thetaindex{5}) + {\sqrt{3}}\,{\sqrt{\Aindex{6}}}\,\sin (\Thetaindex{5})  \right ) \nonumber \\
C_{178-3} &=&
\frac{t}{6}
\, \left ( -J + 12\,t - U + 2\,W + {\sqrt{\Aindex{6}}}\,\cos (\Thetaindex{5}) + {\sqrt{3}}\,{\sqrt{\Aindex{6}}}\,\sin (\Thetaindex{5})  \right ) \nonumber \\
 N_{178} &=& 2\,{\sqrt{{\Cindex{178}{1}}^2 + 2\,\left( {\Cindex{178}{2}}^2 + {\Cindex{178}{3}}^2 \right) }} \nonumber \eeq 
\beq
\ket{\Psi}_{179}& = &\ket{5,- {1 \over 2} , {3 \over 4} ,\Gamma_{4,2}} \nonumber \\ 
& = &
\Cindex{179}{1} \left ( 
\ket{022d} - \ket{20d2} - \ket{2d02} + \ket{d220}\right) 
 \nonumber \\
& + &
\Cindex{179}{2} \left ( 
\ket{02d2} + \ket{0d22} - \ket{202d} - \ket{220d} + \ket{22d0} + \ket{2d20} - \ket{d022} - \ket{d202}\right) 
 \nonumber \\
& + &
\Cindex{179}{3} \left ( 
\ket{2dud} - \ket{2udd} + \ket{d2du} + \ket{dd2u} + \ket{ddu2} - \ket{dud2} - \ket{u2dd} - \ket{ud2d}\right) 
\nonumber \eeq
\beq
C_{179-1} &=&
\frac{1}{18}
\,\left ( -{\Aindex{23}}^2 + 3\,J^2 + 3\,J\,t + 81\,t^2 - 6\,J\,U - 33\,t\,U 
\right .  \nonumber \\
&& \hspace{1cm} 
 \left .
+ 27\,U^2 - 108\,J\,W - 294\,t\,W + 492\,U\,W + 2508\,W^2 
\right . \nonumber \\
&& \hspace{1cm} 
\left .  
- 3\,{\sqrt{\Aindex{6}}}\,J\,\cos (\Thetaindex{5}) 
- 12\,{\sqrt{\Aindex{6}}}\,t\,\cos (\Thetaindex{5}) 
+ 9\,{\sqrt{\Aindex{6}}}\,U\,\cos (\Thetaindex{5}) 
\right .  
\nonumber \\
&& \hspace{1cm} 
\left .
+ 102\,{\sqrt{\Aindex{6}}}\,W\,\cos (\Thetaindex{5})
 + 3\,{\sqrt{3}}\,{\sqrt{\Aindex{6}}}\,J\,\sin (\Thetaindex{5}) 
+ 12\,{\sqrt{3}}\,{\sqrt{\Aindex{6}}}\,t\,\sin (\Thetaindex{5}) 
 \right .  \nonumber \\
&& \hspace{1cm} 
\left .
- 9\,{\sqrt{3}}\,{\sqrt{\Aindex{6}}}\,U\,\sin (\Thetaindex{5}) 
- 102\,{\sqrt{3}}\,{\sqrt{\Aindex{6}}}\,W\,\sin (\Thetaindex{5})
\right )  \nonumber \\
C_{179-2} &=&
\frac{t}{6}
\, \left ( 2\,J + 12\,t + 2\,U - 4\,W + {\sqrt{\Aindex{6}}}\,\cos (\Thetaindex{5}) - {\sqrt{3}}\,{\sqrt{\Aindex{6}}}\,\sin (\Thetaindex{5})  \right ) \nonumber \\
C_{179-3} &=&
\frac{t}{6}
\, \left ( -J + 12\,t - U + 2\,W + {\sqrt{\Aindex{6}}}\,\cos (\Thetaindex{5}) - {\sqrt{3}}\,{\sqrt{\Aindex{6}}}\,\sin (\Thetaindex{5})  \right ) \nonumber \\
 N_{179} &=& 2\,{\sqrt{{\Cindex{179}{1}}^2 + 2\,\left( {\Cindex{179}{2}}^2 + {\Cindex{179}{3}}^2 \right) }} \nonumber \eeq 
\beq
\ket{\Psi}_{180}& = &\ket{5,- {1 \over 2} , {3 \over 4} ,\Gamma_{4,3}} \nonumber \\ 
& = &
\Cindex{180}{1} \left ( 
\ket{022d} + \ket{0d22} - \ket{20d2} + \ket{220d} - \ket{22d0} + \ket{2d02} - \ket{d022} - \ket{d220}\right) 
 \nonumber \\
& + &
\Cindex{180}{2} \left ( 
\ket{02d2} - \ket{202d} - \ket{2d20} + \ket{d202}\right) 
 \nonumber \\
& + &
\Cindex{180}{3} \left ( 
\ket{2ddu} - \ket{2udd} + \ket{d2ud} - \ket{dd2u} - \ket{ddu2} + \ket{du2d} - \ket{u2dd} + \ket{udd2}\right) 
\nonumber \eeq
\beq
C_{180-1} &=&
\frac{t}{3}
\, \left ( J + 6\,t + U - 2\,W - {\sqrt{\Aindex{6}}}\,\cos (\Thetaindex{5})  \right ) \nonumber \\
C_{180-2} &=&
\frac{1}{18}
\,\left ( -{\Aindex{22}}^2 + 3\,\Aindex{22}\,J + 12\,\Aindex{22}\,t - 18\,J\,t\right . \nonumber \\
&& \hspace{1cm} 
 + 
 \left . 45\,t^2 - 9\,\Aindex{22}\,U + 18\,J\,U + 54\,t\,U - 18\,U^2 - 102\,\Aindex{22}\,W + 144\,J\,W
\right .  \nonumber \\
&& \hspace{1cm} 
 + 
 \left . 612\,t\,W - 468\,U\,W - 2592\,W^2
\right )  \nonumber \\
C_{180-3} &=&
\frac{-t}{6}
\, \left ( -J + 12\,t - U + 2\,W - 2\,{\sqrt{\Aindex{6}}}\,\cos (\Thetaindex{5})  \right ) \nonumber \\
 N_{180} &=& 2\,{\sqrt{2\,{\Cindex{180}{1}}^2 + {\Cindex{180}{2}}^2 + 2\,{\Cindex{180}{3}}^2}} \nonumber \eeq 
\beq
\ket{\Psi}_{181}& = &\ket{5,- {1 \over 2} , {3 \over 4} ,\Gamma_{4,3}} \nonumber \\ 
& = &
\Cindex{181}{1} \left ( 
\ket{022d} + \ket{0d22} - \ket{20d2} + \ket{220d} - \ket{22d0} + \ket{2d02} - \ket{d022} - \ket{d220}\right) 
 \nonumber \\
& + &
\Cindex{181}{2} \left ( 
\ket{02d2} - \ket{202d} - \ket{2d20} + \ket{d202}\right) 
 \nonumber \\
& + &
\Cindex{181}{3} \left ( 
\ket{2ddu} - \ket{2udd} + \ket{d2ud} - \ket{dd2u} - \ket{ddu2} + \ket{du2d} - \ket{u2dd} + \ket{udd2}\right) 
\nonumber \eeq
\beq
C_{181-1} &=&
\frac{t}{6}
\, \left ( 2\,J + 12\,t + 2\,U - 4\,W + {\sqrt{\Aindex{6}}}\,\cos (\Thetaindex{5}) + {\sqrt{3}}\,{\sqrt{\Aindex{6}}}\,\sin (\Thetaindex{5})  \right ) \nonumber \\
C_{181-2} &=&
\frac{1}{18}
\,\left ( -{\Aindex{24}}^2 + 3\,J^2 + 3\,J\,t + 81\,t^2 - 6\,J\,U - 33\,t\,U 
\right .  \nonumber \\
&& \hspace{1cm} 
 \left . 
+ 27\,U^2 - 108\,J\,W - 294\,t\,W + 492\,U\,W  + 2508\,W^2 \right . \nonumber \\
&& \hspace{1cm} 
 \left .  - 3\,{\sqrt{\Aindex{6}}}\,J\,\cos (\Thetaindex{5}) 
- 12\,{\sqrt{\Aindex{6}}}\,t\,\cos (\Thetaindex{5}) 
+ 9\,{\sqrt{\Aindex{6}}}\,U\,\cos (\Thetaindex{5}) 
\right .  \nonumber \\
&& \hspace{1cm} 
 \left . 
+ 102\,{\sqrt{\Aindex{6}}}\,W\,\cos (\Thetaindex{5})
-3\,{\sqrt{3}}\,{\sqrt{\Aindex{6}}}\,J\,\sin (\Thetaindex{5}) 
- 12\,{\sqrt{3}}\,{\sqrt{\Aindex{6}}}\,t\,\sin (\Thetaindex{5}) 
\right .  \nonumber \\
&& \hspace{1cm} 
\left . 
+ 9\,{\sqrt{3}}\,{\sqrt{\Aindex{6}}}\,U\,\sin (\Thetaindex{5}) 
+ 102\,{\sqrt{3}}\,{\sqrt{\Aindex{6}}}\,W\,\sin (\Thetaindex{5})
\right )  \nonumber \\
C_{181-3} &=&
\frac{-t}{6}
\, \left ( -J + 12\,t - U + 2\,W + {\sqrt{\Aindex{6}}}\,\cos (\Thetaindex{5}) + {\sqrt{3}}\,{\sqrt{\Aindex{6}}}\,\sin (\Thetaindex{5})  \right ) \nonumber \\
 N_{181} &=& 2\,{\sqrt{2\,{\Cindex{181}{1}}^2 + {\Cindex{181}{2}}^2 + 2\,{\Cindex{181}{3}}^2}} \nonumber \eeq 
\beq
\ket{\Psi}_{182}& = &\ket{5,- {1 \over 2} , {3 \over 4} ,\Gamma_{4,3}} \nonumber \\ 
& = &
\Cindex{182}{1} \left ( 
\ket{022d} + \ket{0d22} - \ket{20d2} + \ket{220d} - \ket{22d0} + \ket{2d02} - \ket{d022} - \ket{d220}\right) 
 \nonumber \\
& + &
\Cindex{182}{2} \left ( 
\ket{02d2} - \ket{202d} - \ket{2d20} + \ket{d202}\right) 
 \nonumber \\
& + &
\Cindex{182}{3} \left ( 
\ket{2ddu} - \ket{2udd} + \ket{d2ud} - \ket{dd2u} - \ket{ddu2} + \ket{du2d} - \ket{u2dd} + \ket{udd2}\right) 
\nonumber \eeq
\beq
C_{182-1} &=&
\frac{t}{6}
\, \left ( 2\,J + 12\,t + 2\,U - 4\,W + {\sqrt{\Aindex{6}}}\,\cos (\Thetaindex{5}) - {\sqrt{3}}\,{\sqrt{\Aindex{6}}}\,\sin (\Thetaindex{5})  \right ) \nonumber \\
C_{182-2} &=&
\frac{1}{18}
\,\left ( -{\Aindex{23}}^2 + 3\,J^2 + 3\,J\,t + 81\,t^2 - 6\,J\,U - 33\,t\,U 
\right . 
\nonumber \\
&& \hspace{1cm} 
\left . 
+ 27\,U^2 - 108\,J\,W - 294\,t\,W + 492\,U\,W  + 2508\,W^2 
\right . 
\nonumber \\
&& \hspace{1cm} 
\left . 
- 3\,{\sqrt{\Aindex{6}}}\,J\,\cos (\Thetaindex{5}) 
- 12\,{\sqrt{\Aindex{6}}}\,t\,\cos (\Thetaindex{5}) 
+ 9\,{\sqrt{\Aindex{6}}}\,U\,\cos (\Thetaindex{5}) 
\right .  \nonumber \\
&& \hspace{1cm} 
 \left . 
+ 102\,{\sqrt{\Aindex{6}}}\,W\,\cos (\Thetaindex{5})
+ 3\,{\sqrt{3}}\,{\sqrt{\Aindex{6}}}\,J\,\sin (\Thetaindex{5}) 
+ 12\,{\sqrt{3}}\,{\sqrt{\Aindex{6}}}\,t\,\sin (\Thetaindex{5}) 
\right .  \nonumber \\
&& \hspace{1cm} 
 \left . 
- 9\,{\sqrt{3}}\,{\sqrt{\Aindex{6}}}\,U\,\sin (\Thetaindex{5}) 
- 102\,{\sqrt{3}}\,{\sqrt{\Aindex{6}}}\,W\,\sin (\Thetaindex{5})
\right )  \nonumber \\
C_{182-3} &=&
\frac{-t}{6}
\, \left ( -J + 12\,t - U + 2\,W + {\sqrt{\Aindex{6}}}\,\cos (\Thetaindex{5}) - {\sqrt{3}}\,{\sqrt{\Aindex{6}}}\,\sin (\Thetaindex{5})  \right ) \nonumber \\
 N_{182} &=& 2\,{\sqrt{2\,{\Cindex{182}{1}}^2 + {\Cindex{182}{2}}^2 + 2\,{\Cindex{182}{3}}^2}} \nonumber \eeq 
\beq
\ket{\Psi}_{183}& = &\ket{5,- {1 \over 2} , {3 \over 4} ,\Gamma_{5,1}} \nonumber \\ 
& = &
\Cindex{183}{1} \left ( 
\ket{022d} - \ket{0d22} - \ket{20d2} - \ket{220d} + \ket{22d0} + \ket{2d02} + \ket{d022} - \ket{d220}\right) 
 \nonumber \\
& + &
\Cindex{183}{2} \left ( 
\ket{2ddu} + \ket{2udd} + \ket{d2ud} + \ket{dd2u} + \ket{ddu2} + \ket{du2d} + \ket{u2dd} + \ket{udd2}\right) 
 \nonumber \\
& + &
\Cindex{183}{3} \left ( 
\ket{2dud} + \ket{d2du} + \ket{dud2} + \ket{ud2d}\right) 
\nonumber \eeq
\beq
C_{183-1} &=&
\frac{{\sqrt{\frac{3}{2}}}\,t}{2} \nonumber \\
C_{183-2} &=&
\frac{1}{4\,{\sqrt{6}}}
\, \left ( {\sqrt{\Aindex{4}}} + J - 2\,t + U - 2\,W  \right ) \nonumber \\
C_{183-3} &=&
\frac{-1}{2\,{\sqrt{6}}}
\, \left ( {\sqrt{\Aindex{4}}} + J - 2\,t + U - 2\,W  \right ) \nonumber \\
 N_{183} &=& 2\,{\sqrt{2\,{\Cindex{183}{1}}^2 + 2\,{\Cindex{183}{2}}^2 + {\Cindex{183}{3}}^2}} \nonumber \eeq 
\beq
\ket{\Psi}_{184}& = &\ket{5,- {1 \over 2} , {3 \over 4} ,\Gamma_{5,1}} \nonumber \\ 
& = &
\Cindex{184}{1} \left ( 
\ket{022d} - \ket{0d22} - \ket{20d2} - \ket{220d} + \ket{22d0} + \ket{2d02} + \ket{d022} - \ket{d220}\right) 
 \nonumber \\
& + &
\Cindex{184}{2} \left ( 
\ket{2ddu} + \ket{2udd} + \ket{d2ud} + \ket{dd2u} + \ket{ddu2} + \ket{du2d} + \ket{u2dd} + \ket{udd2}\right) 
 \nonumber \\
& + &
\Cindex{184}{3} \left ( 
\ket{2dud} + \ket{d2du} + \ket{dud2} + \ket{ud2d}\right) 
\nonumber \eeq
\beq
C_{184-1} &=&
\frac{{\sqrt{\frac{3}{2}}}\,t}{2} \nonumber \\
C_{184-2} &=&
\frac{-1}{4\,{\sqrt{6}}}
\, \left ( {\sqrt{\Aindex{4}}} - J + 2\,t - U + 2\,W  \right ) \nonumber \\
C_{184-3} &=&
\frac{1}{2\,{\sqrt{6}}}
\, \left ( {\sqrt{\Aindex{4}}} - J + 2\,t - U + 2\,W  \right ) \nonumber \\
 N_{184} &=& 2\,{\sqrt{2\,{\Cindex{184}{1}}^2 + 2\,{\Cindex{184}{2}}^2 + {\Cindex{184}{3}}^2}} \nonumber \eeq 
\beq
\ket{\Psi}_{185}& = &\ket{5,- {1 \over 2} , {15 \over 4} ,\Gamma_{5,1}} \nonumber \\ 
&=& \frac{1}{2\,{\sqrt{3}}}
 \left ( \ket{2ddu} + \ket{2dud} + \ket{2udd} + \ket{d2du} + \ket{d2ud} + \ket{dd2u}\right . \nonumber \\
&& \hspace{1em} 
 + 
\left . \ket{ddu2} + \ket{du2d} + \ket{dud2} + \ket{u2dd} + \ket{ud2d} + \ket{udd2}\right ) \nonumber  
\eeq
\beq
\ket{\Psi}_{186}& = &\ket{5,- {1 \over 2} , {3 \over 4} ,\Gamma_{5,2}} \nonumber \\ 
& = &
\Cindex{186}{1} \left ( 
\ket{022d} - \ket{02d2} - \ket{202d} + \ket{20d2} - \ket{2d02} + \ket{2d20} + \ket{d202} - \ket{d220}\right) 
 \nonumber \\
& + &
\Cindex{186}{2} \left ( 
\ket{2ddu} + \ket{2dud} - \ket{d2du} - \ket{d2ud} - \ket{du2d} + \ket{dud2} - \ket{ud2d} + \ket{udd2}\right) 
 \nonumber \\
& + &
\Cindex{186}{3} \left ( 
\ket{2udd} - \ket{dd2u} + \ket{ddu2} - \ket{u2dd}\right) 
\nonumber \eeq
\beq
C_{186-1} &=&
\frac{{\sqrt{\frac{3}{2}}}\,t}{2} \nonumber \\
C_{186-2} &=&
\frac{-1}{4\,{\sqrt{6}}}
\, \left ( {\sqrt{\Aindex{4}}} + J - 2\,t + U - 2\,W  \right ) \nonumber \\
C_{186-3} &=&
\frac{1}{2\,{\sqrt{6}}}
\, \left ( {\sqrt{\Aindex{4}}} + J - 2\,t + U - 2\,W  \right ) \nonumber \\
 N_{186} &=& 2\,{\sqrt{2\,{\Cindex{186}{1}}^2 + 2\,{\Cindex{186}{2}}^2 + {\Cindex{186}{3}}^2}} \nonumber \eeq 
\beq
\ket{\Psi}_{187}& = &\ket{5,- {1 \over 2} , {3 \over 4} ,\Gamma_{5,2}} \nonumber \\ 
& = &
\Cindex{187}{1} \left ( 
\ket{022d} - \ket{02d2} - \ket{202d} + \ket{20d2} - \ket{2d02} + \ket{2d20} + \ket{d202} - \ket{d220}\right) 
 \nonumber \\
& + &
\Cindex{187}{2} \left ( 
\ket{2ddu} + \ket{2dud} - \ket{d2du} - \ket{d2ud} - \ket{du2d} + \ket{dud2} - \ket{ud2d} + \ket{udd2}\right) 
 \nonumber \\
& + &
\Cindex{187}{3} \left ( 
\ket{2udd} - \ket{dd2u} + \ket{ddu2} - \ket{u2dd}\right) 
\nonumber \eeq
\beq
C_{187-1} &=&
\frac{{\sqrt{\frac{3}{2}}}\,t}{2} \nonumber \\
C_{187-2} &=&
\frac{1}{4\,{\sqrt{6}}}
\, \left ( {\sqrt{\Aindex{4}}} - J + 2\,t - U + 2\,W  \right ) \nonumber \\
C_{187-3} &=&
\frac{-1}{2\,{\sqrt{6}}}
\, \left ( {\sqrt{\Aindex{4}}} - J + 2\,t - U + 2\,W  \right ) \nonumber \\
 N_{187} &=& 2\,{\sqrt{2\,{\Cindex{187}{1}}^2 + 2\,{\Cindex{187}{2}}^2 + {\Cindex{187}{3}}^2}} \nonumber \eeq 
\beq
\ket{\Psi}_{188}& = &\ket{5,- {1 \over 2} , {15 \over 4} ,\Gamma_{5,2}} \nonumber \\ 
&=& \frac{1}{2\,{\sqrt{3}}}
 \left ( \ket{2ddu} + \ket{2dud} + \ket{2udd} - \ket{d2du} - \ket{d2ud} - \ket{dd2u}\right . \nonumber \\
&& \hspace{1em} 
 + 
\left . \ket{ddu2} - \ket{du2d} + \ket{dud2} - \ket{u2dd} - \ket{ud2d} + \ket{udd2}\right ) \nonumber  
\eeq
\beq
\ket{\Psi}_{189}& = &\ket{5,- {1 \over 2} , {3 \over 4} ,\Gamma_{5,3}} \nonumber \\ 
& = &
\Cindex{189}{1} \left ( 
\ket{02d2} - \ket{0d22} - \ket{202d} + \ket{220d} - \ket{22d0} + \ket{2d20} + \ket{d022} - \ket{d202}\right) 
 \nonumber \\
& + &
\Cindex{189}{2} \left ( 
\ket{2ddu} + \ket{d2ud} - \ket{du2d} - \ket{udd2}\right) 
 \nonumber \\
& + &
\Cindex{189}{3} \left ( 
\ket{2dud} + \ket{2udd} + \ket{d2du} - \ket{dd2u} - \ket{ddu2} - \ket{dud2} + \ket{u2dd} - \ket{ud2d}\right) 
\nonumber \eeq
\beq
C_{189-1} &=&
\frac{-\left( {\sqrt{\frac{3}{2}}}\,t \right) }{2} \nonumber \\
C_{189-2} &=&
\frac{-1}{2\,{\sqrt{6}}}
\, \left ( {\sqrt{\Aindex{4}}} + J - 2\,t + U - 2\,W  \right ) \nonumber \\
C_{189-3} &=&
\frac{1}{4\,{\sqrt{6}}}
\, \left ( {\sqrt{\Aindex{4}}} + J - 2\,t + U - 2\,W  \right ) \nonumber \\
 N_{189} &=& 2\,{\sqrt{2\,{\Cindex{189}{1}}^2 + {\Cindex{189}{2}}^2 + 2\,{\Cindex{189}{3}}^2}} \nonumber \eeq 
\beq
\ket{\Psi}_{190}& = &\ket{5,- {1 \over 2} , {3 \over 4} ,\Gamma_{5,3}} \nonumber \\ 
& = &
\Cindex{190}{1} \left ( 
\ket{02d2} - \ket{0d22} - \ket{202d} + \ket{220d} - \ket{22d0} + \ket{2d20} + \ket{d022} - \ket{d202}\right) 
 \nonumber \\
& + &
\Cindex{190}{2} \left ( 
\ket{2ddu} + \ket{d2ud} - \ket{du2d} - \ket{udd2}\right) 
 \nonumber \\
& + &
\Cindex{190}{3} \left ( 
\ket{2dud} + \ket{2udd} + \ket{d2du} - \ket{dd2u} - \ket{ddu2} - \ket{dud2} + \ket{u2dd} - \ket{ud2d}\right) 
\nonumber \eeq
\beq
C_{190-1} &=&
\frac{-\left( {\sqrt{\frac{3}{2}}}\,t \right) }{2} \nonumber \\
C_{190-2} &=&
\frac{1}{2\,{\sqrt{6}}}
\, \left ( {\sqrt{\Aindex{4}}} - J + 2\,t - U + 2\,W  \right ) \nonumber \\
C_{190-3} &=&
\frac{-1}{4\,{\sqrt{6}}}
\, \left ( {\sqrt{\Aindex{4}}} - J + 2\,t - U + 2\,W  \right ) \nonumber \\
 N_{190} &=& 2\,{\sqrt{2\,{\Cindex{190}{1}}^2 + {\Cindex{190}{2}}^2 + 2\,{\Cindex{190}{3}}^2}} \nonumber \eeq 
\beq
\ket{\Psi}_{191}& = &\ket{5,- {1 \over 2} , {15 \over 4} ,\Gamma_{5,3}} \nonumber \\ 
&=& \frac{1}{2\,{\sqrt{3}}}
 \left ( \ket{2ddu} + \ket{2dud} + \ket{2udd} + \ket{d2du} + \ket{d2ud} - \ket{dd2u}\right . \nonumber \\
&& \hspace{1em} 
\left . -\ket{ddu2} - \ket{du2d} - \ket{dud2} + \ket{u2dd} - \ket{ud2d} - \ket{udd2}\right ) \nonumber  
\eeq
{\subsection*{\boldmath Unnormalized eigenvectors for ${\rm  N_e}=5$ and   ${\rm m_s}$= $ {1 \over 2} $.}
\beq
\ket{\Psi}_{192}& = &\ket{5, {1 \over 2} , {3 \over 4} ,\Gamma_1} \nonumber \\ 
&=& \frac{1}{2\,{\sqrt{3}}}
 \left ( \ket{022u} + \ket{02u2} + \ket{0u22} + \ket{202u} + \ket{20u2} + \ket{220u}\right . \nonumber \\
&& \hspace{1em} 
 + 
\left . \ket{22u0} + \ket{2u02} + \ket{2u20} + \ket{u022} + \ket{u202} + \ket{u220}\right ) \nonumber  
\eeq
\beq
\ket{\Psi}_{193}& = &\ket{5, {1 \over 2} , {15 \over 4} ,\Gamma_2} \nonumber \\ 
&=& \frac{1}{2\,{\sqrt{3}}}
 \left ( \ket{2duu} + \ket{2udu} + \ket{2uud} - \ket{d2uu} + \ket{du2u} - \ket{duu2}\right . \nonumber \\
&& \hspace{1em} 
\left . -\ket{u2du} - \ket{u2ud} + \ket{ud2u} - \ket{udu2} + \ket{uu2d} - \ket{uud2}\right ) \nonumber  
\eeq
\beq
\ket{\Psi}_{194}& = &\ket{5, {1 \over 2} , {3 \over 4} ,\Gamma_{3,1}} \nonumber \\ 
& = &
\Cindex{194}{1} \left ( 
\ket{022u} + \ket{0u22} + \ket{20u2} + \ket{220u} + \ket{22u0} + \ket{2u02} + \ket{u022} + \ket{u220}\right) 
 \nonumber \\
& + &
\Cindex{194}{2} \left ( 
\ket{02u2} + \ket{202u} + \ket{2u20} + \ket{u202}\right) 
 \nonumber \\
& + &
\Cindex{194}{3} \left ( 
\ket{2duu} - \ket{2uud} - \ket{d2uu} + \ket{duu2} + \ket{u2du} - \ket{ud2u} + \ket{uu2d} - \ket{uud2}\right) 
\nonumber \eeq
\beq
C_{194-1} &=&
\frac{-t}{2\,{\sqrt{2}}} \nonumber \\
C_{194-2} &=&
\frac{t}{{\sqrt{2}}} \nonumber \\
C_{194-3} &=&
\frac{1}{4\,{\sqrt{2}}}
\, \left ( {\sqrt{\Aindex{7}}} + J + 2\,t + U - 2\,W  \right ) \nonumber \\
 N_{194} &=& 2\,{\sqrt{2\,{\Cindex{194}{1}}^2 + {\Cindex{194}{2}}^2 + 2\,{\Cindex{194}{3}}^2}} \nonumber \eeq 
\beq
\ket{\Psi}_{195}& = &\ket{5, {1 \over 2} , {3 \over 4} ,\Gamma_{3,1}} \nonumber \\ 
& = &
\Cindex{195}{1} \left ( 
\ket{022u} + \ket{0u22} + \ket{20u2} + \ket{220u} + \ket{22u0} + \ket{2u02} + \ket{u022} + \ket{u220}\right) 
 \nonumber \\
& + &
\Cindex{195}{2} \left ( 
\ket{02u2} + \ket{202u} + \ket{2u20} + \ket{u202}\right) 
 \nonumber \\
& + &
\Cindex{195}{3} \left ( 
\ket{2duu} - \ket{2uud} - \ket{d2uu} + \ket{duu2} + \ket{u2du} - \ket{ud2u} + \ket{uu2d} - \ket{uud2}\right) 
\nonumber \eeq
\beq
C_{195-1} &=&
\frac{-t}{2\,{\sqrt{2}}} \nonumber \\
C_{195-2} &=&
\frac{t}{{\sqrt{2}}} \nonumber \\
C_{195-3} &=&
\frac{-1}{4\,{\sqrt{2}}}
\, \left ( {\sqrt{\Aindex{7}}} - J - 2\,t - U + 2\,W  \right ) \nonumber \\
 N_{195} &=& 2\,{\sqrt{2\,{\Cindex{195}{1}}^2 + {\Cindex{195}{2}}^2 + 2\,{\Cindex{195}{3}}^2}} \nonumber \eeq 
\beq
\ket{\Psi}_{196}& = &\ket{5, {1 \over 2} , {3 \over 4} ,\Gamma_{3,2}} \nonumber \\ 
& = &
\Cindex{196}{1} \left ( 
\ket{022u} - \ket{0u22} + \ket{20u2} - \ket{220u} - \ket{22u0} + \ket{2u02} - \ket{u022} + \ket{u220}\right) 
 \nonumber \\
& + &
\Cindex{196}{2} \left ( 
\ket{2duu} + \ket{2uud} - \ket{d2uu} - \ket{duu2} - \ket{u2du} + \ket{ud2u} + \ket{uu2d} - \ket{uud2}\right) 
 \nonumber \\
& + &
\Cindex{196}{3} \left ( 
\ket{2udu} + \ket{du2u} - \ket{u2ud} - \ket{udu2}\right) 
\nonumber \eeq
\beq
C_{196-1} &=&
\frac{{\sqrt{\frac{3}{2}}}\,t}{2} \nonumber \\
C_{196-2} &=&
\frac{-1}{4\,{\sqrt{6}}}
\, \left ( {\sqrt{\Aindex{7}}} + J + 2\,t + U - 2\,W  \right ) \nonumber \\
C_{196-3} &=&
\frac{1}{2\,{\sqrt{6}}}
\, \left ( {\sqrt{\Aindex{7}}} + J + 2\,t + U - 2\,W  \right ) \nonumber \\
 N_{196} &=& 2\,{\sqrt{2\,{\Cindex{196}{1}}^2 + 2\,{\Cindex{196}{2}}^2 + {\Cindex{196}{3}}^2}} \nonumber \eeq 
\beq
\ket{\Psi}_{197}& = &\ket{5, {1 \over 2} , {3 \over 4} ,\Gamma_{3,2}} \nonumber \\ 
& = &
\Cindex{197}{1} \left ( 
\ket{022u} - \ket{0u22} + \ket{20u2} - \ket{220u} - \ket{22u0} + \ket{2u02} - \ket{u022} + \ket{u220}\right) 
 \nonumber \\
& + &
\Cindex{197}{2} \left ( 
\ket{2duu} + \ket{2uud} - \ket{d2uu} - \ket{duu2} - \ket{u2du} + \ket{ud2u} + \ket{uu2d} - \ket{uud2}\right) 
 \nonumber \\
& + &
\Cindex{197}{3} \left ( 
\ket{2udu} + \ket{du2u} - \ket{u2ud} - \ket{udu2}\right) 
\nonumber \eeq
\beq
C_{197-1} &=&
\frac{{\sqrt{\frac{3}{2}}}\,t}{2} \nonumber \\
C_{197-2} &=&
\frac{1}{4\,{\sqrt{6}}}
\, \left ( {\sqrt{\Aindex{7}}} - J - 2\,t - U + 2\,W  \right ) \nonumber \\
C_{197-3} &=&
\frac{-1}{2\,{\sqrt{6}}}
\, \left ( {\sqrt{\Aindex{7}}} - J - 2\,t - U + 2\,W  \right ) \nonumber \\
 N_{197} &=& 2\,{\sqrt{2\,{\Cindex{197}{1}}^2 + 2\,{\Cindex{197}{2}}^2 + {\Cindex{197}{3}}^2}} \nonumber \eeq 
\beq
\ket{\Psi}_{198}& = &\ket{5, {1 \over 2} , {3 \over 4} ,\Gamma_{4,1}} \nonumber \\ 
& = &
\Cindex{198}{1} \left ( 
\ket{022u} + \ket{02u2} + \ket{202u} + \ket{20u2} - \ket{2u02} - \ket{2u20} - \ket{u202} - \ket{u220}\right) 
 \nonumber \\
& + &
\Cindex{198}{2} \left ( 
\ket{0u22} - \ket{220u} - \ket{22u0} + \ket{u022}\right) 
 \nonumber \\
& + &
\Cindex{198}{3} \left ( 
\ket{2udu} - \ket{2uud} - \ket{du2u} - \ket{duu2} + \ket{u2du} - \ket{u2ud} + \ket{ud2u} + \ket{udu2}\right) 
\nonumber \eeq
\beq
C_{198-1} &=&
\frac{t}{2\,{\sqrt{2}}}
\, \left ( J + 4\,t + U - 2\,W + 2\,{\sqrt{\Aindex{6}}}\,\cos (\Thetaindex{5})  \right ) \nonumber \\
C_{198-2} &=&
\frac{-t}{3\,{\sqrt{2}}}
\, \left ( J - 12\,t + U - 2\,W + 2\,{\sqrt{\Aindex{6}}}\,\cos (\Thetaindex{5})  \right ) \nonumber \\
C_{198-3} &=&
\frac{-1}{18\,{\sqrt{2}}}
\,\left ( {\Aindex{22}}^2 - 3\,\Aindex{22}\,t - 36\,t^2 + 12\,\Aindex{22}\,U - 18\,t\,U\right . \nonumber \\
&& \hspace{1cm} 
 + 
 \left . 36\,U^2 + 96\,\Aindex{22}\,W - 144\,t\,W + 576\,U\,W + 2304\,W^2
\right  )  \nonumber \\
 N_{198} &=& 2\,{\sqrt{2\,{\Cindex{198}{1}}^2 + {\Cindex{198}{2}}^2 + 2\,{\Cindex{198}{3}}^2}} \nonumber \eeq 
\beq
\ket{\Psi}_{199}& = &\ket{5, {1 \over 2} , {3 \over 4} ,\Gamma_{4,1}} \nonumber \\ 
& = &
\Cindex{199}{1} \left ( 
\ket{022u} + \ket{02u2} + \ket{202u} + \ket{20u2} - \ket{2u02} - \ket{2u20} - \ket{u202} - \ket{u220}\right) 
 \nonumber \\
& + &
\Cindex{199}{2} \left ( 
\ket{0u22} - \ket{220u} - \ket{22u0} + \ket{u022}\right) 
 \nonumber \\
& + &
\Cindex{199}{3} \left ( 
\ket{2udu} - \ket{2uud} - \ket{du2u} - \ket{duu2} + \ket{u2du} - \ket{u2ud} + \ket{ud2u} + \ket{udu2}\right) 
\nonumber \eeq
\beq
C_{199-1} &=&
\frac{t}{2\,{\sqrt{2}}}
\, \left ( J + 4\,t + U - 2\,W - {\sqrt{\Aindex{6}}}\,\cos (\Thetaindex{5}) - {\sqrt{3}}\,{\sqrt{\Aindex{6}}}\,\sin (\Thetaindex{5})  \right ) \nonumber \\
C_{199-2} &=&
\frac{-t}{3\,{\sqrt{2}}}
\, \left ( J - 12\,t + U - 2\,W - {\sqrt{\Aindex{6}}}\,\cos (\Thetaindex{5}) - {\sqrt{3}}\,{\sqrt{\Aindex{6}}}\,\sin (\Thetaindex{5})  \right ) \nonumber \\
C_{199-3} &=&
\frac{-1}{18\,{\sqrt{2}}}
\,\left ( {\Aindex{24}}^2 - 3\,J\,t - 45\,t^2 + 12\,J\,U + 33\,t\,U - 24\,U^2 + 96\,J\,W + 294\,t\,W - 504\,U\,W\right . \nonumber \\
&& \hspace{1cm} 
 \left . -2496\,W^2 + 3\,{\sqrt{\Aindex{6}}}\,t\,\cos (\Thetaindex{5}) - 12\,{\sqrt{\Aindex{6}}}\,U\,\cos (\Thetaindex{5}) - 96\,{\sqrt{\Aindex{6}}}\,W\,\cos (\Thetaindex{5})
\right .  \nonumber \\
&& \hspace{1cm} 
 + 
 \left . 3\,{\sqrt{3}}\,{\sqrt{\Aindex{6}}}\,t\,\sin (\Thetaindex{5}) - 12\,{\sqrt{3}}\,{\sqrt{\Aindex{6}}}\,U\,\sin (\Thetaindex{5}) - 96\,{\sqrt{3}}\,{\sqrt{\Aindex{6}}}\,W\,\sin (\Thetaindex{5})
\right )  \nonumber \\
 N_{199} &=& 2\,{\sqrt{2\,{\Cindex{199}{1}}^2 + {\Cindex{199}{2}}^2 + 2\,{\Cindex{199}{3}}^2}} \nonumber \eeq 
\beq
\ket{\Psi}_{200}& = &\ket{5, {1 \over 2} , {3 \over 4} ,\Gamma_{4,1}} \nonumber \\ 
& = &
\Cindex{200}{1} \left ( 
\ket{022u} + \ket{02u2} + \ket{202u} + \ket{20u2} - \ket{2u02} - \ket{2u20} - \ket{u202} - \ket{u220}\right) 
 \nonumber \\
& + &
\Cindex{200}{2} \left ( 
\ket{0u22} - \ket{220u} - \ket{22u0} + \ket{u022}\right) 
 \nonumber \\
& + &
\Cindex{200}{3} \left ( 
\ket{2udu} - \ket{2uud} - \ket{du2u} - \ket{duu2} + \ket{u2du} - \ket{u2ud} + \ket{ud2u} + \ket{udu2}\right) 
\nonumber \eeq
\beq
C_{200-1} &=&
\frac{t}{2\,{\sqrt{2}}}
\, \left ( J + 4\,t + U - 2\,W - {\sqrt{\Aindex{6}}}\,\cos (\Thetaindex{5}) + {\sqrt{3}}\,{\sqrt{\Aindex{6}}}\,\sin (\Thetaindex{5})  \right ) \nonumber \\
C_{200-2} &=&
\frac{-t}{3\,{\sqrt{2}}}
\, \left ( J - 12\,t + U - 2\,W - {\sqrt{\Aindex{6}}}\,\cos (\Thetaindex{5}) + {\sqrt{3}}\,{\sqrt{\Aindex{6}}}\,\sin (\Thetaindex{5})  \right ) \nonumber \\
C_{200-3} &=&
\frac{-1}{18\,{\sqrt{2}}}
\,\left ( {\Aindex{23}}^2 - 3\,J\,t - 45\,t^2 + 12\,J\,U + 33\,t\,U - 24\,U^2 + 96\,J\,W + 294\,t\,W - 504\,U\,W\right . \nonumber \\
&& \hspace{1cm} 
 \left . -2496\,W^2 + 3\,{\sqrt{\Aindex{6}}}\,t\,\cos (\Thetaindex{5}) - 12\,{\sqrt{\Aindex{6}}}\,U\,\cos (\Thetaindex{5}) - 96\,{\sqrt{\Aindex{6}}}\,W\,\cos (\Thetaindex{5})
\right .  \nonumber \\
&& \hspace{1cm} 
 \left . -3\,{\sqrt{3}}\,{\sqrt{\Aindex{6}}}\,t\,\sin (\Thetaindex{5}) + 12\,{\sqrt{3}}\,{\sqrt{\Aindex{6}}}\,U\,\sin (\Thetaindex{5}) + 96\,{\sqrt{3}}\,{\sqrt{\Aindex{6}}}\,W\,\sin (\Thetaindex{5})
\right )  \nonumber \\
 N_{200} &=& 2\,{\sqrt{2\,{\Cindex{200}{1}}^2 + {\Cindex{200}{2}}^2 + 2\,{\Cindex{200}{3}}^2}} \nonumber \eeq 
\beq
\ket{\Psi}_{201}& = &\ket{5, {1 \over 2} , {3 \over 4} ,\Gamma_{4,2}} \nonumber \\ 
& = &
\Cindex{201}{1} \left ( 
\ket{022u} - \ket{20u2} - \ket{2u02} + \ket{u220}\right) 
 \nonumber \\
& + &
\Cindex{201}{2} \left ( 
\ket{02u2} + \ket{0u22} - \ket{202u} - \ket{220u} + \ket{22u0} + \ket{2u20} - \ket{u022} - \ket{u202}\right) 
 \nonumber \\
& + &
\Cindex{201}{3} \left ( 
\ket{2duu} - \ket{2udu} + \ket{d2uu} + \ket{du2u} - \ket{u2ud} + \ket{udu2} - \ket{uu2d} - \ket{uud2}\right) 
\nonumber \eeq
\beq
C_{201-1} &=&
\frac{1}{18}
\,\left ( -{\Aindex{22}}^2 + 3\,\Aindex{22}\,J + 12\,\Aindex{22}\,t - 18\,J\,t\right . \nonumber \\
&& \hspace{1cm} 
 + 
 \left . 45\,t^2 - 9\,\Aindex{22}\,U + 18\,J\,U + 54\,t\,U - 18\,U^2 - 102\,\Aindex{22}\,W + 144\,J\,W
\right .  \nonumber \\
&& \hspace{1cm} 
 + 
 \left . 612\,t\,W - 468\,U\,W - 2592\,W^2
\right )  \nonumber \\
C_{201-2} &=&
\frac{t}{3}
\, \left ( J + 6\,t + U - 2\,W - {\sqrt{\Aindex{6}}}\,\cos (\Thetaindex{5})  \right ) \nonumber \\
C_{201-3} &=&
\frac{t}{6}
\, \left ( -J + 12\,t - U + 2\,W - 2\,{\sqrt{\Aindex{6}}}\,\cos (\Thetaindex{5})  \right ) \nonumber \\
 N_{201} &=& 2\,{\sqrt{{\Cindex{201}{1}}^2 + 2\,\left( {\Cindex{201}{2}}^2 + {\Cindex{201}{3}}^2 \right) }} \nonumber \eeq 
\beq
\ket{\Psi}_{202}& = &\ket{5, {1 \over 2} , {3 \over 4} ,\Gamma_{4,2}} \nonumber \\ 
& = &
\Cindex{202}{1} \left ( 
\ket{022u} - \ket{20u2} - \ket{2u02} + \ket{u220}\right) 
 \nonumber \\
& + &
\Cindex{202}{2} \left ( 
\ket{02u2} + \ket{0u22} - \ket{202u} - \ket{220u} + \ket{22u0} + \ket{2u20} - \ket{u022} - \ket{u202}\right) 
 \nonumber \\
& + &
\Cindex{202}{3} \left ( 
\ket{2duu} - \ket{2udu} + \ket{d2uu} + \ket{du2u} - \ket{u2ud} + \ket{udu2} - \ket{uu2d} - \ket{uud2}\right) 
\nonumber \eeq
\beq
C_{202-1} &=&
\frac{1}{18}
\,\left ( -{\Aindex{24}}^2 + 3\,J^2 + 3\,J\,t + 81\,t^2 - 6\,J\,U - 33\,t\,U + 27\,U^2 - 108\,J\,W - 294\,t\,W + 492\,U\,W\right . \nonumber \\
&& \hspace{1cm} 
 + 
 \left . 2508\,W^2 - 3\,{\sqrt{\Aindex{6}}}\,J\,\cos (\Thetaindex{5}) - 12\,{\sqrt{\Aindex{6}}}\,t\,\cos (\Thetaindex{5}) + 9\,{\sqrt{\Aindex{6}}}\,U\,\cos (\Thetaindex{5}) + 102\,{\sqrt{\Aindex{6}}}\,W\,\cos (\Thetaindex{5})
\right .  \nonumber \\
&& \hspace{1cm} 
 \left . -3\,{\sqrt{3}}\,{\sqrt{\Aindex{6}}}\,J\,\sin (\Thetaindex{5}) - 12\,{\sqrt{3}}\,{\sqrt{\Aindex{6}}}\,t\,\sin (\Thetaindex{5}) + 9\,{\sqrt{3}}\,{\sqrt{\Aindex{6}}}\,U\,\sin (\Thetaindex{5}) + 102\,{\sqrt{3}}\,{\sqrt{\Aindex{6}}}\,W\,\sin (\Thetaindex{5})
\right )  \nonumber \\
C_{202-2} &=&
\frac{t}{6}
\, \left ( 2\,J + 12\,t + 2\,U - 4\,W + {\sqrt{\Aindex{6}}}\,\cos (\Thetaindex{5}) + {\sqrt{3}}\,{\sqrt{\Aindex{6}}}\,\sin (\Thetaindex{5})  \right ) \nonumber \\
C_{202-3} &=&
\frac{t}{6}
\, \left ( -J + 12\,t - U + 2\,W + {\sqrt{\Aindex{6}}}\,\cos (\Thetaindex{5}) + {\sqrt{3}}\,{\sqrt{\Aindex{6}}}\,\sin (\Thetaindex{5})  \right ) \nonumber \\
 N_{202} &=& 2\,{\sqrt{{\Cindex{202}{1}}^2 + 2\,\left( {\Cindex{202}{2}}^2 + {\Cindex{202}{3}}^2 \right) }} \nonumber \eeq 
\beq
\ket{\Psi}_{203}& = &\ket{5, {1 \over 2} , {3 \over 4} ,\Gamma_{4,2}} \nonumber \\ 
& = &
\Cindex{203}{1} \left ( 
\ket{022u} - \ket{20u2} - \ket{2u02} + \ket{u220}\right) 
 \nonumber \\
& + &
\Cindex{203}{2} \left ( 
\ket{02u2} + \ket{0u22} - \ket{202u} - \ket{220u} + \ket{22u0} + \ket{2u20} - \ket{u022} - \ket{u202}\right) 
 \nonumber \\
& + &
\Cindex{203}{3} \left ( 
\ket{2duu} - \ket{2udu} + \ket{d2uu} + \ket{du2u} - \ket{u2ud} + \ket{udu2} - \ket{uu2d} - \ket{uud2}\right) 
\nonumber \eeq
\beq
C_{203-1} &=&
\frac{1}{18}
\,\left ( -{\Aindex{23}}^2 + 3\,J^2 + 3\,J\,t + 81\,t^2 - 6\,J\,U - 33\,t\,U + 27\,U^2 - 108\,J\,W - 294\,t\,W + 492\,U\,W\right . \nonumber \\
&& \hspace{1cm} 
 + 
 \left . 2508\,W^2 - 3\,{\sqrt{\Aindex{6}}}\,J\,\cos (\Thetaindex{5}) - 12\,{\sqrt{\Aindex{6}}}\,t\,\cos (\Thetaindex{5}) + 9\,{\sqrt{\Aindex{6}}}\,U\,\cos (\Thetaindex{5}) + 102\,{\sqrt{\Aindex{6}}}\,W\,\cos (\Thetaindex{5})
\right .  \nonumber \\
&& \hspace{1cm} 
 + 
 \left . 3\,{\sqrt{3}}\,{\sqrt{\Aindex{6}}}\,J\,\sin (\Thetaindex{5}) + 12\,{\sqrt{3}}\,{\sqrt{\Aindex{6}}}\,t\,\sin (\Thetaindex{5}) - 9\,{\sqrt{3}}\,{\sqrt{\Aindex{6}}}\,U\,\sin (\Thetaindex{5}) - 102\,{\sqrt{3}}\,{\sqrt{\Aindex{6}}}\,W\,\sin (\Thetaindex{5})
\right )  \nonumber \\
C_{203-2} &=&
\frac{t}{6}
\, \left ( 2\,J + 12\,t + 2\,U - 4\,W + {\sqrt{\Aindex{6}}}\,\cos (\Thetaindex{5}) - {\sqrt{3}}\,{\sqrt{\Aindex{6}}}\,\sin (\Thetaindex{5})  \right ) \nonumber \\
C_{203-3} &=&
\frac{t}{6}
\, \left ( -J + 12\,t - U + 2\,W + {\sqrt{\Aindex{6}}}\,\cos (\Thetaindex{5}) - {\sqrt{3}}\,{\sqrt{\Aindex{6}}}\,\sin (\Thetaindex{5})  \right ) \nonumber \\
 N_{203} &=& 2\,{\sqrt{{\Cindex{203}{1}}^2 + 2\,\left( {\Cindex{203}{2}}^2 + {\Cindex{203}{3}}^2 \right) }} \nonumber \eeq 
\beq
\ket{\Psi}_{204}& = &\ket{5, {1 \over 2} , {3 \over 4} ,\Gamma_{4,3}} \nonumber \\ 
& = &
\Cindex{204}{1} \left ( 
\ket{022u} + \ket{0u22} - \ket{20u2} + \ket{220u} - \ket{22u0} + \ket{2u02} - \ket{u022} - \ket{u220}\right) 
 \nonumber \\
& + &
\Cindex{204}{2} \left ( 
\ket{02u2} - \ket{202u} - \ket{2u20} + \ket{u202}\right) 
 \nonumber \\
& + &
\Cindex{204}{3} \left ( 
\ket{2duu} - \ket{2uud} + \ket{d2uu} - \ket{duu2} - \ket{u2du} - \ket{ud2u} + \ket{uu2d} + \ket{uud2}\right) 
\nonumber \eeq
\beq
C_{204-1} &=&
\frac{-t}{2\,{\sqrt{2}}}
\, \left ( J + 4\,t + U - 2\,W + 2\,{\sqrt{\Aindex{6}}}\,\cos (\Thetaindex{5})  \right ) \nonumber \\
C_{204-2} &=&
\frac{t}{3\,{\sqrt{2}}}
\, \left ( J - 12\,t + U - 2\,W + 2\,{\sqrt{\Aindex{6}}}\,\cos (\Thetaindex{5})  \right ) \nonumber \\
C_{204-3} &=&
\frac{-1}{18\,{\sqrt{2}}}
\,\left ( {\Aindex{22}}^2 - 3\,\Aindex{22}\,t - 36\,t^2 + 12\,\Aindex{22}\,U - 18\,t\,U\right . \nonumber \\
&& \hspace{1cm} 
 + 
 \left . 36\,U^2 + 96\,\Aindex{22}\,W - 144\,t\,W + 576\,U\,W + 2304\,W^2
\right  )  \nonumber \\
 N_{204} &=& 2\,{\sqrt{2\,{\Cindex{204}{1}}^2 + {\Cindex{204}{2}}^2 + 2\,{\Cindex{204}{3}}^2}} \nonumber \eeq 
\beq
\ket{\Psi}_{205}& = &\ket{5, {1 \over 2} , {3 \over 4} ,\Gamma_{4,3}} \nonumber \\ 
& = &
\Cindex{205}{1} \left ( 
\ket{022u} + \ket{0u22} - \ket{20u2} + \ket{220u} - \ket{22u0} + \ket{2u02} - \ket{u022} - \ket{u220}\right) 
 \nonumber \\
& + &
\Cindex{205}{2} \left ( 
\ket{02u2} - \ket{202u} - \ket{2u20} + \ket{u202}\right) 
 \nonumber \\
& + &
\Cindex{205}{3} \left ( 
\ket{2duu} - \ket{2uud} + \ket{d2uu} - \ket{duu2} - \ket{u2du} - \ket{ud2u} + \ket{uu2d} + \ket{uud2}\right) 
\nonumber \eeq
\beq
C_{205-1} &=&
\frac{-t}{2\,{\sqrt{2}}}
\, \left ( J + 4\,t + U - 2\,W - {\sqrt{\Aindex{6}}}\,\cos (\Thetaindex{5}) - {\sqrt{3}}\,{\sqrt{\Aindex{6}}}\,\sin (\Thetaindex{5})  \right ) \nonumber \\
C_{205-2} &=&
\frac{t}{3\,{\sqrt{2}}}
\, \left ( J - 12\,t + U - 2\,W - {\sqrt{\Aindex{6}}}\,\cos (\Thetaindex{5}) - {\sqrt{3}}\,{\sqrt{\Aindex{6}}}\,\sin (\Thetaindex{5})  \right ) \nonumber \\
C_{205-3} &=&
\frac{-1}{18\,{\sqrt{2}}}
\,\left ( {\Aindex{24}}^2 - 3\,J\,t - 45\,t^2 + 12\,J\,U + 33\,t\,U - 24\,U^2 + 96\,J\,W + 294\,t\,W - 504\,U\,W\right . \nonumber \\
&& \hspace{1cm} 
 \left . -2496\,W^2 + 3\,{\sqrt{\Aindex{6}}}\,t\,\cos (\Thetaindex{5}) - 12\,{\sqrt{\Aindex{6}}}\,U\,\cos (\Thetaindex{5}) - 96\,{\sqrt{\Aindex{6}}}\,W\,\cos (\Thetaindex{5})
\right .  \nonumber \\
&& \hspace{1cm} 
 + 
 \left . 3\,{\sqrt{3}}\,{\sqrt{\Aindex{6}}}\,t\,\sin (\Thetaindex{5}) - 12\,{\sqrt{3}}\,{\sqrt{\Aindex{6}}}\,U\,\sin (\Thetaindex{5}) - 96\,{\sqrt{3}}\,{\sqrt{\Aindex{6}}}\,W\,\sin (\Thetaindex{5})
\right )  \nonumber \\
 N_{205} &=& 2\,{\sqrt{2\,{\Cindex{205}{1}}^2 + {\Cindex{205}{2}}^2 + 2\,{\Cindex{205}{3}}^2}} \nonumber \eeq 
\beq
\ket{\Psi}_{206}& = &\ket{5, {1 \over 2} , {3 \over 4} ,\Gamma_{4,3}} \nonumber \\ 
& = &
\Cindex{206}{1} \left ( 
\ket{022u} + \ket{0u22} - \ket{20u2} + \ket{220u} - \ket{22u0} + \ket{2u02} - \ket{u022} - \ket{u220}\right) 
 \nonumber \\
& + &
\Cindex{206}{2} \left ( 
\ket{02u2} - \ket{202u} - \ket{2u20} + \ket{u202}\right) 
 \nonumber \\
& + &
\Cindex{206}{3} \left ( 
\ket{2duu} - \ket{2uud} + \ket{d2uu} - \ket{duu2} - \ket{u2du} - \ket{ud2u} + \ket{uu2d} + \ket{uud2}\right) 
\nonumber \eeq
\beq
C_{206-1} &=&
\frac{-t}{2\,{\sqrt{2}}}
\, \left ( J + 4\,t + U - 2\,W - {\sqrt{\Aindex{6}}}\,\cos (\Thetaindex{5}) + {\sqrt{3}}\,{\sqrt{\Aindex{6}}}\,\sin (\Thetaindex{5})  \right ) \nonumber \\
C_{206-2} &=&
\frac{t}{3\,{\sqrt{2}}}
\, \left ( J - 12\,t + U - 2\,W - {\sqrt{\Aindex{6}}}\,\cos (\Thetaindex{5}) + {\sqrt{3}}\,{\sqrt{\Aindex{6}}}\,\sin (\Thetaindex{5})  \right ) \nonumber \\
C_{206-3} &=&
\frac{-1}{18\,{\sqrt{2}}}
\,\left ( {\Aindex{23}}^2 - 3\,J\,t - 45\,t^2 + 12\,J\,U + 33\,t\,U - 24\,U^2 + 96\,J\,W + 294\,t\,W - 504\,U\,W\right . \nonumber \\
&& \hspace{1cm} 
 \left . -2496\,W^2 + 3\,{\sqrt{\Aindex{6}}}\,t\,\cos (\Thetaindex{5}) - 12\,{\sqrt{\Aindex{6}}}\,U\,\cos (\Thetaindex{5}) - 96\,{\sqrt{\Aindex{6}}}\,W\,\cos (\Thetaindex{5})
\right .  \nonumber \\
&& \hspace{1cm} 
 \left . -3\,{\sqrt{3}}\,{\sqrt{\Aindex{6}}}\,t\,\sin (\Thetaindex{5}) + 12\,{\sqrt{3}}\,{\sqrt{\Aindex{6}}}\,U\,\sin (\Thetaindex{5}) + 96\,{\sqrt{3}}\,{\sqrt{\Aindex{6}}}\,W\,\sin (\Thetaindex{5})
\right )  \nonumber \\
 N_{206} &=& 2\,{\sqrt{2\,{\Cindex{206}{1}}^2 + {\Cindex{206}{2}}^2 + 2\,{\Cindex{206}{3}}^2}} \nonumber \eeq 
\beq
\ket{\Psi}_{207}& = &\ket{5, {1 \over 2} , {3 \over 4} ,\Gamma_{5,1}} \nonumber \\ 
& = &
\Cindex{207}{1} \left ( 
\ket{022u} - \ket{0u22} - \ket{20u2} - \ket{220u} + \ket{22u0} + \ket{2u02} + \ket{u022} - \ket{u220}\right) 
 \nonumber \\
& + &
\Cindex{207}{2} \left ( 
\ket{2duu} + \ket{2uud} + \ket{d2uu} + \ket{duu2} + \ket{u2du} + \ket{ud2u} + \ket{uu2d} + \ket{uud2}\right) 
 \nonumber \\
& + &
\Cindex{207}{3} \left ( 
\ket{2udu} + \ket{du2u} + \ket{u2ud} + \ket{udu2}\right) 
\nonumber \eeq
\beq
C_{207-1} &=&
\frac{-\left( {\sqrt{\frac{3}{2}}}\,t \right) }{2} \nonumber \\
C_{207-2} &=&
\frac{1}{4\,{\sqrt{6}}}
\, \left ( {\sqrt{\Aindex{4}}} + J - 2\,t + U - 2\,W  \right ) \nonumber \\
C_{207-3} &=&
\frac{-1}{2\,{\sqrt{6}}}
\, \left ( {\sqrt{\Aindex{4}}} + J - 2\,t + U - 2\,W  \right ) \nonumber \\
 N_{207} &=& 2\,{\sqrt{2\,{\Cindex{207}{1}}^2 + 2\,{\Cindex{207}{2}}^2 + {\Cindex{207}{3}}^2}} \nonumber \eeq 
\beq
\ket{\Psi}_{208}& = &\ket{5, {1 \over 2} , {3 \over 4} ,\Gamma_{5,1}} \nonumber \\ 
& = &
\Cindex{208}{1} \left ( 
\ket{022u} - \ket{0u22} - \ket{20u2} - \ket{220u} + \ket{22u0} + \ket{2u02} + \ket{u022} - \ket{u220}\right) 
 \nonumber \\
& + &
\Cindex{208}{2} \left ( 
\ket{2duu} + \ket{2uud} + \ket{d2uu} + \ket{duu2} + \ket{u2du} + \ket{ud2u} + \ket{uu2d} + \ket{uud2}\right) 
 \nonumber \\
& + &
\Cindex{208}{3} \left ( 
\ket{2udu} + \ket{du2u} + \ket{u2ud} + \ket{udu2}\right) 
\nonumber \eeq
\beq
C_{208-1} &=&
\frac{-\left( {\sqrt{\frac{3}{2}}}\,t \right) }{2} \nonumber \\
C_{208-2} &=&
\frac{-1}{4\,{\sqrt{6}}}
\, \left ( {\sqrt{\Aindex{4}}} - J + 2\,t - U + 2\,W  \right ) \nonumber \\
C_{208-3} &=&
\frac{1}{2\,{\sqrt{6}}}
\, \left ( {\sqrt{\Aindex{4}}} - J + 2\,t - U + 2\,W  \right ) \nonumber \\
 N_{208} &=& 2\,{\sqrt{2\,{\Cindex{208}{1}}^2 + 2\,{\Cindex{208}{2}}^2 + {\Cindex{208}{3}}^2}} \nonumber \eeq 
\beq
\ket{\Psi}_{209}& = &\ket{5, {1 \over 2} , {15 \over 4} ,\Gamma_{5,1}} \nonumber \\ 
&=& \frac{1}{2\,{\sqrt{3}}}
 \left ( \ket{2duu} + \ket{2udu} + \ket{2uud} + \ket{d2uu} + \ket{du2u} + \ket{duu2}\right . \nonumber \\
&& \hspace{1em} 
 + 
\left . \ket{u2du} + \ket{u2ud} + \ket{ud2u} + \ket{udu2} + \ket{uu2d} + \ket{uud2}\right ) \nonumber  
\eeq
\beq
\ket{\Psi}_{210}& = &\ket{5, {1 \over 2} , {3 \over 4} ,\Gamma_{5,2}} \nonumber \\ 
& = &
\Cindex{210}{1} \left ( 
\ket{022u} - \ket{02u2} - \ket{202u} + \ket{20u2} - \ket{2u02} + \ket{2u20} + \ket{u202} - \ket{u220}\right) 
 \nonumber \\
& + &
\Cindex{210}{2} \left ( 
\ket{2duu} - \ket{d2uu} - \ket{uu2d} + \ket{uud2}\right) 
 \nonumber \\
& + &
\Cindex{210}{3} \left ( 
\ket{2udu} + \ket{2uud} - \ket{du2u} + \ket{duu2} - \ket{u2du} - \ket{u2ud} - \ket{ud2u} + \ket{udu2}\right) 
\nonumber \eeq
\beq
C_{210-1} &=&
\frac{{\sqrt{\frac{3}{2}}}\,t}{2} \nonumber \\
C_{210-2} &=&
\frac{-1}{2\,{\sqrt{6}}}
\, \left ( {\sqrt{\Aindex{4}}} + J - 2\,t + U - 2\,W  \right ) \nonumber \\
C_{210-3} &=&
\frac{1}{4\,{\sqrt{6}}}
\, \left ( {\sqrt{\Aindex{4}}} + J - 2\,t + U - 2\,W  \right ) \nonumber \\
 N_{210} &=& 2\,{\sqrt{2\,{\Cindex{210}{1}}^2 + {\Cindex{210}{2}}^2 + 2\,{\Cindex{210}{3}}^2}} \nonumber \eeq 
\beq
\ket{\Psi}_{211}& = &\ket{5, {1 \over 2} , {3 \over 4} ,\Gamma_{5,2}} \nonumber \\ 
& = &
\Cindex{211}{1} \left ( 
\ket{022u} - \ket{02u2} - \ket{202u} + \ket{20u2} - \ket{2u02} + \ket{2u20} + \ket{u202} - \ket{u220}\right) 
 \nonumber \\
& + &
\Cindex{211}{2} \left ( 
\ket{2duu} - \ket{d2uu} - \ket{uu2d} + \ket{uud2}\right) 
 \nonumber \\
& + &
\Cindex{211}{3} \left ( 
\ket{2udu} + \ket{2uud} - \ket{du2u} + \ket{duu2} - \ket{u2du} - \ket{u2ud} - \ket{ud2u} + \ket{udu2}\right) 
\nonumber \eeq
\beq
C_{211-1} &=&
\frac{{\sqrt{\frac{3}{2}}}\,t}{2} \nonumber \\
C_{211-2} &=&
\frac{1}{2\,{\sqrt{6}}}
\, \left ( {\sqrt{\Aindex{4}}} - J + 2\,t - U + 2\,W  \right ) \nonumber \\
C_{211-3} &=&
\frac{-1}{4\,{\sqrt{6}}}
\, \left ( {\sqrt{\Aindex{4}}} - J + 2\,t - U + 2\,W  \right ) \nonumber \\
 N_{211} &=& 2\,{\sqrt{2\,{\Cindex{211}{1}}^2 + {\Cindex{211}{2}}^2 + 2\,{\Cindex{211}{3}}^2}} \nonumber \eeq 
\beq
\ket{\Psi}_{212}& = &\ket{5, {1 \over 2} , {15 \over 4} ,\Gamma_{5,2}} \nonumber \\ 
&=& \frac{1}{2\,{\sqrt{3}}}
 \left ( \ket{2duu} + \ket{2udu} + \ket{2uud} - \ket{d2uu} - \ket{du2u} + \ket{duu2}\right . \nonumber \\
&& \hspace{1em} 
\left . -\ket{u2du} - \ket{u2ud} - \ket{ud2u} + \ket{udu2} - \ket{uu2d} + \ket{uud2}\right ) \nonumber  
\eeq
\beq
\ket{\Psi}_{213}& = &\ket{5, {1 \over 2} , {3 \over 4} ,\Gamma_{5,3}} \nonumber \\ 
& = &
\Cindex{213}{1} \left ( 
\ket{02u2} - \ket{0u22} - \ket{202u} + \ket{220u} - \ket{22u0} + \ket{2u20} + \ket{u022} - \ket{u202}\right) 
 \nonumber \\
& + &
\Cindex{213}{2} \left ( 
\ket{2duu} + \ket{2udu} + \ket{d2uu} - \ket{du2u} + \ket{u2ud} - \ket{udu2} - \ket{uu2d} - \ket{uud2}\right) 
 \nonumber \\
& + &
\Cindex{213}{3} \left ( 
\ket{2uud} - \ket{duu2} + \ket{u2du} - \ket{ud2u}\right) 
\nonumber \eeq
\beq
C_{213-1} &=&
\frac{-\left( {\sqrt{\frac{3}{2}}}\,t \right) }{2} \nonumber \\
C_{213-2} &=&
\frac{-1}{4\,{\sqrt{6}}}
\, \left ( {\sqrt{\Aindex{4}}} + J - 2\,t + U - 2\,W  \right ) \nonumber \\
C_{213-3} &=&
\frac{1}{2\,{\sqrt{6}}}
\, \left ( {\sqrt{\Aindex{4}}} + J - 2\,t + U - 2\,W  \right ) \nonumber \\
 N_{213} &=& 2\,{\sqrt{2\,{\Cindex{213}{1}}^2 + 2\,{\Cindex{213}{2}}^2 + {\Cindex{213}{3}}^2}} \nonumber \eeq 
\beq
\ket{\Psi}_{214}& = &\ket{5, {1 \over 2} , {3 \over 4} ,\Gamma_{5,3}} \nonumber \\ 
& = &
\Cindex{214}{1} \left ( 
\ket{02u2} - \ket{0u22} - \ket{202u} + \ket{220u} - \ket{22u0} + \ket{2u20} + \ket{u022} - \ket{u202}\right) 
 \nonumber \\
& + &
\Cindex{214}{2} \left ( 
\ket{2duu} + \ket{2udu} + \ket{d2uu} - \ket{du2u} + \ket{u2ud} - \ket{udu2} - \ket{uu2d} - \ket{uud2}\right) 
 \nonumber \\
& + &
\Cindex{214}{3} \left ( 
\ket{2uud} - \ket{duu2} + \ket{u2du} - \ket{ud2u}\right) 
\nonumber \eeq
\beq
C_{214-1} &=&
\frac{-\left( {\sqrt{\frac{3}{2}}}\,t \right) }{2} \nonumber \\
C_{214-2} &=&
\frac{1}{4\,{\sqrt{6}}}
\, \left ( {\sqrt{\Aindex{4}}} - J + 2\,t - U + 2\,W  \right ) \nonumber \\
C_{214-3} &=&
\frac{-1}{2\,{\sqrt{6}}}
\, \left ( {\sqrt{\Aindex{4}}} - J + 2\,t - U + 2\,W  \right ) \nonumber \\
 N_{214} &=& 2\,{\sqrt{2\,{\Cindex{214}{1}}^2 + 2\,{\Cindex{214}{2}}^2 + {\Cindex{214}{3}}^2}} \nonumber \eeq 
\beq
\ket{\Psi}_{215}& = &\ket{5, {1 \over 2} , {15 \over 4} ,\Gamma_{5,3}} \nonumber \\ 
&=& \frac{1}{2\,{\sqrt{3}}}
 \left ( \ket{2duu} + \ket{2udu} + \ket{2uud} + \ket{d2uu} - \ket{du2u} - \ket{duu2}\right . \nonumber \\
&& \hspace{1em} 
 + 
\left . \ket{u2du} + \ket{u2ud} - \ket{ud2u} - \ket{udu2} - \ket{uu2d} - \ket{uud2}\right ) \nonumber  
\eeq
{\subsection*{\boldmath Unnormalized eigenvectors for ${\rm  N_e}=5$ and   ${\rm m_s}$= $ {3 \over 2} $.}
\beq
\ket{\Psi}_{216}& = &\ket{5, {3 \over 2} , {15 \over 4} ,\Gamma_2} \nonumber \\ 
&=& \frac{1}{2}
 \left ( \ket{2uuu} - \ket{u2uu} + \ket{uu2u} - \ket{uuu2}\right) \nonumber 
\eeq
\beq
\ket{\Psi}_{217}& = &\ket{5, {3 \over 2} , {15 \over 4} ,\Gamma_{5,1}} \nonumber \\ 
&=& \frac{1}{2}
 \left ( \ket{2uuu} + \ket{u2uu} + \ket{uu2u} + \ket{uuu2}\right) \nonumber 
\eeq
\beq
\ket{\Psi}_{218}& = &\ket{5, {3 \over 2} , {15 \over 4} ,\Gamma_{5,2}} \nonumber \\ 
&=& \frac{1}{2}
 \left ( \ket{2uuu} - \ket{u2uu} - \ket{uu2u} + \ket{uuu2}\right) \nonumber 
\eeq
\beq
\ket{\Psi}_{219}& = &\ket{5, {3 \over 2} , {15 \over 4} ,\Gamma_{5,3}} \nonumber \\ 
&=& \frac{1}{2}
 \left ( \ket{2uuu} + \ket{u2uu} - \ket{uu2u} - \ket{uuu2}\right) \nonumber 
\eeq
{\subsection*{\boldmath Unnormalized eigenvectors for ${\rm  N_e}=6$ and   ${\rm m_s}$= $-1$.}
\beq
\ket{\Psi}_{220}& = &\ket{6,-1,2,\Gamma_{4,1}} \nonumber \\ 
&=& \frac{1}{2}
 \left ( \ket{2d2d} + \ket{2dd2} + \ket{d22d} + \ket{d2d2}\right) \nonumber 
\eeq
\beq
\ket{\Psi}_{221}& = &\ket{6,-1,2,\Gamma_{4,2}} \nonumber \\ 
&=& \frac{1}{2}
 \left ( \ket{22dd} + \ket{2d2d} - \ket{d2d2} - \ket{dd22}\right) \nonumber 
\eeq
\beq
\ket{\Psi}_{222}& = &\ket{6,-1,2,\Gamma_{4,3}} \nonumber \\ 
&=& \frac{1}{2}
 \left ( \ket{22dd} - \ket{2dd2} + \ket{d22d} + \ket{dd22}\right) \nonumber 
\eeq
\beq
\ket{\Psi}_{223}& = &\ket{6,-1,2,\Gamma_{5,1}} \nonumber \\ 
&=& \frac{1}{2}
 \left ( \ket{22dd} + \ket{2dd2} - \ket{d22d} + \ket{dd22}\right) \nonumber 
\eeq
\beq
\ket{\Psi}_{224}& = &\ket{6,-1,2,\Gamma_{5,2}} \nonumber \\ 
&=& \frac{1}{2}
 \left ( \ket{2d2d} - \ket{2dd2} - \ket{d22d} + \ket{d2d2}\right) \nonumber 
\eeq
\beq
\ket{\Psi}_{225}& = &\ket{6,-1,2,\Gamma_{5,3}} \nonumber \\ 
&=& \frac{1}{2}
 \left ( \ket{22dd} - \ket{2d2d} + \ket{d2d2} - \ket{dd22}\right) \nonumber 
\eeq
{\subsection*{\boldmath Unnormalized eigenvectors for ${\rm  N_e}=6$ and   ${\rm m_s}$= $0$.}
\beq
\ket{\Psi}_{226}& = &\ket{6,0,0,\Gamma_1} \nonumber \\ 
& = &
\Cindex{226}{1} \left ( 
\ket{0222} + \ket{2022} + \ket{2202} + \ket{2220}\right) 
 \nonumber \\
& + &
\Cindex{226}{2} \left ( 
\ket{22du} - \ket{22ud} + \ket{2d2u} + \ket{2du2} - \ket{2u2d} - \ket{2ud2}\right . \nonumber \\
&& \hspace{3em} 
 + 
\left . \ket{d22u} + \ket{d2u2} + \ket{du22} - \ket{u22d} - \ket{u2d2} - \ket{ud22}\right ) 
\nonumber \eeq
\beq
C_{226-1} &=&
{\sqrt{3}}\,t \nonumber \\
C_{226-2} &=&
\frac{1}{4\,{\sqrt{3}}}
\, \left ( {\sqrt{\Aindex{8}}} + J + 4\,t + U - 2\,W  \right ) \nonumber \\
 N_{226} &=& 2\,{\sqrt{{\Cindex{226}{1}}^2 + 3\,{\Cindex{226}{2}}^2}} \nonumber \eeq 
\beq
\ket{\Psi}_{227}& = &\ket{6,0,0,\Gamma_1} \nonumber \\ 
& = &
\Cindex{227}{1} \left ( 
\ket{0222} + \ket{2022} + \ket{2202} + \ket{2220}\right) 
 \nonumber \\
& + &
\Cindex{227}{2} \left ( 
\ket{22du} - \ket{22ud} + \ket{2d2u} + \ket{2du2} - \ket{2u2d} - \ket{2ud2}\right . \nonumber \\
&& \hspace{3em} 
 + 
\left . \ket{d22u} + \ket{d2u2} + \ket{du22} - \ket{u22d} - \ket{u2d2} - \ket{ud22}\right ) 
\nonumber \eeq
\beq
C_{227-1} &=&
{\sqrt{3}}\,t \nonumber \\
C_{227-2} &=&
\frac{-1}{4\,{\sqrt{3}}}
\, \left ( {\sqrt{\Aindex{8}}} - J - 4\,t - U + 2\,W  \right ) \nonumber \\
 N_{227} &=& 2\,{\sqrt{{\Cindex{227}{1}}^2 + 3\,{\Cindex{227}{2}}^2}} \nonumber \eeq 
\beq
\ket{\Psi}_{228}& = &\ket{6,0,0,\Gamma_{3,1}} \nonumber \\ 
& = &
\Cindex{228}{1} \left ( 
\ket{22du} - \ket{22ud} + \ket{2du2} - \ket{2ud2} + \ket{d22u} + \ket{du22} - \ket{u22d} - \ket{ud22}\right) 
 \nonumber \\
& + &
\Cindex{228}{2} \left ( 
\ket{2d2u} - \ket{2u2d} + \ket{d2u2} - \ket{u2d2}\right) 
\nonumber \eeq
\beq
C_{228-1} &=&
\frac{-1}{2\,{\sqrt{6}}} \nonumber \\
C_{228-2} &=&
\frac{1}{{\sqrt{6}}} \nonumber \\
 N_{228} &=& 2\,{\sqrt{2\,{\Cindex{228}{1}}^2 + {\Cindex{228}{2}}^2}} \nonumber \eeq 
\beq
\ket{\Psi}_{229}& = &\ket{6,0,0,\Gamma_{3,2}} \nonumber \\ 
&=& \frac{1}{2\,{\sqrt{2}}}
 \left ( \ket{22du} - \ket{22ud} - \ket{2du2} + \ket{2ud2} - \ket{d22u} + \ket{du22} + \ket{u22d} - \ket{ud22}\right) \nonumber 
\eeq
\beq
\ket{\Psi}_{230}& = &\ket{6,0,0,\Gamma_{4,1}} \nonumber \\ 
& = &
\Cindex{230}{1} \left ( 
\ket{0222} + \ket{2022} - \ket{2202} - \ket{2220}\right) 
 \nonumber \\
& + &
\Cindex{230}{2} \left ( 
\ket{22du} - \ket{22ud} - \ket{du22} + \ket{ud22}\right) 
\nonumber \eeq
\beq
C_{230-1} &=&
-t \nonumber \\
C_{230-2} &=&
\frac{1}{4}
\, \left ( {\sqrt{\Aindex{1}}} + J + U - 2\,W  \right ) \nonumber \\
 N_{230} &=& 2\,{\sqrt{{\Cindex{230}{1}}^2 + {\Cindex{230}{2}}^2}} \nonumber \eeq 
\beq
\ket{\Psi}_{231}& = &\ket{6,0,0,\Gamma_{4,1}} \nonumber \\ 
& = &
\Cindex{231}{1} \left ( 
\ket{0222} + \ket{2022} - \ket{2202} - \ket{2220}\right) 
 \nonumber \\
& + &
\Cindex{231}{2} \left ( 
\ket{22du} - \ket{22ud} - \ket{du22} + \ket{ud22}\right) 
\nonumber \eeq
\beq
C_{231-1} &=&
-t \nonumber \\
C_{231-2} &=&
\frac{1}{4}
\, \left ( -{\sqrt{\Aindex{1}}} + J + U - 2\,W  \right ) \nonumber \\
 N_{231} &=& 2\,{\sqrt{{\Cindex{231}{1}}^2 + {\Cindex{231}{2}}^2}} \nonumber \eeq 
\beq
\ket{\Psi}_{232}& = &\ket{6,0,2,\Gamma_{4,1}} \nonumber \\ 
&=& \frac{1}{2\,{\sqrt{2}}}
 \left ( \ket{2d2u} + \ket{2du2} + \ket{2u2d} + \ket{2ud2} + \ket{d22u} + \ket{d2u2} + \ket{u22d} + \ket{u2d2}\right) \nonumber 
\eeq
\beq
\ket{\Psi}_{233}& = &\ket{6,0,0,\Gamma_{4,2}} \nonumber \\ 
& = &
\Cindex{233}{1} \left ( 
\ket{0222} - \ket{2022} - \ket{2202} + \ket{2220}\right) 
 \nonumber \\
& + &
\Cindex{233}{2} \left ( 
\ket{2du2} - \ket{2ud2} - \ket{d22u} + \ket{u22d}\right) 
\nonumber \eeq
\beq
C_{233-1} &=&
-t \nonumber \\
C_{233-2} &=&
\frac{1}{4}
\, \left ( {\sqrt{\Aindex{1}}} + J + U - 2\,W  \right ) \nonumber \\
 N_{233} &=& 2\,{\sqrt{{\Cindex{233}{1}}^2 + {\Cindex{233}{2}}^2}} \nonumber \eeq 
\beq
\ket{\Psi}_{234}& = &\ket{6,0,0,\Gamma_{4,2}} \nonumber \\ 
& = &
\Cindex{234}{1} \left ( 
\ket{0222} - \ket{2022} - \ket{2202} + \ket{2220}\right) 
 \nonumber \\
& + &
\Cindex{234}{2} \left ( 
\ket{2du2} - \ket{2ud2} - \ket{d22u} + \ket{u22d}\right) 
\nonumber \eeq
\beq
C_{234-1} &=&
-t \nonumber \\
C_{234-2} &=&
\frac{1}{4}
\, \left ( -{\sqrt{\Aindex{1}}} + J + U - 2\,W  \right ) \nonumber \\
 N_{234} &=& 2\,{\sqrt{{\Cindex{234}{1}}^2 + {\Cindex{234}{2}}^2}} \nonumber \eeq 
\beq
\ket{\Psi}_{235}& = &\ket{6,0,2,\Gamma_{4,2}} \nonumber \\ 
&=& \frac{1}{2\,{\sqrt{2}}}
 \left ( \ket{22du} + \ket{22ud} + \ket{2d2u} + \ket{2u2d} - \ket{d2u2} - \ket{du22} - \ket{u2d2} - \ket{ud22}\right) \nonumber 
\eeq
\beq
\ket{\Psi}_{236}& = &\ket{6,0,0,\Gamma_{4,3}} \nonumber \\ 
& = &
\Cindex{236}{1} \left ( 
\ket{0222} - \ket{2022} + \ket{2202} - \ket{2220}\right) 
 \nonumber \\
& + &
\Cindex{236}{2} \left ( 
\ket{2d2u} - \ket{2u2d} - \ket{d2u2} + \ket{u2d2}\right) 
\nonumber \eeq
\beq
C_{236-1} &=&
-t \nonumber \\
C_{236-2} &=&
\frac{1}{4}
\, \left ( {\sqrt{\Aindex{1}}} + J + U - 2\,W  \right ) \nonumber \\
 N_{236} &=& 2\,{\sqrt{{\Cindex{236}{1}}^2 + {\Cindex{236}{2}}^2}} \nonumber \eeq 
\beq
\ket{\Psi}_{237}& = &\ket{6,0,0,\Gamma_{4,3}} \nonumber \\ 
& = &
\Cindex{237}{1} \left ( 
\ket{0222} - \ket{2022} + \ket{2202} - \ket{2220}\right) 
 \nonumber \\
& + &
\Cindex{237}{2} \left ( 
\ket{2d2u} - \ket{2u2d} - \ket{d2u2} + \ket{u2d2}\right) 
\nonumber \eeq
\beq
C_{237-1} &=&
-t \nonumber \\
C_{237-2} &=&
\frac{1}{4}
\, \left ( -{\sqrt{\Aindex{1}}} + J + U - 2\,W  \right ) \nonumber \\
 N_{237} &=& 2\,{\sqrt{{\Cindex{237}{1}}^2 + {\Cindex{237}{2}}^2}} \nonumber \eeq 
\beq
\ket{\Psi}_{238}& = &\ket{6,0,2,\Gamma_{4,3}} \nonumber \\ 
&=& \frac{1}{2\,{\sqrt{2}}}
 \left ( \ket{22du} + \ket{22ud} - \ket{2du2} - \ket{2ud2} + \ket{d22u} + \ket{du22} + \ket{u22d} + \ket{ud22}\right) \nonumber 
\eeq
\beq
\ket{\Psi}_{239}& = &\ket{6,0,2,\Gamma_{5,1}} \nonumber \\ 
&=& \frac{1}{2\,{\sqrt{2}}}
 \left ( \ket{22du} + \ket{22ud} + \ket{2du2} + \ket{2ud2} - \ket{d22u} + \ket{du22} - \ket{u22d} + \ket{ud22}\right) \nonumber 
\eeq
\beq
\ket{\Psi}_{240}& = &\ket{6,0,2,\Gamma_{5,2}} \nonumber \\ 
&=& \frac{1}{2\,{\sqrt{2}}}
 \left ( \ket{2d2u} - \ket{2du2} + \ket{2u2d} - \ket{2ud2} - \ket{d22u} + \ket{d2u2} - \ket{u22d} + \ket{u2d2}\right) \nonumber 
\eeq
\beq
\ket{\Psi}_{241}& = &\ket{6,0,2,\Gamma_{5,3}} \nonumber \\ 
&=& \frac{1}{2\,{\sqrt{2}}}
 \left ( \ket{22du} + \ket{22ud} - \ket{2d2u} - \ket{2u2d} + \ket{d2u2} - \ket{du22} + \ket{u2d2} - \ket{ud22}\right) \nonumber 
\eeq
{\subsection*{\boldmath Unnormalized eigenvectors for ${\rm  N_e}=6$ and   ${\rm m_s}$= $1$.}
\beq
\ket{\Psi}_{242}& = &\ket{6,1,2,\Gamma_{4,1}} \nonumber \\ 
&=& \frac{1}{2}
 \left ( \ket{2u2u} + \ket{2uu2} + \ket{u22u} + \ket{u2u2}\right) \nonumber 
\eeq
\beq
\ket{\Psi}_{243}& = &\ket{6,1,2,\Gamma_{4,2}} \nonumber \\ 
&=& \frac{1}{2}
 \left ( \ket{22uu} + \ket{2u2u} - \ket{u2u2} - \ket{uu22}\right) \nonumber 
\eeq
\beq
\ket{\Psi}_{244}& = &\ket{6,1,2,\Gamma_{4,3}} \nonumber \\ 
&=& \frac{1}{2}
 \left ( \ket{22uu} - \ket{2uu2} + \ket{u22u} + \ket{uu22}\right) \nonumber 
\eeq
\beq
\ket{\Psi}_{245}& = &\ket{6,1,2,\Gamma_{5,1}} \nonumber \\ 
&=& \frac{1}{2}
 \left ( \ket{22uu} + \ket{2uu2} - \ket{u22u} + \ket{uu22}\right) \nonumber 
\eeq
\beq
\ket{\Psi}_{246}& = &\ket{6,1,2,\Gamma_{5,2}} \nonumber \\ 
&=& \frac{1}{2}
 \left ( \ket{2u2u} - \ket{2uu2} - \ket{u22u} + \ket{u2u2}\right) \nonumber 
\eeq
\beq
\ket{\Psi}_{247}& = &\ket{6,1,2,\Gamma_{5,3}} \nonumber \\ 
&=& \frac{1}{2}
 \left ( \ket{22uu} - \ket{2u2u} + \ket{u2u2} - \ket{uu22}\right) \nonumber 
\eeq
{\subsection*{\boldmath Unnormalized eigenvectors for ${\rm  N_e}=7$ and   ${\rm m_s}$= $- {1 \over 2} $.}
\beq
\ket{\Psi}_{248}& = &\ket{7,- {1 \over 2} , {3 \over 4} ,\Gamma_1} \nonumber \\ 
&=& \frac{1}{2}
 \left ( \ket{222d} + \ket{22d2} + \ket{2d22} + \ket{d222}\right) \nonumber 
\eeq
\beq
\ket{\Psi}_{249}& = &\ket{7,- {1 \over 2} , {3 \over 4} ,\Gamma_{4,1}} \nonumber \\ 
&=& \frac{1}{2}
 \left ( \ket{222d} + \ket{22d2} - \ket{2d22} - \ket{d222}\right) \nonumber 
\eeq
\beq
\ket{\Psi}_{250}& = &\ket{7,- {1 \over 2} , {3 \over 4} ,\Gamma_{4,2}} \nonumber \\ 
&=& \frac{1}{2}
 \left ( \ket{222d} - \ket{22d2} - \ket{2d22} + \ket{d222}\right) \nonumber 
\eeq
\beq
\ket{\Psi}_{251}& = &\ket{7,- {1 \over 2} , {3 \over 4} ,\Gamma_{4,3}} \nonumber \\ 
&=& \frac{1}{2}
 \left ( \ket{222d} - \ket{22d2} + \ket{2d22} - \ket{d222}\right) \nonumber 
\eeq
{\subsection*{\boldmath Unnormalized eigenvectors for ${\rm  N_e}=7$ and   ${\rm m_s}$= $ {1 \over 2} $.}
\beq
\ket{\Psi}_{252}& = &\ket{7, {1 \over 2} , {3 \over 4} ,\Gamma_1} \nonumber \\ 
&=& \frac{1}{2}
 \left ( \ket{222u} + \ket{22u2} + \ket{2u22} + \ket{u222}\right) \nonumber 
\eeq
\beq
\ket{\Psi}_{253}& = &\ket{7, {1 \over 2} , {3 \over 4} ,\Gamma_{4,1}} \nonumber \\ 
&=& \frac{1}{2}
 \left ( \ket{222u} + \ket{22u2} - \ket{2u22} - \ket{u222}\right) \nonumber 
\eeq
\beq
\ket{\Psi}_{254}& = &\ket{7, {1 \over 2} , {3 \over 4} ,\Gamma_{4,2}} \nonumber \\ 
&=& \frac{1}{2}
 \left ( \ket{222u} - \ket{22u2} - \ket{2u22} + \ket{u222}\right) \nonumber 
\eeq
\beq
\ket{\Psi}_{255}& = &\ket{7, {1 \over 2} , {3 \over 4} ,\Gamma_{4,3}} \nonumber \\ 
&=& \frac{1}{2}
 \left ( \ket{222u} - \ket{22u2} + \ket{2u22} - \ket{u222}\right) \nonumber 
\eeq
{\subsection*{\boldmath Unnormalized eigenvectors for ${\rm  N_e}=8$ and   ${\rm m_s}$= $0$.}
\beq
\ket{\Psi}_{256}& = &\ket{8,0,0,\Gamma_1} \nonumber \\ 
&=& 1
 \left ( \ket{2222}\right) \nonumber 
\eeq

%% file: Parts/figure01.tex
\begin{figure}
 \begin{minipage}[c]{0.3\textwidth}
 \centering  
 \includegraphics[height=32mm]{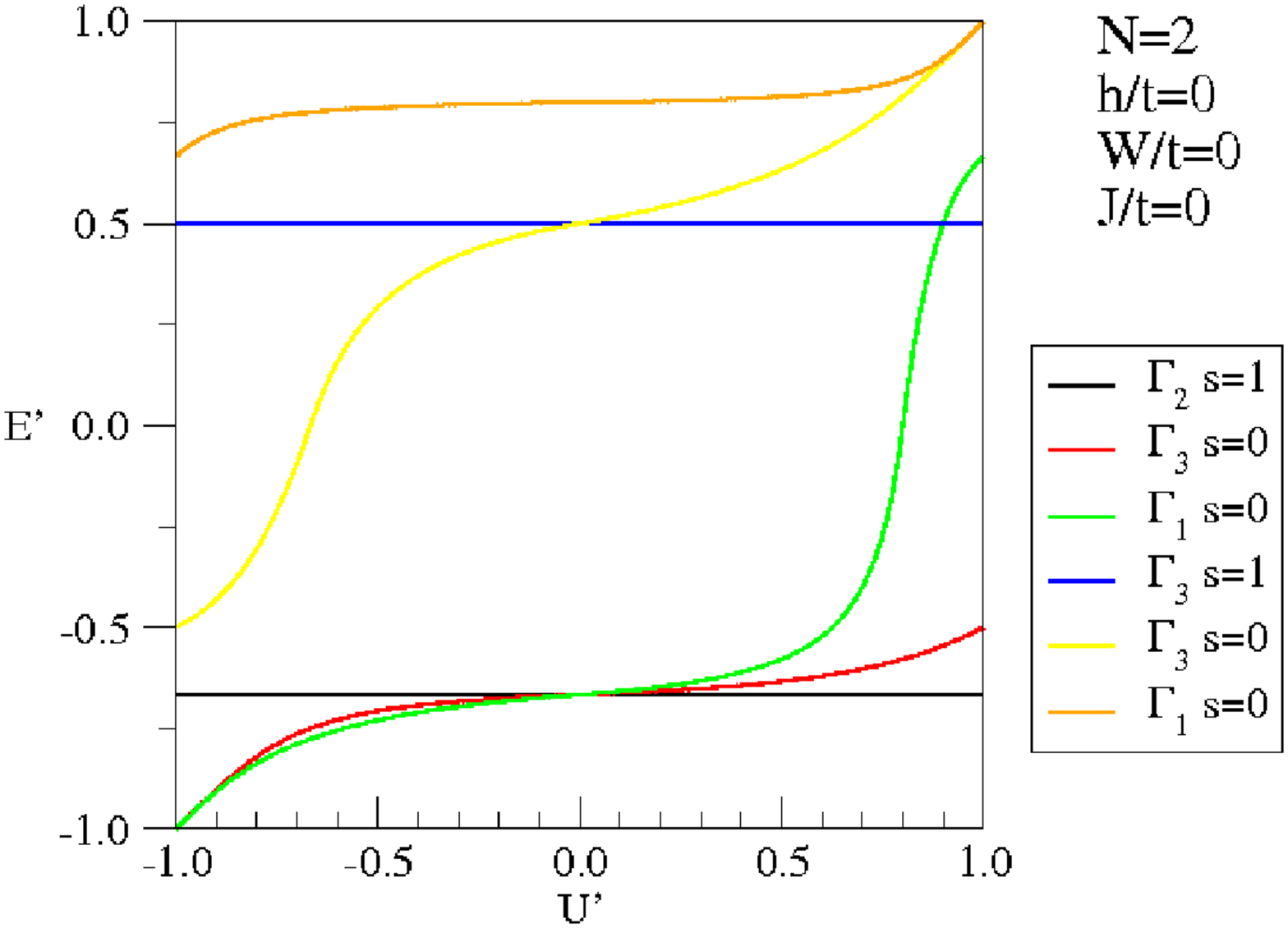}
 \includegraphics[height=32mm]{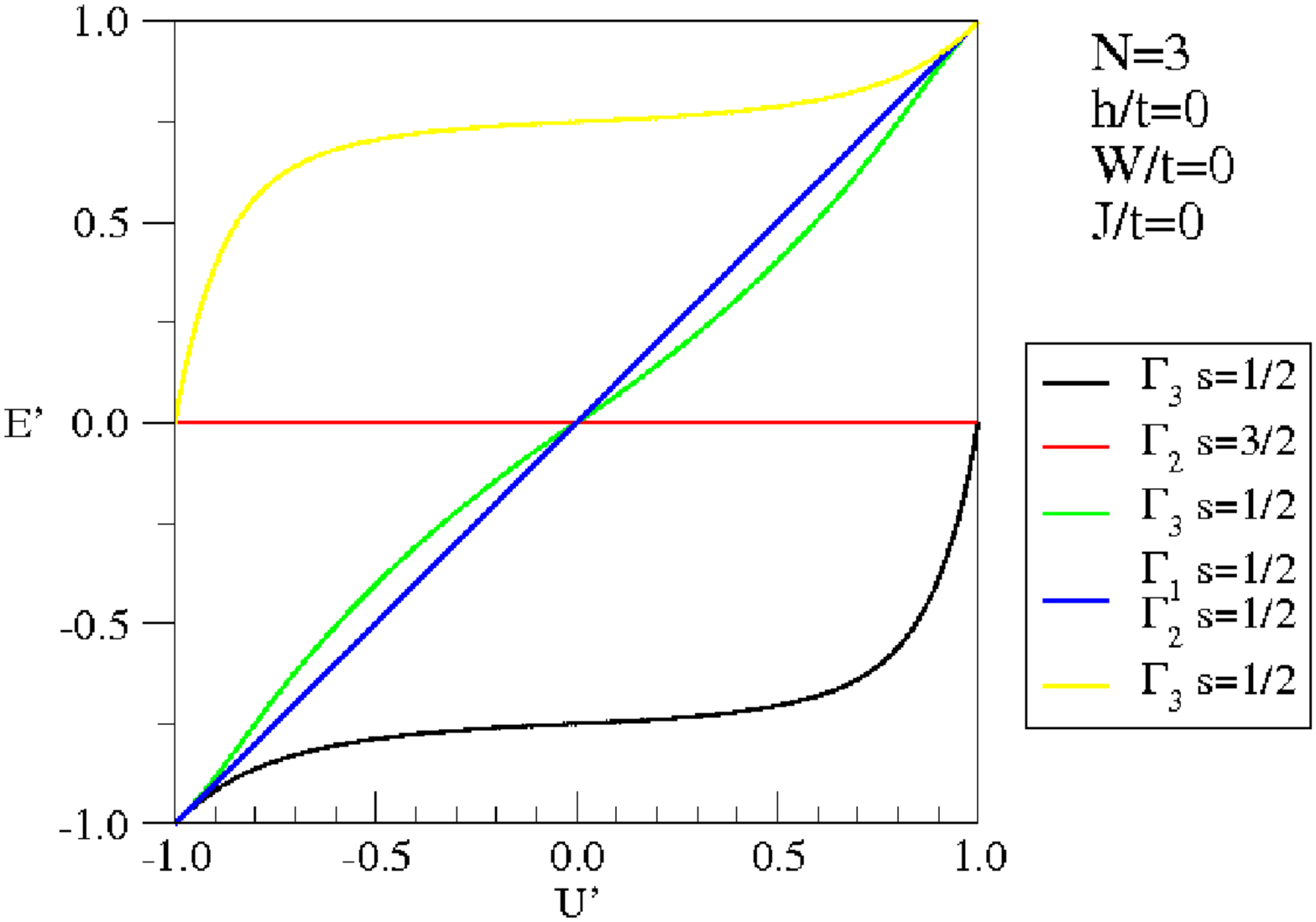}
 \includegraphics[height=32mm]{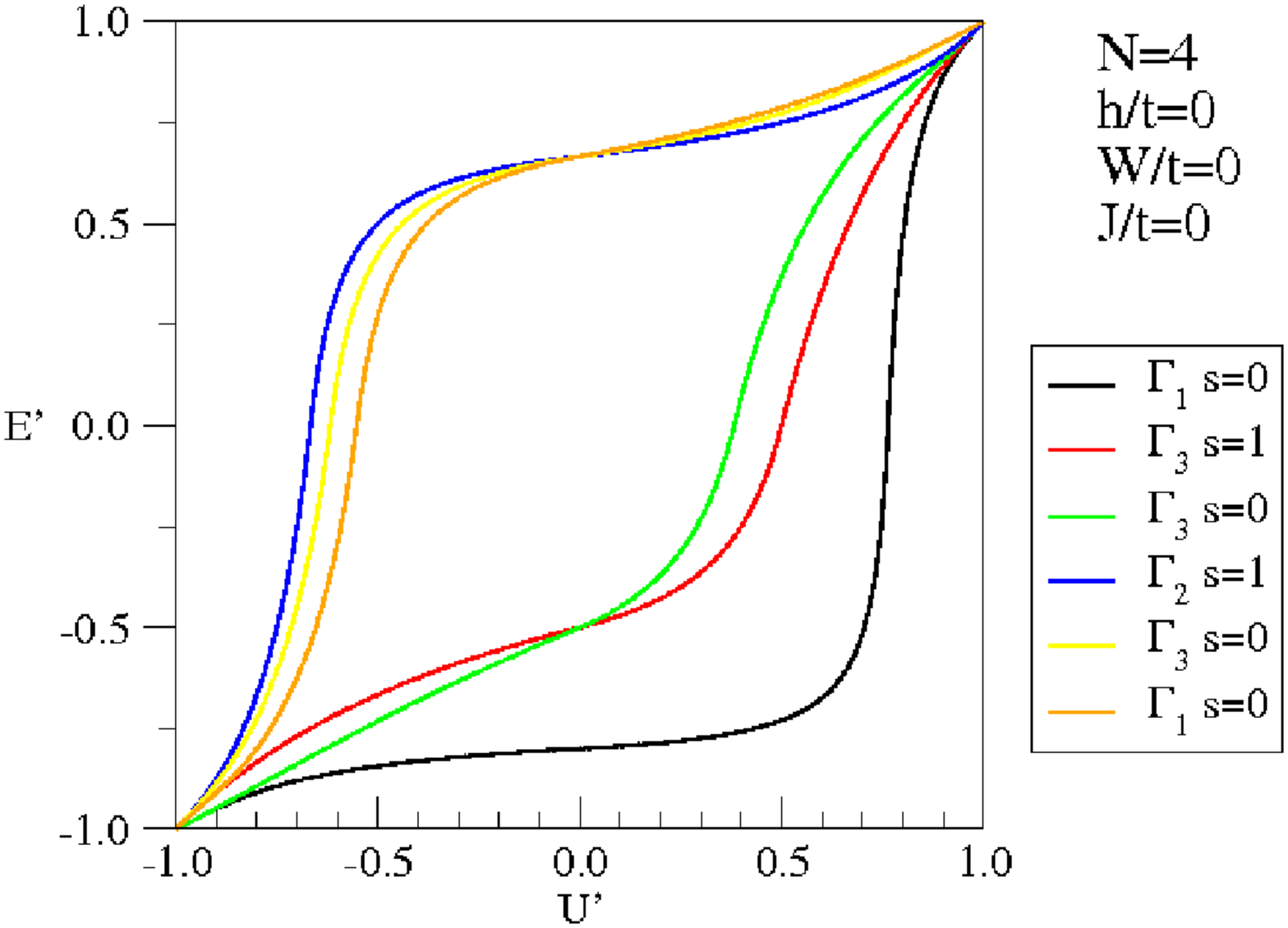}
 \end{minipage}%
 \hfill
 \begin{minipage}[c]{0.3\textwidth}
 \centering 
 \includegraphics[height=32mm]{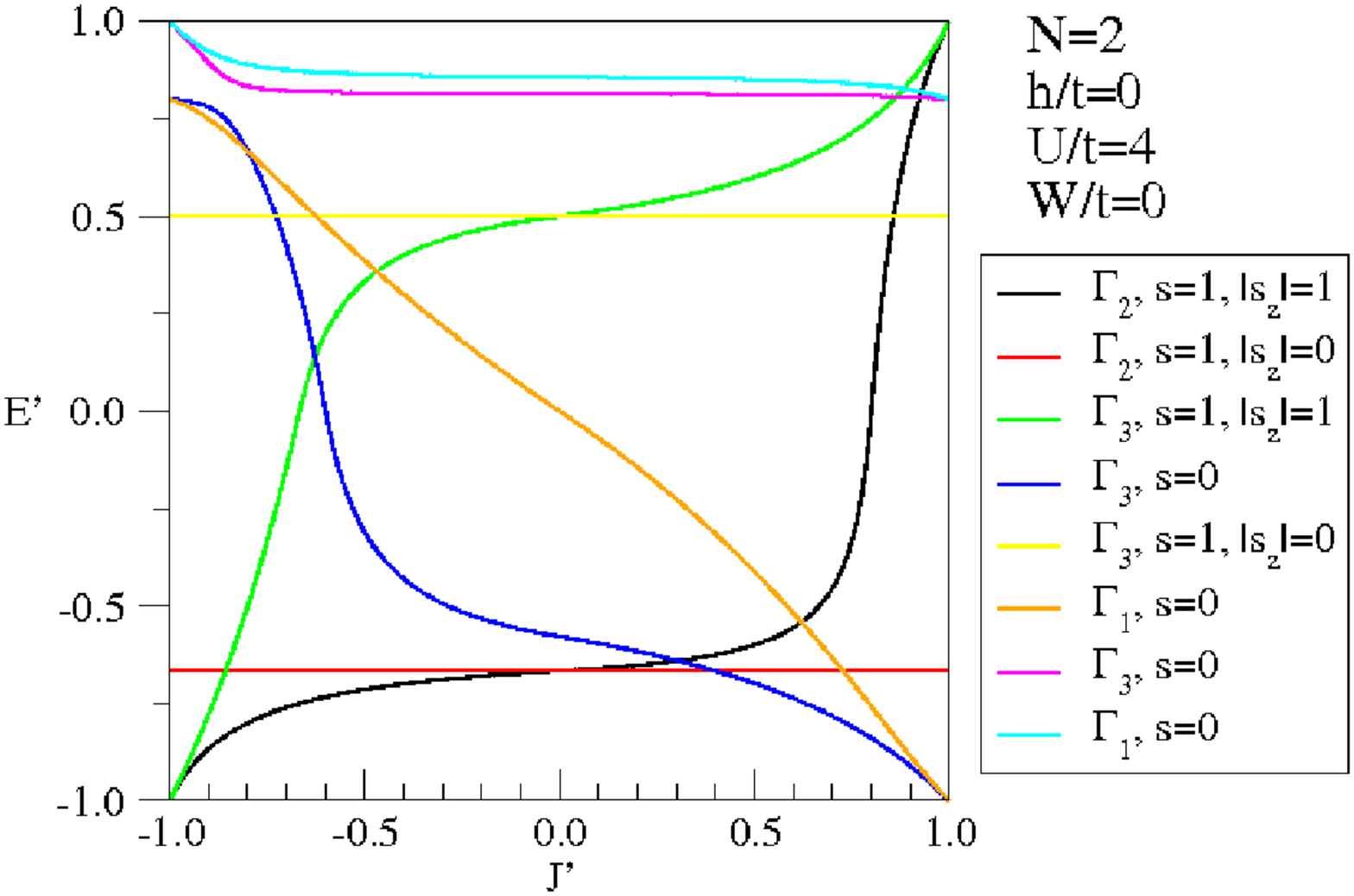}
 \includegraphics[height=32mm]{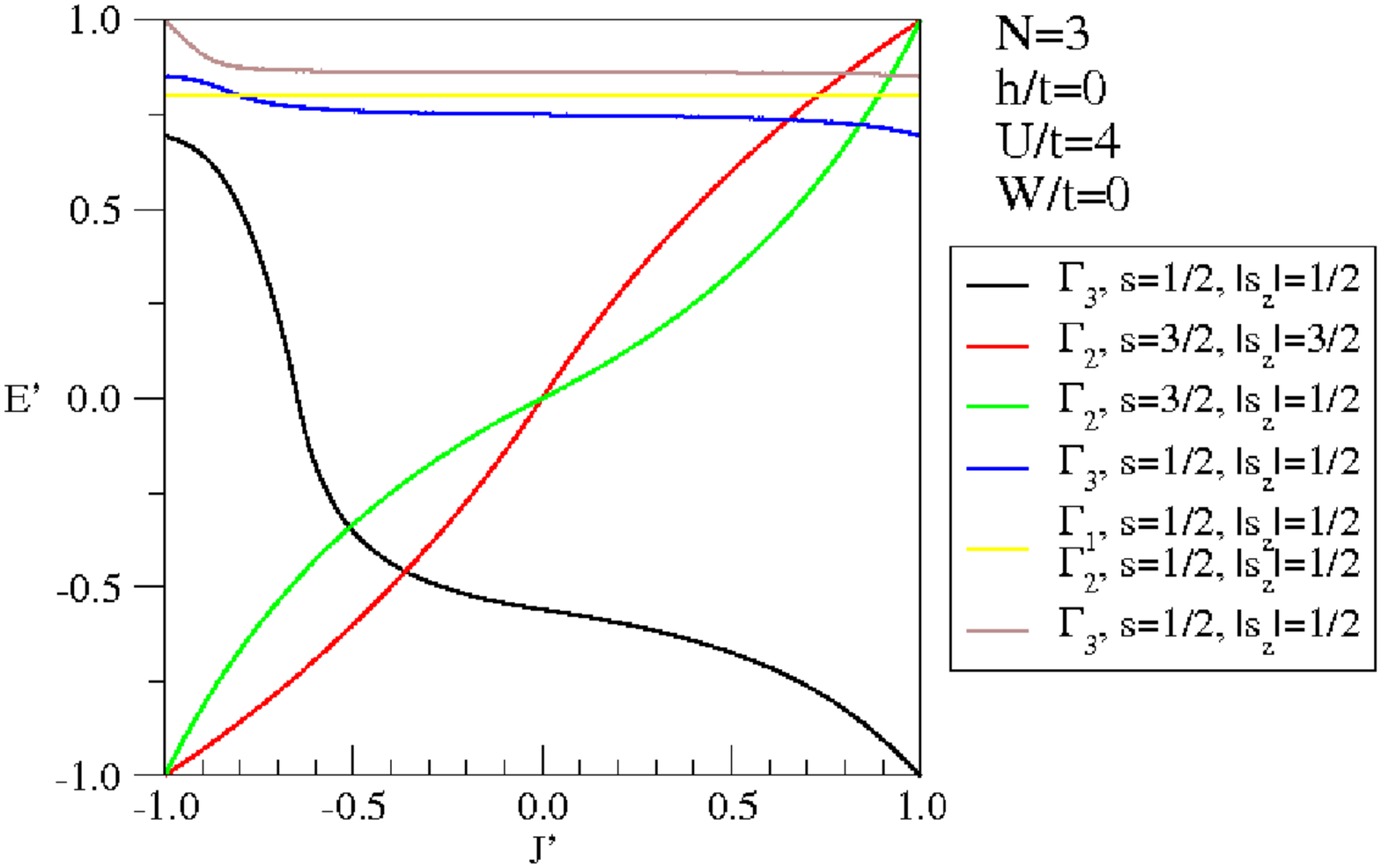}
 \includegraphics[height=32mm]{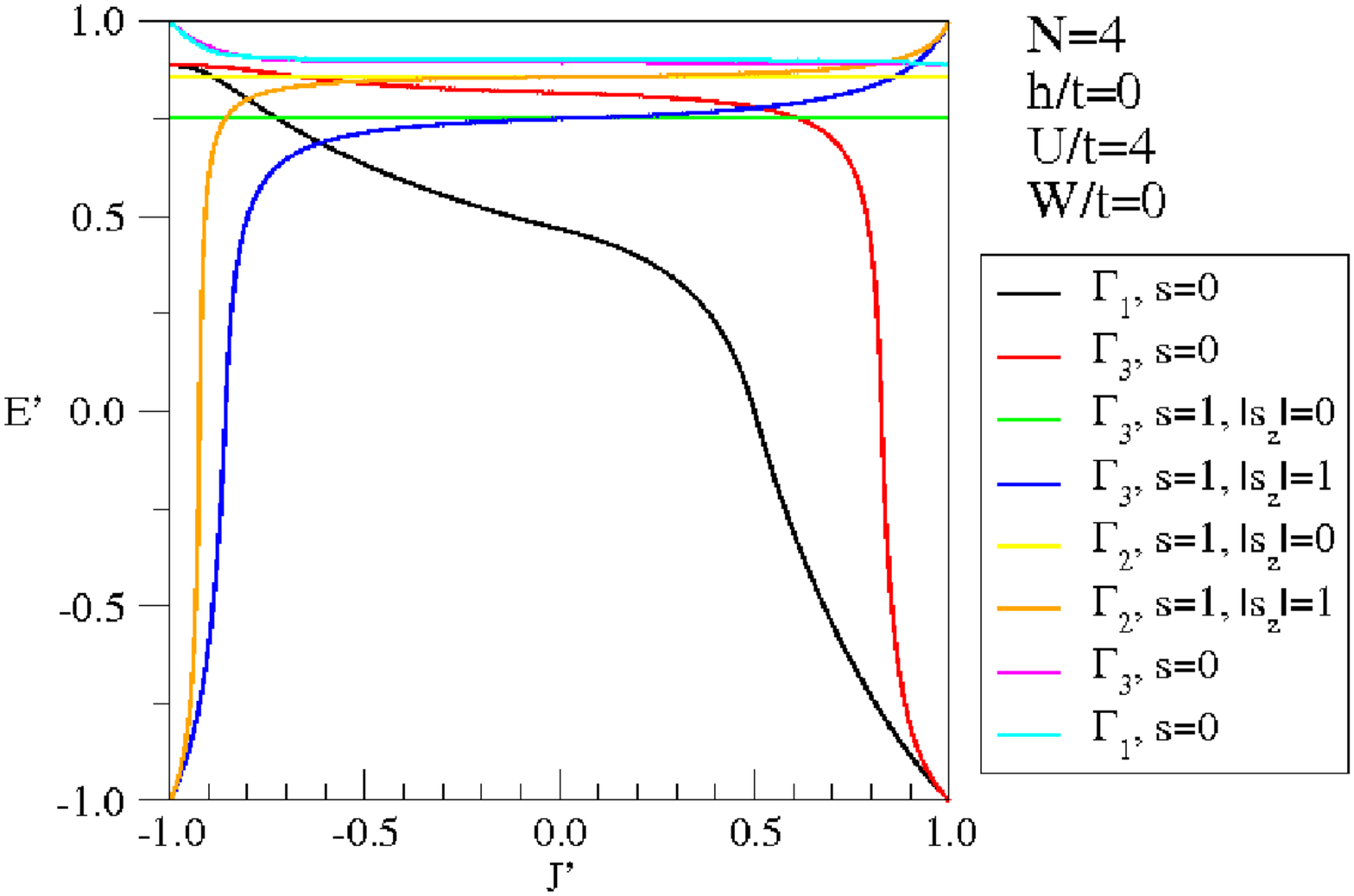}
 \end{minipage} 
\hfill
 \begin{minipage}[c]{0.3\textwidth}
 \centering 
 \includegraphics[height=32mm]{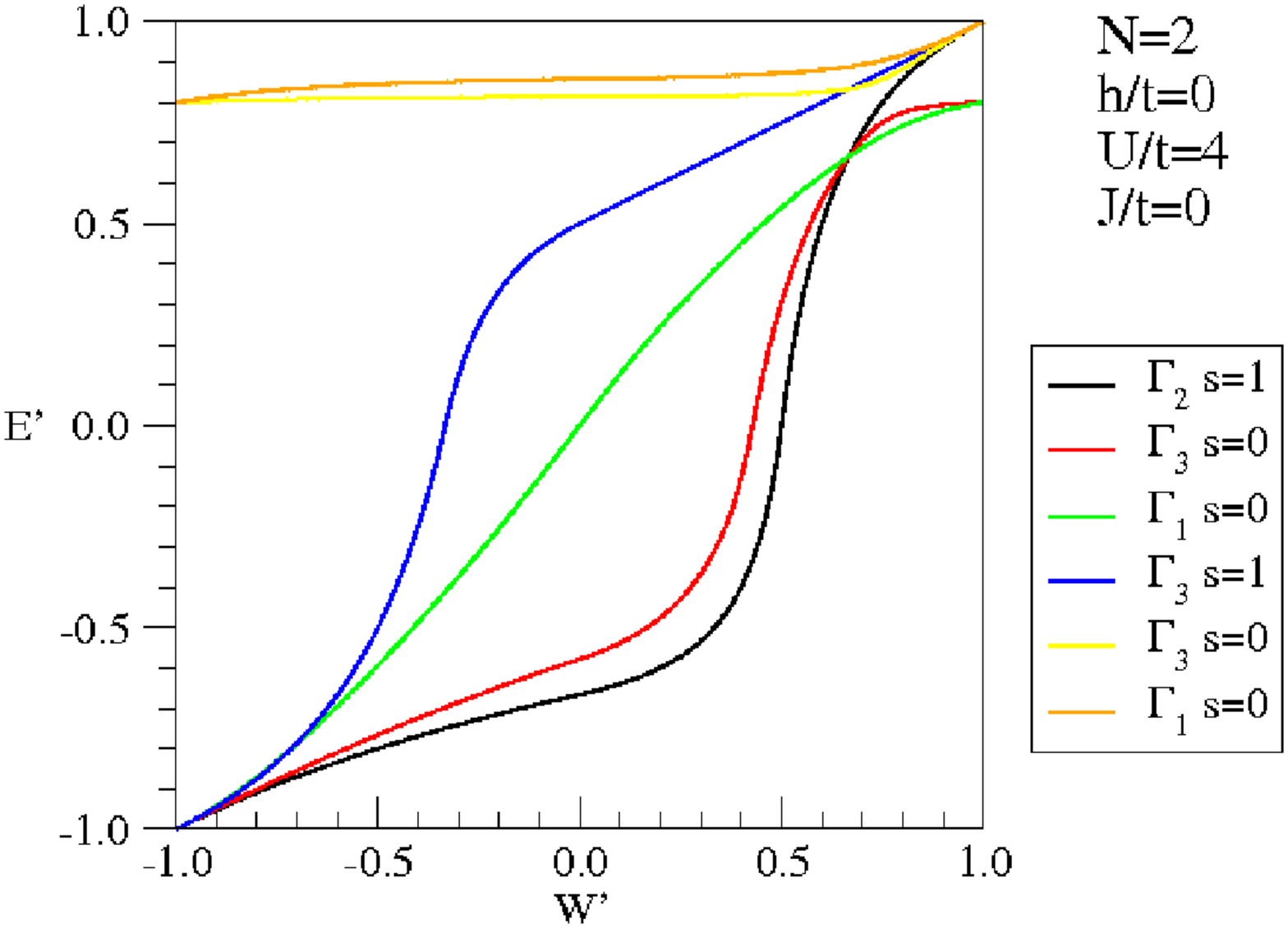}
 \includegraphics[height=32mm]{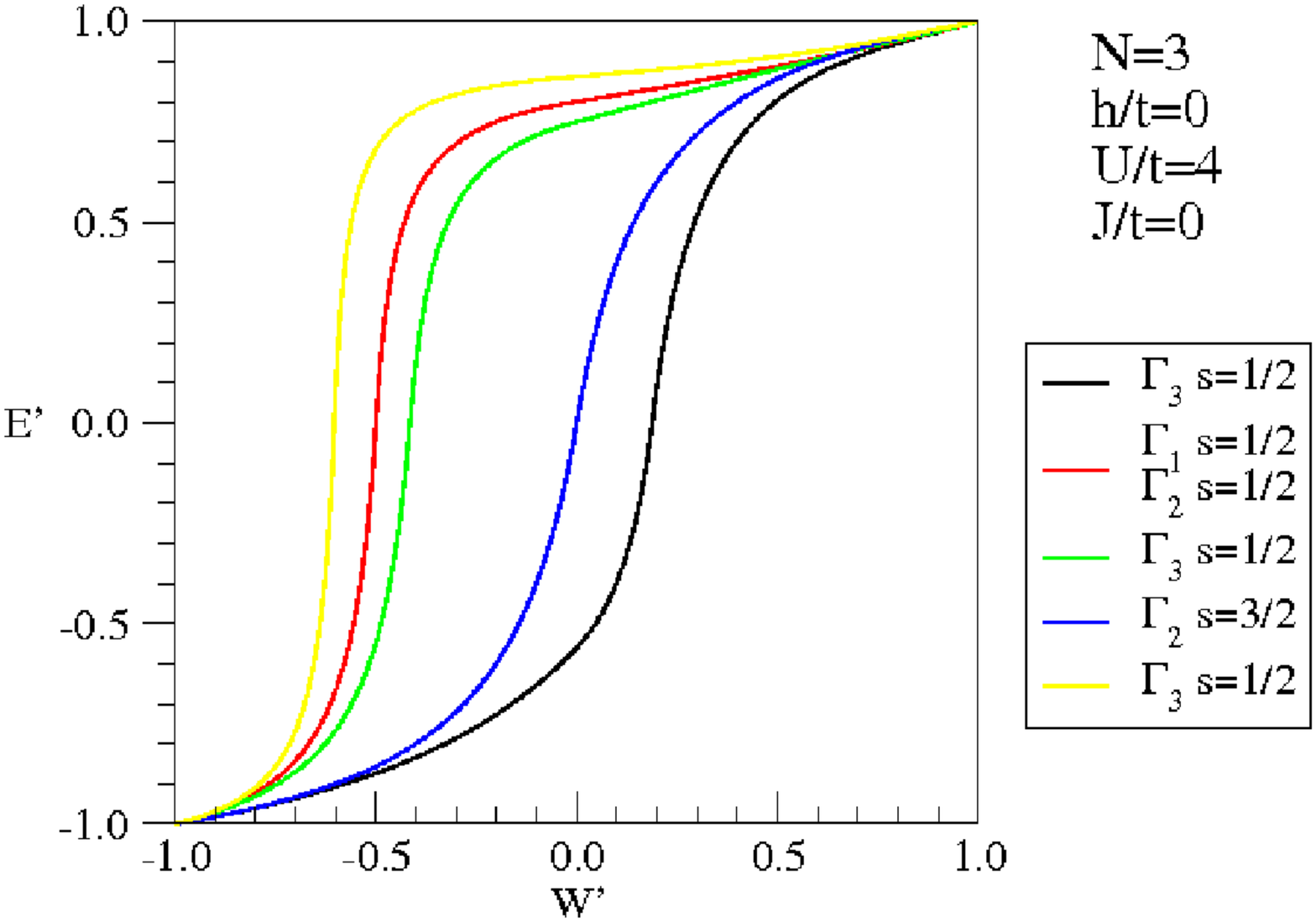}
 \includegraphics[height=32mm]{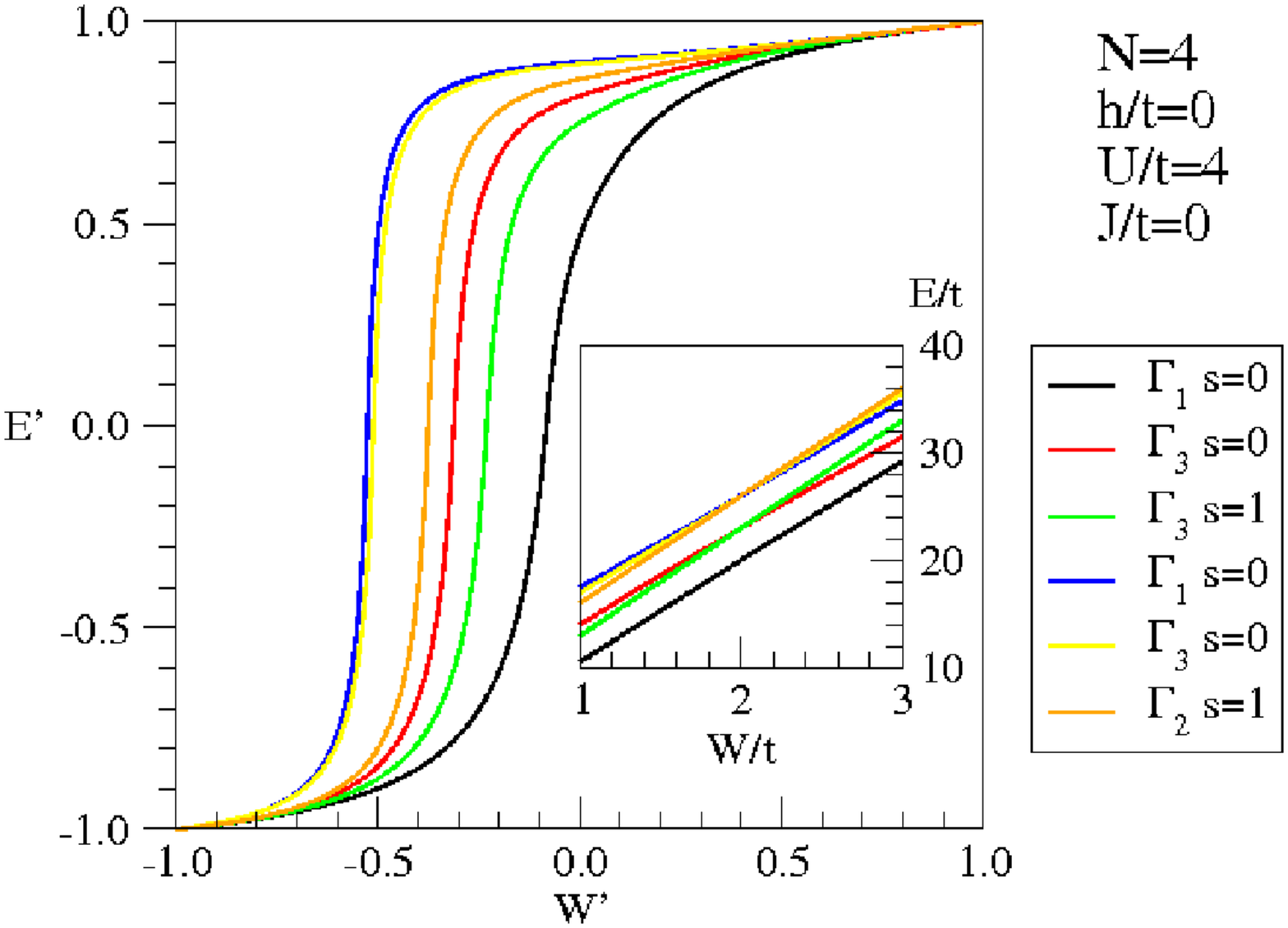}
 \end{minipage} 
\caption{The canonical spectrum (scaled to primed values) 
   of the triangle in dependence on $U'$ (left panel), 
   $J'$ (middle panel), and $W'$ (right panel) for two, three, 
   and four respectively electrons on the cluster. 
   The values of the other parameters are indicated within the figures.
   }
\label{triangleCanonicalSpectra}
\end{figure}

%% file: Parts/figure02.tex
\begin{figure}
 \begin{minipage}[c]{0.3\textwidth}
 \centering 
 \includegraphics[width=45mm]{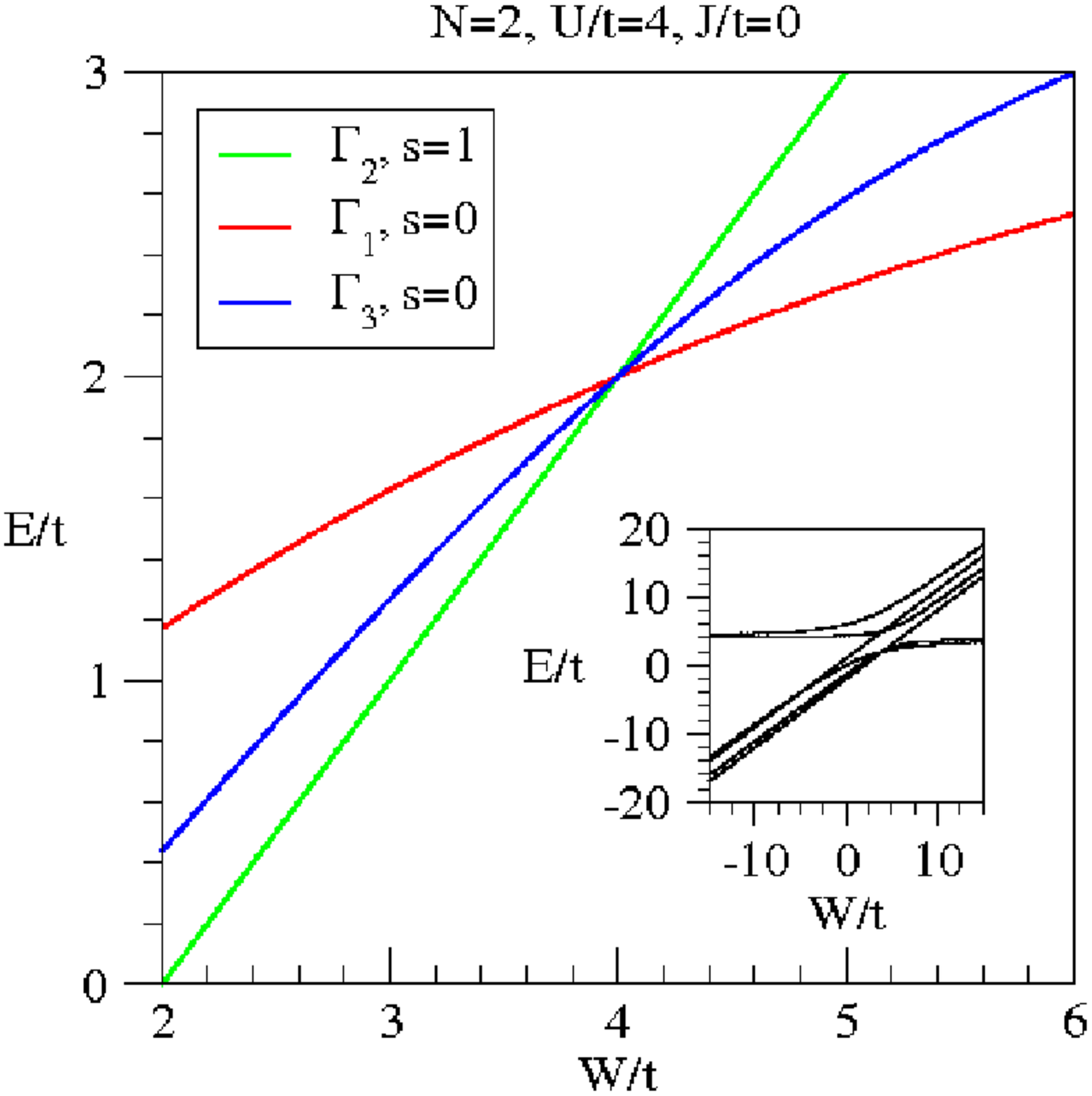}
 \end{minipage}%
 \hfill
 \begin{minipage}[c]{0.3\textwidth}
 \centering 
 \includegraphics[width=45mm]{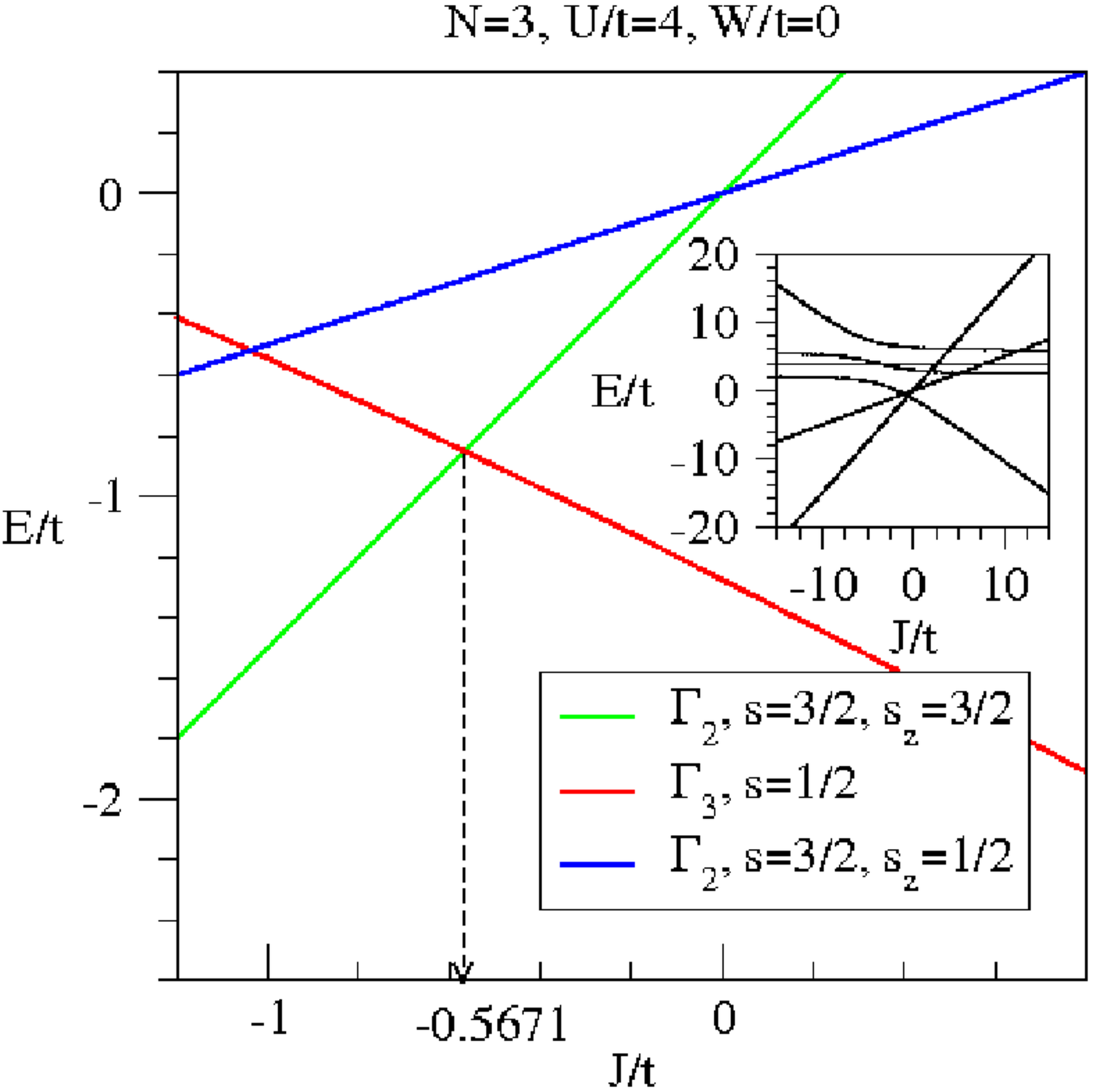}
 \end{minipage} 
 \hfill
 \begin{minipage}[c]{0.3\textwidth}
 \centering 
 \includegraphics[width=45mm]{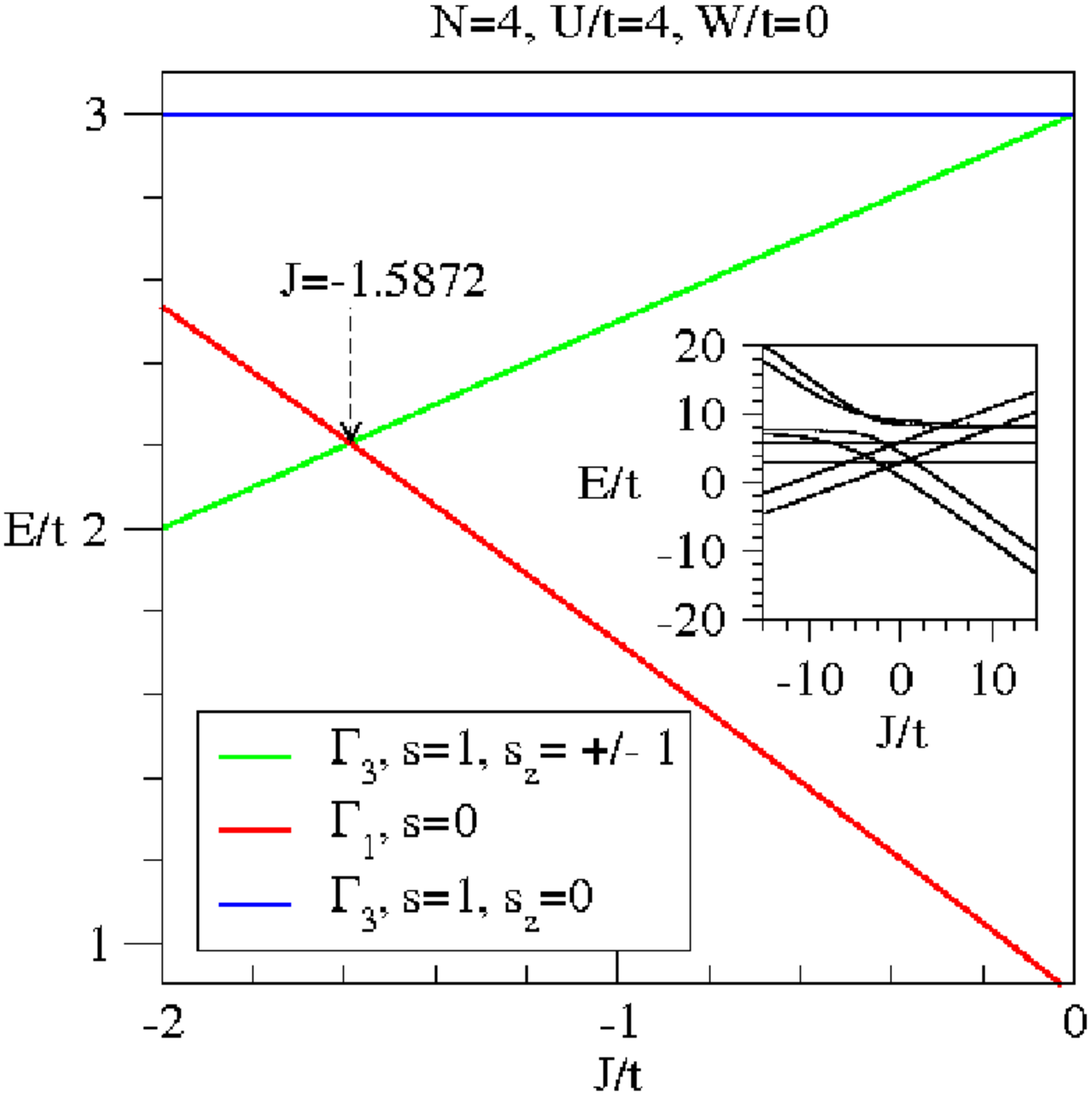}
 \end{minipage} 
\caption{The crossings of the canonical groundstates in dependence of the exchange parameter $J$ for the triangular cluster. The insets sketch the $J$-dependencies for the complete canonical spectra. For details see explicit formula in appendix \ref{appendix1}.}
\label{triangleLevelCrossings}
\end{figure}

%% file: Parts/figure03.tex
\begin{figure}
  \begin{minipage}[c]{0.5\textwidth}
    \includegraphics[height=60mm]{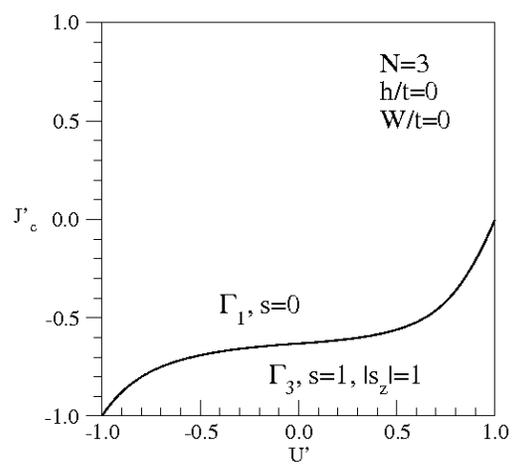}

\end{minipage}  
\caption{$J'_c$ as a function of $U'$ for three electrons on the 
            triangular cluster. }
\label{figureJcvonUN3}
\end{figure}

%% file: Parts/figure04_01.tex
\begin{figure} 
 \begin{minipage}[c]{0.4\textwidth}
 \centering 
 \includegraphics[height=60mm]{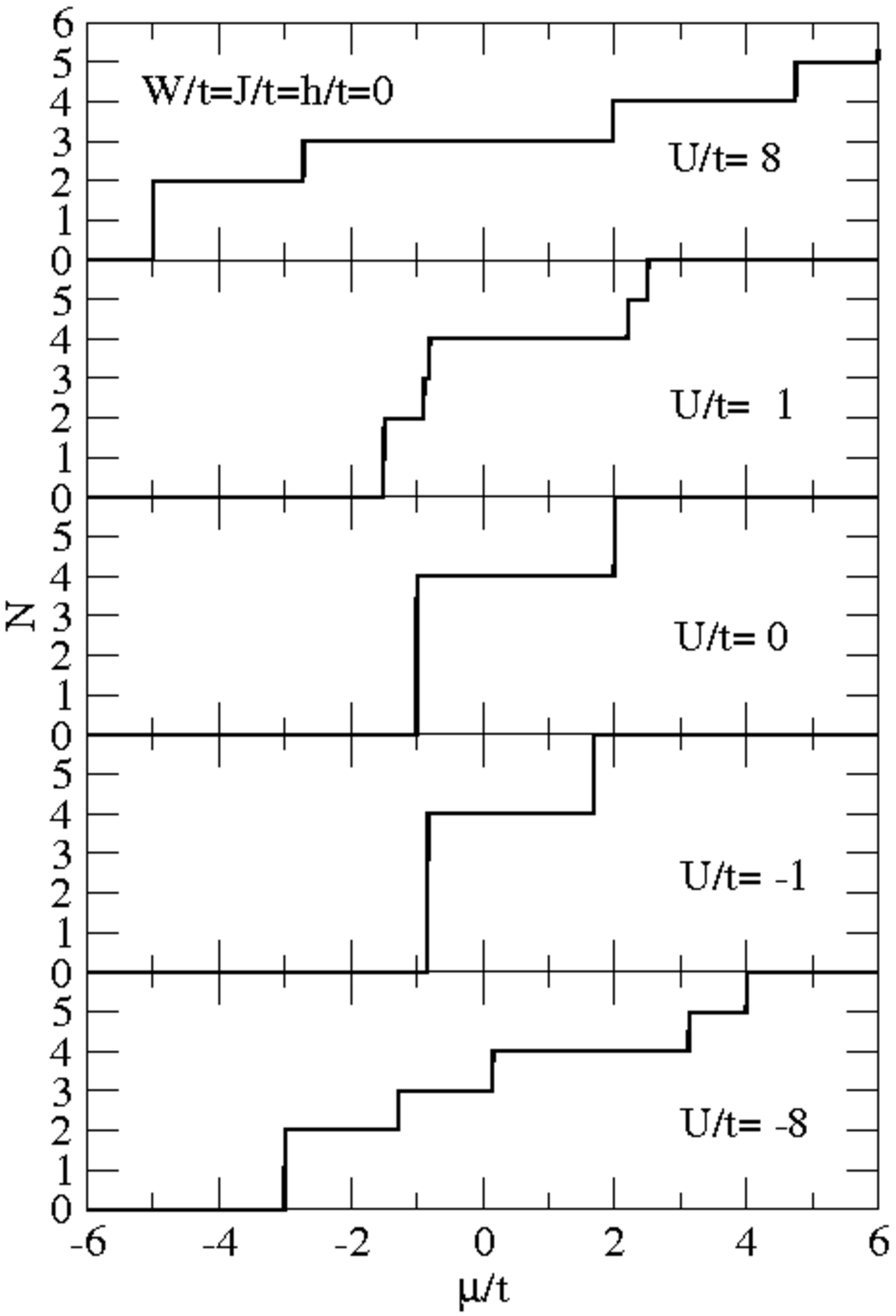}
 \label{figure1a}
 \end{minipage}%
 \hfill
 \begin{minipage}[c]{0.6\textwidth}
 \centering  
 \includegraphics[height=60mm]{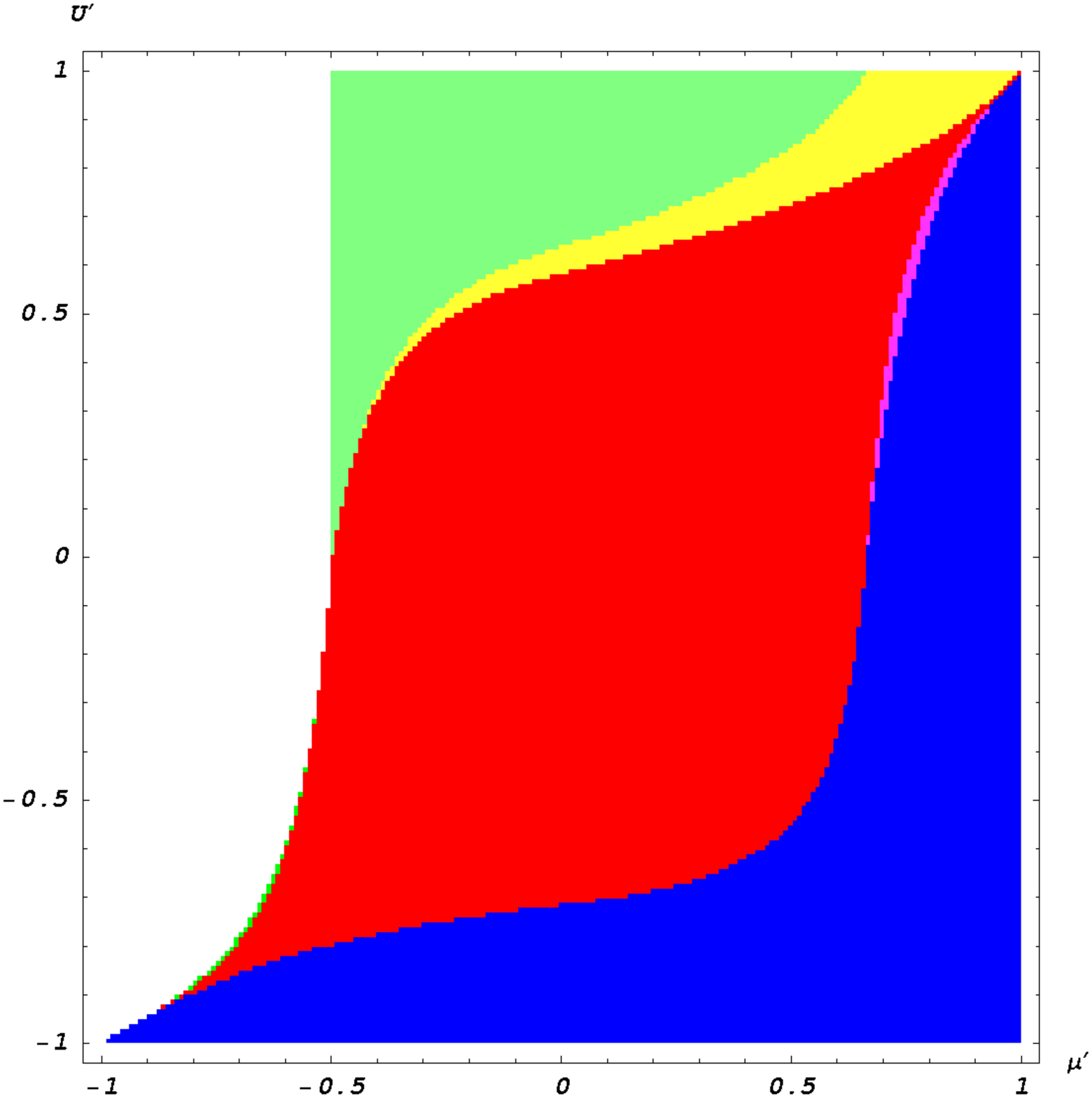}
 \end{minipage}  
 \caption{
 Left panel: $N(\mu)$ of the triangle for different values of the on-site correlation $U$. 
 Right panel: The electron occupation  for the related
 $\mu$-$U$ plane  scaled to primed variables. The meaning of the colours is
according to the palette given in Fig. \ref{palette3site} 
 }
 \label{triangleNvonMueprimeUprime}
\vspace{\floatsep}
 \begin{minipage}[c]{0.5\textwidth}
 \centering  
  \includegraphics[height=60mm]{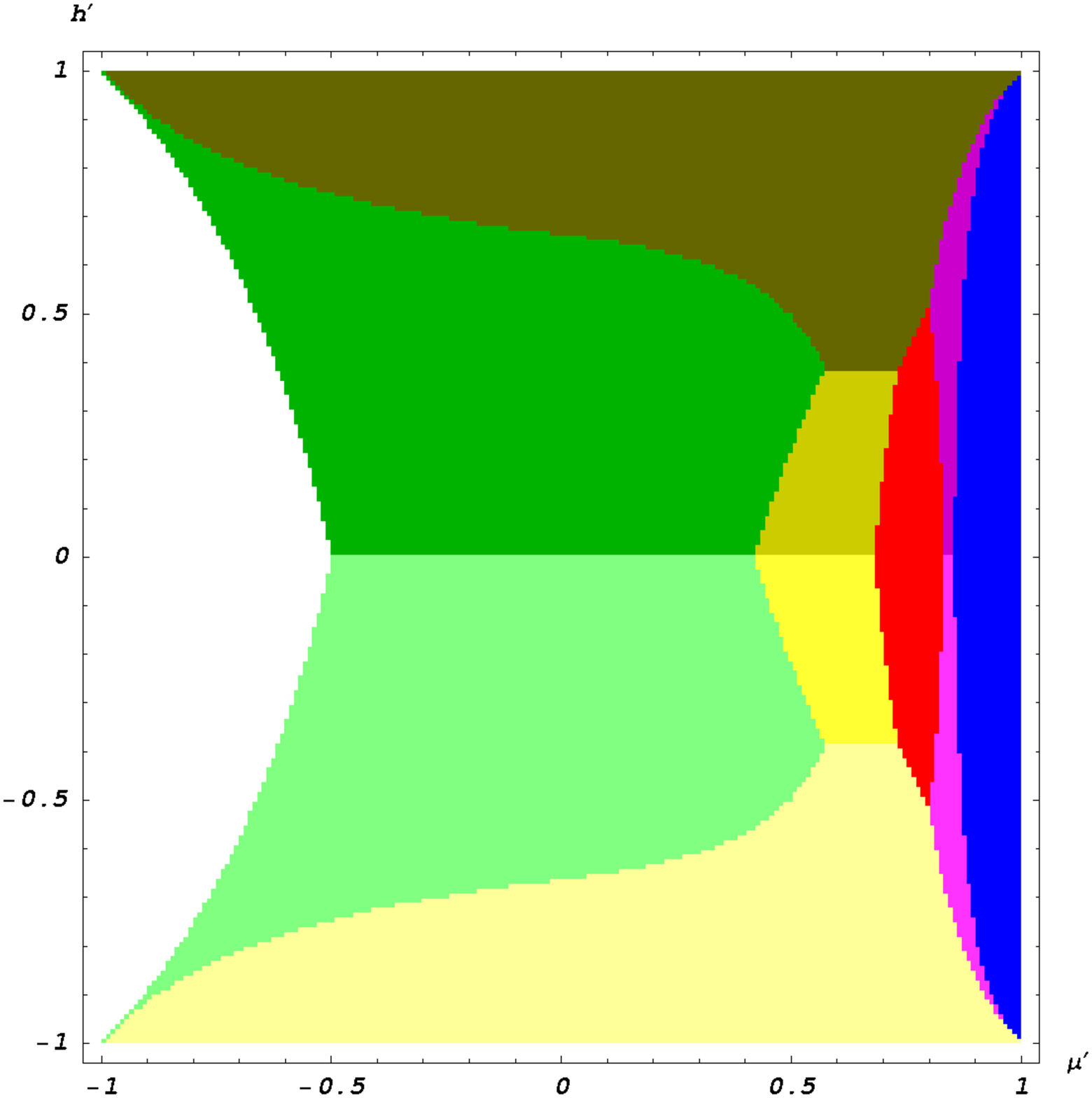}
 \end{minipage}    
 \label{triangleNvonMueh}
 \hfill
 \begin{minipage}[c]{0.5\textwidth}
 \centering 
 \includegraphics[width=60mm]{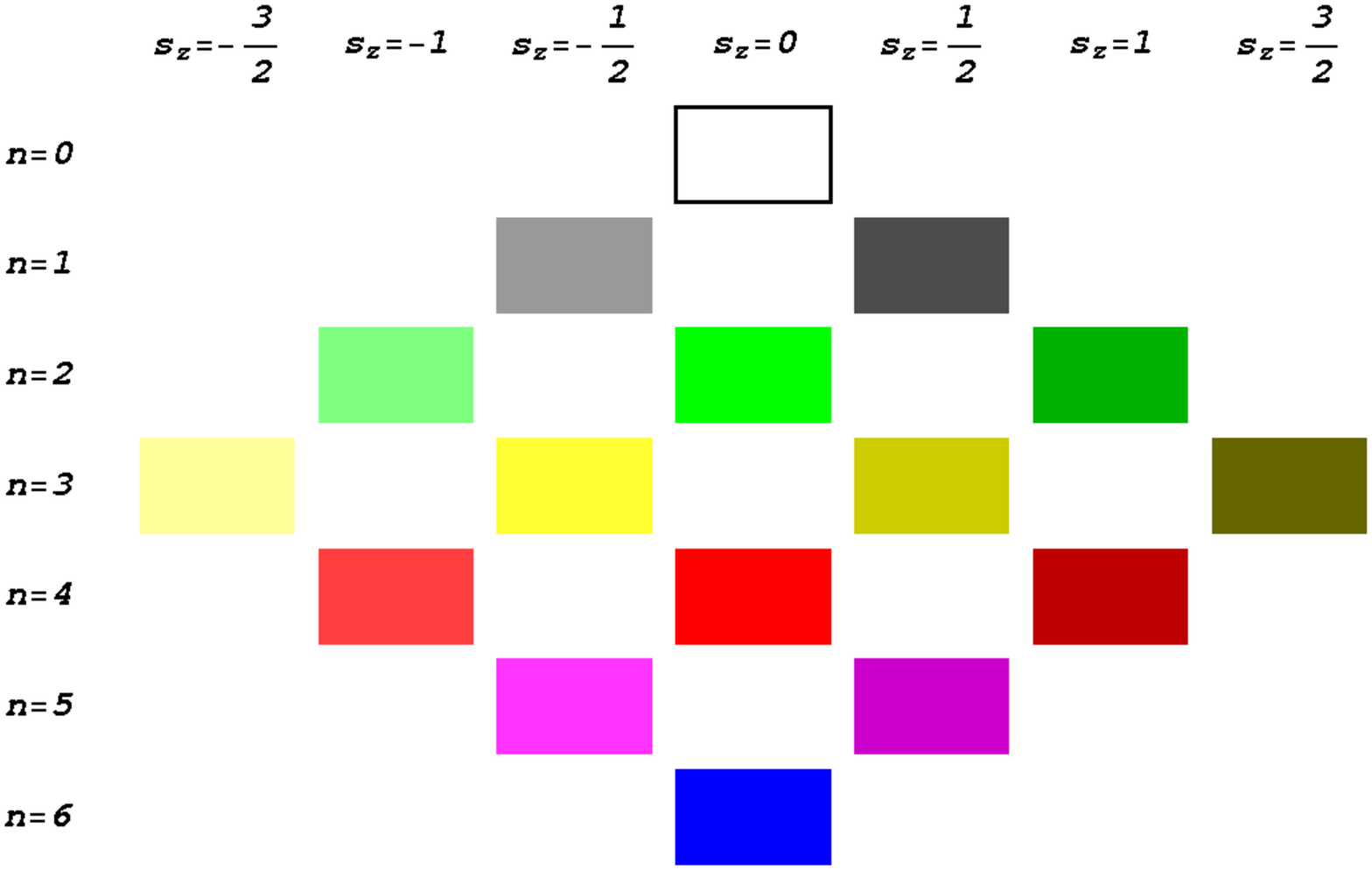}
 \end{minipage}%
\caption{
Left panel: The electron occupation and $S_z$ eigenvalues of the groundstate 
 of the triangular cluster for the complete $\mu$-$h$ plane 
(scaled to primed variables). 
 Right panel: The palette used here and in Figs. 
\ref{triangleNvonMueprimeUprime}, \ref{triangleU4NvonMueprimeJprime}, and \ref{triangleU4NvonMueprimeWprime}
to differ the groundstates with respect of their electron occupation and $S_z$ eigenvalues.
}
\label{palette3site}
\end{figure}

%% file: Parts/figure04_23.tex
\begin{figure} 
\begin{minipage}[c]{0.4\textwidth}
 \centering 
 \includegraphics[height=60mm]{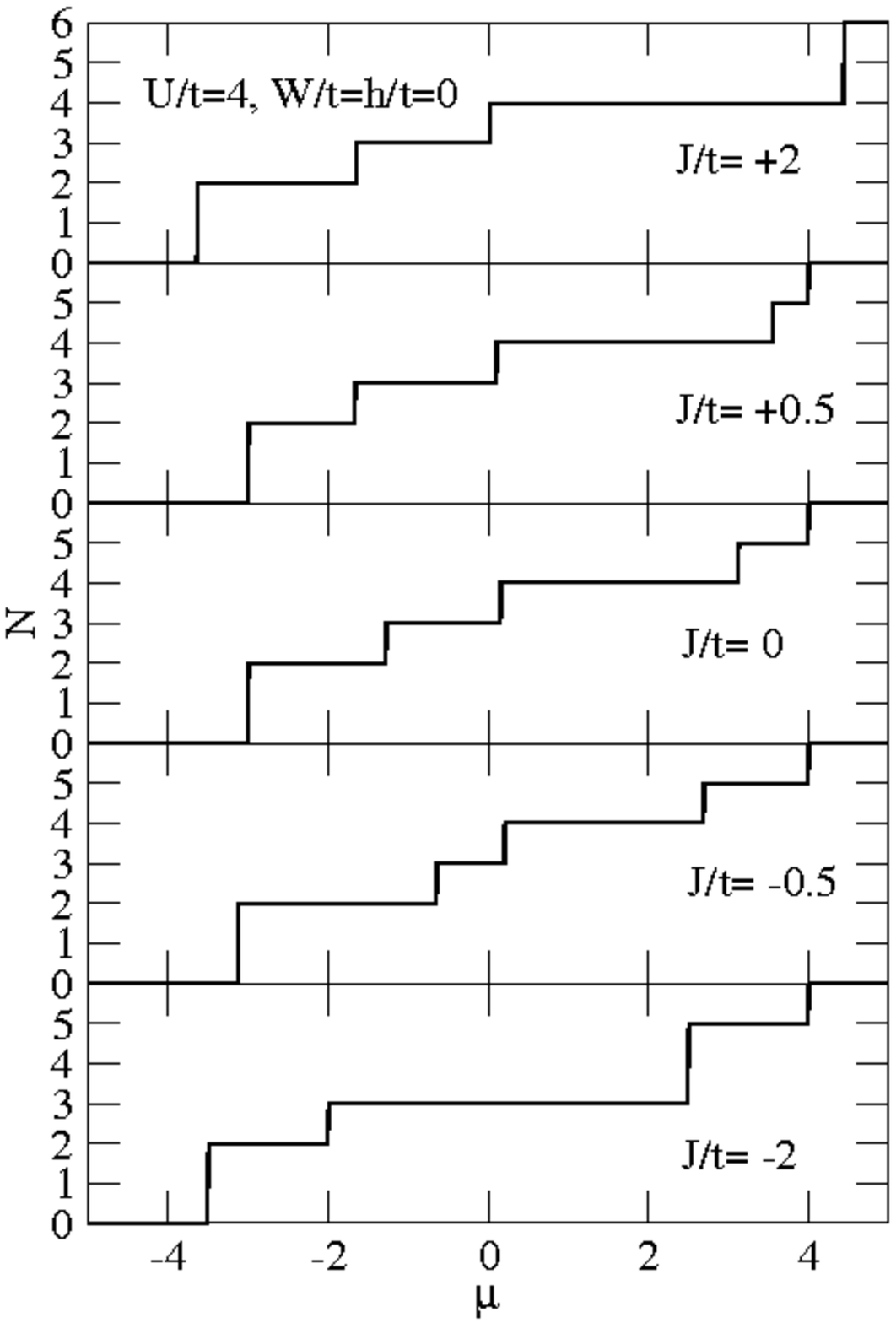}
 \end{minipage}%
 \hfill
 \begin{minipage}[c]{0.6\textwidth}
 \centering  
  \includegraphics[height=60mm]{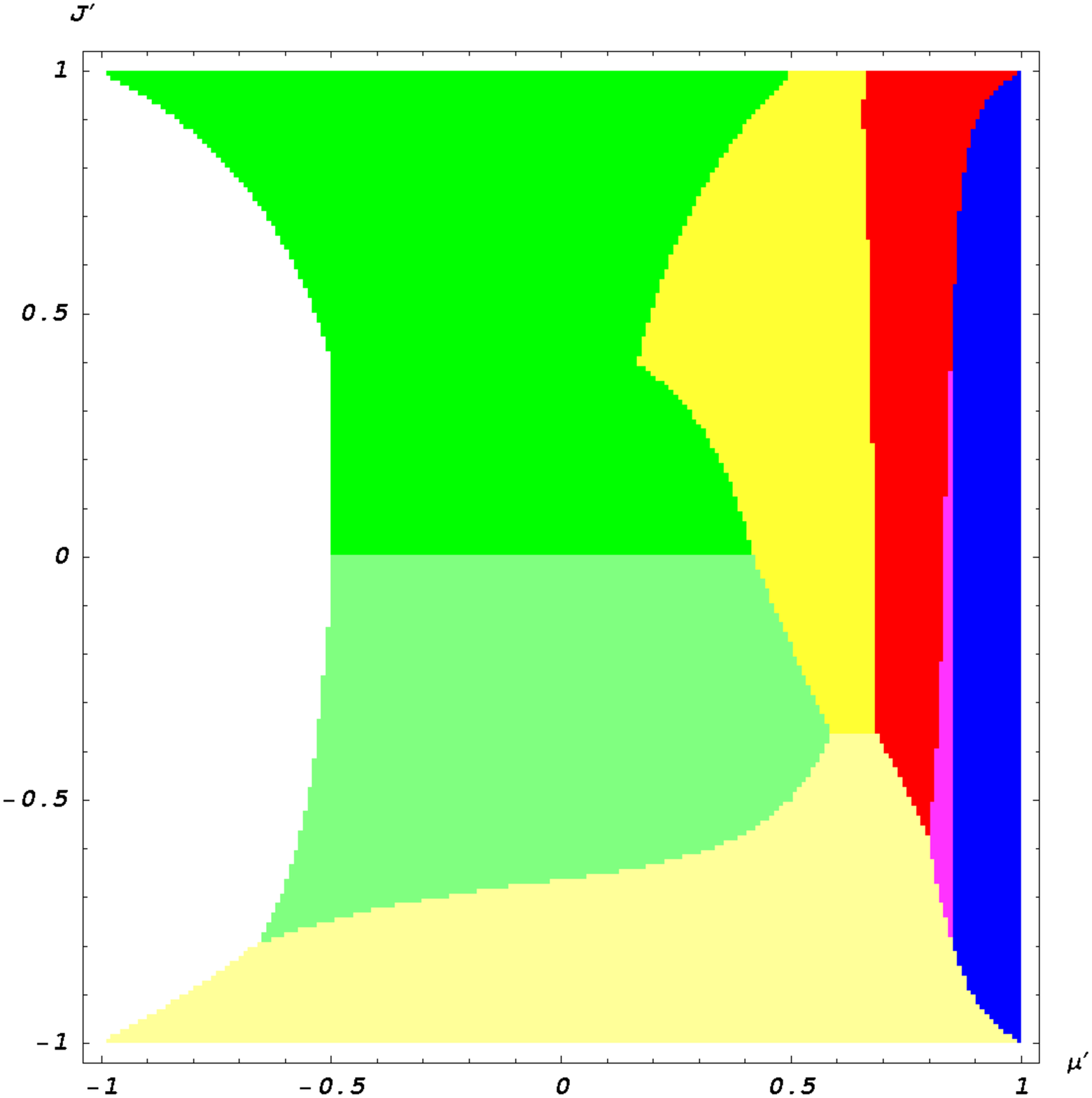}
 \end{minipage}    
 \caption{
 Left panel: $N(\mu)$ of the triangle for different values of the exchange 
 parameter $J$ for $U/t=4$
 Right panel: The electron occupation and $S_z$ eigenvalues of the groundstate 
 for the related $\mu$-$J$ plane scaled to primed variables. 
 The meaning of the colours is the same as in Fig. \ref{palette3site}. 
 } 
 \label{triangleU4NvonMueprimeJprime}
\vspace{\floatsep}
 \begin{minipage}[c]{0.4\textwidth}
  \centering 
  \includegraphics[height=60mm]{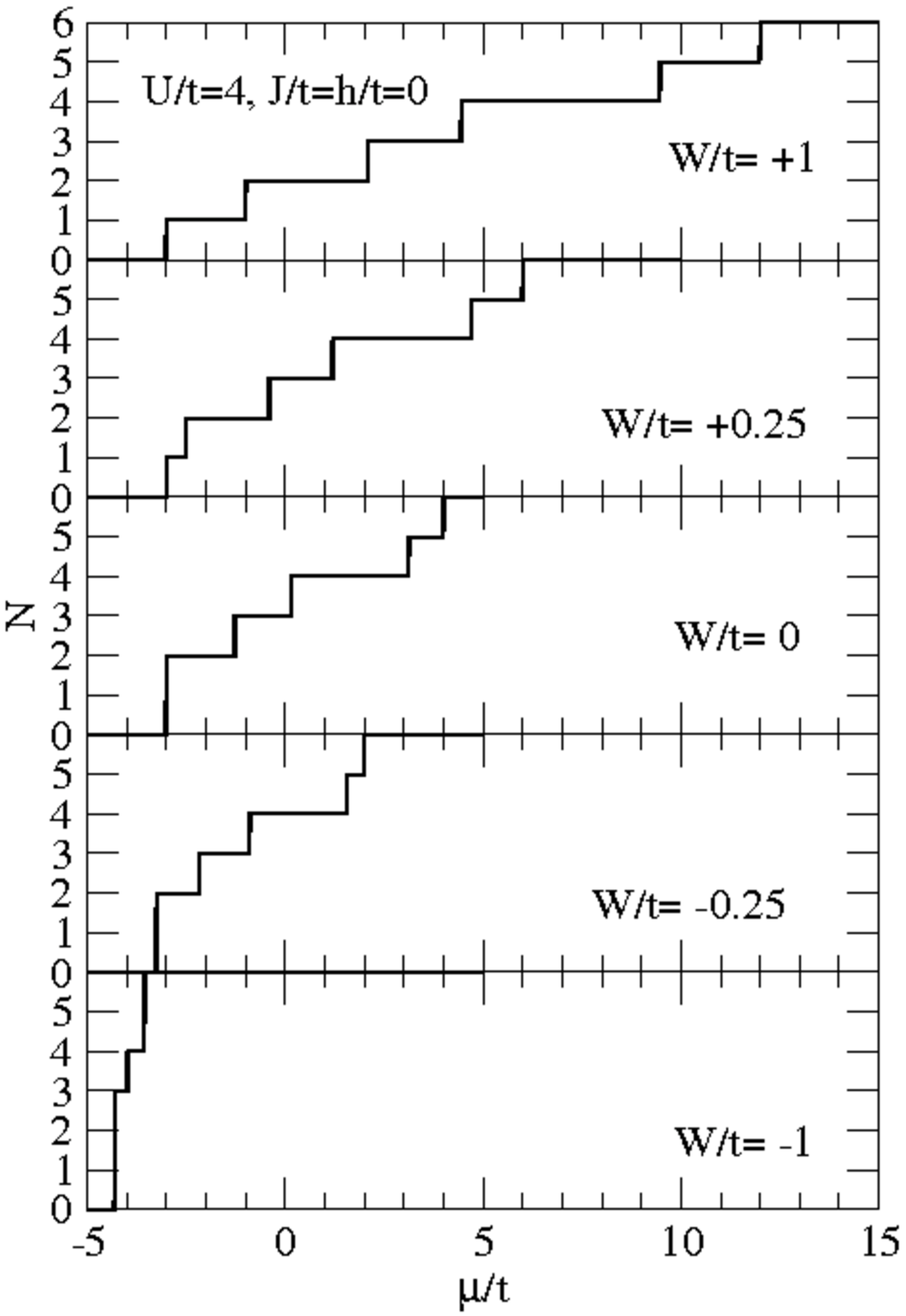}
 \end{minipage}%
 \hfill
 \begin{minipage}[c]{0.6\textwidth}
 \centering  
 \includegraphics[height=60mm]{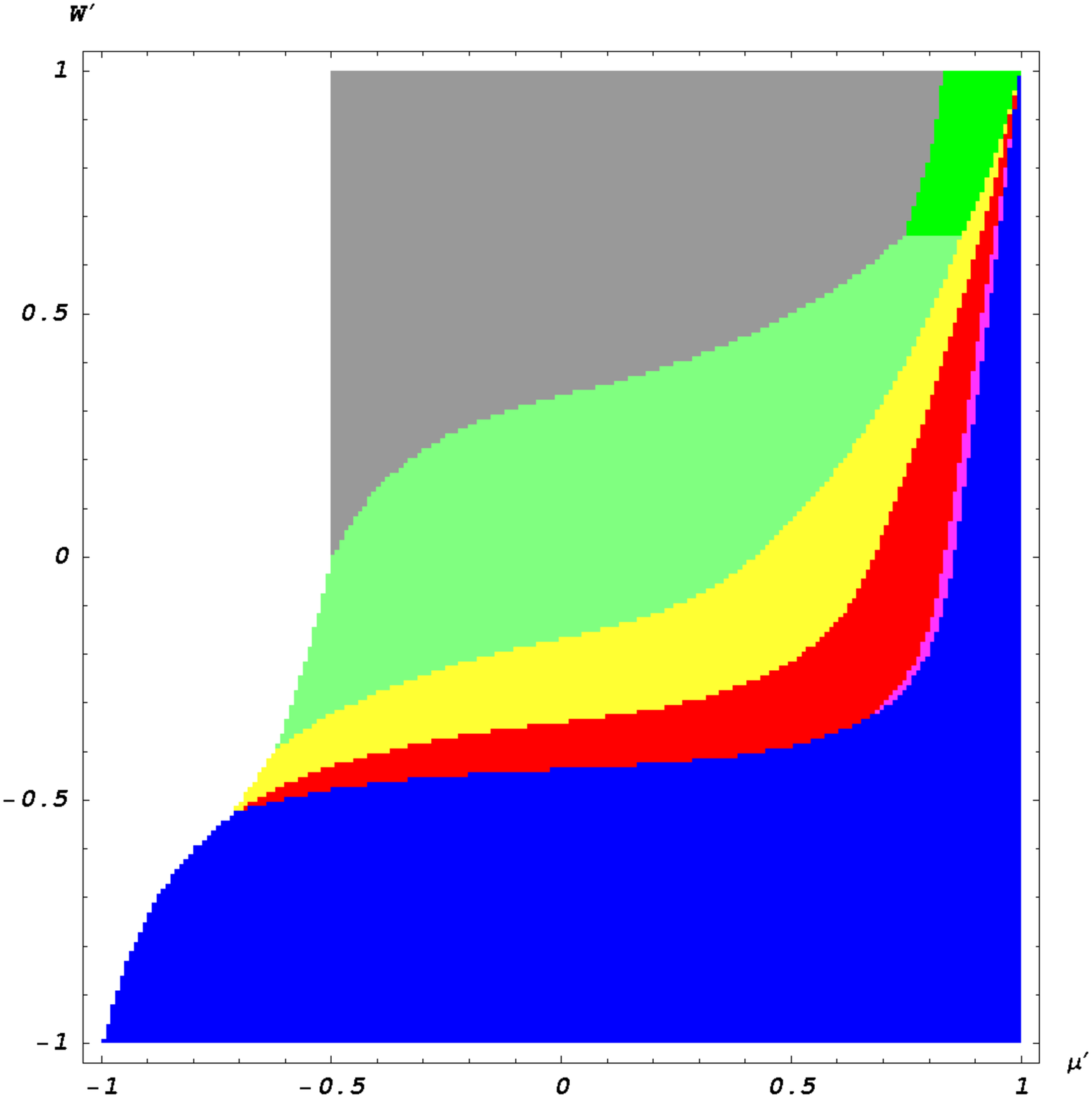}
 \end{minipage}  
 \caption{
  Left panel: $N(\mu)$ of the triangle for different values of the 
  nearest-neighbour Coulomb interaction $W$ for $U/t=4$. 
  Right panel: The electron occupation and $S_z$ eigenvalues 
  of the groundstate 
  for the related  $\mu$-$W$ plane scaled to primed variables.
  The meaning of the colours is the same as in Fig. \ref{palette3site}. 
  }
 \label{triangleU4NvonMueprimeWprime}
\end{figure}

%% file: Parts/figure05.tex
\begin{figure}
 \begin{minipage}[c]{0.3\textwidth}
 \centering 
 \includegraphics[height=50mm]{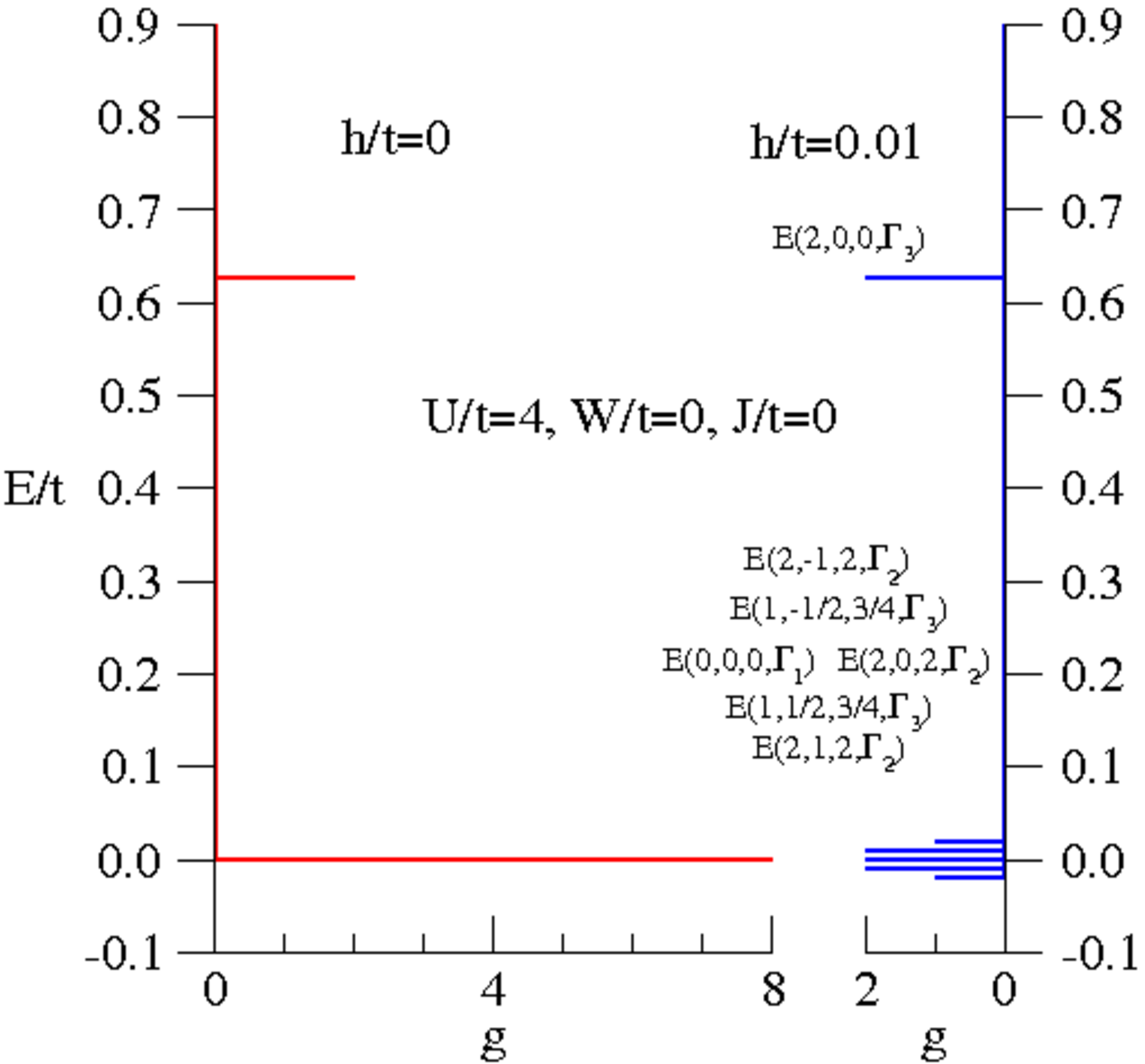}
 \end{minipage}%
 \hfill
 \begin{minipage}[c]{0.6\textwidth}
 \centering 
 \includegraphics[height=50mm]{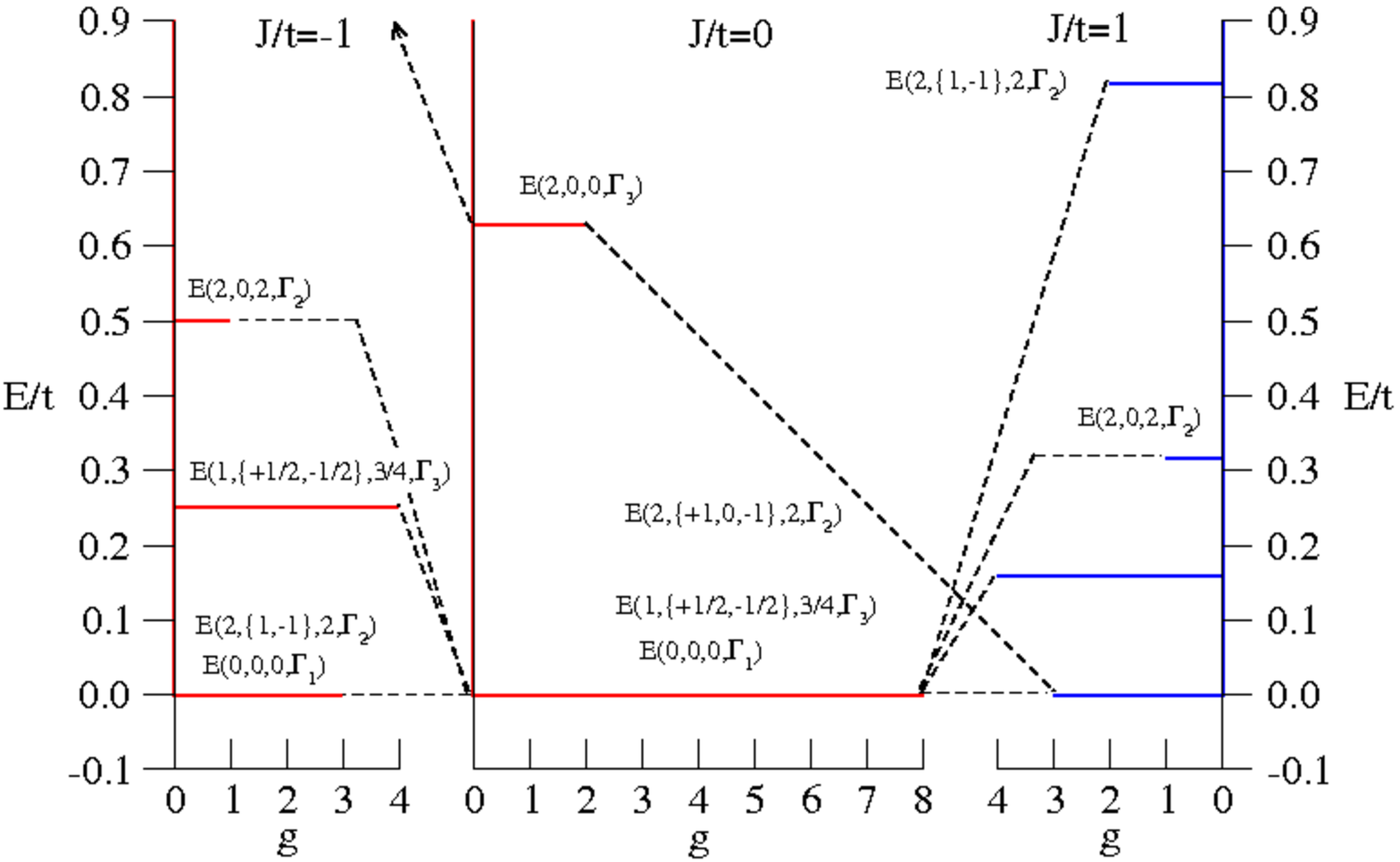}
  \end{minipage} 
\caption{The lowest part of the grand canonical spectrum for the triangle with and without a magnetic
field (left panel) and with ferromagnetic and antiferromagnetic exchange taken into account (right panel). The energylevels are labeled by the magnetic quantum numbers and the irreducible representation of the point group.}
\label{triangleLowestGrandCanonicalLevels}
\end{figure}

%% file: Parts/figure06.tex
\begin{figure}
 \begin{minipage}[t]{0.5\textwidth}
 \centering  
 \includegraphics[height=40mm]{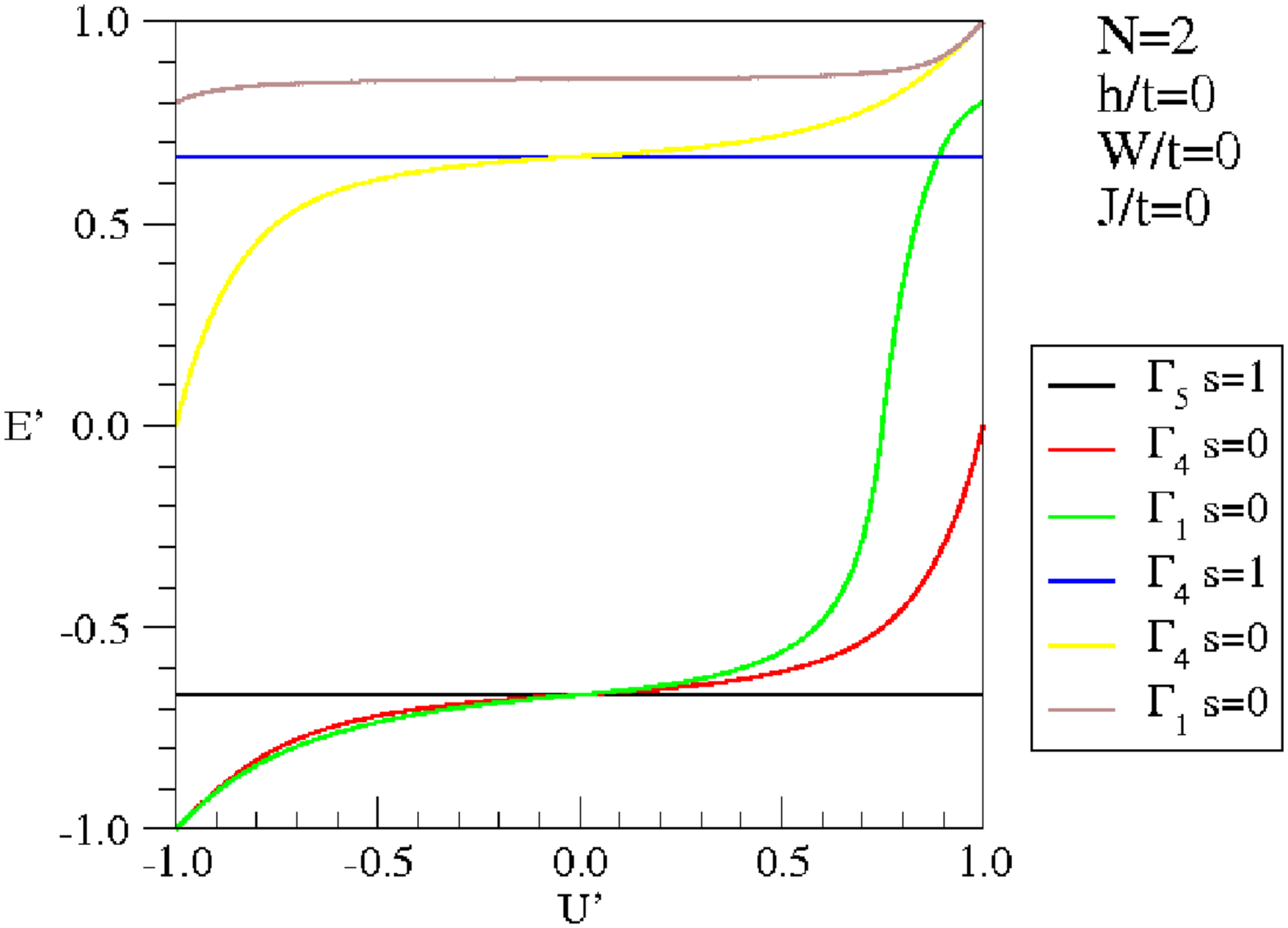}
 \includegraphics[height=40mm]{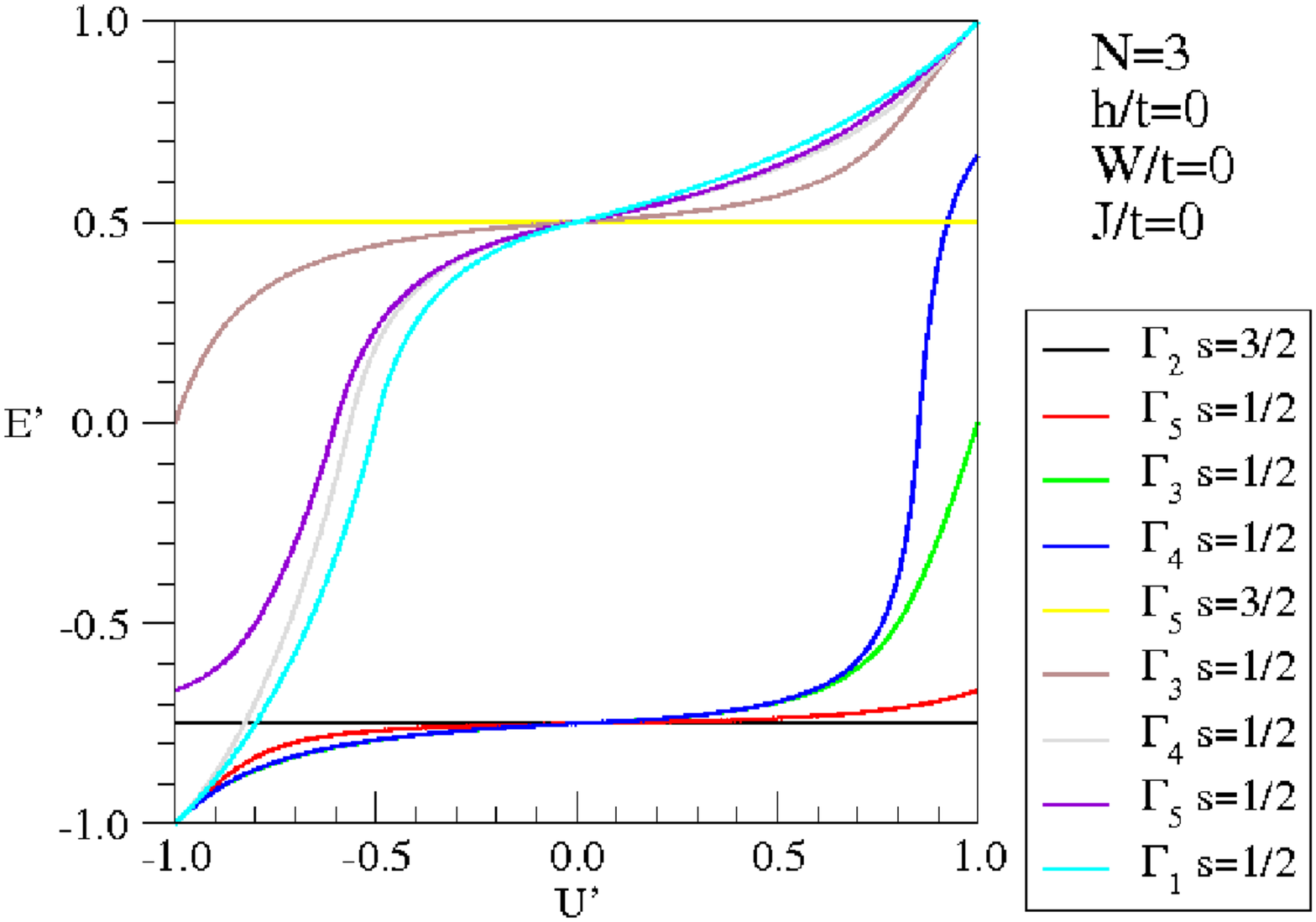}
 \includegraphics[height=40mm]{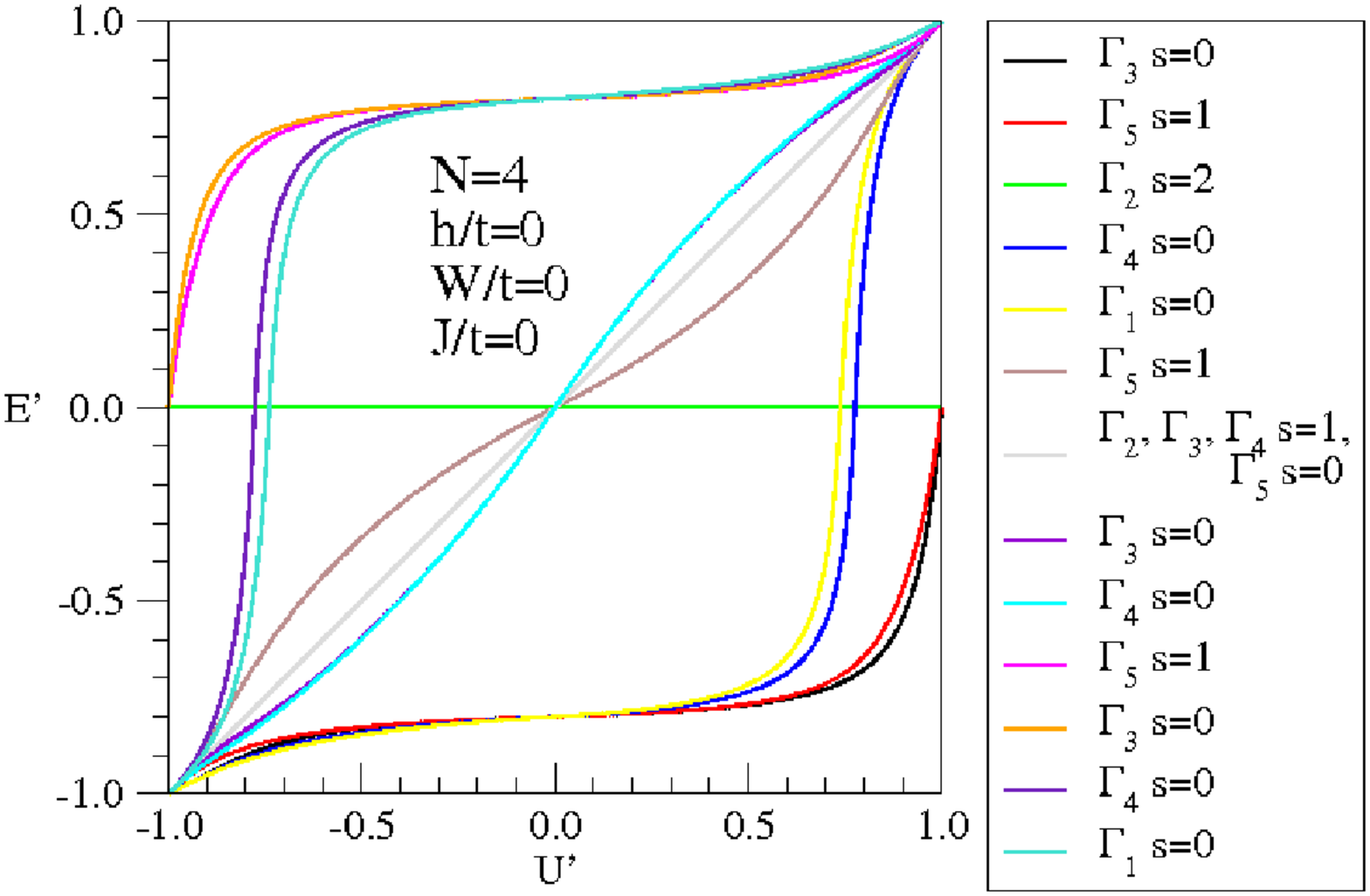}
 \includegraphics[height=40mm]{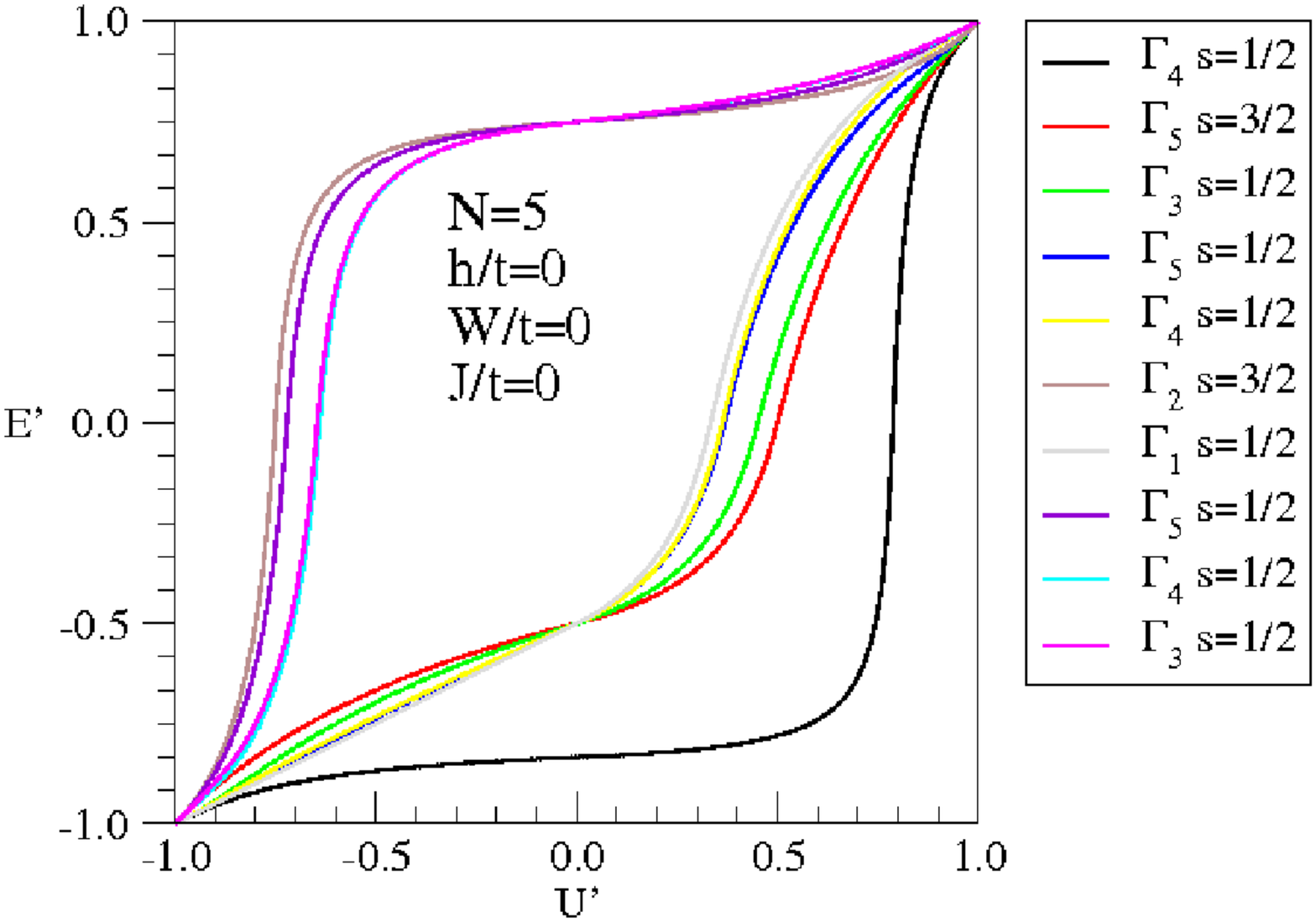}
 \includegraphics[height=40mm]{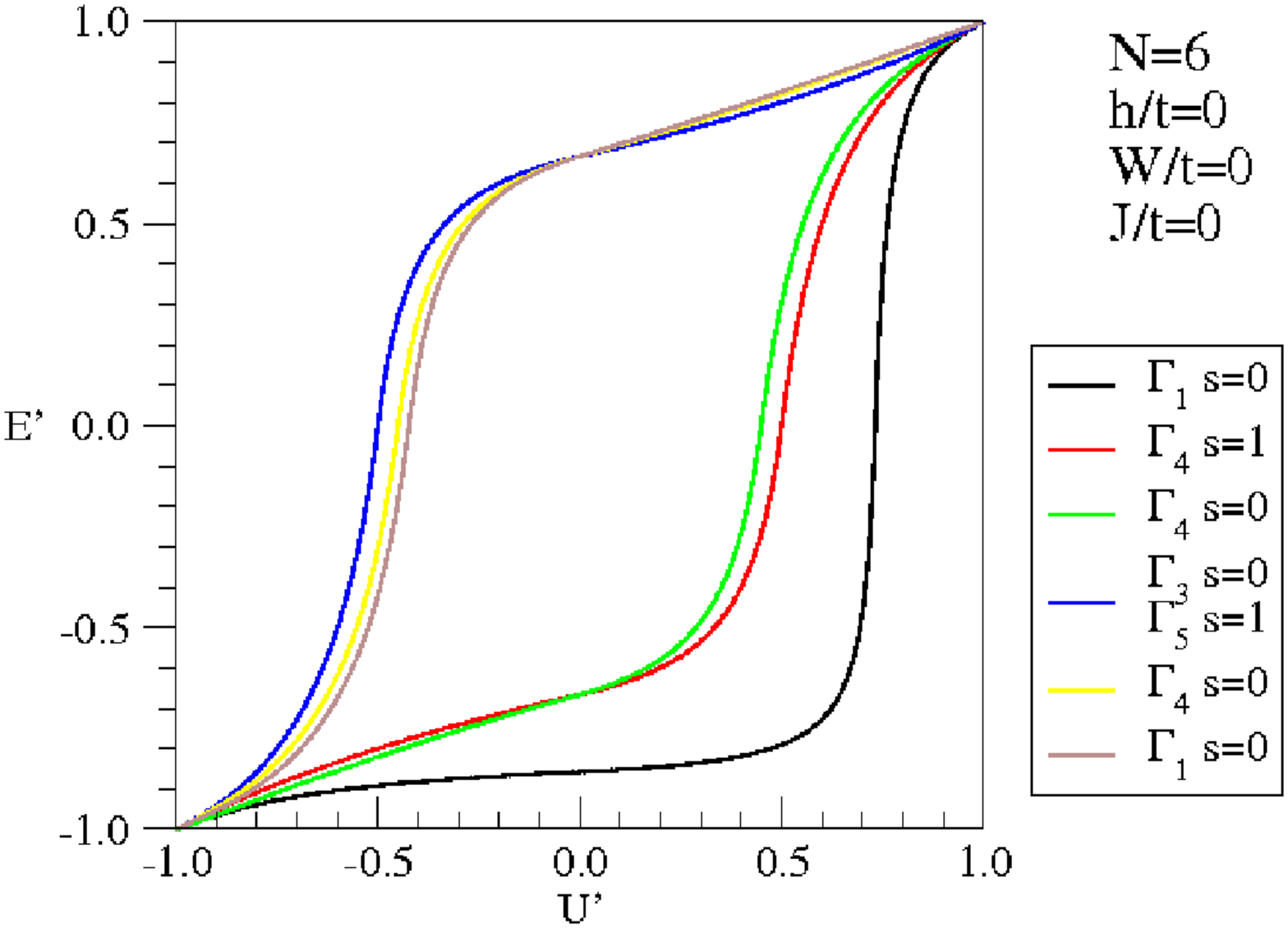}
\end{minipage}  
\begin{minipage}[t]{0.5\textwidth}
 \centering 
 \includegraphics[height=40mm]{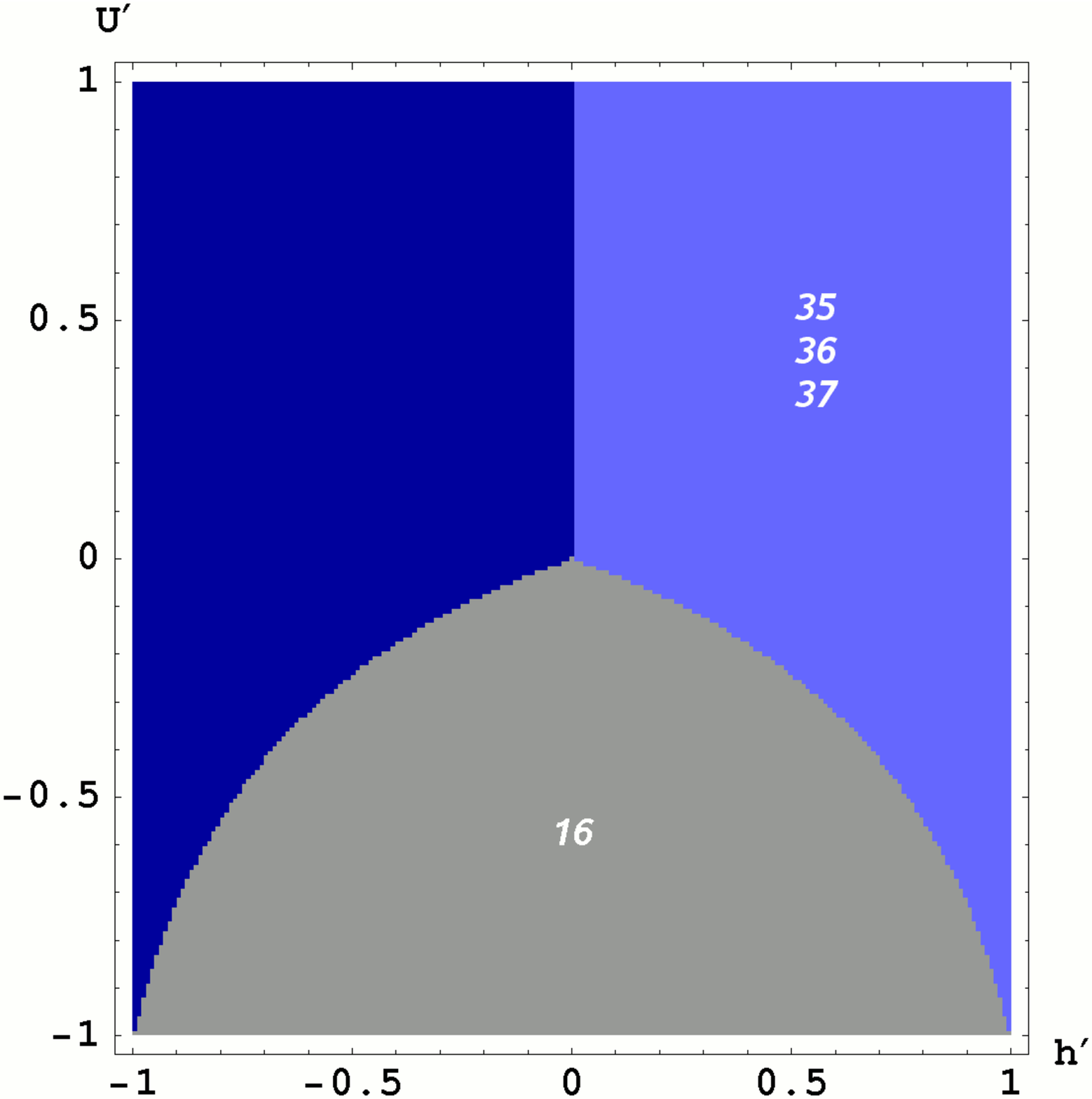}
 \includegraphics[height=40mm]{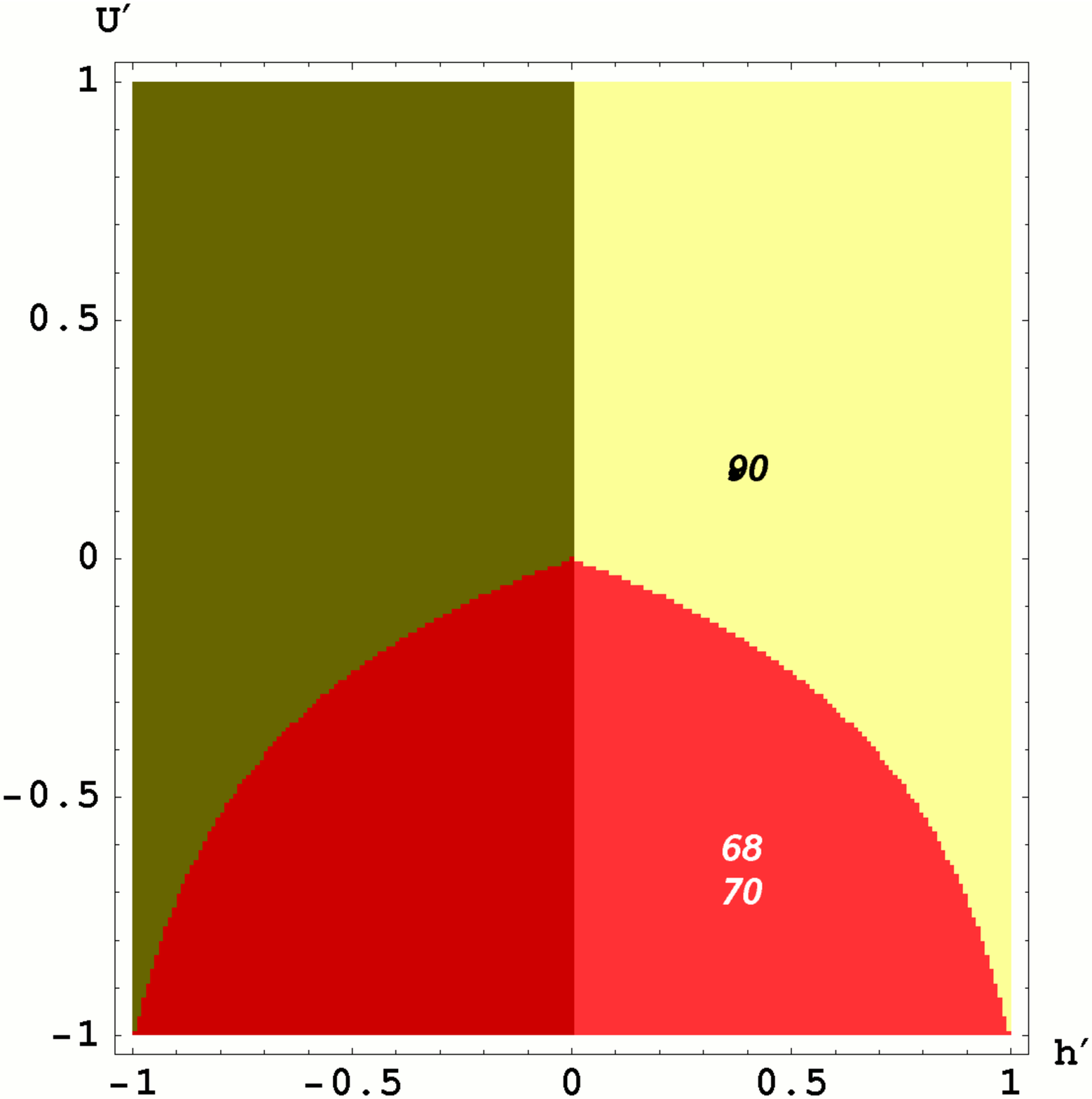}
 \includegraphics[height=40mm]{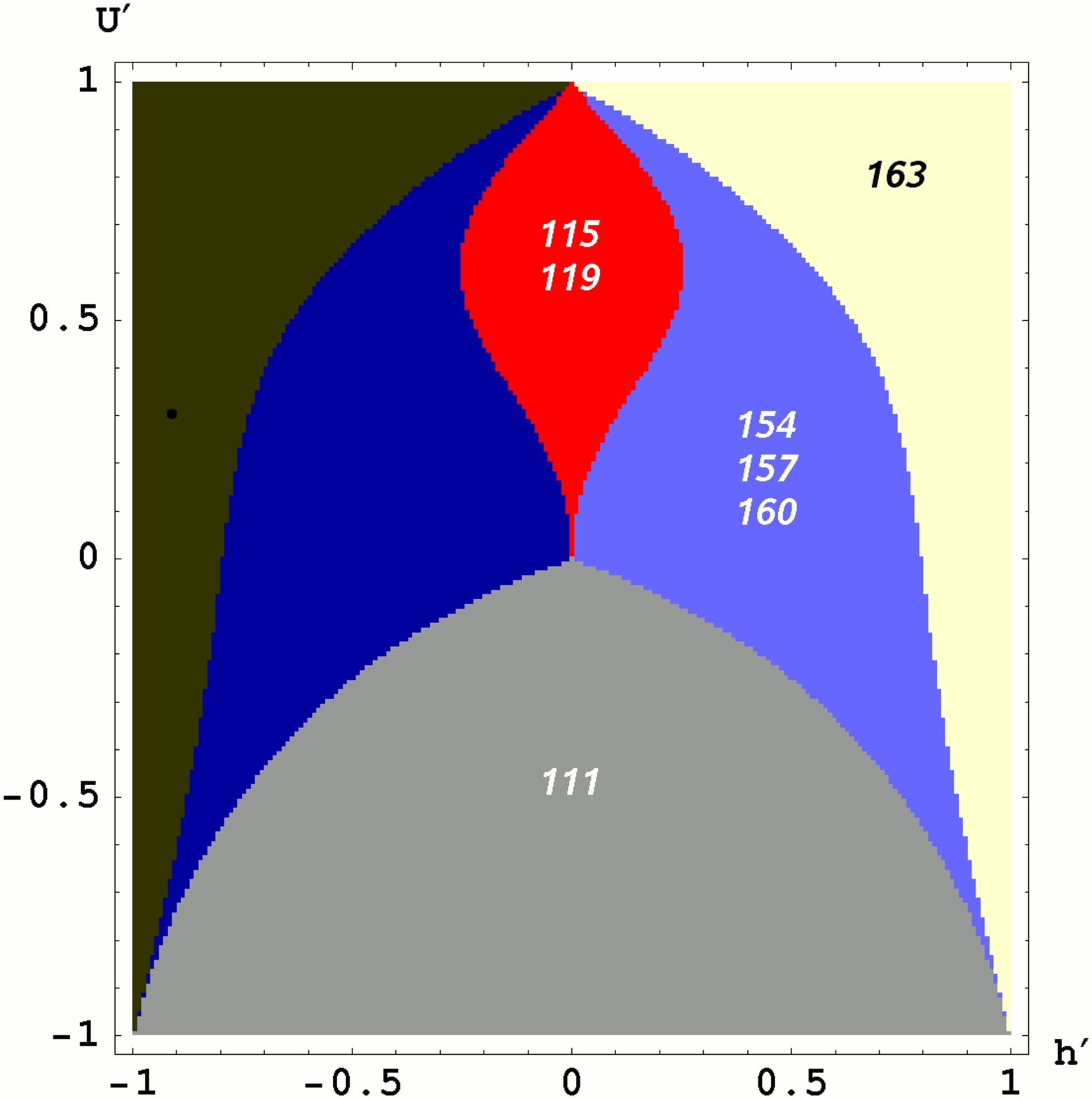}
 \includegraphics[height=40mm]{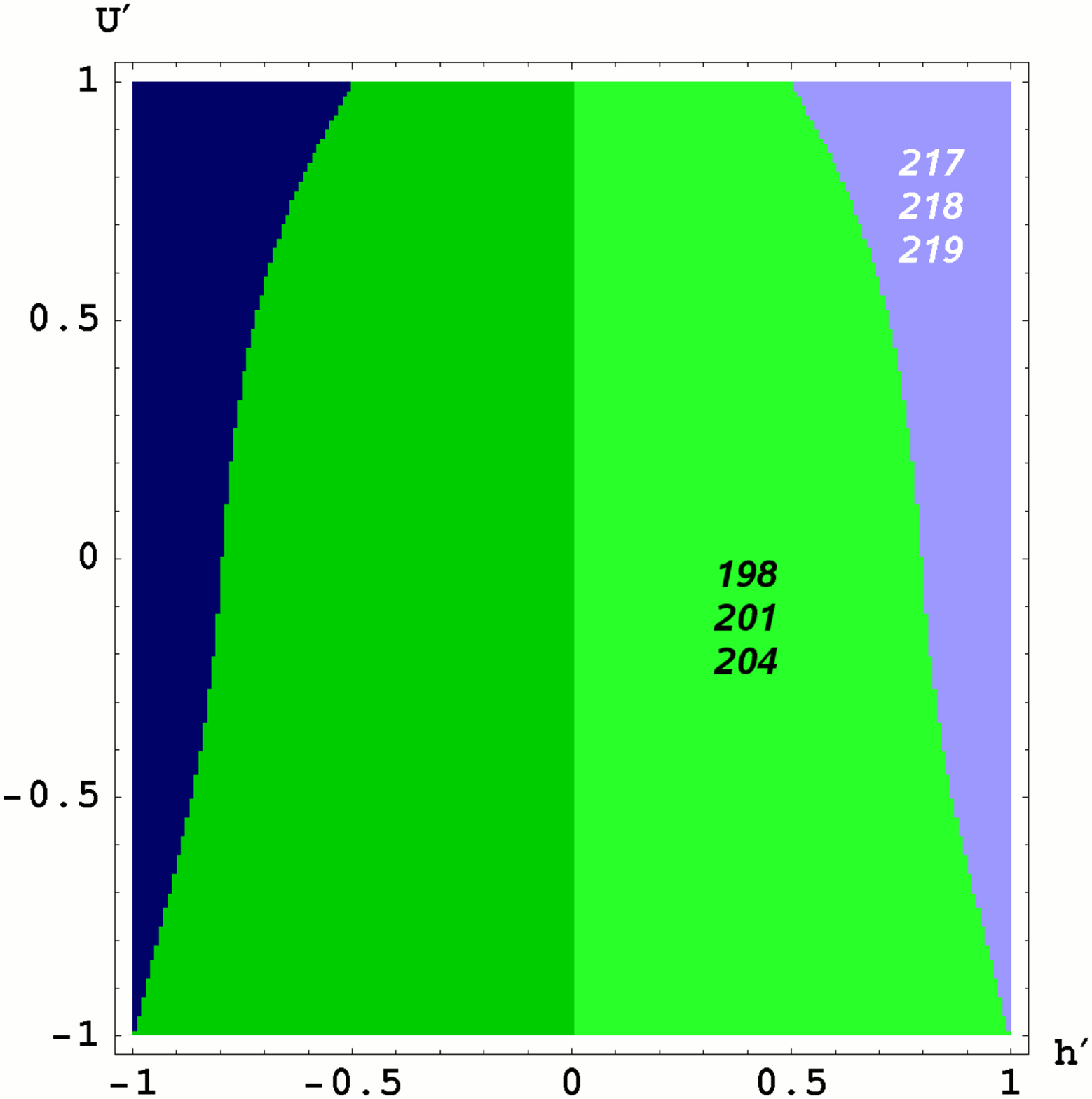}
 \includegraphics[height=40mm]{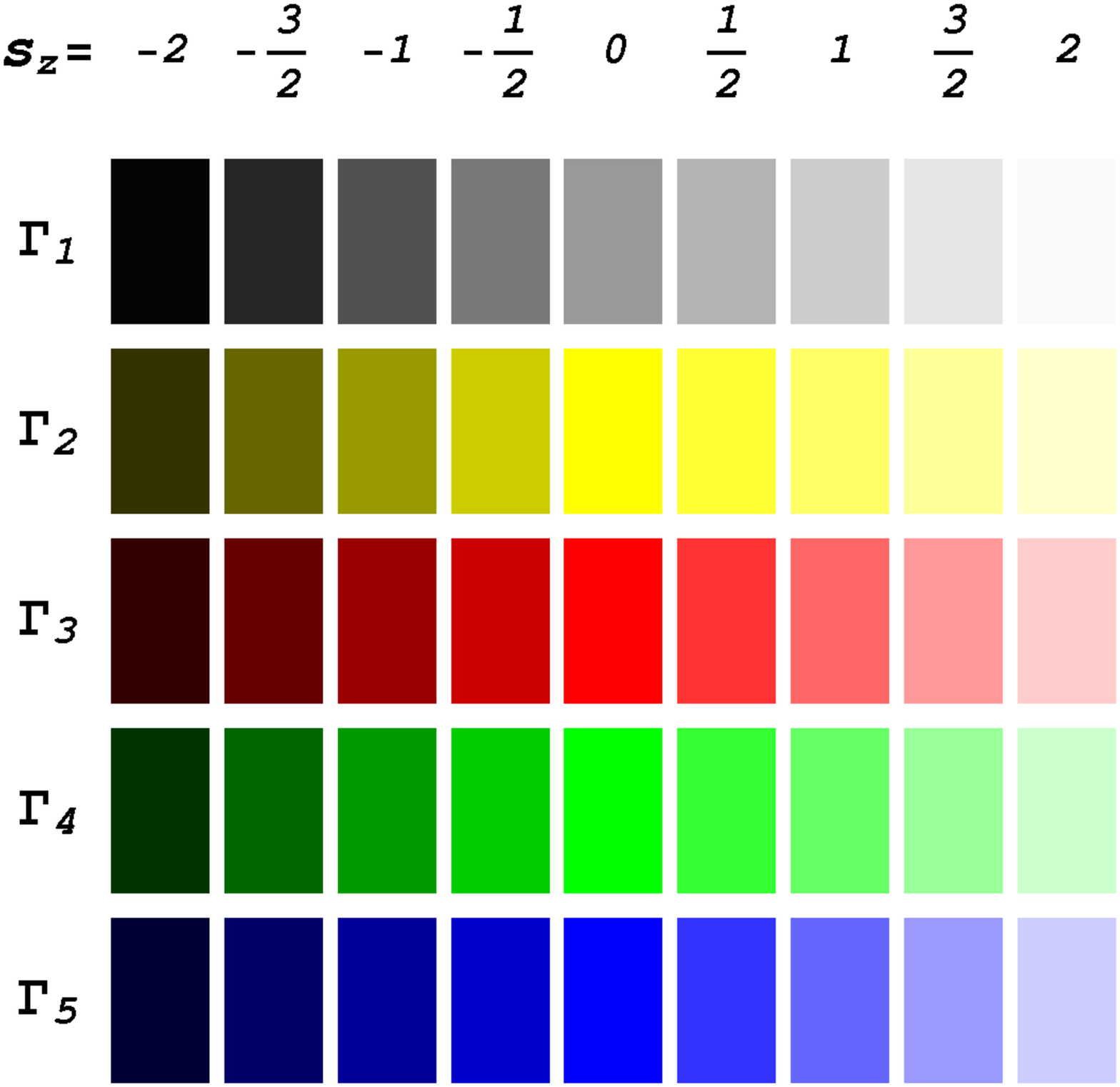}
 
 \end{minipage}%
 \caption{Left: The complete canonical spectrum for the tetrahedron 
 (scaled to primed values) in dependence on $U'$ for 
 $h/t=0$, $W/t=0$ and $J/t=0$. 
 Right: The groundstates for the complete $h-U$-plane 
 (scaled to primed values) for $W/t=0$ and $J/t=0$. 
 The indicated numbers, referring to
 the appendix \ref{appendix2a} show which states are degenerated. 
 The colours contain the
 information about the irr. representation of the point group and their intensity 
 shows the value of $S_z$ as depicted in the palette in the left lower corner. 
 Since the system is symmetrically with respect to a simultaneous
 sign change of magnetic field and spin projection we gave the numbers of the 
 right half-plane only.}
\label{primedewsTetraU}
\end{figure}

%% file: Parts/figure07.tex
\begin{figure}
 \begin{minipage}[t]{0.5\textwidth}
 \centering 
 \includegraphics[height=40mm]{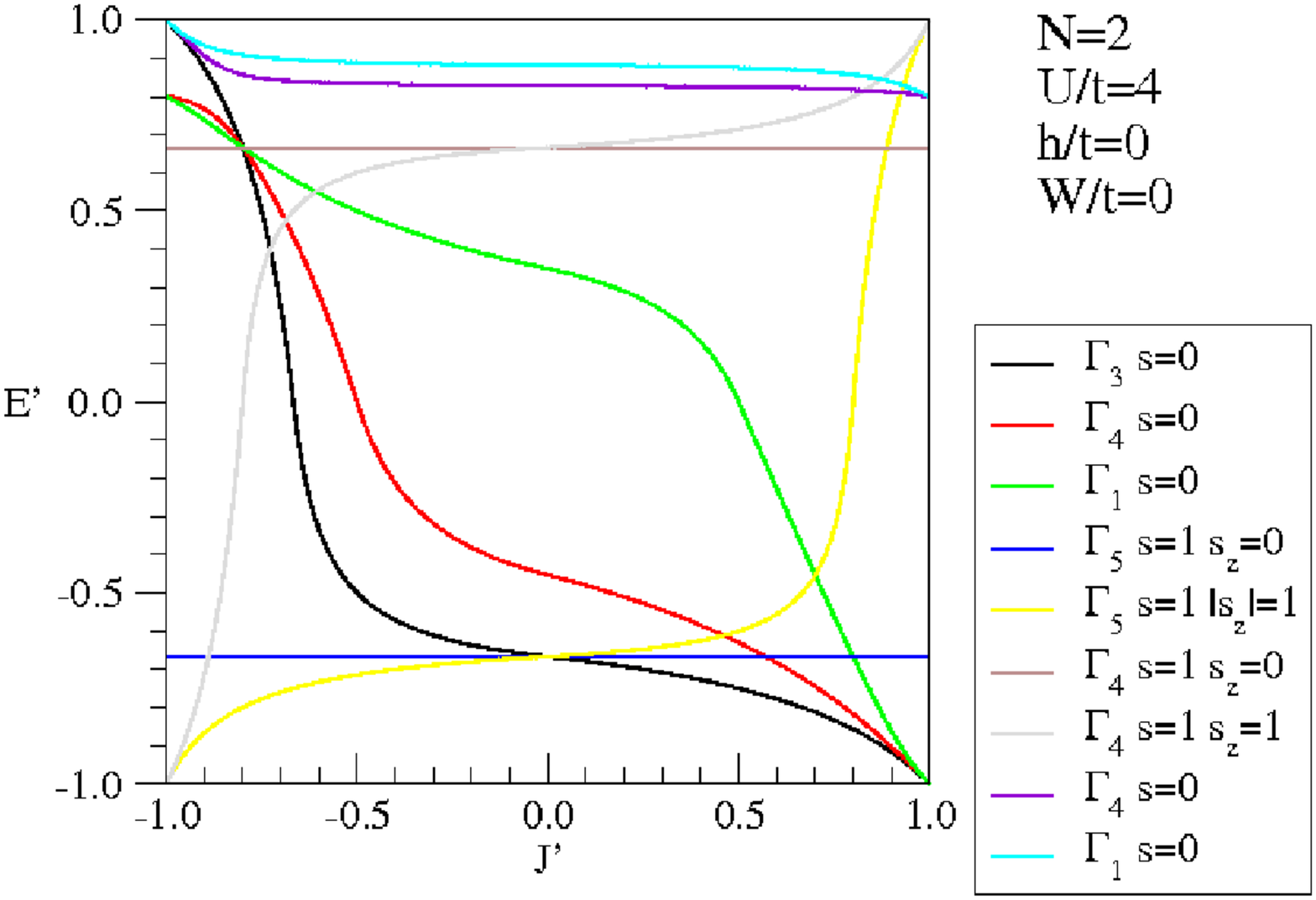}
 \includegraphics[height=40mm]{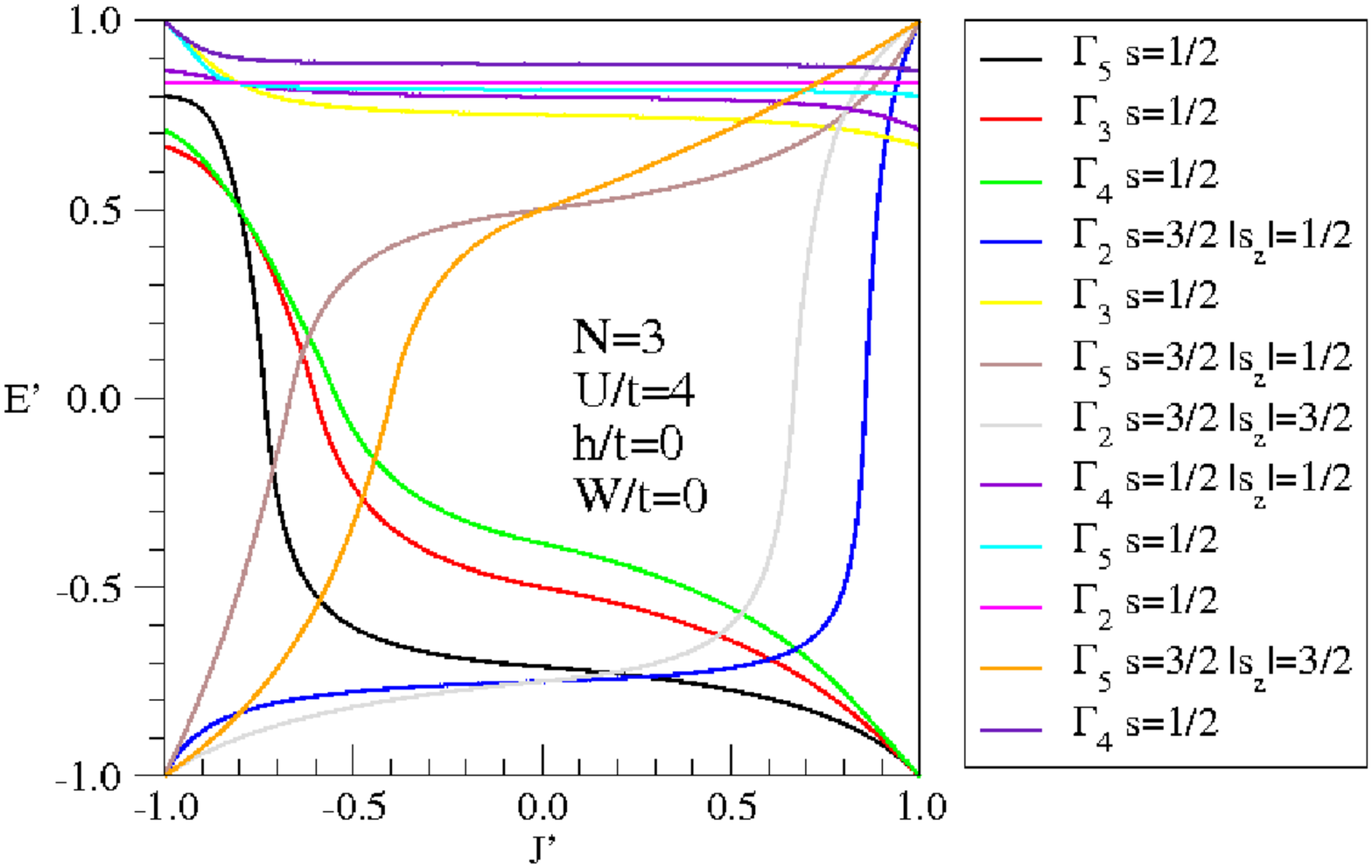}
\hspace*{2.5mm}
\includegraphics[height=41mm]{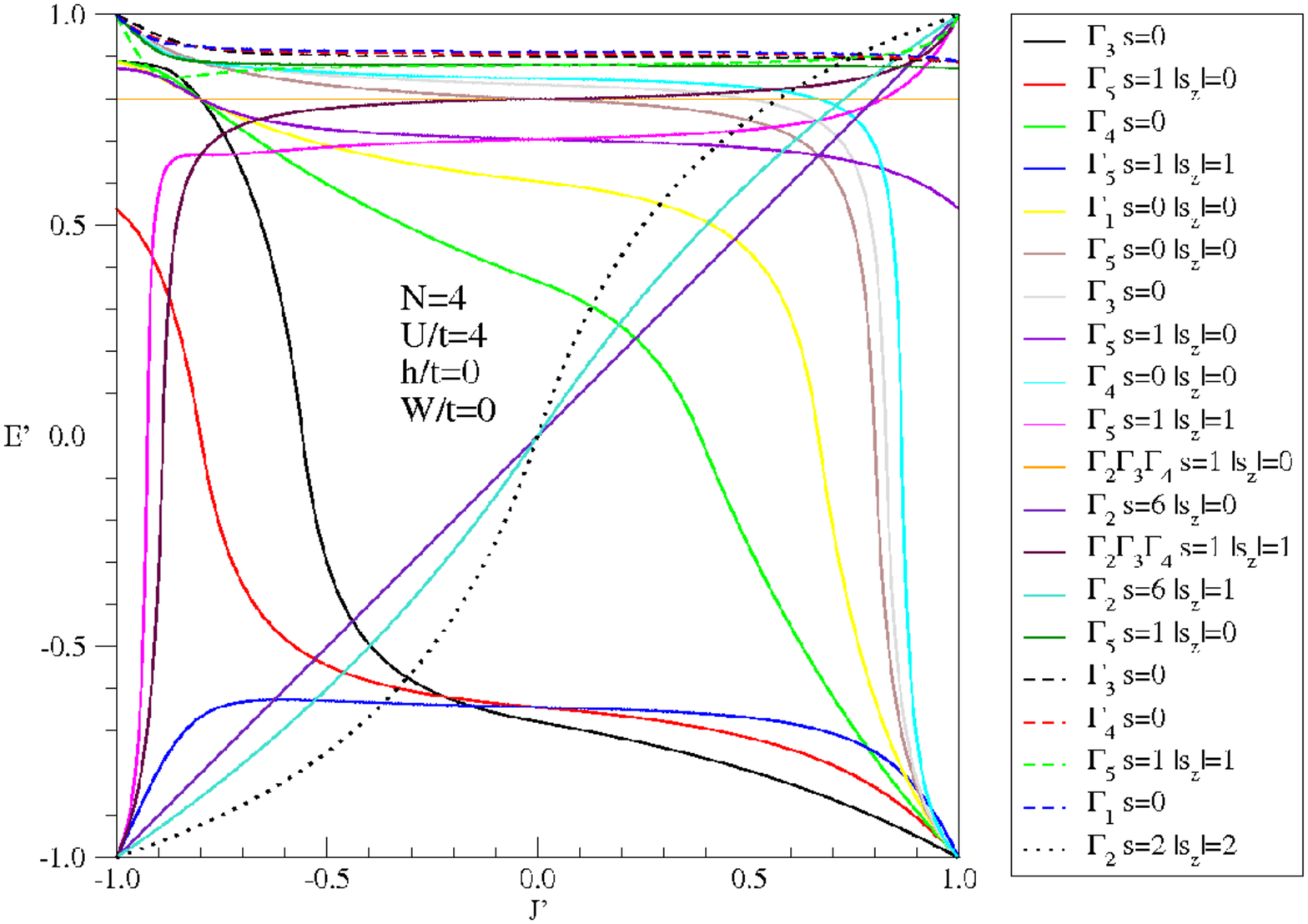}  \includegraphics[height=40mm]{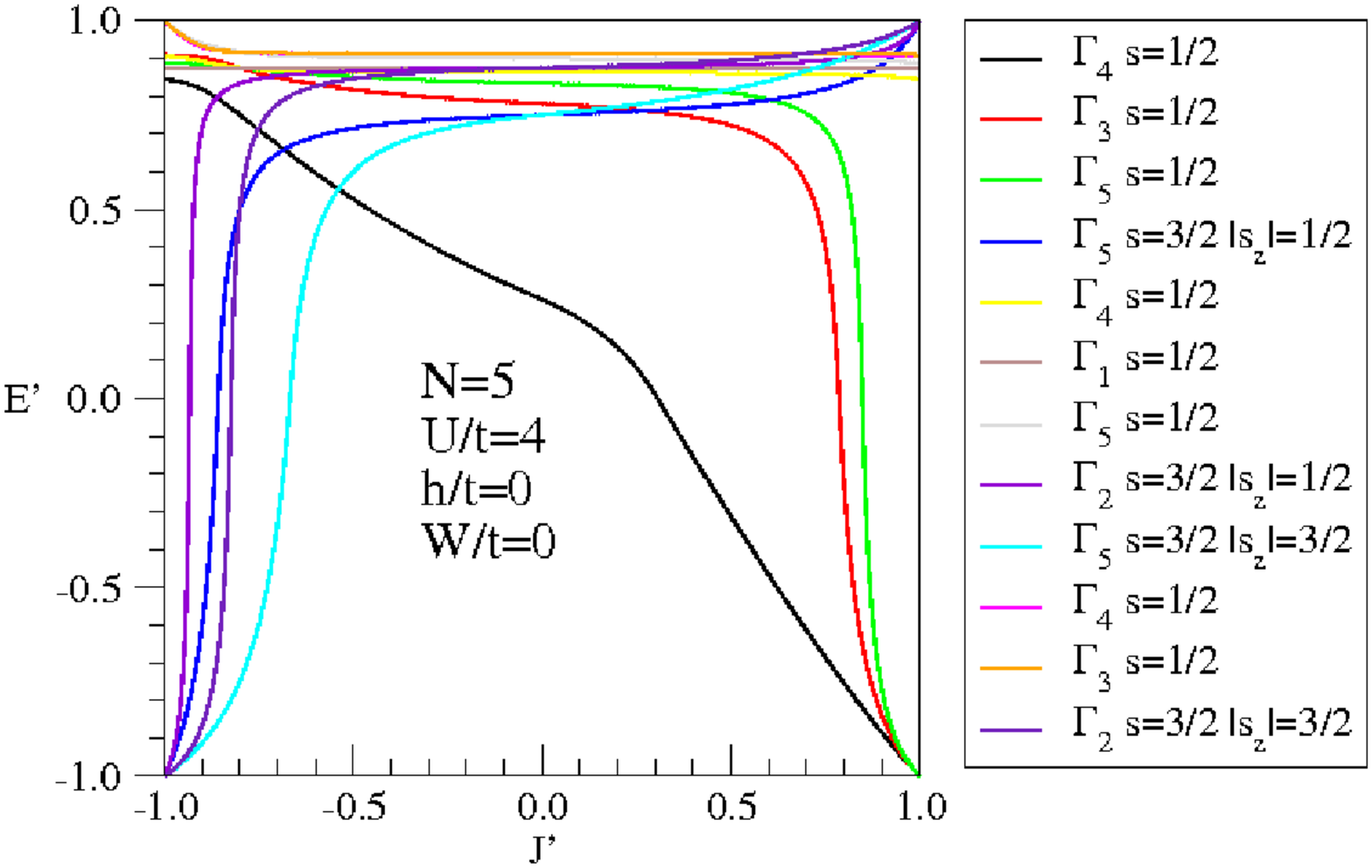}
\includegraphics[height=40mm]{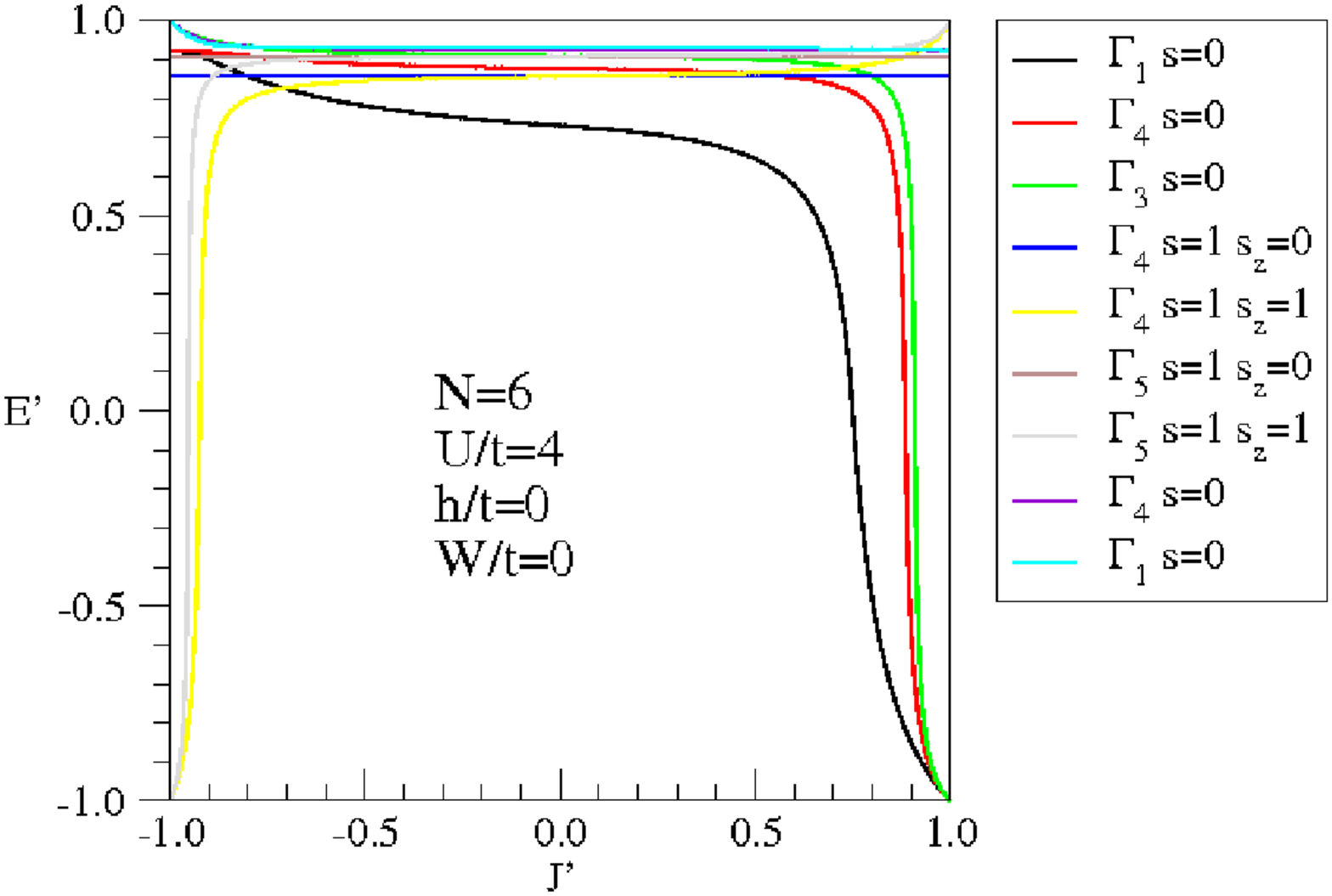}
\end{minipage} 
\hfill
\begin{minipage}[t]{0.5\textwidth}
 \centering 
 \includegraphics[height=40mm]{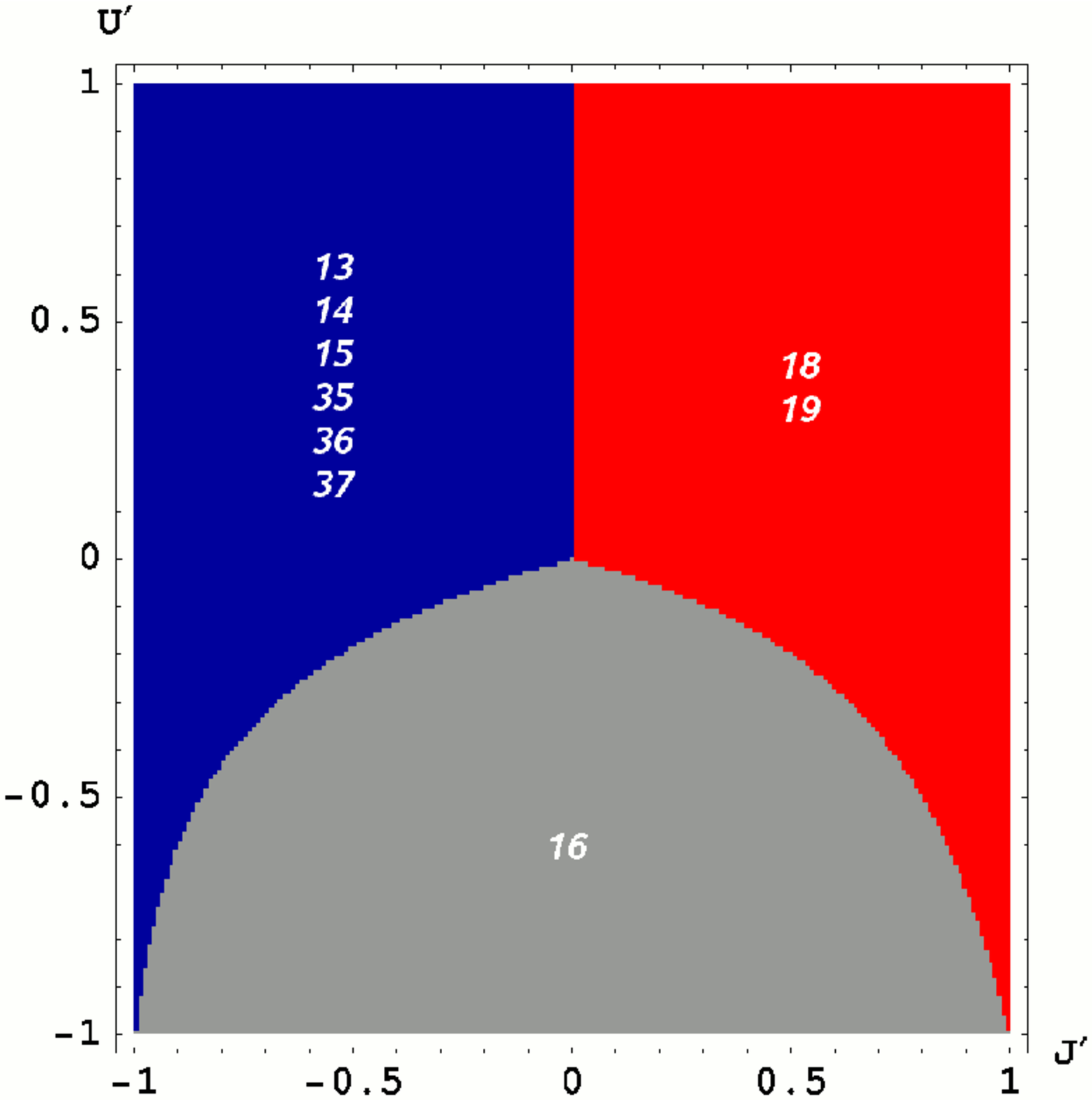}
 \includegraphics[height=40mm]{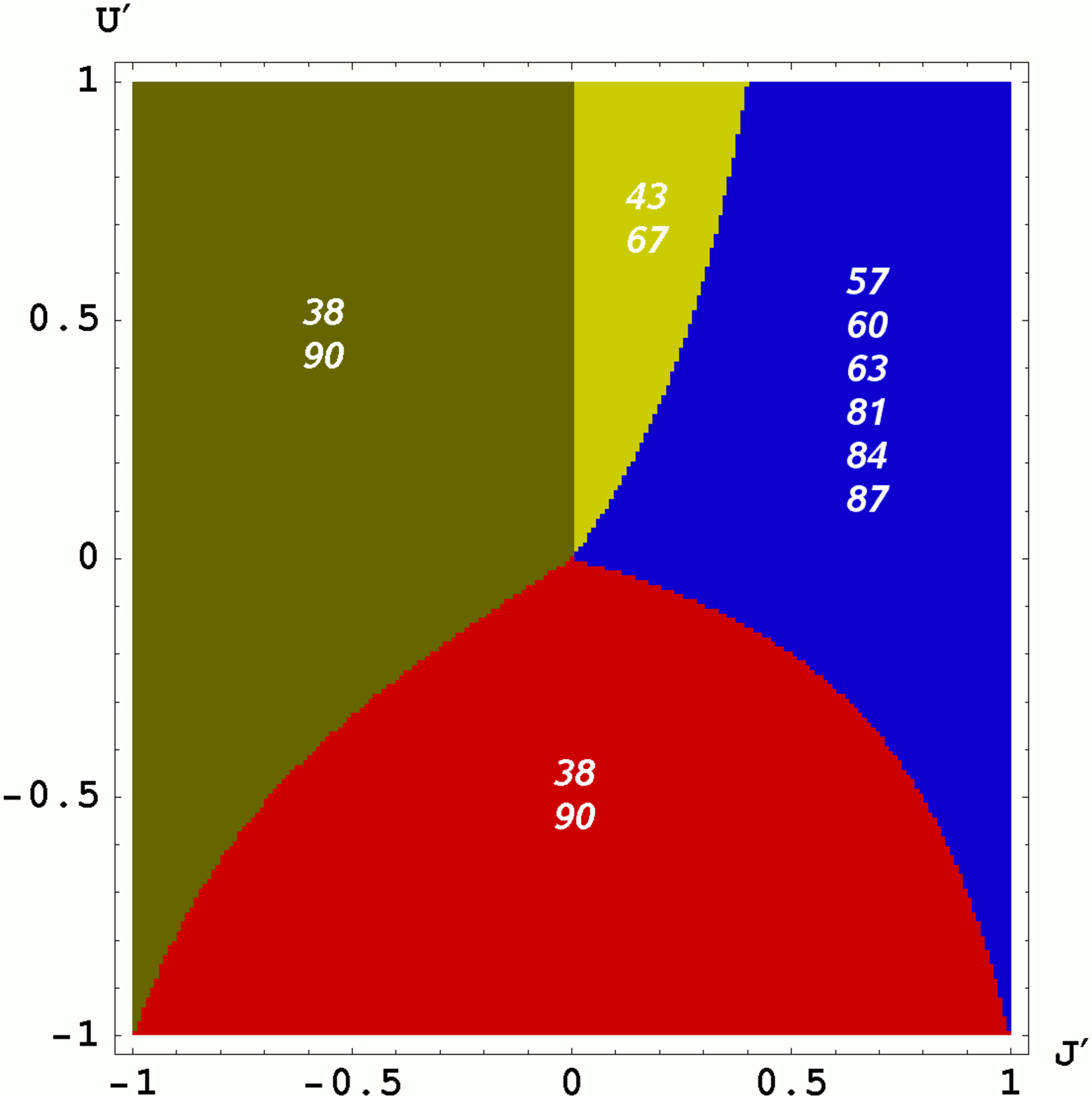}
 \includegraphics[height=40mm]{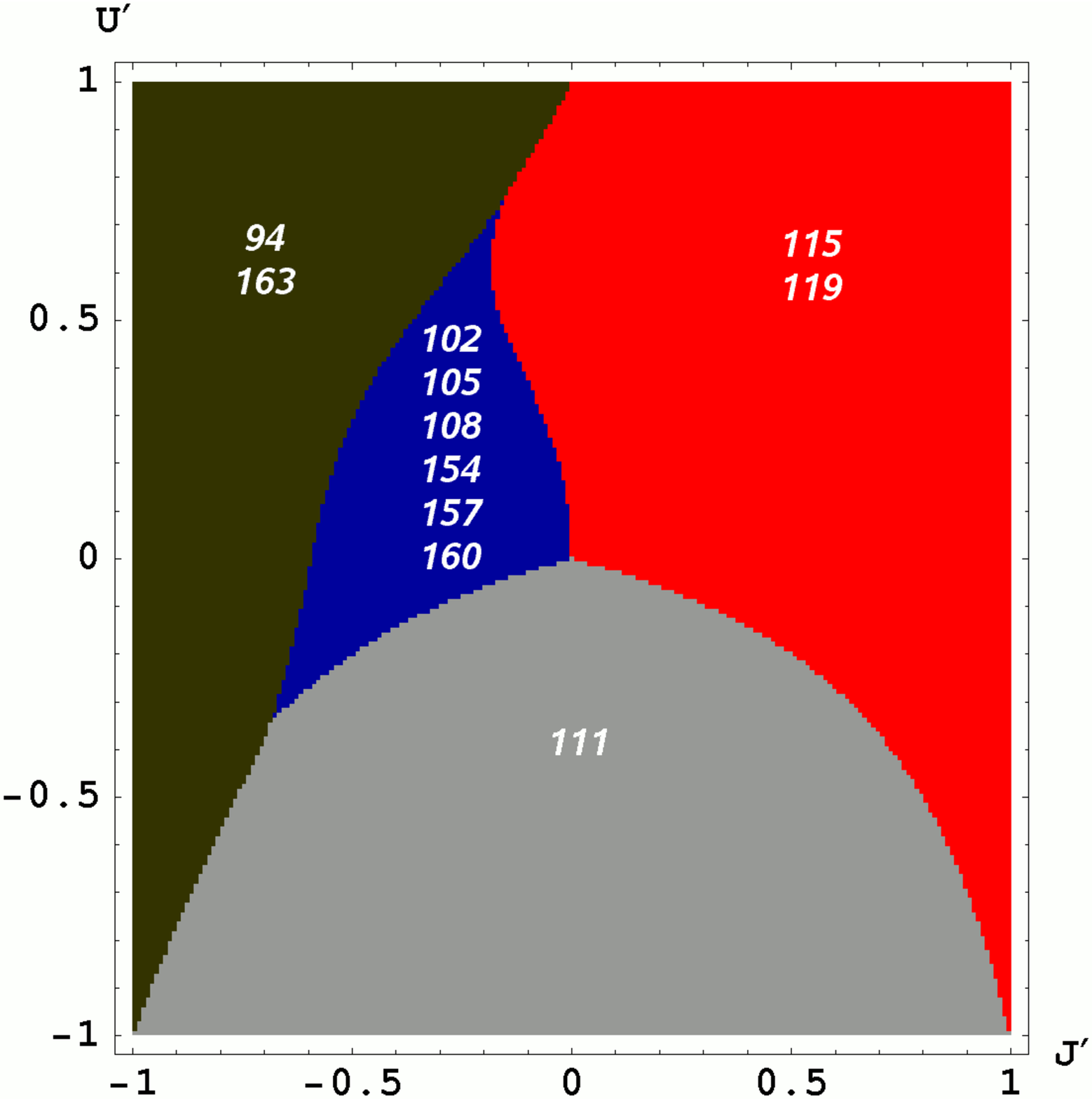}
 \includegraphics[height=40mm]{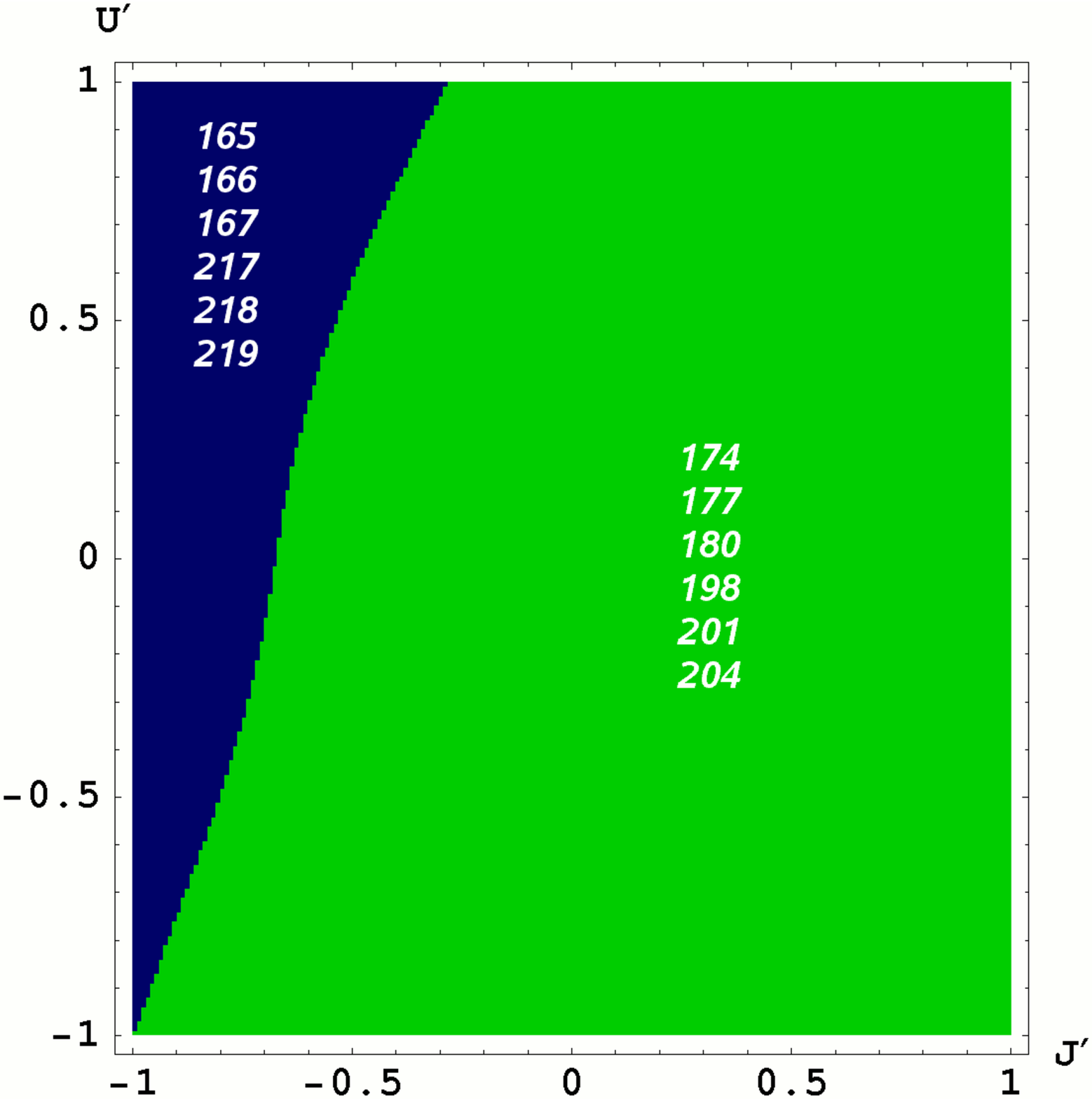}
 \includegraphics[height=40mm]{./Figures/canonicalPalette.eps}
\end{minipage}
 \caption{Left: The complete canonical spectrum for the tetrahedron
  (scaled to primed values) in dependence on $J'$ for $h/t=0$, $U/t=4$ 
  and $W/t=0$. Right: The groundstates for the complete $J-U$-plane 
  (scaled to primed values) 
  for $W/t=0$ and $h/t=0$.
  The numbers and colours have the same 
  meaning as in Fig.\ref{primedewsTetraU}.}
\label{primedewsTetraJ}
\end{figure}

%% file: Parts/figure0809.tex
\begin{figure}
 \begin{minipage}[t]{0.48\textwidth}
 \centering 
 \includegraphics[height=45mm]{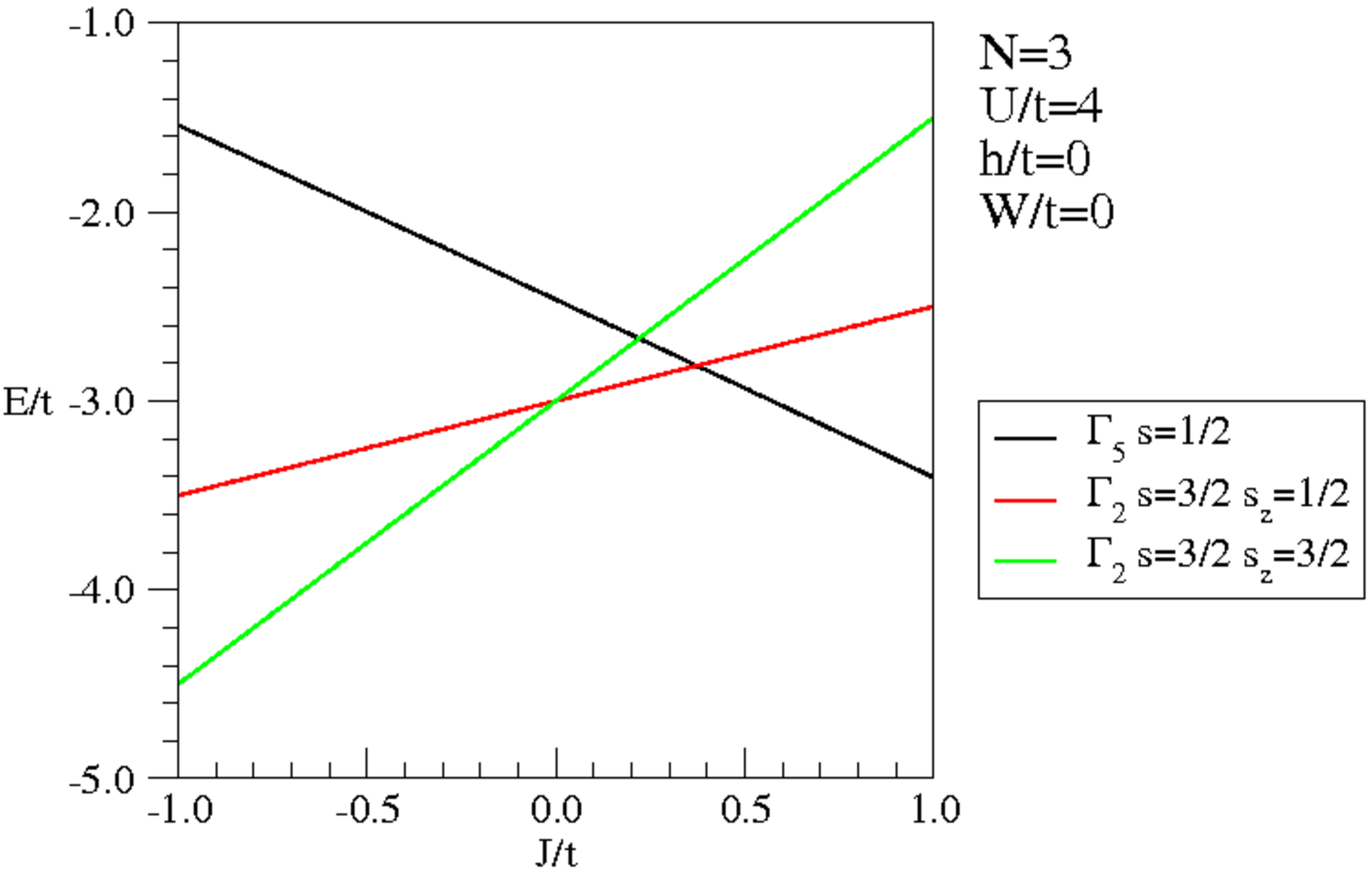}
\caption{The ground state levels for three electrons on the tetrahedron in dependence of the exchange parameter $J$ for $h/t=0$, $U/t=4$ and $W/t=0$.}
\label{lowEwsU4N3J}
\end{minipage} 
\hfill
 \begin{minipage}[t]{0.48\textwidth}
 \centering 
 \includegraphics[height=45mm]{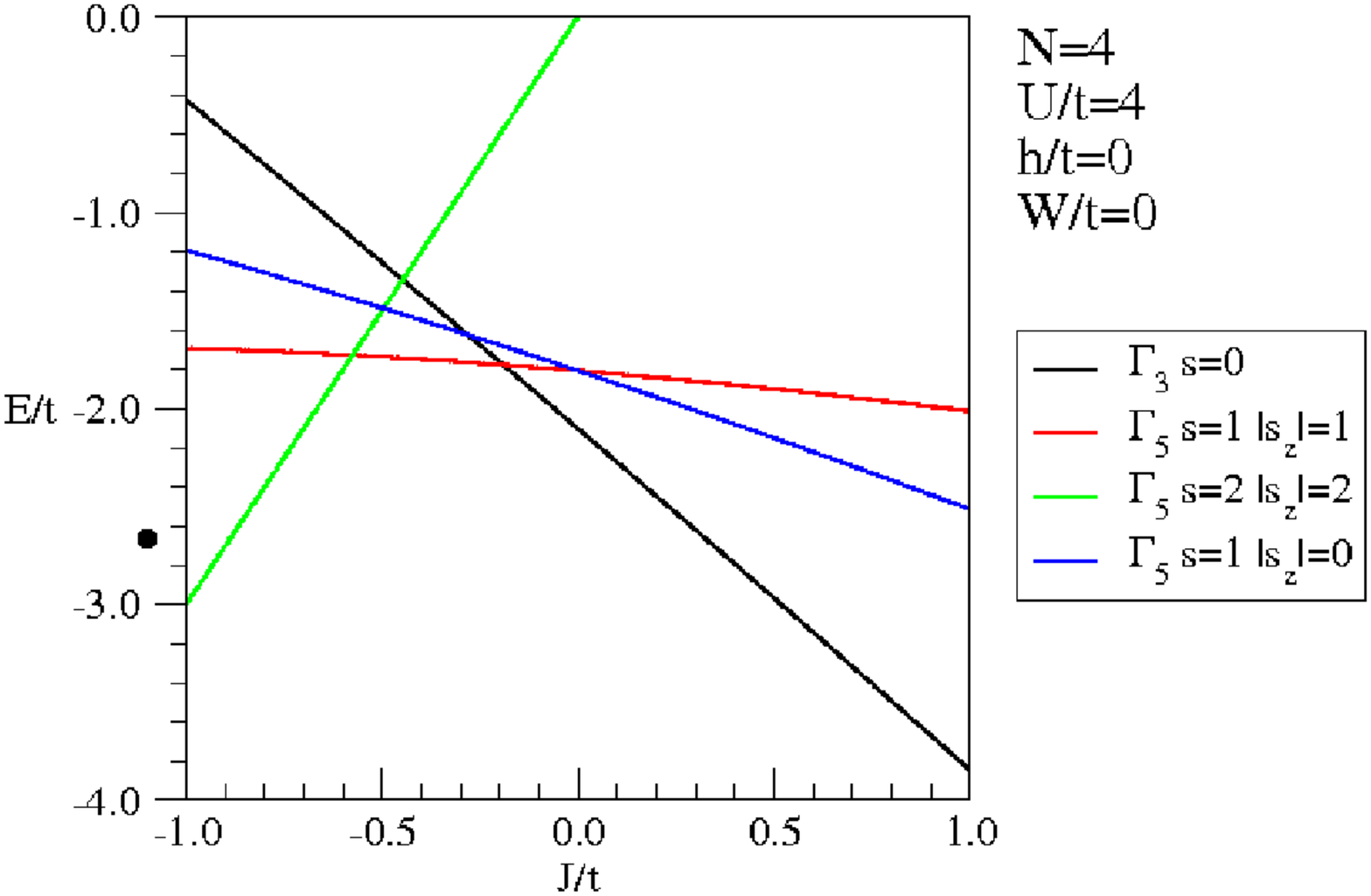}
\caption{The ground state levels for four electrons on the tetrahedron in dependence of the exchange parameter $J$ for $h/t=0$, $U/t=4$ and $W/t=0$.}
\label{lowEwsU4N4J}
\end{minipage} 
\end{figure}

%% file: Parts/figure10.tex
\begin{figure}
 \begin{minipage}[t]{0.5\textwidth}
 \centering 
 \includegraphics[height=40mm]{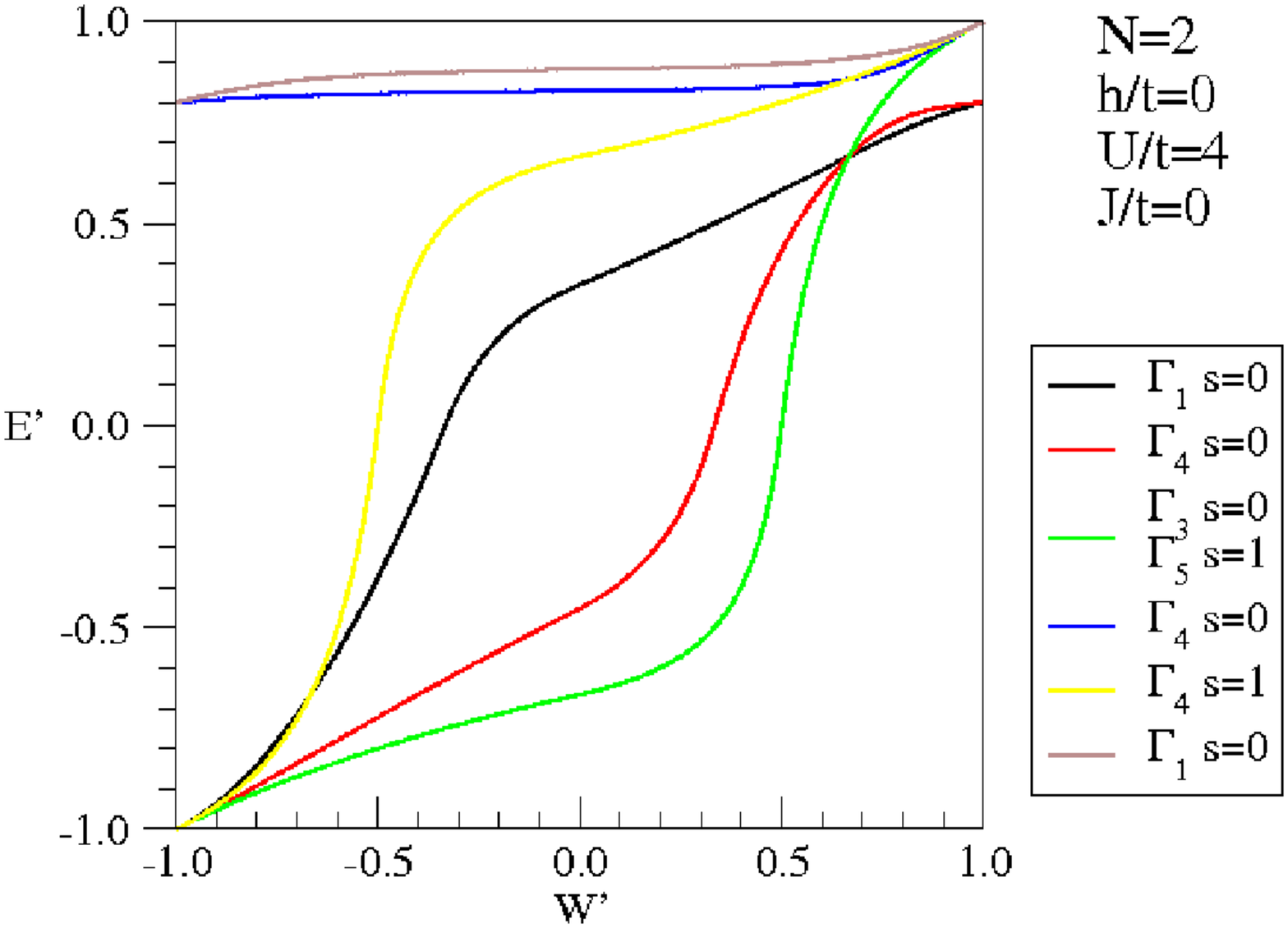}
 \includegraphics[height=40mm]{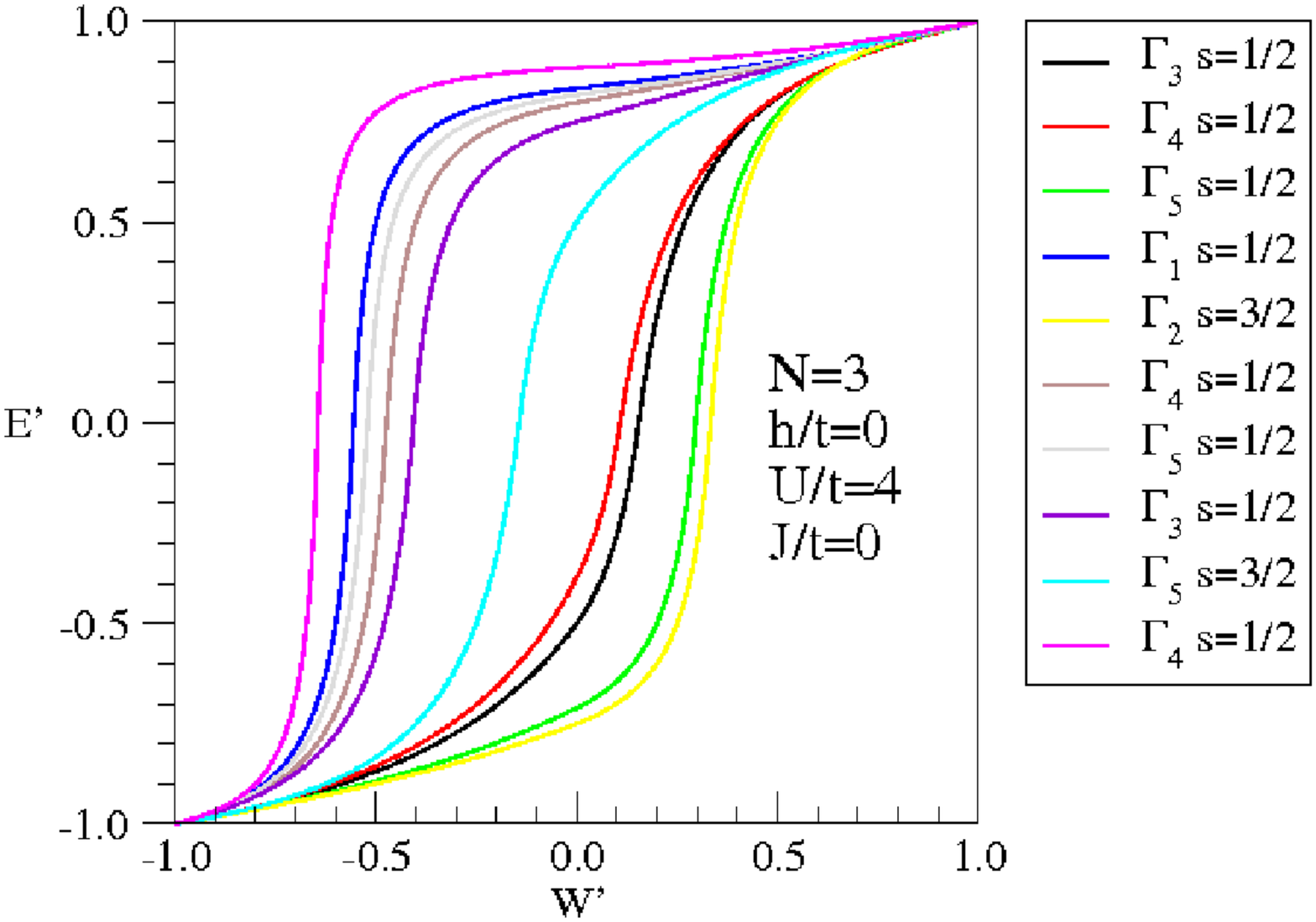}
 \includegraphics[height=40mm]{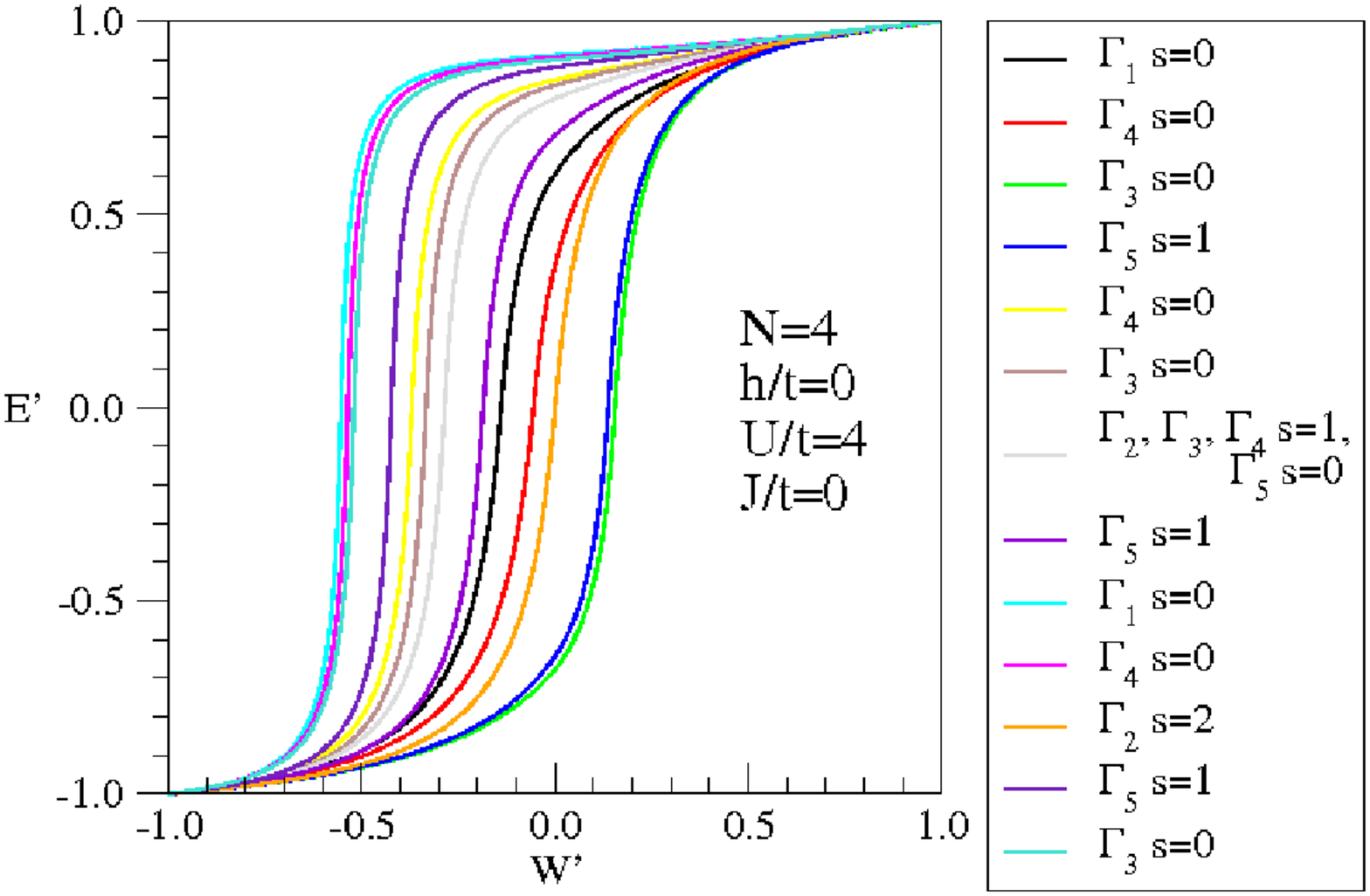}
 \includegraphics[height=40mm]{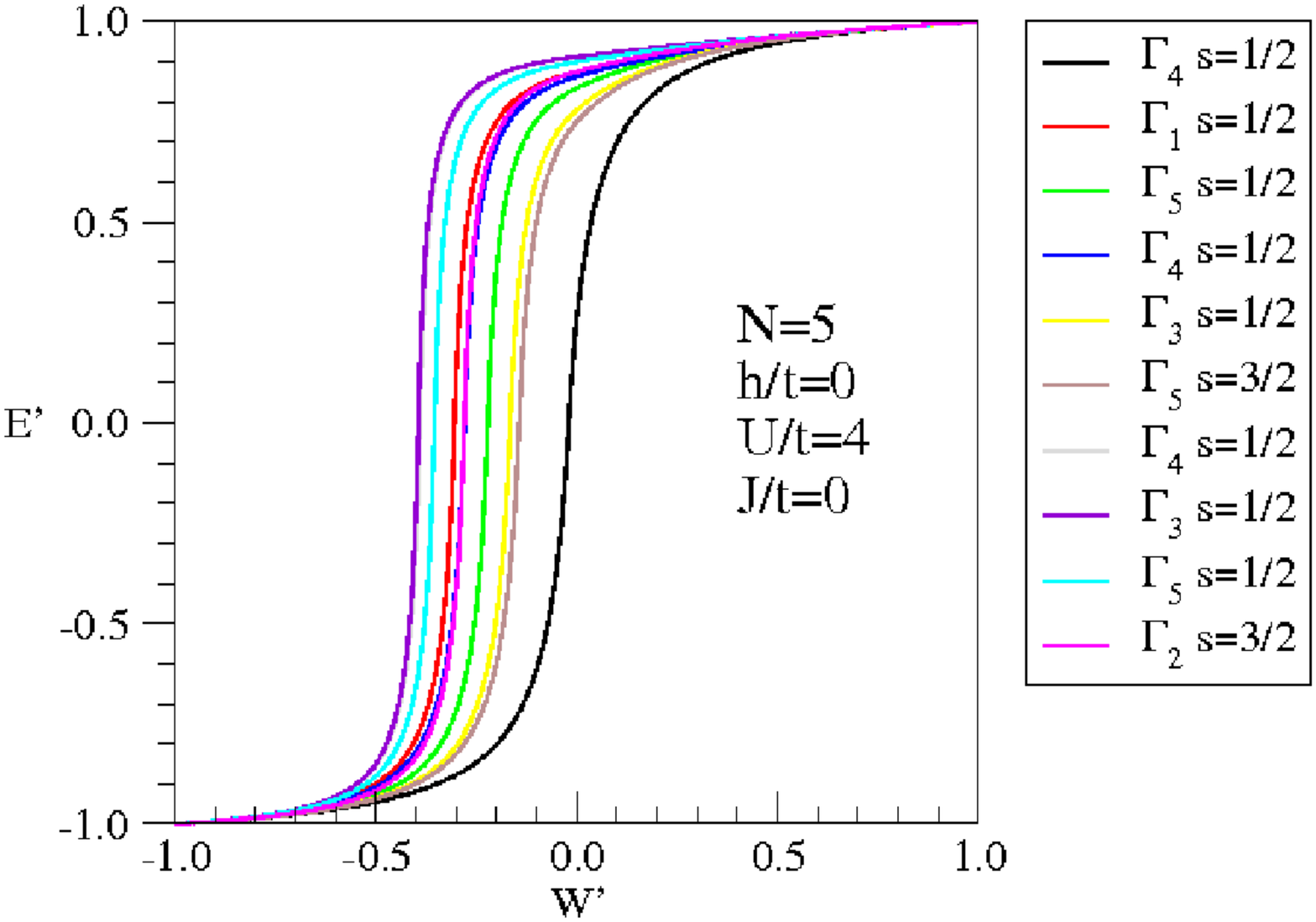}
 \end{minipage} 
 \begin{minipage}[t]{0.5\textwidth}
 \centering 
 \includegraphics[height=40mm]{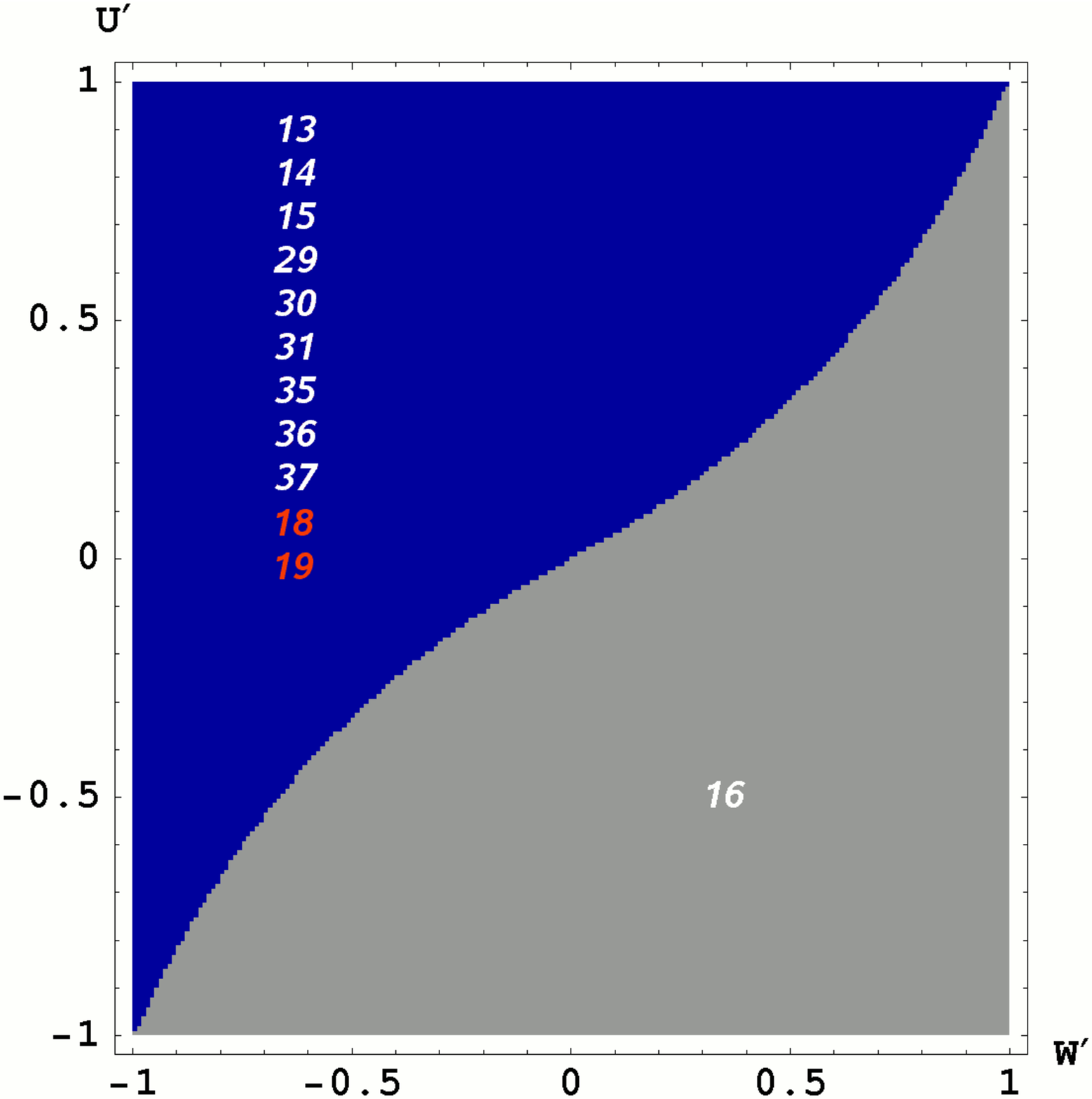}
 \includegraphics[height=40mm]{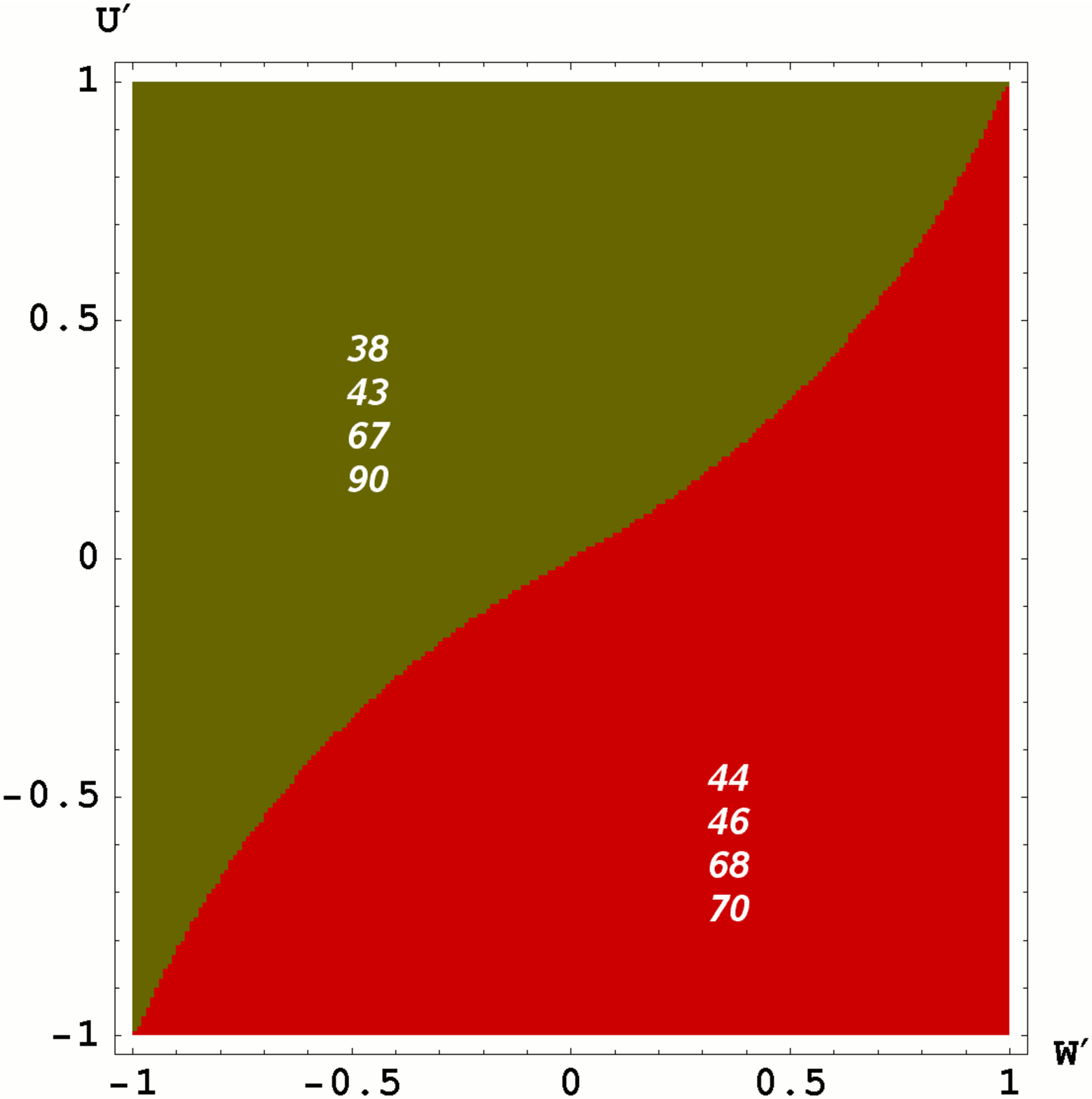}
 \includegraphics[height=40mm]{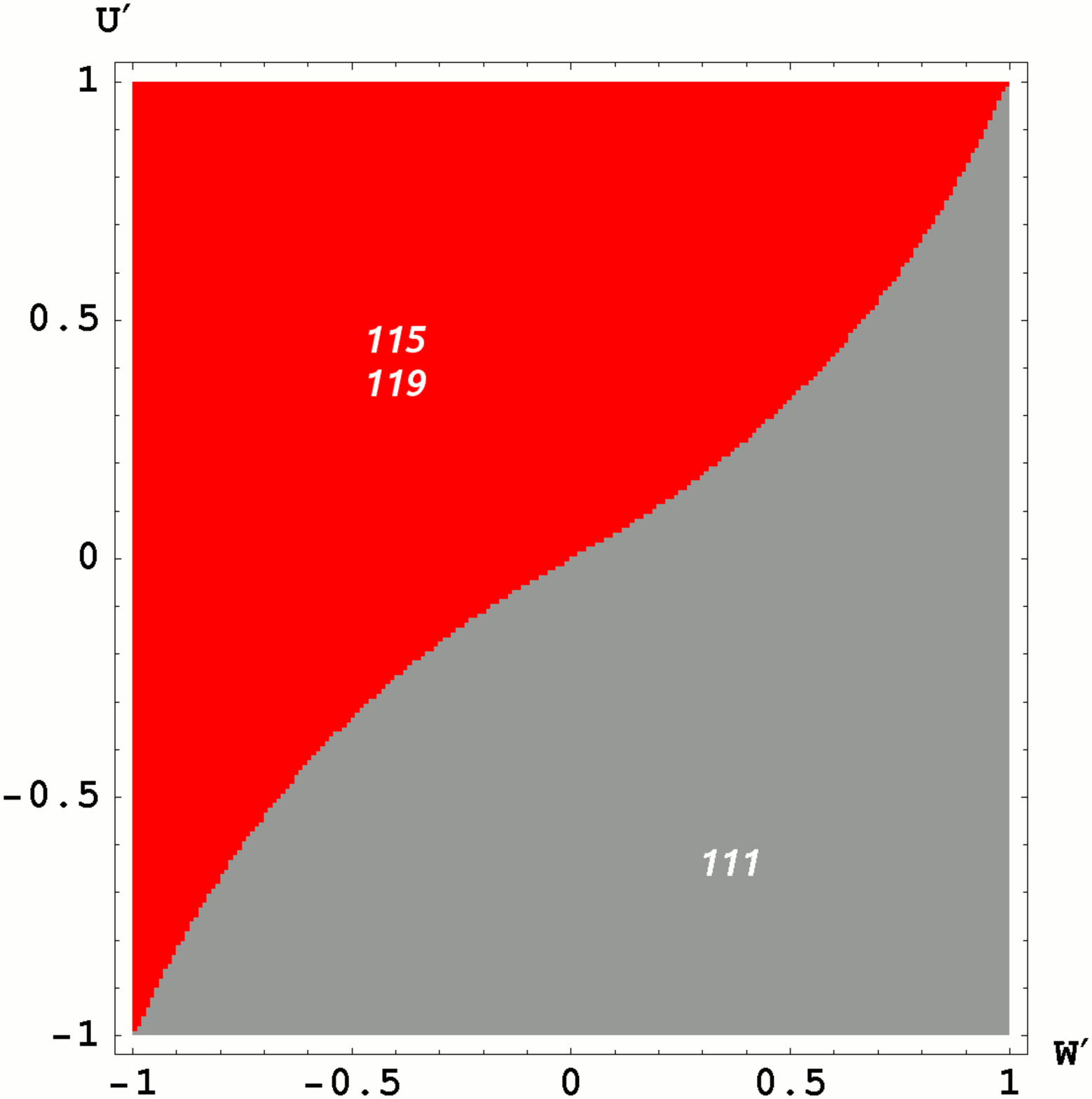}
 \includegraphics[height=40mm]{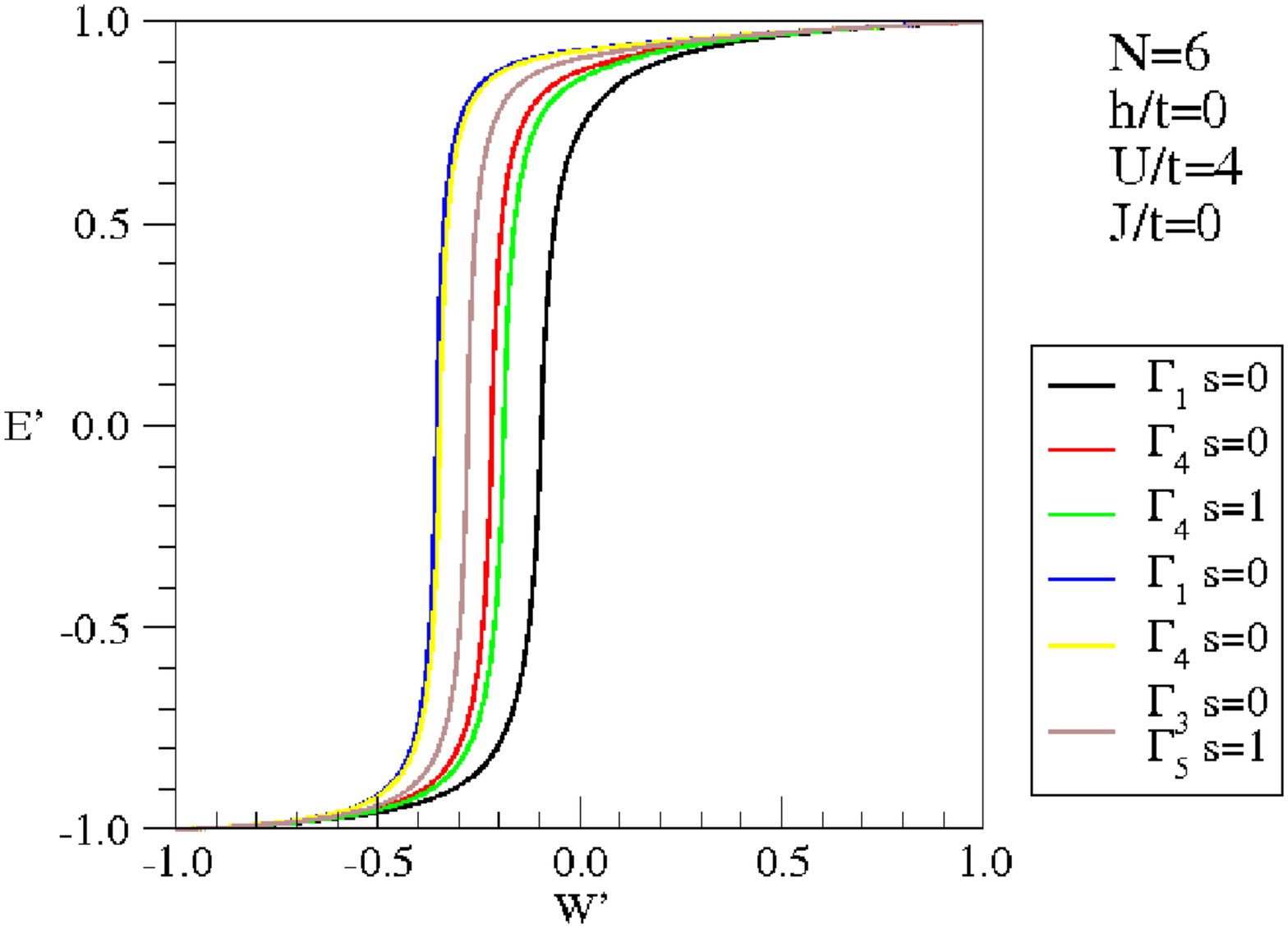}
 \end{minipage} 
 \caption{Left: The complete canonical spectrum for the tetrahedron 
          (scaled to primed values) in dependence on $W'$. The 
           values of the other parameters are indicated in the 
          figures. Right: The groundstates for the complete $W-U$-plane 
  (scaled to primed values) 
  for $J/t=0$ and $h/t=0$.
  The numbers and colours have the same 
  meaning as in Fig. \ref{primedewsTetraU}}
\label{primedewsTetraW}
\end{figure}

%% file: Parts/figure11.tex
\begin{figure}
 \begin{minipage}[c]{0.5\textwidth}
 \centering 
 \includegraphics[height=55mm]{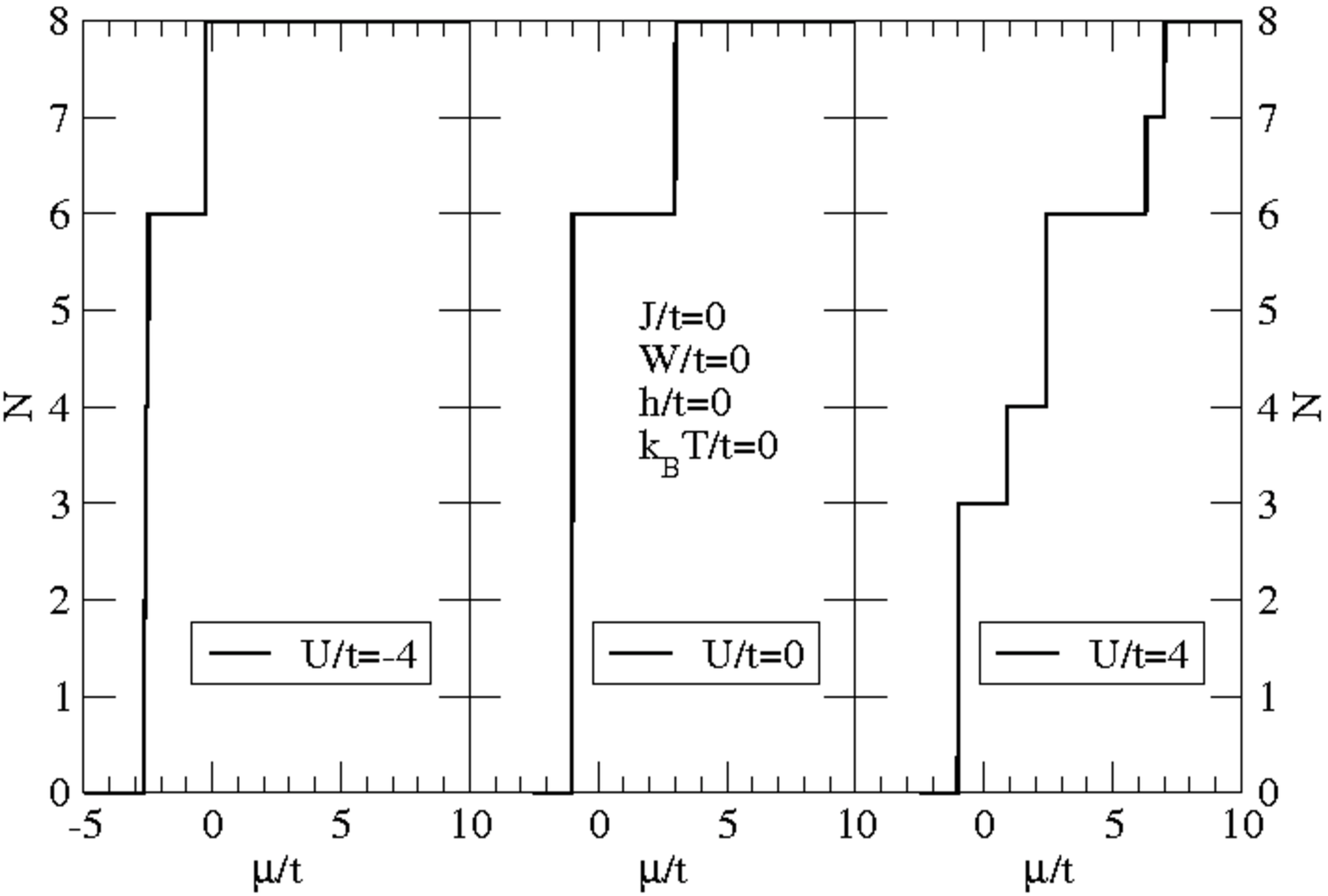}
 \end{minipage}%
 \hfill
 \begin{minipage}[c]{0.5\textwidth}
 \centering 
 \includegraphics[height=60mm]{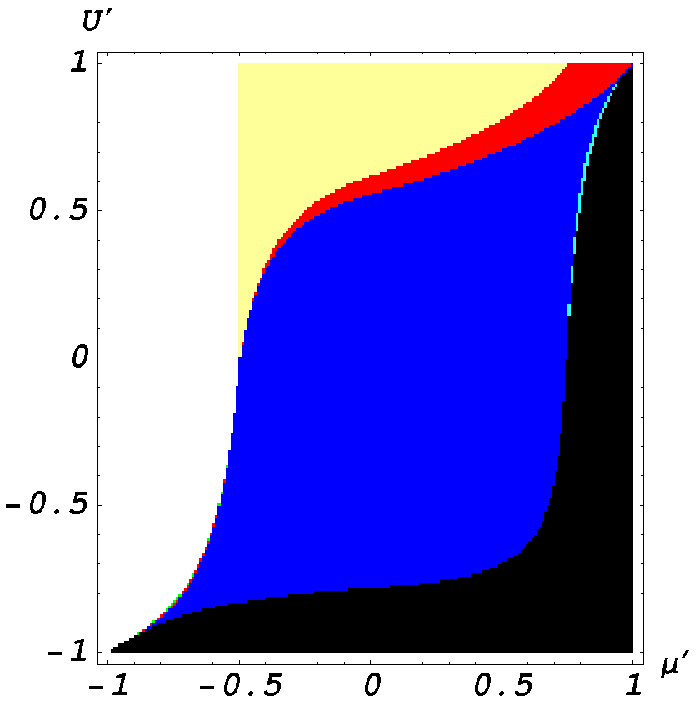}
 \end{minipage} 
\caption{
Left: $N(\mu)$ for the tetrahedron for different values of the on-site correlation $U$. Right: The electron occupation  for the whole
$\mu$-$U$ plane (scaled to primed values).
The curves with $U/t=-4$, $U/t=0$, and $U/T=4$ in the left picture 
correspond to horizontal cuts across the right one for $U'=-0.5$, $U'=0$, and $U'=0.5$ respectively.}
\label{NvonMueU}
\end{figure}

%% file: Parts/figure11a.tex
\begin{figure}
 \begin{minipage}[c]{0.3\textwidth}
 \centering 
 \includegraphics[height=60mm]{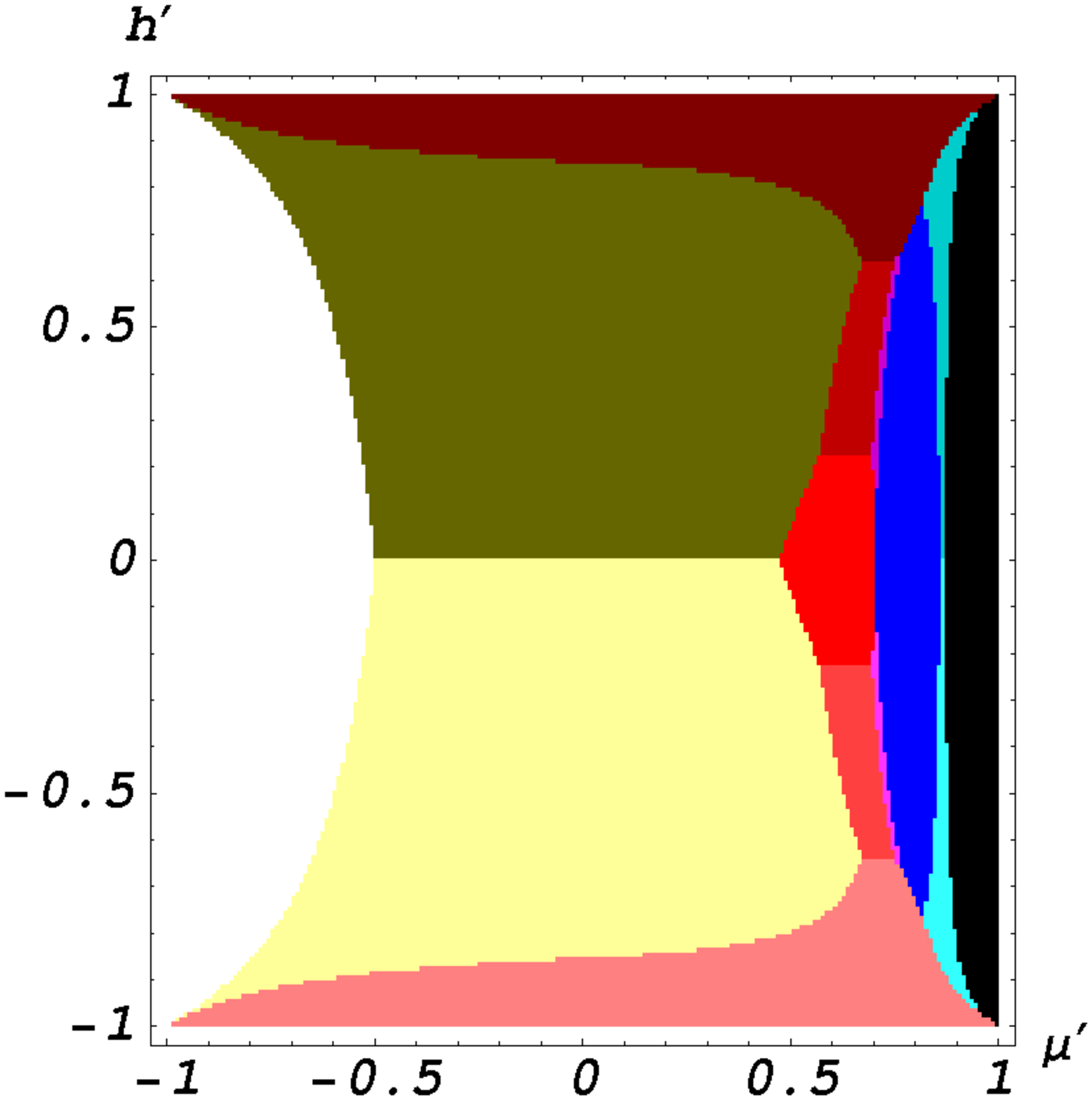}
 \end{minipage}%
 \hfill
 \begin{minipage}[c]{0.6\textwidth}
 \centering 
 \includegraphics[height=55mm]{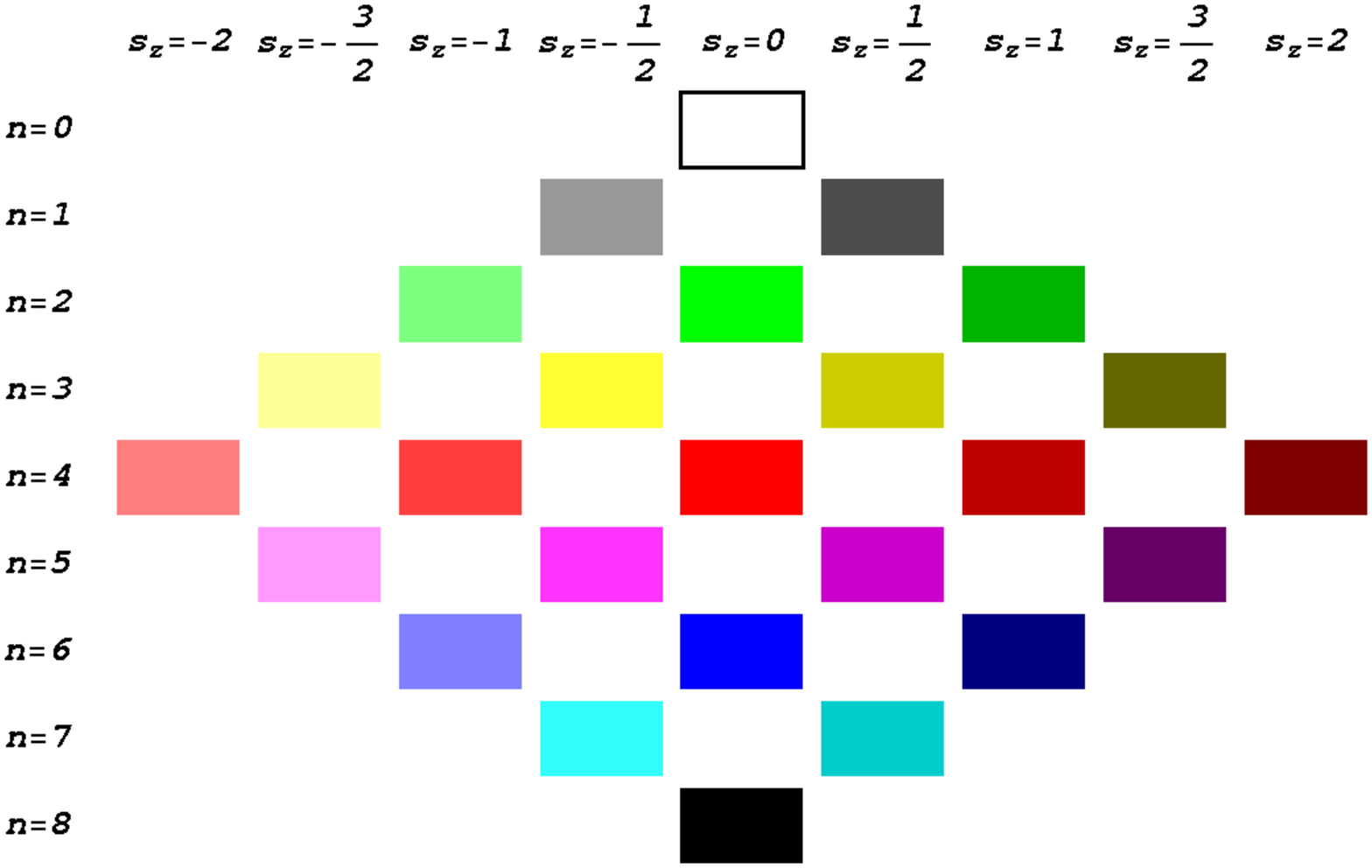}
 \end{minipage} 
\caption{
Left panel: The electron occupation $N$ and the $S_z$-eigenvalue for the ground state in dependence on the chemical potential and applied magnetic field
(both scaled to primed values)
for the tetrahedron and $U/t=4$, $J/t=0$, and $W/t=0$.
Right panel. The palette used here and in Figs. 
\ref{NvonMueU},\ref{NvonMueJU4}, and \ref{NvonMueWU4}
to differ the groundstates with respect of their electron occupation and $S_z$ eigenvalues.
}
\label{palette4site}
\end{figure}

%% file: Parts/figure12.tex
\begin{figure}
\centering
\includegraphics[width=0.8\textwidth]{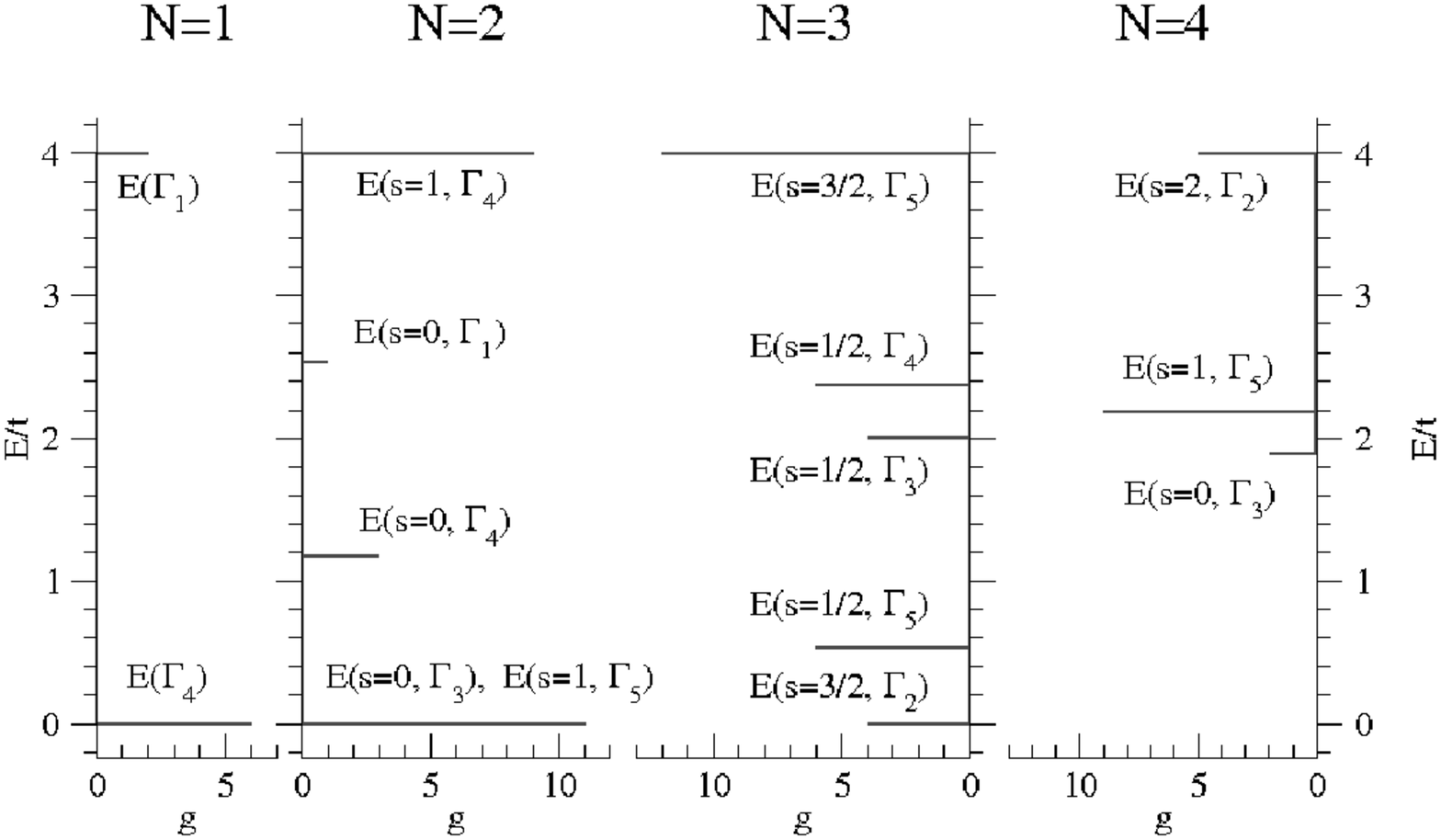}
\caption{The grand-canonical spectrum for the tetrahedron at $\mu=-t$. 
The other parameters are $U/t=4$, $J/t=0$, $W/t=0$, and $h/t=0$.} 
\label{lowlevelsN0N3}
\end{figure}

%% file: Parts/figure13.tex
\begin{figure}
  \centering
  \includegraphics[width=0.8\textwidth]{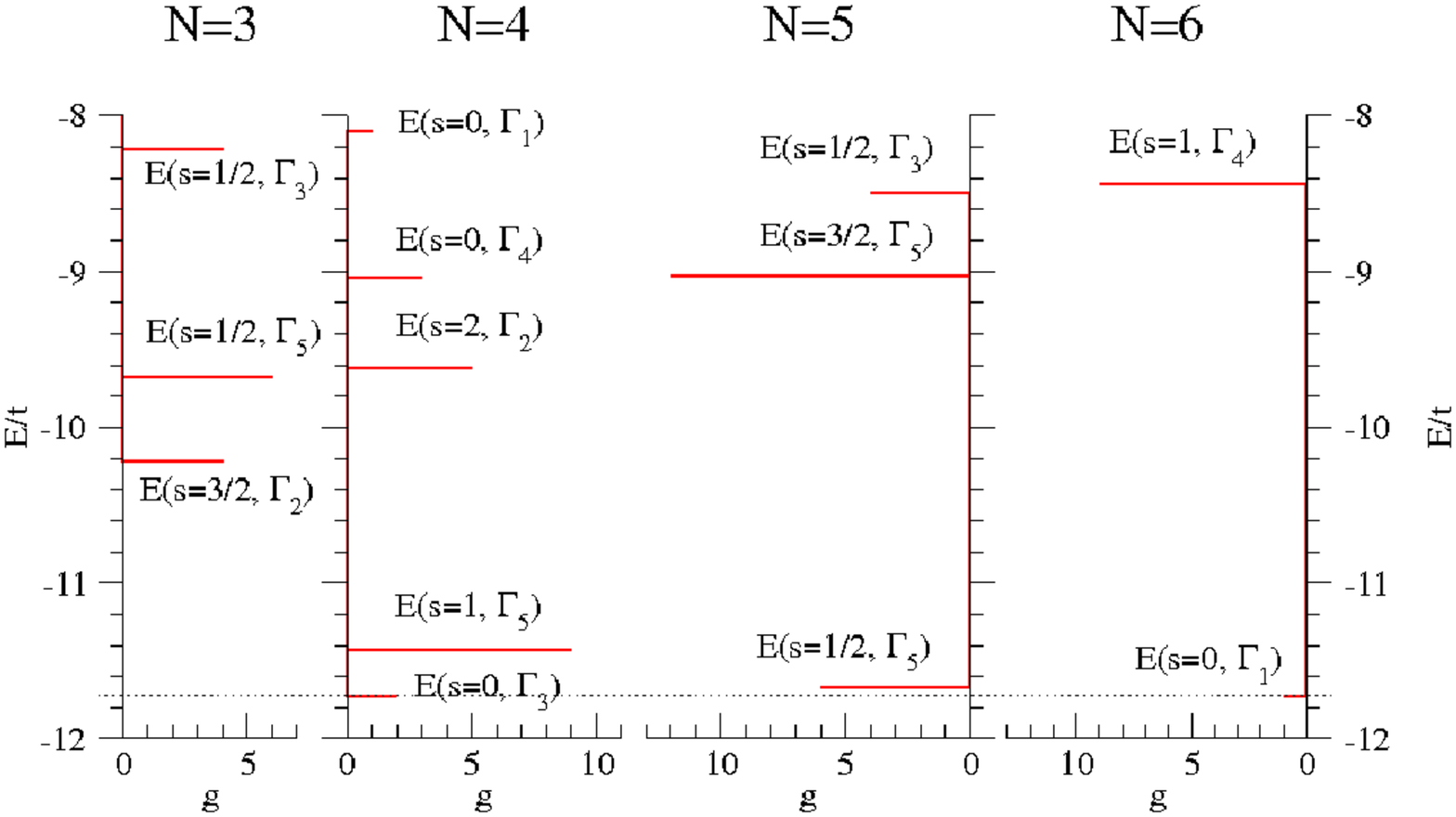}
  \caption{The grand-canonical spectrum for the tetrahedron at $\mu=2.4056\,t$. 
     The other parameters are $U/t=4$, $J/t=0$, $W/t=0$, 
     and $h/t=0$.
     }
  \label{lowlevelsN4N6}
\end{figure}

%% file: Parts/figure14.tex
\begin{figure}
\begin{minipage}[t]{0.5\textwidth}
\centering 
\includegraphics[height=55mm]{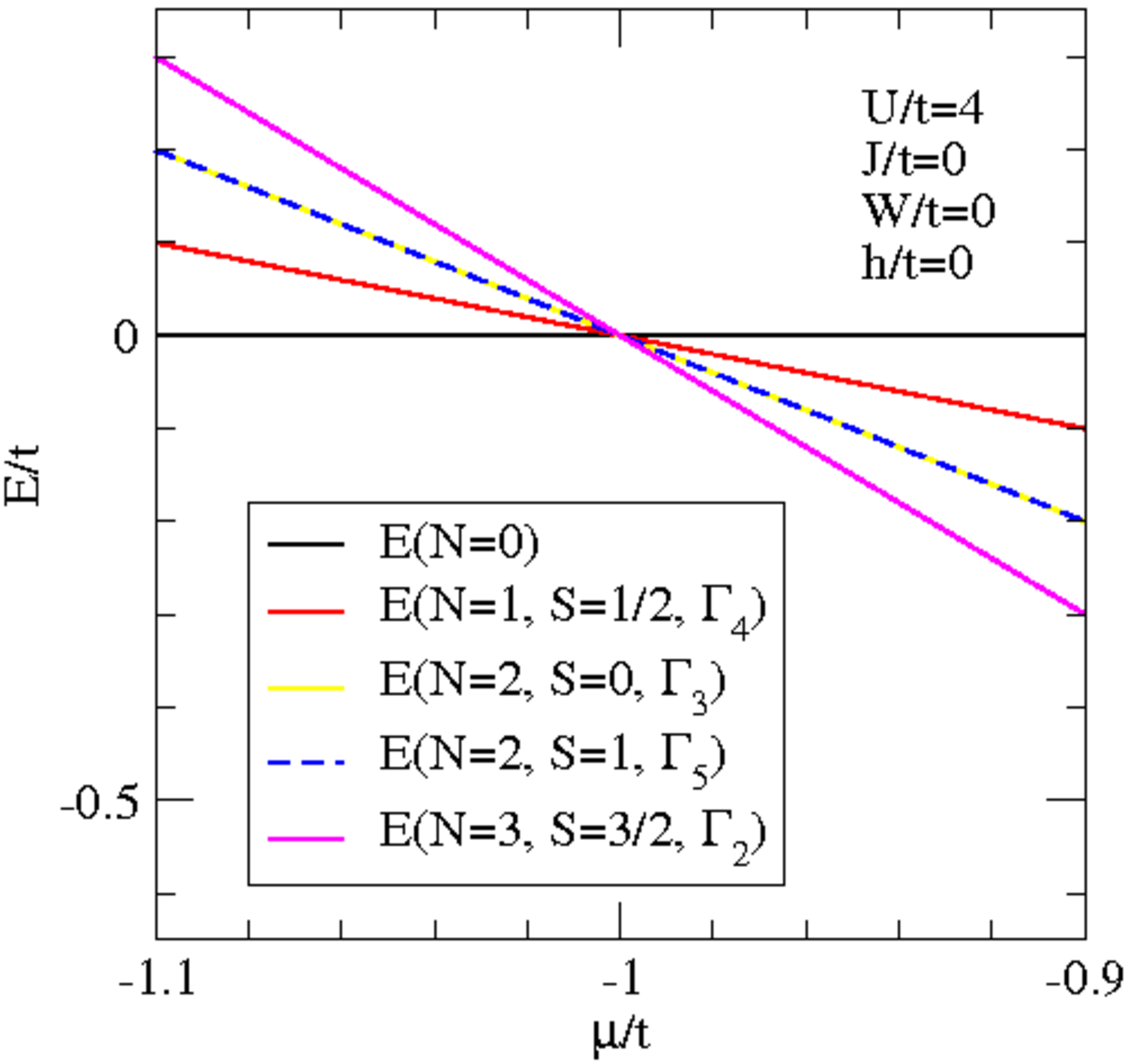}
\end{minipage}%
\hfill
\begin{minipage}[t]{0.5\textwidth}
\centering 
\includegraphics[height=55mm]{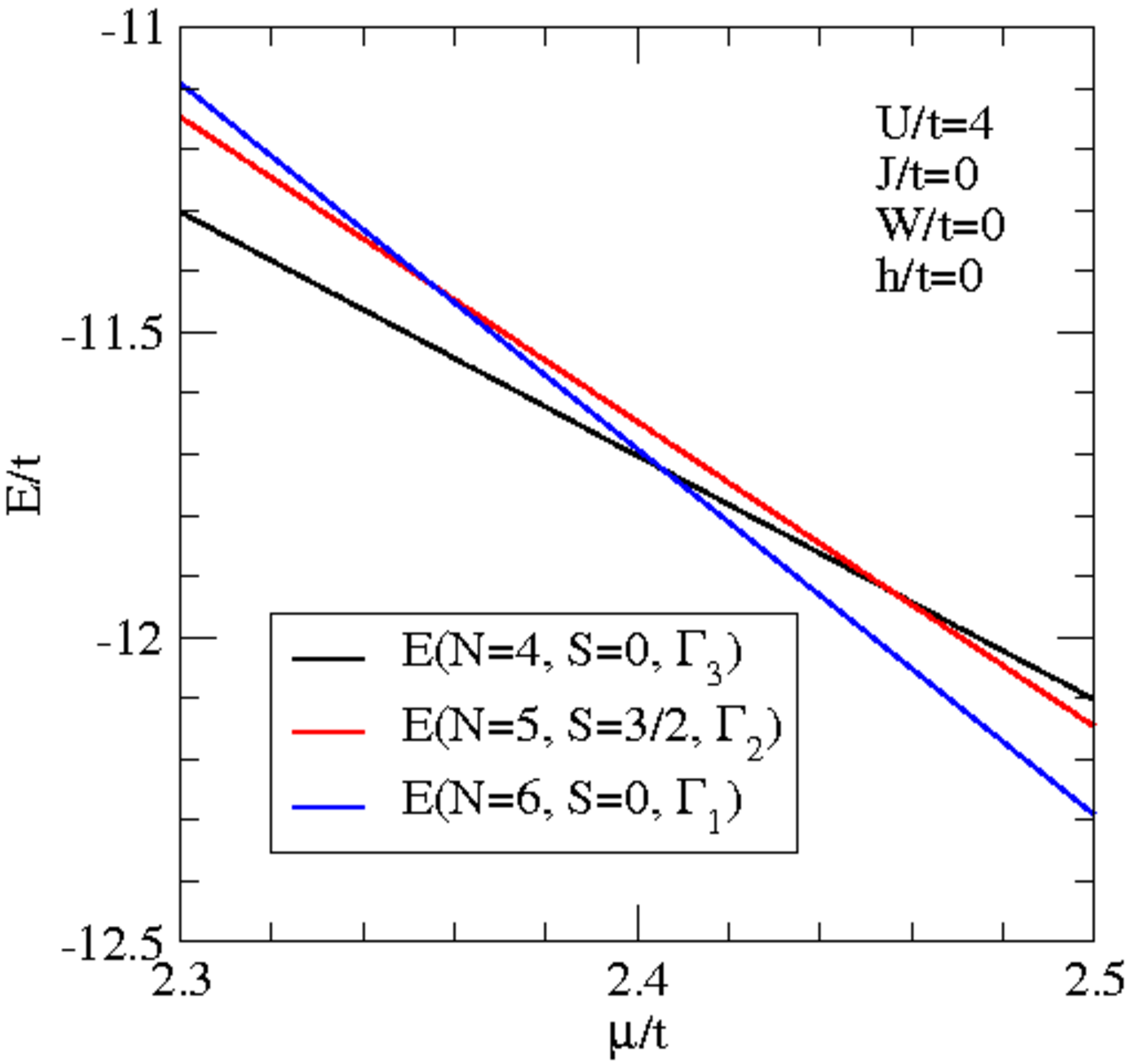}
\end{minipage}%
\caption{The grand-canonical ground states for the tetrahedon in dependence on the chemical potential in the
vicinity of the degeneration points.}
\label{gcLevelsU4}
\end{figure} 

%% file: Parts/figure15.tex
\begin{figure}
 \begin{minipage}[c]{0.5\textwidth}
 \centering 
 \includegraphics[height=55mm]{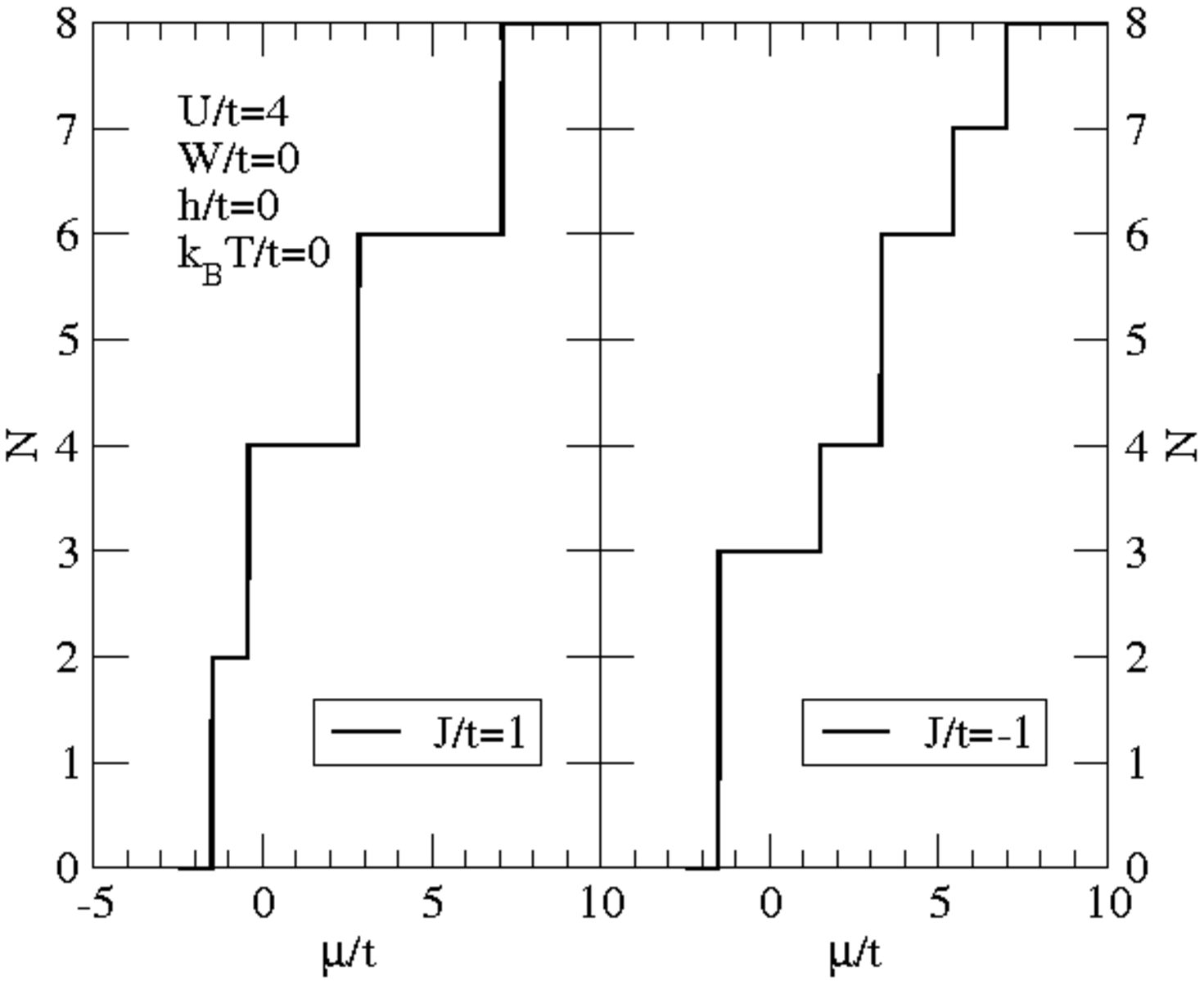}
 \end{minipage}%
 \hfill
 \begin{minipage}[c]{0.5\textwidth}
 \centering 
 \includegraphics[height=60mm]{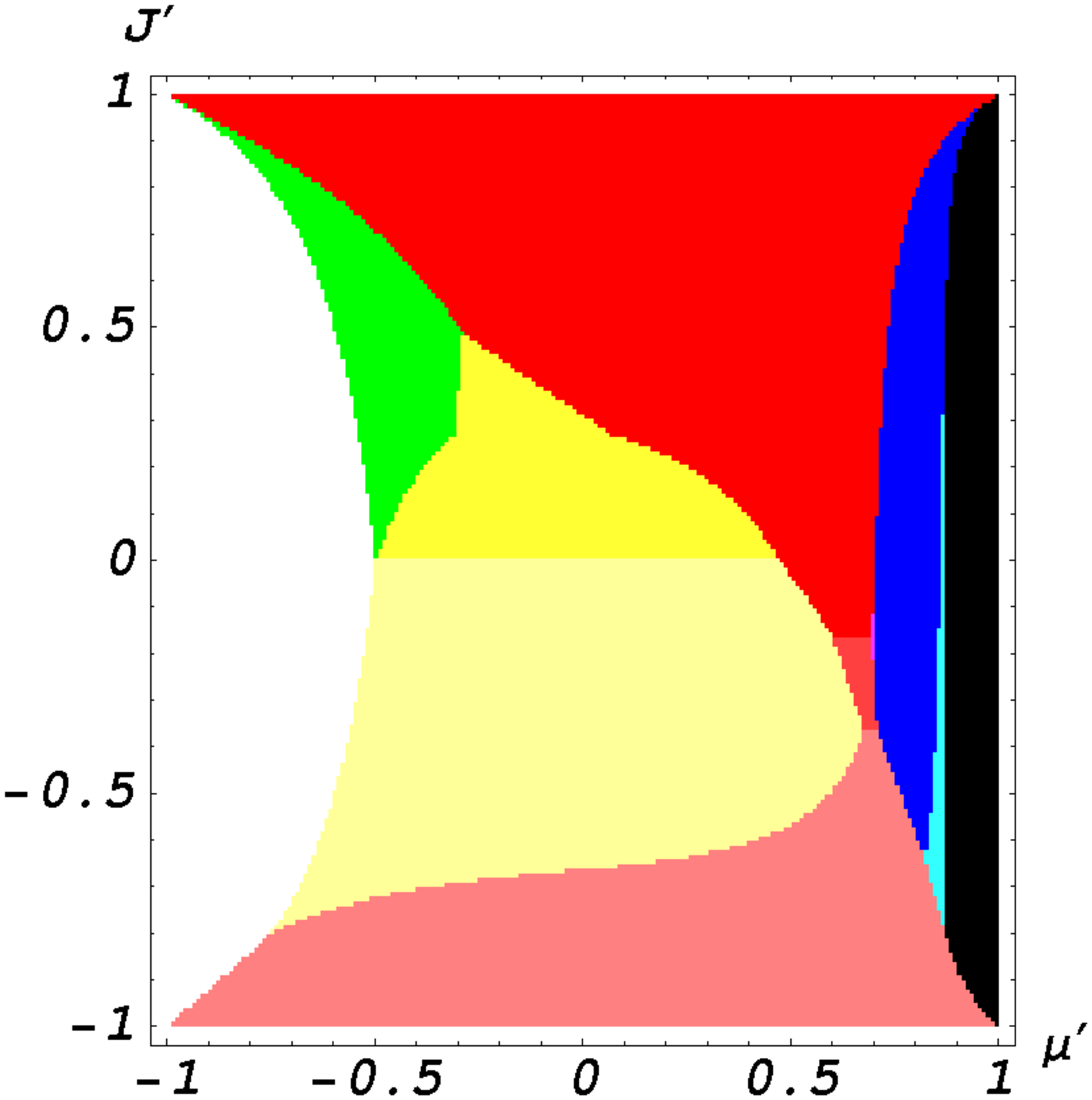}
 \end{minipage} 
\caption{
Left panel: $N(\mu)$ of the tetrahedron for the on-site correlation fixed to $U=4\,t$ for an
antiferromagnetic ($J/t=1$) and a ferromagnetic exchange ($J=-1\,t$). 
Right panel: The electron occupation $N$ and the $S_z$-eigenvalue for the ground state in dependence on the chemical potential and exchange parameter $J$
(both scaled to primed values)
for the tetrahedron and $U/t=4$, $h/t=0$, and $W/t=0$. 
The curves with $J=-\,t$ and $J=t$ in the left picture 
correspond to horizontal cuts across the right one for
$J'=-0.5$ and $J'=0.5$ respectively. 
The meaning of the colours is
according to the palette given in Fig. \ref{palette4site}.
}
\label{NvonMueJU4}
\end{figure}

%% file: Parts/figure16.tex
\begin{figure}
\begin{minipage}[c]{0.5\textwidth}
 \centering 
 \includegraphics[height=55mm]{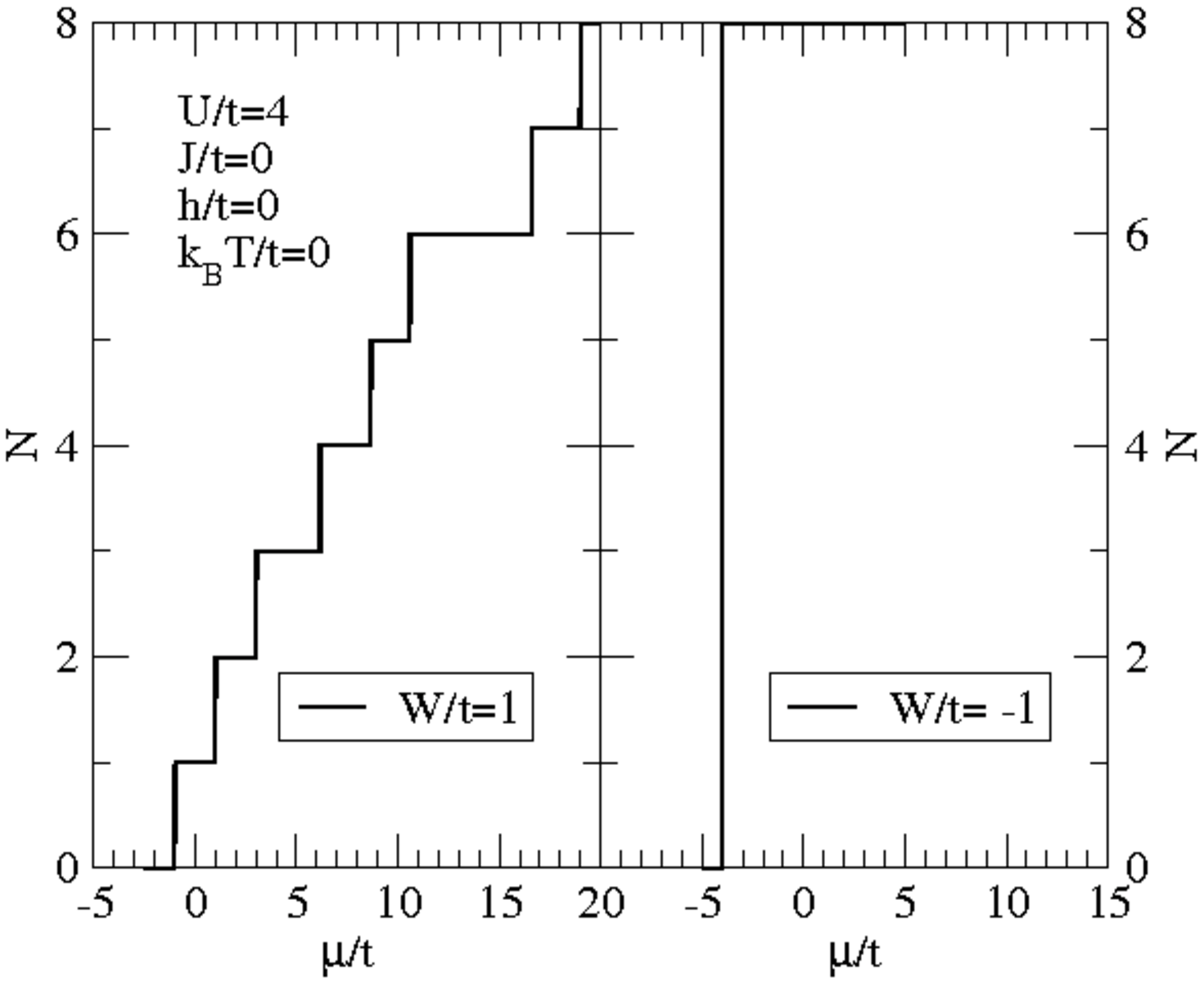}
 \label{NvonMueU4W1W1mDouble}
 \end{minipage}%
 \hfill
 \begin{minipage}[c]{0.5\textwidth}
 \centering 
 \includegraphics[height=60mm]{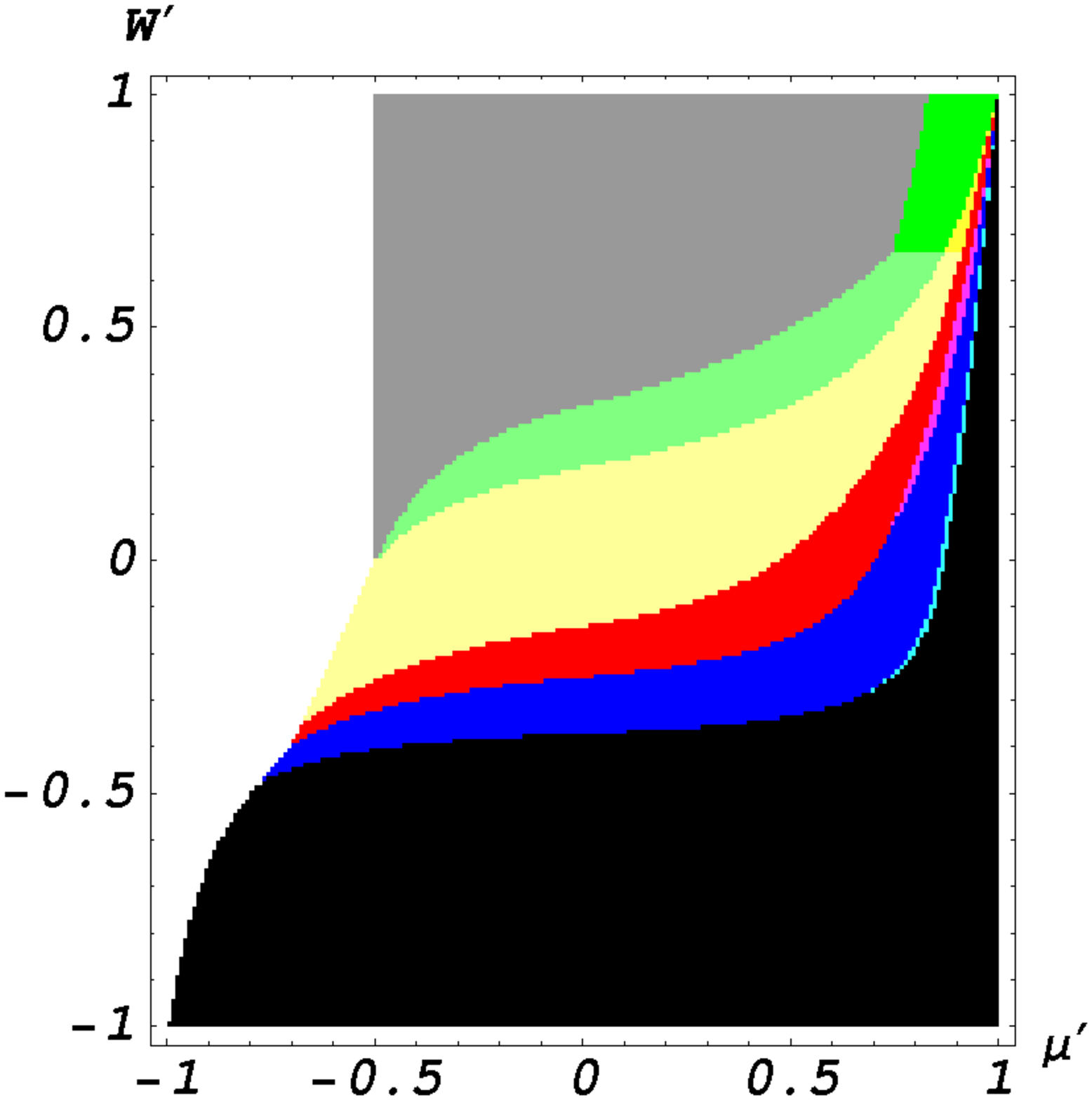}
 \label{NvonMueU4mueWContour}
 \end{minipage} 
\caption{
Left panel: $N(\mu)$ of the tetrahedron for the on-site correlation fixed to $U/t=4$ for a repulsive ($W/t=1$) and an attractive ($W/t=-0.1$) nn Coulomb
interaction.
Right panel: The electron occupation $N$ and the $S_z$-eigenvalue for the ground state in dependence on the chemical potential and nn-Coulomb interaction parameter $W$
(both scaled to primed values)
for the tetrahedron and $U/t=4$, $h/t=0$, and $J/t=0$. 
The curves with $W/t=1$ and $W/t=-1$ in the left picture 
correspond to horizontal cuts across the right one for
$W'=0.5$ and $W'=-0.5$ respectively. 
The meaning of the colours is
according to the palette given in Fig. \ref{palette4site}.
}
\label{NvonMueWU4}
\end{figure}

%% file: tetrahedronCondMat.bbl
\begin{thebibliography}{00}
\bibitem{Senechal00}
 S. Senechal, D. Perez and M. Pioro-Ladriere, Phys. Rev. Lett. 84 (2000) 522
\bibitem{Maier05}  T. Maier, M. Jarrell, T. Pruschke and M. H. Hettler, Rev. Mod. Phys. 77, 1027 (2005).
\bibitem{Wang05}  W. Z. Wang, Phys. Rev. B72 (2005) 125116
\bibitem{Kikoin06} T. Kuzmenko, K. Kikoin, and Y. Avishai, Phys. Rev. Lett. 96 (2006)046601
\bibitem{Lieb68} E.H. Lieb and F.Y. Wu, Phys. Rev. Lett. 20 (1968) 1445 
 \bibitem{Korepin00}Deguchi, F.H.L. Essler, F. G\"ohmann, A. Kl\"umper, 
 V.E. Korepin, and K. Kusakabe, Phys. Rep. 331 (2000) 197
\bibitem{Falicov84} L. M. Falicov and R.M. Victora, Phys. Rev. B 30 (1984) 1696,
\bibitem{Schumann02} R. Schumann, Ann. Phys. (Leipzig) 11, (2002) 49
\bibitem{Merino06} J. Merino, B.J. Powell, and Ross H. McKenzie, 
Phys. Rev. B 73 (2006) 235107
\bibitem{Perez96} J. A. Perez, O. Navarro, and C. Wang, 
Phys. Rev. B 53 (1996)15389
\bibitem{Nakano06} T. Nakano and K. Kuroki,
Phys. Rev. B 74 (2006) 174502
\bibitem{Kocharian05} A. N. Kocharian, G. W. Fernando, K. Palandage, and J. W. Davenport, cond-mat/0510609 
\bibitem{Kocharian06}A.M. Kocharian, G.W. Fernando, K. Palandage, and J.W. Davenport, Phys. Rev. B 74, 024511 (2006) 
\bibitem{Tsai06} Wai Feng Tsai and Steven A. Kivelson, Phys. Rev. B 73, 214510 (2006)
\bibitem{Zhang89} Y. Zhang and J. Callaway, Phys. Rev. B 39 (2006) 9397
\bibitem{Schumann06} R. Schumann, Proceedings of the M2S-HTSC VIII Dresden, to appear in Physica C 
\bibitem{FuldeBuch} P. Fulde. {\em Electron Correlations in Molecules and
Solids}, p. 243, Springer, Berlin Heidelberg New York Tokyo, 1997
\bibitem{Davoudi06}  B. Davoudi and A.-M. S. Tremblay, 
Phys. Rev. B 74 (2006) 035113
\bibitem{CornwellBook}J.F. Cornwell, Group Theory in Physics, Academic Press, London, 1984
\bibitem{Tremblay05}  D. Senechal, P.-L. Lavertu, M.-A, Marois, and A.-M. S. Tremblay, 
Phys. Rev. Lett. 94 (2005) 156404
\bibitem{Maier06}  T. A. Maier, M. Jarrell, and D. Scalapino, Phys. Rev. B74 (2006) 094513
\end{thebibliography}
